\documentstyle[epsfig]{osudissert96}

\newcommand {\half}{\frac{1}{2}}
\newcommand {\gp}{\gamma^{+}}
\newcommand {\gm}{\gamma^{-}}
\newcommand {\gz}{\gamma^{0}}
\newcommand {\gmp}{\gamma^{\mp}}
\newcommand {\gpm}{\gamma^{\pm}}
\newcommand {\Lp}{\Lambda_{+}}
\newcommand {\Lm}{\Lambda_{-}}
\newcommand {\Lpm}{\Lambda_{\pm}}
\newcommand {\Lmp}{\Lambda_{\mp}}
\newcommand {\beq}{\begin{eqnarray}}
\newcommand {\eeq}{\end{eqnarray}}
\newcommand {\barr}{\begin{array}}
\newcommand {\earr}{\end{array}}

\newcommand {\psl}{p\hspace{-.17cm}\slash}
\newcommand {\cH}{{\cal H}}
\newcommand {\bT}{{\bf T}}
\newcommand {\chp}{\chi^{\hspace{.05cm}\prime}}
\newcommand {\bd}{b^{\dagger}}
\newcommand {\ad}{a^{\dagger}}
\newcommand {\dd}{d^{\dagger}}

\newcommand {\inpar}{\left( \frac{1}{\partial^{+}}\right) }
\newcommand {\CD}{\mathrm{C\hspace{.2mm}D}}

\newcommand {\n}{\nonumber\\}

\newcommand {\p}{{\cal P}}
\newcommand {\la}{\Lambda}

\newcommand {\bx}[2]{\Delta^{#1}_{#2}}
\newcommand {\Bx}[2]{\Delta^{#1}_{#2}}

\newcommand {\M}{{\cal M}^2(\la)}

\newcommand {\dVome}[1]{\left< F \right| \delta V^{(#1)} \left| I \right>}

\newcommand {\Vcdme}[3]{\langle {#1}\vert V_{\CD}^{(#3)}(\la)\vert{#2}\rangle}

\newcommand {\qb}{\bar{q}}

\newcommand {\T}[2]{{\bf T}^{#1}_{#2}}
\newcommand {\hone}{\hspace{.1cm}}
\newcommand {\mol}{$\frac{m}{\Lambda}$ }

\newcommand {\Vme}[2]{\langle {#1}\vert V^{(1)}\vert{#2}\rangle}

\newcommand {\bsplpath}{bspline-plots/}
\newcommand {\gluepath}{glueball-plots/}

\newcommand {\diagrampath}{diagrams/}
\newcommand {\flowpath}{flowcharts/}
\newcommand {\mesonpath}{meson-plots/}

\def\bbox#1{%
\relax\ifmmode
\mathchoice
{{\hbox{\boldmath$\displaystyle#1$}}}%
{{\hbox{\boldmath$\textstyle#1$}}}%
{{\hbox{\boldmath$\scriptstyle#1$}}}%
{{\hbox{\boldmath$\scriptscriptstyle#1$}}}%
\else
\mbox{#1}%
\fi
}

\begin{document}

\graduationyear{2001}
\author{Roger Donald Kylin}
\title{THE MESON MASS SPECTRUM FROM A SYSTEMATICALLY RENORMALIZED LIGHT-FRONT HAMILTONIAN}
\authordegrees{B.S., M.S.}  
\unit{Department of Physics}

\advisorname{Professor Robert J. Perry}
\member{Professor Michael A.~Lisa, Professor Greg Kilcup, Professor Richard J.~Furnstahl, Professor Henri Moscovici}
\maketitle

%
%



\begin{abstract}
  We extend a systematic renormalization procedure for quantum field theory to include particle masses and present several 
applications.  We use a Hamiltonian formulation and light-front quantization because this may produce a convergent 
Fock-space expansion.  The QCD Hamiltonian is systematically renormalized to second order in the strong coupling and 
the Fock-space expansion is truncated  to lowest order to produce a finite-dimensional Hamiltonian matrix.  The 
renormalized Hamiltonian is used to calculate the spectra of the $b\bar{b}$ and $c\bar{c}$ mesons as a lowest-order 
test of our procedure for full QCD.  

The analytic determination of the renormalized Hamiltonian matrix generates expressions that must be numerically 
integrated to generate quantitative results.  The efficiency of the numerical calculation depends on how well the 
basis functions can approximate the real state.  
We examine the effectiveness of using Basis-Splines (B-Splines) to represent QCD 
states.  After briefly describing these functions, we test them using the one- and two-dimensional harmonic oscillator 
problems.  We test their ability to represent realistic wavefunctions by using them to find the glueball mass 
spectrum.

An efficient algorithm for numerically calculating the matrix elements in the glueball and meson problems is 
necessary because the calculation is numerically intensive.  We describe our algorithm and discuss its 
parallel-cpu implementation.

\end{abstract}

%
%


\dedication{For Nikki and David}

%
%

\begin{acknowledgments}

I want to thank my advisor Robert Perry, who has been a wonderful teacher and inspiration.  

Brent Allen has been a great source of knowledge, logic and experience.  He continues to be a wonderful friend.

I am grateful to Rick Mohr who has been a good sounding board and has quietly kept our Linux cluster working 
beautifully.

The other members of our research group deserve credit for their help, support and cpu time, especially Dick Furnstahl, 
Negussie Tirfessa, Ulrich Heinz, Hans-Werner Hammer, Thomas Mehen, Stephen Wong and Peter Kolb.

My understanding of parallel computing and the Message Passing Interface was made possible only through the help 
of many individuals.  John Wilkins encouraged me to begin parallel programming and helped get my first 
supercomputing accounts.  Greg Kilcup and Jeongnim Kim helped me better understand parallel 
programming and Lars Jonsson spent a lot of time helping to optimize my code. 

I would like to thank Dick Furnstahl, Greg Kilcup, Mike Lisa, and Henri Moscovici for serving on my committee.

I am indebted to my wife Nikki who supported me, and encouraged me to reach my potential over the last ten years.  Finally, I 
would like to thank my son, David, who adds a bit of sunshine to every day.

This work was partially supported by the National Computational Science Alliance
under PHY990016N and PHY000009N, utilizing the NCSA SGI/CRAY Origin2000.
This work was also supported by the National Science Foundation under grant numbers PHY-9511923 and 
PHY-9800964.

\end{acknowledgments}

\begin{vita}

\dateitem{March 15, 1973}{Born -- Mayfield Heights, Ohio}
\dateitem{1995}{B.S. Physics, The University of California, Davis}
\dateitem{1998}{M.S. Physics, The Ohio State University}
\dateitem{1995--present}{Graduate Teaching and Research Associate, The Ohio State University}

\begin{publist}
    \researchpubs
    
    \pubitem{``Optimization of Fullerene Yields in a Plasma Arc Reactor;'' P.E. Anderson, T. T. Anderson,
    P. L. Dyer, J. W. Dykes, S. H. Irons, C. A. Smith, R. D. Kylin, P. Klavins, J. Z. Liu, and R. N.
    Shelton, in the proceedings on Recent Advances in the Chemistry and Physics of
    Fullerenes and Related Materials, Karl M. Kadish and Rodney S. Ruoff, eds. (The
    Electrochemical Society, Inc., 1994)}
    \pubitem{``Systematic Renormalization in Hamiltonian Light-Front Field Theory: The Massive 
    Generalization;''  Roger D. Kylin, Brent H. Allen, and Robert J. Perry  Phys. Rev. D {\bf 60} 
    (1999) 067704, hep-th/9812080.}
\end{publist}
\newpage
\begin{fieldsstudy}
    
    \majorfield{Physics}
    
\end{fieldsstudy}
\end{vita}

%
%

\tableofcontents
\listoffigures
\listoftables

%
%

\chapter{Introduction}

The physics of quarks and gluons is described by quantum chromodynamics (QCD), which is universally believed to be 
the fundamental theory of the strong interaction.  High energy QCD systems 
(small spatial separations) behave as though the particles interact weakly (asymptotic 
freedom).  Low energy systems (large spatial separations) interact so strongly that it is not possible to isolate a 
quark or gluon (confinement).  The physics in the confining and asymptotically free systems is so different that it is 
has not yet been possible to write down a solution to QCD that is valid in both regimes.  Historically, difficult 
theories like QCD are first solved for simple systems.  Solving these systems provides an 
initial test of the theory before more complex systems are attacked and a development ground for new techniques.

Heavy quark bound states are relatively simple because the large quark mass dominates the spectrum and the average 
separation of quarks and gluons is small, so that dynamics are largely described by asymptotic freedom.  This also allows the 
interactions to be treated nonrelativistically \cite{heavy-new,quark-model,martinaA,martinaB,martina}.
Techniques that have been applied to the heavy quark 
system include nonrelativistic potential models and nonrelativistic QCD (NRQCD).  Potential models 
\cite{potnrqcd,pot-model-quigg} obtain good predictive power by approximating the interaction between particles 
using nonrelativistic potentials, although these potentials are not directly obtained from QCD.
Predictions include the $\Upsilon$/$\Upsilon'$ splitting \cite{e-and-g}, 
spin-splitting, transition rates and hadronic transitions \cite{pot-model-quigg}.  

NRQCD uses a nonrelativistic expansion of QCD operators to generate an approximate Lagrangian that can be 
systematically improved \cite{latticeB}.  Heavy-quark spectroscopy can be determined on the lattice 
\cite{latticeA}, which can then be used to determine the strong coupling constant, which is important because 
the coupling can be compared to perturbative results\footnote{There are also perturbative models of NRQCD.  See 
\cite{eftnrqcd}.} and could reveal the existence of new physics 
\cite{nrqcd-review}.  Both potential models and NRQCD have proven to be predictive theories for nonrelativistic systems.  Despite both 
theories being improvable, a relativistic treatment of QCD is necessary to treat systems of lighter quarks. 

There are many problems one must solve (or circumvent) to solve full QCD.  In Euclidean or equal-time coordinates 
vacuum fluctuations prevent using a constituent picture for low energy hadrons.  Also, since QCD is a field theory, 
there are divergences that must be regulated, often breaking symmetries.  Any solution to QCD will require the use of 
renormalization to remove the dependence on the regulator and to restore broken symmetries.  


Hadrons are composed of at least two (meson) or three (baryon) quarks.\footnote{In this overview we do not 
distinguish between quarks and antiquarks.}  The 
quarks are bound by the strong force through the exchange of color-charged gluons, which are in turn bound together by the 
same mechanism.  Hadronic bound 
states (and glueball states) are color singlets; there is no net color.  As quarks in a baryon are separated, the 
energy of the system increases, eventually allowing the creation of new quark-antiquark pairs.  These pairs form 
bound states with 
the separated quarks, preventing the isolation of any individual quark.  If we consider low-energy bound states, 
the particles interact weakly and we may be able to make a constituent approximation.  However in a standard 
equal-time picture Heisenberg's uncertainty principle allows quark-antiquark pairs to fluctuate out 
of the vacuum.  These quarks can then interact with any hadronic bound states, preventing the state from being 
successfully represented by two or three quarks.  Moreover, these vacuum fluctuations are usually thought to be 
critical for strong interactions.

We would like to derive a constituent picture for mesons from QCD, with mesonic states and masses determined by a 
Schr\"{o}dinger equation:
\beq
H \vert \Psi \rangle = E \vert \Psi \rangle,
\eeq
where we can approximate the eigenstate with a truncated Fock-space expansion,
\beq
\vert \Psi \rangle = \phi_{q\bar{q}} \vert q \bar{q} \rangle + \phi_{q\bar{q}g} \vert q \bar{q} g \rangle + 
\cdot\cdot\cdot .
\eeq
$E$ is the energy eigenvalue of the state $\vert \Psi \rangle$ and we use shorthand notation for the Fock states 
where $q$ is a quark, $\bar{q}$ an antiquark, and $g$ is a gluon.  Unfortunately a Fock-space expansion is not 
usually thought to be
practical, because the complicated equal-time vacuum forces bound-states to contain an infinite number 
of particles that are part of the physical vacuum on which hadrons are built.  It may be possible, however, to obtain a convergent Fock-space 
expansion if we work in light-front coordinates because the free energy of a state increases at least like the number of 
particles squared (Sec.~\ref{lf-overview}).  This means all states with many particles are high-energy states.
Inspired by the work of Dyson 
\cite{dyson}, Wilson \cite{wilson}, G{\l}azek and Wilson \cite{glazek and wilson}, and Wegner \cite{wegner}, significant 
work has been done to perturbatively derive light-front Hamiltonians in the full Fock-space, neglecting zero modes 
\cite{martinaA,martinaB,martina,long paper,robert,billy,stan,zhang,zhang2,elana,walhout}.

Local interactions in a field theory couple states with arbitrarily large differences in free energy, which will invalidate our Fock-space 
expansion if we can not decouple high-energy states from low-energy states.
One way we can force high-energy states to decouple from low-energy states is to simply truncate the Fock-space 
expansion.  The truncated states can be important, and their removal may discard important 
physics.  We can retain all of the physics if we place a 
cutoff on change in invariant mass between states, and then use a similarity transformation \cite{glazek 
and wilson,wegner} to map 
the effects from mixing of low- and high-energy states to interactions involving only low-energy states.

The importance of high-energy states can be appreciated by considering the second-order perturbation theory 
correction to a free energy:
\beq
\label{energy-shift}
\delta E_{n}=\sum_{m\neq n}^{N}\frac{\left|\langle 
m\left|V\right|n\rangle\right|^{2}}{\langle m\left|H_{0}\right|m\rangle-\langle n\left|H_{0}\right|n\rangle}.
\eeq
The sum can diverge as $N\rightarrow \infty$ if the matrix elements between states with large differences in energy 
do not fall off sufficiently rapidly.  We regularize the theory by implementing a gaussian cutoff on interactions 
between states with differing free
invariant masses\footnote{The invariant mass can be thought of as the energy of the state after subtracting the 
energy due to translations.}.  This cutoff serves two purposes.  It makes perturbative corrections like Eq.~(\ref{energy-shift})
finite and it reduces the coupling between states with large differences in invariant mass, 
which further helps justify our Fock-space expansion by reducing the importance of high-energy physics for low-energy states.

The cutoff we use violates Lorentz covariance and gauge invariance, so we can not
renormalize the Hamiltonian exclusively through the redefinition of masses and canonical couplings.  Renormalization 
must be completed by requiring the Hamiltonian to produce cutoff-independent physical quantities and by requiring it 
to obey the physical principles of the theory that are not violated by the cutoff.  These requirements, and the 
assumption that this can be done perturbatively,  
are sufficient to determine the Hamiltonian so that it will give results consistent with all the physical principles of the theory, even those 
violated by the cutoff.  The most powerful characteristic of this approach is that it systematically ``repairs'' the 
theory and requires only the fundamental parameters of the canonical Hamiltonian.

Recently, Allen \cite{brentA,brentB,brentD} derived recursion relations that systematically determine the 
Hamiltonian for massless theories order by order in the coupling and he applied this method to pure-glue QCD.  In this dissertation we extend the
recursion relations to full QCD and present three applications.  We first show how matrix elements of the 
renormalized Hamiltonian are calculated with our extension of the method, which has been published in \cite{kylinA}. 
Secondly we apply new numerical methods to verify 
Allen's previous glueball spectrum calculation \cite{brentB,brentD}.  The final application uses these numerical methods to 
calculate heavy quarkonia ($q\bar{q}$) spectra.

In light-front field theory the Hamiltonian is trivially related to the invariant mass operator (IMO).  The IMO is 
a natural operator to use because it is boost and rotationally invariant (although rotational invariance is 
realized dynamically).  We calculate the IMO in this dissertation, but refer to it 
as the Hamiltonian throughout since it governs the time evolution of all states.

The Hamiltonian matrix is calculated analytically to ${\cal O}(\alpha_{s})$ in a plane-wave basis.  
We then calculate the matrix elements in a basis we choose to approximate the real meson states (i.e., 
eigenstates of the Hamiltonian).
The real meson states are approximated by the first term in a Fock-space expansion, quark-antiquark pairs.  Solving the 
eigensystem gives the invariant masses (eigenvalues) of the system and the coefficients (eigenvectors) needed to build 
the eigenstates from the basis functions.

The numerical calculation of the Hamiltonian matrix requires us to choose a set of basis functions to approximate the 
real states.  
We use Basis-splines (B-splines) to represent the longitudinal and transverse momentum states. 
Although their derivation from mathematical first principles is non-trivial, their benefit is easily 
understood.  First, B-splines do not oscillate, so they do not add extra oscillations to the function that must be 
integrated to determine each matrix element.\footnote{The matrix elements are approximated using Monte-Carlo 
integration.  Monte-Carlo is a powerful method to approximate many-dimensional integrals, but it does not work well 
with oscillating integrands.}  Second, each 
spline has only a limited spatial overlap with the other splines; thus, there can be a significant number of matrix 
elements that vanish if the interactions are local because the basis functions do not overlap.

The most substantial drawback to using B-splines is that they are non-orthogonal.  
This means that an overlap matrix must be calculated to solve Schr\"{o}dinger's equation.  The main inconvenience 
is that a general eigensystem problem must be 
solved and efficient methods that compute only low-lying eigenvalues can not be used.

The numerical calculation of the Hamiltonian matrix is very cpu-intensive because each matrix element includes a five-dimensional integral.  Unfortunately one can not know ahead of time 
how accurately each matrix element must be calculated.  For instance, if all matrix elements are calculated to the 
same precision, much computer time is wasted because all elements are not equally important when determining the 
low-lying eigenvalues.
With no unambiguous method to determine the relative importance of various matrix elements, there is also no 
obvious algorithm to solve the problem efficiently.  Fortunately each matrix element can be calculated 
independently, so most algorithms can be executed in a parallel computing environment. 

In Chapter \ref{chapt:renormalization} we discuss our motivation for using light-front coordinates and our choice of 
regularization and renormalization techniques.  We then review a recent technique that builds on previous work, 
systematically determining the renormalized Hamiltonian for massless theories, and discuss how this method is 
generalized to theories with massive particles.   We conclude Chapter \ref{chapt:renormalization} by listing 
which terms in the Hamiltonian are needed for an ${\cal O}(\alpha)$ calculation, followed by a discussion of how 
these terms are combined to explicitly cancel light-front divergences.

We present B-splines and show some simple applications in Chapter \ref{section:bsplines}.  In Chapter \ref{physical-states}
we introduce the basis used to represent the real meson states.  This basis is tested by computing 
the glueball mass spectrum in Section \ref{chapt:glue-calc}.

The matrix elements of the Hamiltonian are presented in Chapters \ref{chapt:ham-me}, \ref{chapt:me-renorm-ham}, and \ref{chapt:me-physical}.
Chapter \ref{chapt:ham-me} presents some matrix elements of the full Hamiltonian in a plane wave basis.  The 
matrix elements of the second-order renormalized Hamiltonian in a plane wave basis are derived in Chapter 
\ref{chapt:me-renorm-ham}.  Finally the matrix elements of the Hamiltonian in the approximate-state basis are given in 
Chapter \ref{chapt:me-physical}.

Although there is a lot of analytic work required to solve this problem, quantitative results can only be obtained after applying numerical 
methods to attack the problem.  We discuss some of the numerical considerations when deriving the matrix elements 
and describe the general algorithm and how it is parallelized in Chapter \ref{chapt:numerical-parallel}.

Finally, we present our analysis of our numerical methods and the calculation of the meson spectrum in Chapter 
\ref{results}.  We make concluding remarks in Chapter \ref{conclusions}.

This dissertation builds upon previous work and applies it to a familiar problem using numerical techniques that may 
be similar to other calculations unknown to the author.  We believe the following parts of this calculation are 
original work:
\begin{list}{$\bullet$}{}
    \item We included masses in Renormalization Group calculations using the method  in \cite{robert}.
    \item Calculation of the renormalized light-front QCD Hamiltonian to ${\cal O}(\alpha)$ using the renormalization 
    methods from \cite{long paper, robert, brentA}.
    \item Use of B-spline basis function to calculate the glueball mass spectrum.
    \item Numerical calculation of the heavy meson mass spectrum for our renormalized QCD Hamiltonian.
    \item Development of an efficient algorithm to calculate the Hamiltonian matrix.
\end{list}

\chapter{Motivations and Formalism Overview}
\label{chapt:renormalization}

In this chapter we motivate and give an overview of our method.  This includes a discussion of light-front 
coordinates as well as discussions about regularization and renormalization.  We then discuss a recent 
development that systematically determines the renormalized Hamiltonian by requiring it to produce cutoff-independent results 
and to obey all physical principles not violated by the cutoff \cite{brentA,brentB,brentD}.  We then review the generalization
to include massive particles from \cite{kylinA}.  In Section \ref{mass-o2-begin} we list all of the diagrams that 
contribute to the second-order Hamiltonian.  Finally we describe how these diagrams are combined to explicitly 
cancel the light-front divergences.

This chapter is intended to give a good understanding of the entire approach without 
saturating the reader with too many details, which can be found in the references.

\section{Light-Front Coordinates}
\label{lf-overview}
\begin{figure}
    \centerline{\epsfig{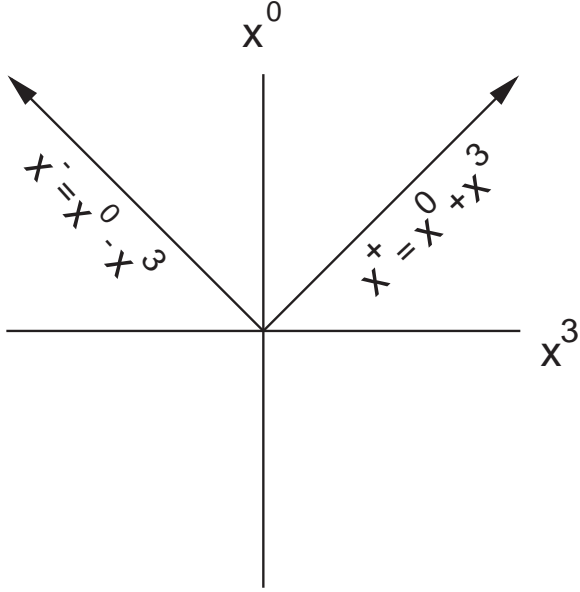}}
    \caption[Light-front coordinates.]{\label{lf-3plus1}Light-from coordinates in $3+1$ dimensions.  $x^{+}$ and $x^{-}$, the light-front time and 
    longitudinal space components are a linear combination of the equal-time 3-direction and time.}
\end{figure}

Light-front coordinates may be considered a rotation of equal-time coordinates such that light-front time is along 
the forward light-cone (See Fig \ref{lf-3plus1}).
Explicitly we write the light-front time coordinate:
\beq
x^{+}=x^{0}+x^{3},
\eeq
and the light-front longitudinal space coordinate as:
\beq
x^{-}=x^{0}-x^{3}.
\eeq
The scalar product of light-front vectors is:
\beq
a \cdot b = \half a^{+}b^{-} + \half a^{-}b^{+} - {\mathrm {\bf a}}_{\perp}\cdot{\mathrm {\bf b}}_{\perp},
\eeq
where ${\mathrm {\bf a}}_{\perp}$ and ${\mathrm {\bf b}}_{\perp}$ are the transverse components.  This shows that the 
light-front energy, which is conjugate to light-front time, is $p^{-}$ and the longitudinal momentum is $p^{+}$.  An 
on-mass-shell particle obeys 
$p^{0} \ge p^{3}$ which implies $p^{+} \ge 0$.  Finally a free particle of mass $m$ has the light-front energy
\beq
p^{-}=\frac{{\mathrm {\bf p}}^{2}_{\perp}+m^{2}}{p^{+}}.
\eeq
The light-front energy is finite unless the longitudinal momentum is zero.  The study of these infinite-energy 
zero modes is rich.\footnote{Reference \cite{brodsky} is a review of theories on the light-front
that includes a discussion of zero-modes and many references to zero-mode calculations.}
Since they are infinite energy modes we discard them in our approach, although their effects will need to be 
added to the Hamiltonian.  Using the light-front dispersion relation we 
can write the free energy of a state as:
\beq
P^{-}=\frac{1}{P^{+}}\sum_{i}^{\mathrm{N}} \frac{\mathrm{{\bf p}}^{2}_{i \perp} +m_{i}^{2}}{x_{i}},
\eeq
where $P^{+}$ is the total longitudinal momentum of the state, N is the number of particles in the state, the index $i$ 
refers to each particle in the state, and $x_{i}$ is the $i^{th}$ particle's longitudinal momentum fraction.  If we 
assume all particles in the state have similar masses and transverse momenta, we can write the energy as:
\beq
P^{-}\sim\frac{(\mathrm{{\bf p}}^{2}_{\perp} +m^{2})_{typical}}{P^{+}}\sum_{i}^{\mathrm{N}} \frac{1}{x_{i}}.
\eeq
The free energy is minimized if the particles equally share the longitudinal momentum ($x_{i}=\frac{1}{N}$).  This 
means the minimum free energy of a light-front state behaves like:
\beq
P^{-}_{\mathrm{min.}}\sim N^{2} (\mathrm{{\bf p}}^{2}_{\perp} +m^{2})_{typical}.
\eeq
Since the energy of the state increases like the number of particles squared, a Fock-space expansion may be 
justified. 

See appendix \ref{lf-formalism} for a complete presentation of the light-front formalism used in this paper.

\section{Regularization}
\label{regulate}

In the next two sections we discuss the procedure we use to regulate and renormalize the Hamiltonian.  Our 
discussion follows the one in \cite{robert} and we do not cite each instance an argument from this reference is 
used.

If particle number is not conserved in a theory, there can be arbitrarily many particles in a state.  In addition each state can be 
coupled to every other state through a matrix element of the Hamiltonian, yielding a Hamiltonian with an infinite number of matrix elements.  
We must regularize the Hamiltonian so it is finite dimensional.  We could simply limit the number of particles in 
a state.  This type of truncation is called a Tamm-Dancoff truncation \cite{tamm,dancoff,td-robert} and suffers from 
divergent sensitivity to the precise form of the truncation, and removing this sensitivity can be as hard as 
solving the full theory.

Consider a Hamiltonian that can be written as;
\begin{equation}
H=H_{0}+V,
\end{equation}
where V can be considered a perturbation.  To second order in perturbation theory, the energy of a state $|n\rangle$, is given by:
\begin{equation}
\label{intermediate}
E_{n}=\langle n\left|H_{0}\right|n\rangle+\langle n\left|V\right|n\rangle-
\sum_{m\neq n}\frac{\left|\langle 
m\left|V\right|n\rangle\right|^{2}}{\langle m\left|H_{0}\right|m\rangle-
\langle n\left|H_{0}\right|n\rangle}.
\end{equation}
If this sum is over a finite number of states, and the matrix elements are finite, the sum will be finite.  However, if the sum is 
over an infinite number of states, the sum may not converge if the matrix elements between states do not fall off 
fast enough.  One way we can ensure this sum is finite is to place a cutoff on the Hamiltonian that reduces the 
coupling between states with arbitrarily large differences in free energy. 

Since perturbative calculations can only lead to 
reasonable results when the cutoff is a function of the free energy variables and we want a cutoff 
that will also prevent small energy denominators, we choose to place a smooth cutoff on change in invariant-mass 
(free energy) between the final and initial states [Eq.~(\ref{regulation})].  Like all cutoffs, this violates Lorentz covariance and gauge invariance.  
However, it is invariant under boosts and rotations about the 3-axis.

The light-front dispersion relation shows that the energy of a state rapidly increases with the number of particles, and 
the cutoff reduces the coupling between high- and low-energy states.  These two facts support the use of a Fock-space expansion.  

\section{Renormalization}
\label{renorm}

We regulate the Hamiltonian by applying a smooth cutoff on change in invariant-mass between the final and initial 
states.  This introduces cutoff dependence in the Hamiltonian, and it breaks Lorentz covariance and gauge 
invariance.  We restrict the form of the Hamiltonian by requiring it to obey all physical principles unviolated by 
the cutoff and use the similarity renormalization scheme to perturbatively remove the cutoff dependence.
An ${\cal O}(\alpha)$ perturbative renormalization of the regularized Hamiltonian does three things.  It removes 
cutoff dependence and restores broken symmetries to ${\cal O}(\alpha)$.  

For notational consistency\footnote{Some of the notation is this section is shorthand to simplify the 
discussion.  The full notation is introduced in Section \ref{sec:formal}.} with Ref.~\cite{brentD} 
(which we follow in this section) we remind the reader that if we neglect the total transverse momentum of 
a system,\footnote{Transverse boosts are kinematic on the light-front, so all of the interactions in the 
Hamiltonian are in the IMO.} 
the Hamiltonian is trivially related to the Invariant Mass Operator (IMO) by:
\beq
{\cal M}^{2}={\cal P}^{+} {\cal H}.
\eeq
We split the IMO into free and interacting parts:
\beq
{\cal M}^{2}(\Lambda)={\cal M}^{2}_{\mathrm{free}}+{\cal M}^{2}_{\mathrm{int}}(\Lambda).
\eeq
The IMO will produce cutoff independent results if it is unitarily equivalent to itself with an infinite cutoff and 
satisfies the relation:\footnote{We assume the $\Lambda \rightarrow \infty$ limit exists.}
\beq
\label{similarity-trans}
{\cal M}^{2}(\Lambda)=U(\Lambda,\Lambda') {\cal M}^{2}(\Lambda') U^{\dagger}(\Lambda,\Lambda'),
\eeq
where $U$ is a unitary operator that reduces the cutoff from $\Lambda'$ to $\Lambda$.  The transformation, which is 
a simplified version of the one introduced by Wegner \cite{wegner}, is 
determined by the differential equation:
\beq
\label{sim-diffeq}
\frac{dU(\Lambda,\Lambda')}{d(\Lambda^{-4})}=T(\Lambda)U(\Lambda,\Lambda'),
\eeq
with the boundary condition:
\beq
U(\Lambda,\Lambda)=1.
\eeq
We treat the free part of the IMO (${\cal M}^{2}_{\mathrm{free}}$) nonperturbatively and use:
\beq
T(\Lambda)=\left[ {\cal M}^{2}_{\mathrm{free}} , {\cal M}^{2}(\Lambda) \right] .
\eeq
If we can determine $T(\Lambda)$, and therefore $U(\Lambda,\Lambda')$, then we can use the perturbative transformation repeatedly 
to lower the cutoff as long as the couplings are not too large.  We can choose the initial cutoff to be as large as 
we want assuming the couplings in the large cutoff limit do not grow, which is true in an asymptotically free theory.  This implies 
that ${\cal M}^{2}(\Lambda)$ is unitarily equivalent to $\lim_{\Lambda \rightarrow \infty}{\cal M}^{2}(\Lambda)$, 
the IMO with no cutoff.  Thus ${\cal M}^{2}(\Lambda)$ will give cutoff-independent results.

We continue with the discussion of the exact form of the transformation and how we add restrictions from unviolated 
physical principles to fix the form of the renormalized Hamiltonian after more notation has been introduced in 
Section \ref{sec:formal}.

\section{Review of the Systematic Approach}
\label{sec:formal}
In this section we introduce more of the notation developed in Refs.~\cite{brentA,brentB,brentD} and outline the method 
to systematically determine the renormalized Hamiltonian.
Formalism that is necessary for a detailed understanding of this method but that we do not repeat in this paper 
can be found in this earlier work.
 
We want to find the regulated invariant-mass operator, ${\cal M}^{2}(g_{_{\Lambda}},m,\Lambda)$, which is trivially related to the 
Hamiltonian.  It can be split into a free part (which contains implicit mass dependence) and an interacting part:
\beq
\label{separation}
{\cal M}^{2}(g_{_{\Lambda}},m,\Lambda)={\cal M}^{2}_{\mathrm{free}}(m)+{\cal M}_{\mathrm{int}}^{2}(g_{_{\Lambda}},m,\Lambda).
\eeq
Since the method treats ${\cal M}_{\mathrm{int}}^{2}(g_{_{\Lambda}},m,\Lambda)$ perturbatively, we put the particle-mass term in ${\cal 
M}_{\mathrm{free}}^{2}(m)$, to treat it non-perturbatively; however, ${\cal M}_{\mathrm{int}}^{2}(g_{_{\Lambda}},m,\Lambda)$ will still have mass dependence.
The matrix elements of ${\cal M}^{2}(g_{_{\Lambda}},m,\Lambda)$ are written
\begin{eqnarray}
\label{regulation}
\langle F \vert {\cal M}^{2}(g_{_{\Lambda}},m,\Lambda)\vert I\rangle&=&\langle F\vert {\cal 
M}_{\mathrm{free}}^{2}(m)\vert I\rangle+
\langle F\vert {\cal M}_{\mathrm{int}}^{2}(g_{_{\Lambda}},m,\Lambda)\vert I\rangle\nonumber\\
&=&M^{2}_{F}\langle F\vert I\rangle +e^{-\frac{\Delta^{2}_{FI}}{\Lambda^{4}}}\langle F\vert V(g_{_{\Lambda}},m,\Lambda)\vert 
I\rangle,
\end{eqnarray}
where $\vert F\rangle$ and $\vert I\rangle$ are eigenstates of the free invariant-mass operator with eigenvalues 
$M_{F}^{2}$ and $M_{I}^{2}$, and $\Delta_{FI}$ is the difference of these eigenvalues.  $V(g_{_{\Lambda}},m,\Lambda)$ is the 
interacting part of the invariant-mass operator with the Gaussian cutoff factor removed  
and is called the ``reduced interaction.''  The Gaussian cutoff on change in invariant-mass is what regulates the 
theory (Sec.~\ref{regulate}) and it has no effect on the free part of the Hamiltonian.

We expand $V(g_{_{\Lambda}},m,\Lambda)$ in powers of the running coupling,
 $g_{_{\Lambda}}$:
\beq
\label{gseries}
V\hspace{-.07cm}(g_{_{\Lambda}},m,\Lambda)=\sum_{r=1}^{\infty}g_{_{\Lambda}}^{r}V^{(r)}\hspace{-.07cm}(m,\Lambda),
\eeq
where $V^{(1)}$ is the canonical interaction and the $V^{(r\ge 2)}\hspace{-.07cm}(m,\Lambda)$'s are non-canonical 
interactions in the scalar theory.  These non-canonical operators can be thought of as counterterms in a traditional approach.  
Note that $g_{_{\Lambda}}$ implicitly depends on $m$.  The assumption that this expansion exists is equivalent to a 
restricted coupling coherence \cite{perry-wilson}.  We now continue the discussion of the unitary transformation 
described in Sec.~\ref{renorm}.

Allen shows in \cite{brentD} that the perturbative version of the unitary transformation that lowers the cutoff 
[derived from Eq.~(\ref{similarity-trans})] is:
\beq
\label{pert-trans}
V(\Lambda)-V(\Lambda')=\delta V,
\eeq
where $\delta V$ is the change in the reduced reaction and is a function of both $\Lambda$ and $\Lambda'$.  Since 
the free part of the Hamiltonian is independent of the cutoff, calculating $\delta V$ gives us the change in the 
Hamiltonian when lowering the cutoff from $\Lambda'$ to $\Lambda$.  The matrix elements of $\delta V$ are found to 
be\footnote{We do not derive this here as it is done in earlier work.  Simply stated, we use 
Eqns.~(\ref{similarity-trans}) and (\ref{sim-diffeq}) and match powers of the interaction.}:
\beq
\left< F \right| \delta V \left| I \right> &&= \frac{1}{2} \sum_K
\left< F \right| V(\Lambda') \left| K \right> \left< K \right| V(\Lambda') \left| I \right> T_2^{(\Lambda,\Lambda')}(F,K,I) \nonumber\\
&&\hspace{-.5cm}+ \frac{1}{4} \sum_{K,L}  \left< F \right| V(\Lambda') \left| K \right> \left< K \right| V(\Lambda') \left| L \right>
\left< L \right| V(\Lambda') \left| I \right> T_3^{(\Lambda,\Lambda')}(F,K,L,I) \nonumber\\
&&\hspace{-.5cm}+ {\cal O}\left( \left[ V(\Lambda') \right] ^4\right)  ,
\eeq
where the sums are over complete sets of states.  $T_2^{(\la,\la')}(F,K,I)$ and $T_3^{(\la,\la')}(F,K,L,I)$ are:
\beq
T_2^{(\la,\la')}(F,K,I) &=& \left(  \frac{1}{\Bx{}{FK}} - \frac{1}{\Bx{}{KI}} \right)  \left(  
e^{2 \la'^{-4} \bx{}{FK} \bx{}{KI}} - e^{2 \la^{-4} \bx{}{FK} \bx{}{KI}} \right) ,
\eeq
and
\beq
&&T_3^{(\la,\la')}(F,K,L,I) = \nonumber\\
&& \hspace{1cm} \left(  \frac{1}{\Bx{}{KL}} - \frac{1}{\Bx{}{LI}} \right)  \left(  \frac{1}{\Bx{}{KI}} -
\frac{1}{\Bx{}{FK}} \right) 
\n && \hspace{2cm} \times 
e^{2 \la'^{-4} \bx{}{KL} \bx{}{LI}} \left(  e^{2 \la^{-4} \bx{}{FK} \bx{}{KI}} -
e^{2 \la'^{-4} \bx{}{FK} \bx{}{KI}} \right)  \nonumber\\
&&\hspace{1cm}+ \left(  \frac{1}{\Bx{}{KL}} - \frac{1}{\Bx{}{LI}} \right)  
\left( \frac{\Bx{}{FK} + \Bx{}{IK}}{\Bx{}{KL} \Bx{}{LI} + \Bx{}{FK} \Bx{}{KI}} \right)
\n && \hspace{2cm} \times
\left(   e^{2 \la'^{-4} (\bx{}{FK}
\bx{}{KI}
+ \bx{}{KL} \bx{}{LI} )} - 
e^{2 \la^{-4} (\bx{}{FK} \bx{}{KI} + \bx{}{KL} \bx{}{LI} )} \right)  \nonumber\\
&&\hspace{1cm}+ \left(  \frac{1}{\Bx{}{FK}} -
\frac{1}{\Bx{}{KL}} \right)  \left(  \frac{1}{\Bx{}{LI}} -
\frac{1}{\Bx{}{FL}} \right)
\n && \hspace{2cm} \times
  e^{2 \la'^{-4} \bx{}{FK} \bx{}{KL}}  \left(  e^{2 \la^{-4} \bx{}{FL} \bx{}{LI}} -
e^{2 \la'^{-4} \bx{}{FL} \bx{}{LI}} \right)  \nonumber\\
&&\hspace{1cm} + \left(  \frac{1}{\Bx{}{FK}} - \frac{1}{\Bx{}{KL}} \right)  
\left( \frac{\Bx{}{FL} + \Bx{}{IL}}{\Bx{}{FK} \Bx{}{KL} + \Bx{}{FL} \Bx{}{LI}} \right) 
\n && \hspace{2cm} \times
\left(   e^{2 \la'^{-4} (\bx{}{FK} \bx{}{KL} + \bx{}{FL} \bx{}{LI} )} - 
e^{2 \la^{-4} (\bx{}{FK} \bx{}{KL} + \bx{}{FL} \bx{}{LI} )}  \right)  , 
\eeq
where $\bx{}{AB}$ is the difference in the invariant-mass of states $A$ and $B$.  

Next we want to be able to solve for the reduced interaction order by order in the coupling, 
$g_{\Lambda}$.  However the reduced interaction contains operators and couplings.  This means the change in the 
reduced interaction when lowering the cutoff is due to the renormalization of the operators and the coupling.  
If we want to see order by order in the coupling how the reduced interaction changes as the cutoff is 
lowered due to the renormalization of the operators we need to remove the change due to the renormalization 
of the coupling.  This is done next.

Expand the reduced interaction and $\delta V$ in powers of $g_{\Lambda}$ and $g_{\Lambda'}$, respectively:
\beq
V(\Lambda) &=& \sum_{t=1}^{\infty} g_{\Lambda}^{t} V^{(t)}(\Lambda), \n
\delta V &=& \sum_{t=2}^{\infty} g_{\Lambda'}^{t} \delta V^{(t)},
\eeq
where $V^{(t)}(\Lambda)$ is the ${\cal O}(g_{\Lambda}^{t})$ contribution to the reduced interaction and $\delta 
V^{(t)}$ is the ${\cal O}(g_{\Lambda'}^{t})$ contribution to $\delta V$.  The expansion of $\delta V$ starts at 
second order because the first order reduced interaction is the canonical interaction which is unchanged by the 
cutoff.  We can now expand Eq.~(\ref{pert-trans}) in powers of $g_{\Lambda}$ and $g_{\Lambda'}$:
\beq
\label{pert-trans-series}
\sum_{t=1}^{\infty} g_{\Lambda}^{t}V^{(t)}(\Lambda)-\sum_{t=1}^{\infty}g_{\Lambda'}^{t}V^{(t)}(\Lambda')
=\sum_{t=2}^{\infty}g_{\Lambda'}^{t}\delta V^{(t)}.
\eeq
This equation contains the coupling at different cutoffs, so we expand $g_{\Lambda}$ in powers of 
$g_{\Lambda'}$:
\beq
\label{g-series}
g_{\Lambda}=g_{\Lambda'}+\sum_{s=3}^{\infty} g_{\Lambda'}^{s}C_{s}(\Lambda,\Lambda').
\eeq
We can use Eq.~(\ref{g-series}) to 
determine $g_{\Lambda}$ raised to the power $t\ge 1$ in powers of $g_{\Lambda}$:
\beq
\label{gt-series}
g_{\Lambda}^{t}=g_{\Lambda'}^{t}+\sum_{s=2}^{\infty} g_{\Lambda'}^{t+s}B_{t,s}(\Lambda,\Lambda'),
\eeq
where the $B_{t,s}$'s can be determined in terms of the $C$'s by raising Eq.~(\ref{g-series}) to the $t^{th}$ 
power.  We now substitute Eq.~(\ref{gt-series}) into Eq.~(\ref{pert-trans-series}) and match powers of 
$g_{\Lambda'}$ which yields:
\beq
\label{diff}
V^{(r)}\hspace{-.07cm}(m,\Lambda)-V^{(r)}\hspace{-.12cm}\left(m,\Lambda^{\prime}\right)=
{\delta V^{(r)}\hspace{-.07cm}(m,\Lambda,\Lambda^{\prime})-\sum_{s=2}^{r-1}B_{r-s,s}V^{(r-s)}\hspace{-.07cm}(m,\Lambda)},
\eeq 
where the $B_{r-s,s}$'s are functions of $m$, $\Lambda$, and $\Lambda^{\prime}$ that contain information on the scale dependence 
of the coupling.  Since the scale dependence of the reduced interaction comes from $g_{_{\Lambda}}$ and the 
$V^{(r)}\hspace{-.07cm}(m,\Lambda)$'s [See Eq.~(\ref{gseries})], Eq.~(\ref{diff}) simply states that if we subtract 
from $\delta V^{(r)}\hspace{-.07cm}(m,\Lambda,\Lambda^{\prime})$ the contribution due to the scale dependence of the coupling, then we are left with the 
contribution due to the scale dependence of the $V^{(r)}\hspace{-.07cm}(m,\Lambda)$'s.

If there is a part of $V^{(r)}\hspace{-.07cm}(m,\Lambda)$ that is independent of the cutoff, it will cancel on the 
left-hand-side of Eq.~(\ref{diff}).  For this reason, we split $V^{(r)}\hspace{-.07cm}(m,\Lambda)$ into a part that 
depends on the cutoff, $V^{(r)}_{\mathrm{CD}}\hspace{-.07cm}(m,\Lambda)$, and a part that is independent of the cutoff, 
$V^{(r)}_{\mathrm{CI}}\hspace{-.07cm}(m)$:
\beq
V^{(r)}\hspace{-.07cm}(m,\Lambda)=
V^{(r)}_{\mathrm{CD}}\hspace{-.07cm}(m,\Lambda)+V^{(r)}_{\mathrm{CI}}\hspace{-.07cm}(m).
\eeq
We must solve for both $V_{\mathrm{CD}}^{(r)}\hspace{-.07cm}(m,\Lambda)$ and $V_{\mathrm{CI}}^{(r)}\hspace{-.07cm}(m)$ 
to find the IMO.  The recursion relations for 
$V_{\mathrm{CD}}^{(r)}\hspace{-.07cm}(m,\Lambda)$ and $V_{\mathrm{CI}}^{(r)}\hspace{-.07cm}(m)$ are given in 
Sections \ref{dependent} and \ref{independent}, respectively.

\section{Addition of Particle Masses}
\label{sec:difference}

This method of renormalization has been generalized \cite{kylinA} to include particle masses and was illustrated using massive $\phi^{3}$ 
theory in 5+1 dimensions .  This theory is asymptotically free and its diagrammatic structure is similar to QCD, which make it a good 
perturbative development ground.  It is straightforward to extend the method for massless theories developed in Ref. 
\cite{brentA} 
to calculate QCD quantities for which particle masses are unimportant, such as the low-lying glueball spectrum 
\cite{brentB,brentD}.  In this section, we show how to incorporate particle masses non-perturbatively as a necessary step toward 
a treatment of full QCD.

In our renormalized scalar theory $m$ is the physical particle mass.  In a confining theory $m$ is the particle mass in the 
zero-coupling limit, although other definitions are possible.  Since the mass is being treated non-perturbatively, it must 
be included in the free part of 
${\cal M}^{2}(g_{_{\Lambda}},m,\Lambda)$ in Eq.~(\ref{separation}).  This alters the unitary transformation and 
leads to fundamental changes in the renormalization procedure.  

The changes in the procedure are discussed in the next three subsections.  The redefinition 
of the coupling (Sec.~\ref{coupling}) is straightforward.  In Sections \ref{dependent} and \ref{independent}, 
we present the expressions for the matrix elements of $V_{\mathrm{CD}}^{(r)}\hspace{-.07cm}(m,\Lambda)$ and 
$V_{\mathrm{CI}}^{(r)}\hspace{-.07cm}(m)$, respectively.  We also qualitatively discuss the additional steps that are required to 
interpret and use them in a massive theory.  In Section \ref{gen-full-qcd} we 
present the changes to the procedure when generalizing from massive $\phi^{3}$ to full QCD.

\subsection{Coupling}
\label{coupling}
The coupling gives the strength of the three-point canonical interaction in the theory, thus the canonical definition 
of the coupling is
\beq
g = \left[ 64 \pi^{5} p_1^+ \delta^{(5)}(p_1 - p_2 - p_3) \right]^{-1} \left< \phi_2 \phi_3 \right| 
{\cal M}^2_{\mathrm{can}}
\left| \phi_1 \right>|_{p_{2}=p_{3}} .
\eeq
In the massive theory, we choose
\begin{eqnarray}
g_{_{\Lambda}}&=&\left[64\pi^{5}p_{1}^{+}\delta^{(5)}\left(p_{1}-p_{2}-p_{3}
\right)\right]^{-1}
\mathrm{exp}\left(9\frac{m^{4}}{\Lambda^{4}}\right)\langle 
\phi_{2}\phi_{3}\vert{\cal{M}}^{2}(g_{_{\Lambda}},m,\Lambda)
\vert\phi_{1}\rangle|_{p_{2}=p_{3}}\nonumber\\
&=&\left[64\pi^{5}p_{1}^{+}\delta^{(5)}\left(p_{1}-p_{2}-p_{3}
\right)\right]^{-1}
\langle \phi_{2}\phi_{3}\vert 
V(g_{_{\Lambda}},m,\Lambda)\vert\phi_{1}\rangle|_{p_{2}=p_{3}},
\end{eqnarray}
which differs from the definition in the massless theory by the factor 
$\mathrm{exp}\left(9\frac{m^{4}}{\Lambda^{4}}\right)$.  This choice of coupling cancels the added mass dependence 
in the regulator [Eq.~(\ref{regulation})] and allows us to closely follow the formalism developed in the massless theory. 

\subsection{Cutoff-Dependent Contributions to $V^{(r)}\hspace{-.07cm}(m,\Lambda)$}
\label{dependent}

Momentum conservation implies that any matrix element of $V^{(r)}\hspace{-.07cm}(m,\Lambda)$ contains a sum of terms, 
each with a unique product of momentum-conserving delta functions.  Assuming that approximate transverse locality is 
maintained, the coefficient of each product of delta functions can be written as an 
expansion in powers of transverse momenta.  In massive $\phi^{3}$ theory, we can also make a generalized expansion in 
powers and logarithms of $m$.  For a theory in $5+1$ dimensions, the Hamiltonian has dimensions $[\Lambda^{6}]$, 
where $\Lambda$ is the cutoff on change in invariant mass.  For the dimensions to work out 
properly, the scalar fields must have a dimension of $[\Lambda^{2}]$.  Thus, the scale dependence of any term in this expansion has the form
\beq
\label{fcdi}
\Lambda^{6-2N_{\mathrm{int}}}\left(\frac{m}{\Lambda}\right)^{\alpha}
\left[\log\frac{m}{\Lambda}\right]^{\beta}
\left(\frac{p_{\bot}}{\Lambda}\right)^{\gamma},
\eeq
where $N_{\mathrm{int}}$ is the total number of particles in the final and initial states that participate in the 
interaction.  Also $\alpha$, $\beta$, 
and $\gamma$ are non-negative integers.  
For simplicity we display one component of transverse momentum, $p_{\bot}$; however, the general form includes a 
product of all transverse components from all particles.  In principle, the introduction of a particle mass allows 
any function of $\frac{m}{\Lambda}$ to appear.  However, to ${\cal O}(g_{_{\Lambda}}^{3})$ the only extra scale 
dependence comes in the form $\left(\frac{m}{\Lambda}\right)^{\alpha}\left[\log\frac{m}{\Lambda}\right]^{\beta}$.  
If $\beta=0$ and
\beq
\label{condition}
6-2N_{\mathrm{int}}-\alpha-\gamma=0,
\eeq
the term is independent of the cutoff and is referred to as a ``cutoff-independent'' contribution.  These 
contributions are discussed in the next subsection.  
\newpage
The expression for a matrix element of $V^{(r)}_{\mathrm{CD}}\hspace{-.07cm}(m,\Lambda)$ is derived from 
Eq.~(\ref{diff}):
\beq
\label{fcd}
\left<F \right| V^{(r)}_{\mathrm{CD}}(m,\Lambda) \left| I \right> &=& \left[ \langle F\vert
\delta V^{(r)}\hspace{-.07cm}(m,\Lambda,\Lambda^{\prime})\vert I\rangle
\right. \n && \hspace{1cm} \left.
 - \sum_{s=2}^{r-1}  B_{r-s,s}
\left<F \vert V^{(r-s)}(m,\Lambda) \vert I \right>\right]_{\Lambda \hspace{.1cm}\mathrm{terms}} .
\eeq
``$\Lambda$ terms'' means the terms in the momentum and mass expansion that contain $\Lambda^{\prime}$ are to be removed 
from the expression in brackets.  In terms that depend on 
positive powers of $\Lambda^{\prime}$, we do this by letting $\Lambda^{\prime}\rightarrow 0$, and in terms that 
depend on negative powers of $\Lambda^{\prime}$, we let $\Lambda^{\prime}\rightarrow \infty$.

\subsection{Cutoff-Independent Contributions to $V^{(r)}\hspace{-.07cm}(m,\Lambda)$}
\label{independent}

Considering the condition in Eq.~(\ref{condition}), 
only two-point and three-point interactions can have cutoff-independent contributions.  The lowest-order cutoff-independent 
three-point interaction is $V_{\mathrm{CI}}^{(3)}\hspace{-.07cm}(m)$ and has not been explicitly computed in the 
massless and massive theories.  However,  $V_{\mathrm{CI}}^{(2)}\hspace{-.07cm}(m)$ is the lowest-order 
cutoff-independent two-point interaction and must be calculated before anything is calculated to third order.

Due to boost invariance, $V_{\mathrm{CI}}^{(2)}\hspace{-.07cm}(m)$ must be independent of the interacting particles 
transverse momentum, implying $\gamma$ can only be zero.\footnote{In light-front coordinates, a transverse boost 
shifts all transverse momenta.  This means to ensure boost invariance $V_{\mathrm{CI}}^{(2)}\hspace{-.05cm}(m)$ 
must be independent of the interacting particle's transverse momentum.}  This means, for example, the 
cutoff-independent part of a self-energy contribution will be proportional to $m^{2}$.  Thus, 
to isolate the cutoff-independent part of a matrix element we must expand it in powers of  
$\frac{p_{\bot}}{\Lambda}$ and in powers of $\log \left( \frac{m}{\Lambda} \right) $.  Then the term that 
is independent of the cutoff obeys the relation:
\beq
6-2N_{\mathrm{int}}-\alpha=0.
\eeq

The matrix elements of $V_{\mathrm{CI}}^{(r)}\hspace{-.07cm}(m)$ are divided into 2-point and 3-point 
contributions, and are given by the expression
\beq
\label{cime}
&&\langle F\vert V^{(r)}_{\mathrm{CI}}\hspace{-.07cm}(m)\vert I\rangle= \n
&&\hspace{3cm}
\frac{1}{B_{r,2}}\left[ \langle F\vert\delta V^{(r+2)}\hspace{-.07cm}(m,\Lambda,\Lambda^{\prime})\vert I\rangle 
\right. \n && \hspace{5cm} \left.
-\sum_{s=3}^{r+1}B_{r+2-s,s}\langle F\vert V^{(r+2-s)}\hspace{-.07cm}(m)\vert 
I\rangle\right] ^{3-\mathrm{point}}_{m^{0}\vec{p}_{\bot}^{\hspace{.05cm}0}\hspace{.1cm}\mathrm{term}} \n
&&\hspace{3cm}+\frac{1}{B_{r,2}}\left[\langle F\vert\delta V^{(r+2)}\hspace{-.07cm}(m,\Lambda,\Lambda^{\prime})\vert I\rangle
\right. \n && \hspace{5cm} \left.
-\sum_{s=3}^{r+1}B_{r+2-s,s}\langle F\vert V^{(r+2-s)}\hspace{-.07cm}(m)\vert I\rangle
\right] ^{\mathrm{2-point}}_{m^{2}\hspace{.1cm}\mathrm{term}}.
\eeq
Here, ``$m^{0}\vec{p}_{\bot}^{\hspace{.1cm}0}\hspace{.1cm}\mathrm{term}$'' and ``$m^{2}\hspace{.1cm}\mathrm{term}$''
mean expand the term in brackets in powers of external transverse momenta and in powers and logs of $m$, and keep only 
the term that is proportional to $m^{0}\vec{p}_{\bot}^{\hspace{.1cm}0}$ or $m^{2}$, respectively.
The removal of $\Lambda$ and $\Lambda'$ dependence is guaranteed by construction.

Initially Eq.~(\ref{cime}) looks useless because $V_{\mathrm{CI}}^{(r)}\hspace{-.07cm}(m)$ depends on 
$V^{(r+1)}_{\mathrm{CI}}\hspace{-.07cm}(m)$ [which is inside an integral in $\delta V^{(r+2)}$], suggesting the theory 
must be solved to all orders simultaneously.  However, contributions to the reduced interaction from three-point 
interactions can only appear at odd orders, and contributions from two-point interactions can appear only at even orders.  
Thus, in the massless theory, this apparent problem does not manifest itself because there are no cutoff-independent two-point 
interactions.   In the massive theory, although there are cutoff-independent two-point interactions, it is possible to 
solve for $V^{(2)}_{\mathrm{CI}}\hspace{-.07cm}(m)$ and $V^{(3)}_{\mathrm{CI}}\hspace{-.07cm}(m)$ simultaneously, without 
considering higher orders.  This even-order/odd-order pattern can be extended to all orders.  

Including self-energy contributions, the theory we want to describe contains particles of mass $m$.  We can simplify 
the problem by using this fact instead of using Eq.~(\ref{cime}) to solve for the even-order 
$V_{\mathrm{CI}}^{(r)}\hspace{-.07cm}(m)$'s.  We do this by forcing the completely disconnected 
parts of the forward $T$-matrix elements to be zero.  
(This part of a $T$-matrix element contains initial and final states that have the same number of particles, 
$n$, and $n$ momentum-conserving delta functions.)  This fixes the even-order 
$V_{\mathrm{CI}}^{(r)}\hspace{-.07cm}(m)$'s since they only involve interactions on single particle lines.  
This allows us to calculate 
$V^{(2)}_{\mathrm{CI}}\hspace{-.07cm}(m)$ independently of $V^{(3)}_{\mathrm{CI}}\hspace{-.07cm}(m)$.  This extra 
condition can be used to fix all even-order $V_{\mathrm{CI}}^{(r)}\hspace{-.07cm}(m)$'s.

\subsection{Results}
\label{phi3-results}
\begin{figure}
\unitlength1in
\begin{center}
\begin{picture}(4,4)
\put(-0.05,.1){\epsfig{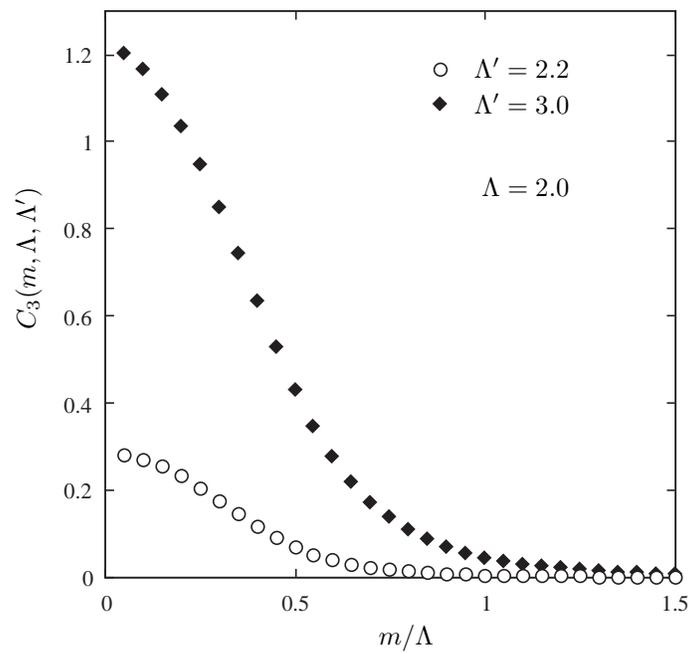}}
\put(2.4,2.92){{\footnotesize $\Lambda^{\prime}=2.2$}}
\put(2.4,2.73){{\footnotesize $\Lambda^{\prime}=3.0$}}
\put(1.9,-.05){{\footnotesize $m/\Lambda$}}
\put(0,1.6){\rotatebox{90}{{\footnotesize $C_{3}(m,\Lambda,\Lambda^{\prime})$}}}
\put(2.44,2.3){{\footnotesize $\Lambda=2.0$}}
\end{picture}
\end{center}

\caption[The third-order coefficient of the running coupling as a function of the particle mass.]{The third-order coefficient of the running coupling as a function of the particle mass.  Curves for various 
upper cutoffs with fixed lower cutoff show the coupling is exponentially damped with increasing mass.}
\label{fig:c3}
\end{figure}

The coupling in massive $\phi^{3}$ theory runs at third order.  We can compare the coupling at two different 
scales, $\Lambda$ and $\Lambda^{\prime}$:
\begin{equation}
g_{_{\Lambda}}=g_{_{\Lambda^{\prime}}}+\sum_{s=3}^{\infty}g_{_{\Lambda^{\prime}}}^{s}\hspace{.2cm}
\hspace{-.12cm}C_{s}(m,\Lambda,\Lambda^{\prime}).
\end{equation}
We can determine how the coupling runs at third order by solving for $C_{3}(m,\Lambda,\Lambda^{\prime})$ 
(which is proportional to the matrix element 
$\langle\phi_{2}\phi_{3}\vert\delta V^{(3)}\hspace{-.07cm}(m,\Lambda,\Lambda^{\prime})\vert\phi_{1}\rangle\vert_{p_{2}=
p_{3}}$).  Figure \ref{fig:c3} shows how $C_{3}(m,\Lambda,\Lambda^{\prime})$ depends on the mass. The running of the 
coupling is exponentially damped as the mass grows since the cutoff inhibits production of 
intermediate particles.  The difference between the values of the running coupling at two different scales increases 
as the two scales are separated.  This is shown by the 
larger magnitude of $C_{3}(m,\Lambda,\Lambda^{\prime})$ as the separation between $\Lambda$ and $\Lambda^{\prime}$ 
grows. 

\begin{figure}
\unitlength1in
\begin{center}
\begin{picture}(4,4)
\put(0,.1){\epsfig{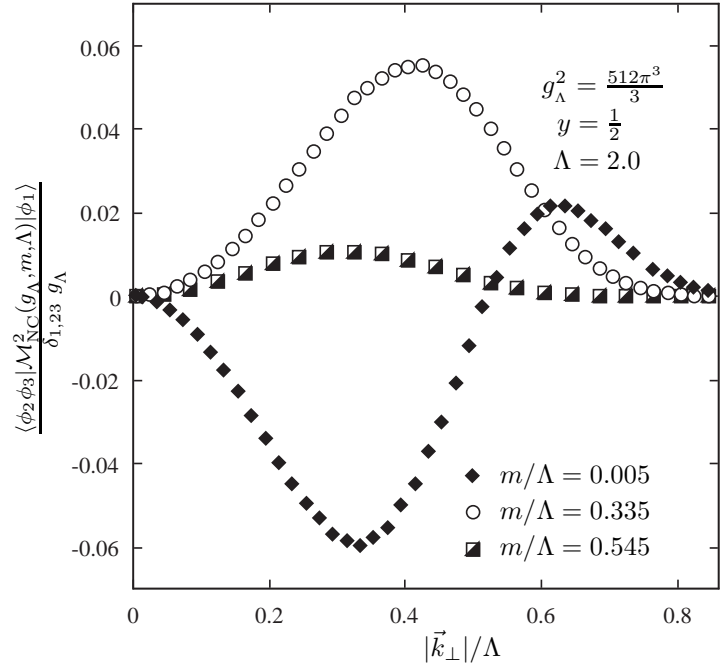}}
\put(-.15,1.1){ \rotatebox{90}{{\small $\frac{\langle \phi_{2} \phi_{3}\vert {\cal 
 M}^{2}_{\mathrm{NC}}(g_{_{\Lambda}},m,\Lambda)\vert\phi_{1}\rangle}{\delta_{1,23}
 \hspace{.1cm}g_{_{\Lambda}}}$}}}
\put(2.45,.86){{\footnotesize $m/\Lambda=0.005$}}
\put(2.45,.67){{\footnotesize $m/\Lambda=0.335$}}
\put(2.45,.48){{\footnotesize $m/\Lambda=0.545$}}
\put(2.75,2.7){{\footnotesize $y=\frac{1}{2}$}}
\put(2.67,2.9){{\footnotesize $g_{_{\Lambda}}^{2}=\frac{512 \pi^{3}}{3}$}}
\put(2.05,-.05){{\footnotesize $\vert\vec{k}_{\bot}\vert/\Lambda$}}
\put(2.73,2.5){{\footnotesize $\Lambda=2.0$}}
\end{picture}
\end{center}

\caption[The matrix element of the non-canonical part of the invariant-mass operator for 
$\phi_{1}\rightarrow\phi_{2}\phi_{3}$.]{The matrix element of the non-canonical part of the invariant-mass operator for 
$\phi_{1}\rightarrow\phi_{2}\phi_{3}$ versus the magnitude of the relative transverse momentum of particles 2 and 3 in the center-of-momentum frame.  $y$ is the longitudinal momentum fraction carried by particle 2.}
 \label{v3}
\end{figure}

Determining $V_{\mathrm{CI}}^{(3)}\hspace{-.07cm}(m)$ requires a fifth-order calculation and is not attempted.  However, 
calculating the matrix element $\langle\phi_{2}\phi_{3}\vert V_{\mathrm{CD}}^{(3)}\hspace{-.07cm}(m,\Lambda)\vert\phi_{1}\rangle$ 
gives the relative sizes of the non-canonical interactions and the canonical interaction.  Their relative magnitudes 
are similar to those in Ref. \cite{brentA}, suggesting that an expansion of the reduced interaction in powers of the 
running coupling is valid through third order.

Figure \ref{v3} shows how the non-canonical part of the matrix element of the invariant-mass operator for the 
interaction $\phi_{1}\rightarrow\phi_{2}\phi_{3}$ depends on the magnitude of the relative transverse momentum in the 
center-of-momentum frame.  Increasing the relative transverse momentum in the center-of-momentum frame increases the free mass of the 
system.

\subsection{Generalizing to Full QCD}
\label{gen-full-qcd}

As in the $\phi^{3}$ theory, momentum conservation implies that any matrix element of 
$V^{(r)}\hspace{-.07cm}(m,\Lambda)$ in full QCD contains terms that can be written as an expansion in powers of 
$m$ and powers of logarithms 
of $m$.  However, in QCD we work in 3+1 dimensions, so the Hamiltonian has dimension $ \left[ \Lambda^{4} 
\right] $ and the 
particle fields have dimension $\left[ \Lambda \right] $.  So for QCD, the scale dependence of any term in this expansion has 
the form
\beq
\Lambda^{4-N_{\mathrm{int}}}\left(\frac{m}{\Lambda}\right)^{\alpha}
\left[\log\frac{m}{\Lambda}\right]^{\beta}
\left(\frac{p_{\bot}}{\Lambda}\right)^{\gamma}.
\eeq
This is the analogous to Eq.~\ref{fcdi} in $\phi^{3}$ theory.  Thus if $\beta=0$ and
\beq
\label{qcd-cd-ci}
4-N_{\mathrm{int}}-\alpha-\gamma=0,
\eeq
the term is independent of the cutoff.  Thus for QCD there are cutoff independent interactions for two, three 
and four interacting particles.  However, since we are approximating mesons as a color singlet $q \bar{q}$ pair, there can only be 
either two or four interacting particles.  So the expression for the cutoff independent interactions is given by:
\newpage
\beq
&&\langle F\vert V^{(r)}_{\mathrm{CI}}\hspace{-.07cm}(m)\vert I\rangle= \n
&&\hspace{3cm}
\frac{1}{B_{r,2}}\left[ \langle F\vert\delta V^{(r+2)}\hspace{-.07cm}(m,\Lambda,\Lambda^{\prime})\vert I\rangle 
\right. \n && \hspace{5cm} \left.
-\sum_{s=3}^{r+1}B_{r+2-s,s}\langle F\vert V^{(r+2-s)}\hspace{-.07cm}(m)\vert 
I\rangle\right] ^{4-\mathrm{point}}_{m^{0}\vec{p}_{\bot}^{\hspace{.05cm}0}\hspace{.1cm}\mathrm{term}} \n
&&\hspace{3cm}+\frac{1}{B_{r,2}}\left[\langle F\vert\delta V^{(r+2)}\hspace{-.07cm}(m,\Lambda,\Lambda^{\prime})\vert I\rangle
\right. \n && \hspace{5cm} \left.
-\sum_{s=3}^{r+1}B_{r+2-s,s}\langle F\vert V^{(r+2-s)}\hspace{-.07cm}(m)\vert I\rangle
\right] ^{\mathrm{2-point}}_{m^{2}\hspace{.1cm}\mathrm{term}}.
\eeq

The expression for $\left<F \right| V^{(r)}_{\mathrm{CD}}(m,\Lambda) \left| I \right>$ is the same in full QCD as it is in 
$\phi^{3}$ theory.

\section{Mass-Squared to Second Order}
\label{mass-o2-begin}
We want to find which matrix elements we will need to calculate to find:
\begin{eqnarray}
\langle F \vert {\cal M}^{2}(g_{_{\Lambda}},m,\Lambda)\vert I\rangle=M^{2}_{F}\langle F\vert I\rangle +
e^{-\frac{\Delta^{2}_{FI}}{\Lambda^{4}}}\langle F\vert V (g_{_{\Lambda}},m,\Lambda)\vert I\rangle
\end{eqnarray}
to second order in $g_{\Lambda}$.  Thus, from Section \ref{sec:formal} we must determine
\beq
\langle F\vert V^{(2)}(g_{_{\Lambda}},m,\Lambda)\vert I\rangle =
\langle F\vert V_{\mathrm{CD}}^{(2)}(m,\Lambda)\vert I\rangle +
\langle F\vert V_{\mathrm{CI}}^{(2)}(m)\vert I\rangle ,
\eeq
where the second-order reduced interaction is divided into the part that depends on the cutoff (CD) and the part that is 
independent of the cutoff (CI)\footnote{See Eq.~(\ref{qcd-cd-ci}) for the rule to determine cutoff dependence or 
independence.}.  

\begin{figure}
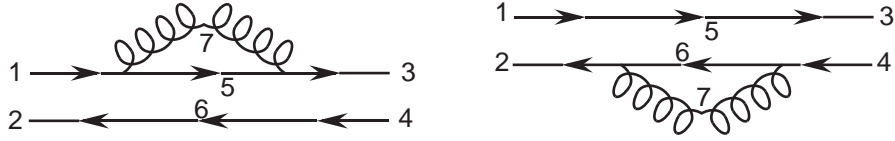

    \centerline{\epsfig{file=\diagrampath self-q.epsf}\hspace{1cm}\epsfig{file=\diagrampath self-qbar.epsf}}
    \caption{\label{se-diagrams}The quark and antiquark self-energy diagrams.}
\end{figure}

\begin{figure}
    \centerline{\epsfig{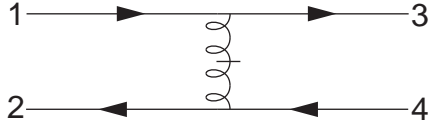}}
    \caption{\label{in-diagram}The instantaneous exchange diagram.}
\end{figure}

\begin{figure}
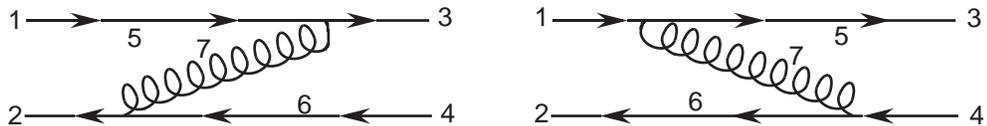

    \centerline{\epsfig{file=\diagrampath exch-qbq.epsf}\hspace{1cm}\epsfig{file=\diagrampath exch-qqb.epsf}}
    \caption{\label{exch-diagrams}The gluon exchange diagrams.}
\end{figure}

There are only three types of diagrams we must consider when calculating the 
reduced interaction to ${\cal O}(g^{2})$.  The self-energy (SE) is a one-body operator that acts on both quarks and 
antiquarks.  These 
diagrams are illustrated in Figure \ref{se-diagrams}.
There are two types of two-body exchange diagrams.  The first is the instantaneous gluon exchange (IN) and is shown 
in Figure \ref{in-diagram}.
The second type of exchange is a single gluon exchange (EX) from quark to antiquark or vice-versa.  These exchange 
diagrams are shown in Figure \ref{exch-diagrams}.

The self-energy (SE) interaction has a cutoff-dependent and cutoff-independent part, so we split it 
into these two parts:
\beq
\langle F \vert V^{(2)}_{\mathrm{SE}}(m,\Lambda)\vert I\rangle =
\langle F \vert V^{(2)}_{\mathrm{SE_{CD}}}(m,\Lambda)\vert I\rangle +
\langle F \vert V^{(2)}_{\mathrm{SE_{CI}}}(m)\vert I\rangle .
\eeq
The second-order cutoff-dependent reduced interaction can be written:
\beq
\left<F \right| V^{(2)}_{\CD}(m,\la) \left| I \right> =\dVome{2}\vert_{\Lambda\mathrm{\hspace{.05cm} terms}}.
\eeq
For two-particle final and initial states, the matrix elements of $\dVome{2}$ are given by:
\beq
\label{deltav}
\dVome{2}&=&\langle F \vert V^{(2)}_{\mathrm{SE_{CD}}}(\Lambda)\vert I\rangle 
T_{2}^{(\Lambda,\Lambda')}(F,K_{\mathrm{SE}},I) \vert_{\Lambda\mathrm{\hspace{.05cm} terms}}\n
&&\hspace{1.5cm}+\langle F \vert V^{(2)}_{\mathrm{EX}}(\Lambda)\vert I\rangle 
T_{2}^{(\Lambda,\Lambda')}(F,K_{\mathrm{EX}},I) \vert_{\Lambda\mathrm{\hspace{.05cm} terms}} ,
\eeq
where $K_{\mathrm{SE}}$, and $K_{\mathrm{EX}}$, are the intermediate states associated with the 
quark self-energy and gluon exchange, respectively.

The cutoff-independent contribution comes from the instantaneous gluon exchange and the cutoff-independent part 
of the self-energy.  The instantaneous exchange is cutoff-independent since there are four interacting particles, 
and there are no terms proportional to any nonzero power of 
the quark mass or external transverse momentum.  Finally we write the cutoff-independent part of the reduced 
interaction as:
\beq
\langle F\vert V^{(2)}_{\mathrm{CI}}(m)\vert I\rangle &=& 
\n &&\hspace{-1cm}
\langle F \vert V^{(2)}_{\mathrm{IN}}(m)\vert I\rangle + \langle F 
\vert V^{(2)}_{\mathrm{SE_{CI}}}(m)\vert I\rangle T_{2}^{(\Lambda,\Lambda')}(F,K_{\mathrm{SE}},I) 
\vert_{\Lambda\mathrm{ terms}} .
\eeq

\section{Combining the Interactions}
\label{comb-inter}
This section describes how the interactions are combined to explicitly cancel divergences in the context of the meson 
calculation (quark-antiquark states), however the general discussion is true for the glueball calculation 
(glue-glue states) with the quark and antiquark replaced with gluons.  We list how the interactions are 
divided and recombined which is simply the overview of Perry's work in \cite{robert}.  

The self-energy, gluon exchange and instantaneous gluon exchange diagrams produce divergences when the longitudinal 
momentum of the exchanged gluon vanishes.  These `infrared' divergences are an artifact of the light-front quantization and 
therefore should cancel.  Although the divergences cancel, Perry uses the QCD Hamiltonian to ${\cal 
O}(\alpha)$ to show how a logarithmic confining mechanism arises after the cancellation.

We regulate the divergences with a cutoff on longitudinal momentum.  However, after the matrix elements are 
combined, as described below, this cutoff can be taken to zero and all matrix elements in the renormalized 
Hamiltonian are finite.  The combination of matrix elements cancels the apparent divergences so no renormalization 
is needed to remove divergences from small longitudinal momentum.  

The divergent parts of the exchange interaction and the self-energy are cancelled by different parts of the instantaneous 
exchange.  We can make this cancellation explicit by dividing the 
instantaneous interaction into two parts, one above the cutoff and one below the cutoff:
\beq
\langle q_{3} \bar{q}_{4}\vert {\cal M}^{2}(\Lambda)\vert q_{1} \bar{q}_{2}\rangle_{\mathrm{IN}}
&=&\langle q_{3} \bar{q}_{4}\vert {\cal M}^{2}(\Lambda)\vert q_{1} \bar{q}_{2}\rangle_{\mathrm{IN}}^{\mathrm{A}}
+\langle q_{3} \bar{q}_{4}\vert {\cal M}^{2}(\Lambda)\vert q_{1} \bar{q}_{2}\rangle_{\mathrm{IN}}^{\mathrm{B}} ,
\eeq
where
\beq
\langle q_{3} \bar{q}_{4}\vert {\cal M}^{2}(\Lambda)\vert q_{1} \bar{q}_{2}\rangle_{\mathrm{IN}}^{\mathrm{A}}
&=&\left( 1-e^{2\Lambda^{-4}\Delta_{FK}\Delta_{KI}}\right) 
\langle q_{3} \bar{q}_{4}\vert {\cal M}^{2}(\Lambda)\vert q_{1} \bar{q}_{2}\rangle_{\mathrm{IN}} , \n
\langle q_{3} \bar{q}_{4}\vert {\cal M}^{2}(\Lambda)\vert q_{1} \bar{q}_{2}\rangle_{\mathrm{IN}}^{\mathrm{B}}
&=&e^{2\Lambda^{-4}\Delta_{FK}\Delta_{KI}} 
\langle q_{3} \bar{q}_{4}\vert {\cal M}^{2}(\Lambda)\vert q_{1} \bar{q}_{2}\rangle_{\mathrm{IN}} .
\eeq
Next break up the self-energy and exchange terms into finite and divergent pieces:
\beq
\langle q_{3} \bar{q}_{4}\vert {\cal M}^{2}(\Lambda)\vert q_{1} \bar{q}_{2}\rangle_{\mathrm{SE}}
&=&\langle q_{3} \bar{q}_{4}\vert {\cal M}^{2}(\Lambda)\vert q_{1} \bar{q}_{2}\rangle_{\mathrm{SE}}^{\mathrm{D}}
+\langle q_{3} \bar{q}_{4}\vert {\cal M}^{2}(\Lambda)\vert q_{1} \bar{q}_{2}\rangle_{\mathrm{SE}}^{\mathrm{F}} , \n
\langle q_{3} \bar{q}_{4}\vert {\cal M}^{2}(\Lambda)\vert q_{1} \bar{q}_{2}\rangle_{\mathrm{EX}}
&=&\langle q_{3} \bar{q}_{4}\vert {\cal M}^{2}(\Lambda)\vert q_{1} \bar{q}_{2}\rangle_{\mathrm{EX}}^{\mathrm{D}}
+\langle q_{3} \bar{q}_{4}\vert {\cal M}^{2}(\Lambda)\vert q_{1} \bar{q}_{2}\rangle_{\mathrm{EX}}^{\mathrm{F}} .
\eeq
We combine the divergent part of the exchange interaction and the instantaneous 
interaction above the cutoff into the finite expression:
\beq
\langle q_{3} \bar{q}_{4}\vert {\cal M}^{2}(\Lambda)\vert q_{1} \bar{q}_{2}\rangle_{\mathrm{IN+EX}}=
\langle q_{3} \bar{q}_{4}\vert {\cal M}^{2}(\Lambda)\vert q_{1} \bar{q}_{2}\rangle_{\mathrm{EX}}^{\mathrm{D}}+
\langle q_{3} \bar{q}_{4}\vert {\cal M}^{2}(\Lambda)\vert q_{1} \bar{q}_{2}\rangle_{\mathrm{IN}}^{\mathrm{A}} .
\eeq
We also combine the instantaneous interaction below the cutoff with the divergent part of the self-energy 
because the divergences in the two terms cancel.  Thus we have,
\beq
\langle q_{3} \bar{q}_{4}\vert {\cal M}^{2}(\Lambda)\vert q_{1} \bar{q}_{2}\rangle_{\mathrm{IN}}^{\mathrm{B,F}}=
\langle q_{3} \bar{q}_{4}\vert {\cal M}^{2}(\Lambda)\vert q_{1} \bar{q}_{2}\rangle_{\mathrm{IN}}^{\mathrm{B}}+
\langle q_{3} \bar{q}_{4}\vert {\cal M}^{2}(\Lambda)\vert q_{1} \bar{q}_{2}\rangle_{\mathrm{SE}}^{\mathrm{D}} ,
\eeq
leading to
\beq
\langle q_{3} \bar{q}_{4}\vert {\cal M}^{2}(\Lambda)\vert q_{1} \bar{q}_{2}\rangle =
\langle q_{3} \bar{q}_{4}\vert {\cal M}^{2}(\Lambda)\vert q_{1} \bar{q}_{2}\rangle_{\mathrm{KE}}&+&
\langle q_{3} \bar{q}_{4}\vert {\cal M}^{2}(\Lambda)\vert q_{1} \bar{q}_{2}\rangle_{\mathrm{SE}}^{\mathrm{F}} \n +
\langle q_{3} \bar{q}_{4}\vert {\cal M}^{2}(\Lambda)\vert q_{1} \bar{q}_{2}\rangle_{\mathrm{EX}}^{\mathrm{F}} &+&
\langle q_{3} \bar{q}_{4}\vert {\cal M}^{2}(\Lambda)\vert q_{1} \bar{q}_{2}\rangle_{\mathrm{IN+EX}} \n &+&
\langle q_{3} \bar{q}_{4}\vert {\cal M}^{2}(\Lambda)\vert q_{1} \bar{q}_{2}\rangle_{\mathrm{IN}}^{\mathrm{B,F}}
\eeq

\chapter{Basis Functions:  B-Splines}
\label{section:bsplines}

In this Chapter we give some background on the variational method as discussed in most quantum mechanical texts.  
We then discuss the approximation of functions using basis functions.  Next we introduce B-splines, and motivate 
their use for a set of efficient basis functions.  We conclude the introduction to B-splines with some simple 
examples.  The first example is simple function 
approximation and the last two are the one and two-dimensional harmonic oscillator.

This Chapter introduces the basis functions we use to represent the longitudinal- and transverse-momentum degrees 
of freedom in our 
bound-state calculations.  However, a good understanding of B-splines is not required to follow the meson 
calculation.  If a detailed understanding of the B-splines is not needed, Section \ref{bspline-intro} can be 
skipped.  The most important facts about B-splines for this calculation are that they have a finite nonzero 
range and that each B-spline only has a spatial overlap with a limited number of other B-splines.

\section{Eigenstate Approximation}
\label{variational-method}

In this section we show that when using a particular set of basis functions to approximate eigenstates,
the lowest eigenvalue of the Hamiltonian in the approximate basis is never 
less than the real ground state of the system.  If more basis functions are used, giving a better approximation of 
the real ground state, 
the eigenvalue of the lowest approximate state should decrease and converge to the real ground state eigenvalue.  
Finally, as the number of basis functions increases the lowest 
eigenvalue should converge to the ground state eigenvalue.  The following discussion is derived from \cite{QMtext}.

The expectation value of the operator $H$ in the state $\vert \psi \rangle $ is given by:
\beq
\langle H \rangle = \frac{ \langle \psi \vert H \vert \psi \rangle }{ \langle \psi \vert \psi \rangle } \ge E_{0},
\eeq
where $E_{0}$ is the smallest eigenvalue of $H$.
If $\vert \psi \rangle$ is expanded in the set of eigenfunctions of $H$,
\beq
\vert \psi \rangle = \sum_{n} c_{n} \vert \phi_{n} \rangle ,
\eeq
the expectation value can be written:
\beq
\label{var-princ}
\langle \psi \vert H \vert \psi \rangle = \sum_{n} \vert c_{n} \vert^{2} E_{n} \ge E_{0} \sum_{n} \vert c_{n} 
\vert^{2},
\eeq
with
\beq
\langle \psi \vert \psi \rangle = \sum_{n} \vert c_{n}\vert^{2}.
\eeq
The only way for the equality in Eq. (\ref{var-princ}) to be true is if all of the $c_{n}$'s are zero except for 
$c_{0}$, indicating $\vert \phi_{0} \rangle$ is the ground state.  We want to approximate each eigenstate of $H$ 
with a finite set of basis functions $\vert B_{i} \rangle$:
\beq
\vert \phi_{n} \rangle = \sum_{i=1}^{N_{f}} a_{i}^{(n)} \vert B_{i} \rangle ,
\eeq
where $a_{i}^{(n)}$ is the coefficient of the $i^{th}$ basis function when approximating the 
$n^{th}$ eigenstate and $N_{f}$ is the number of basis functions.

If one more basis function is added to the set $\{B_{i}\}$, $\vert \phi_{n} \rangle$ will be a better 
(at worst the same) approximation to the eigenstate.  If this extra function could produce a worse approximation, 
its coefficient, $a_{i}^{(n)}$ will be zero.  Thus, as more basis functions are used to approximate the 
ground state, the lowest eigenvalue will converge to the ground state eigenvalue.  However, if 
the original set of basis functions is altered when adding the new function, the lowest eigenvalue may not 
decrease each time a new function is added.

\section{Motivation for B-Splines}

The success of a Hamiltonian approach will depend on the choice of basis functions.  The basis functions are used 
to approximate the real state of the system, so if these functions are very 
different from the real states, it will take a large number of functions to approximate the real state.  
Thus, the convergence of the approximate state to the real state is slow if a poor basis is used.
A large number of basis functions leads to a Hamiltonian with many matrix elements.\footnote{The number of matrix 
elements is proportional to the number of basis functions, for each degree of freedom, squared.}
In this calculation, the matrix elements are determined by numerically calculating 
five-dimensional integrals, which is cpu intensive.  Therefore, a choice of basis functions that limits the 
number of integrals that need to be calculated is important.

We can approximate the function $f(x)$ by using a finite set of basis functions:
\beq
\label{func-approx}
f(x) \approx \sum_{i} a_{i} g_{i}(x).
\eeq
Since $f(x)$ and $g_{i}(x)$ are known, the coefficients $a_{i}$ are determined by multiplying both sides of Eq. 
(\ref{func-approx}) by $g_{j}(x)$, integrating over $x$, and solving:
\beq
\label{small-overlap}
\int dx f(x) g_{j}(x) = a_{i} \int dx g_{i}(x)g_{j}(x),
\eeq
where the sum over $i$ is implied.  If the $g_{i}(x)$ are orthogonal, then the right hand side is nonzero only for 
$i=j$.  

To determine the matrix elements of the operator, ${\cal O}(x)$, in the approximate basis [using the $g_{i}(x)$'s], 
it is necessary to compute integrals that look like:
\beq
\int dx \hspace{.1cm}{\cal O}(x) g_{j}(x) g_{i}(x).
\eeq
These integrals are generally nonzero even if $i \ne j$.  The number of integrals that need to be determined can be 
reduced by choosing a set of basis functions that are non-zero over different ranges of $x$.

The basis functions known as B-splines (Basis Splines) have the property that they are non-zero only over a finite range of their 
argument, and this range is different for each spline in the set.  However, they are non-orthogonal, which means the 
right hand side of Eq. (\ref{small-overlap}) has non-zero off-diagonal terms.

\section{Introduction to B-Splines}
\label{bspline-intro}
In this section we introduce B-spline functions and outline their derivation by N\"{u}rnberger \cite{splineA} 
and de Boor \cite{splineB}.  We begin with a discussion of the ``knots'' or ``control points'' that determine 
each spline's shape and then discuss a few of their basic properties.  Then we state the recurrence relation for 
the B-splines as well as their polynomial generators, discussing subtleties that should be understood.  Note all of 
the equations and notation are introduced in reference \cite{splineA}.  When using these 
two references, there are subtle differences in the notation that can cause confusion when comparing derivations.

Since the goal of this section is to give a basic understanding of B-splines, we write the 
definition of a B-spline and then return to the derivation.  The $i^{th}$ B-spline can be written:  
\beq
\label{bdefn}
B^{m}_{i}(t)=\sum_{j=i}^{i+m+1}a_{j}(t-x_{j})^{m}\theta(t-x_{j})\theta(x_{j+m+1}-t).
\eeq
$t$ is the argument of the 
B-spline and $x_{j}$ is the $j^{th}$ knot.  $a_{j}$ is a numerical coefficient for which we must solve.  
This B-spline is made of $m^{th}$ order polynomials, thus the index $m$ gives the order of the 
B-spline.  The index of the B-spline, $i$, is associated with the knots that are discussed in Section 
\ref{knot-section}.  The range of $i$ is $-m < i < k$ where $k$ controls how many B-splines make up the basis, 
giving $m+k+1$ B-spline functions in the set.
\subsection{Knots}
\label{knot-section}
The knots in a given knot sequence are labeled:
\beq
x_{-m}<\ldots <x_{-1}<a=x_{0}<x_{1}<\ldots <x_{k}<x_{k+1}=b<\ldots <x_{k+m+1},
\eeq
where $a$ and $b$ define the range of $t$ for which the B-splines will form a basis.  $m$ is 
the order of the B-spline and $k$ allows us to choose how many individual splines we use to form our 
basis.  The knots do not need to be equally spaced but must be non-decreasing.\footnote{If there are regions of 
phase space that are known a priori to be more important (the functions being approximated may have more structure 
in these regions) than other regions, clustering knots in the important region produces a set of B-splines that 
will approximate the real functions with fewer B-splines.}  It is possible to place multiple 
knots at one point, but we do not discuss this until section \ref{degen-knots}, because it complicates the 
following recurrence relations. 
\subsection{Basic Properties}
One of the benefits of using B-splines for basis states is that they are non-negative and have a spatial overlap with a 
limited number of other B-splines (the number of overlapping splines depends on the order).  
The $i^{th}$ B-spline of order $m$, $B^{m}_{i}(t)$ is 
positive in the range $[x_{i},x_{i+m+1}]$, and zero outside.  Explicitly:
\beq
B^{m}_{i}(t)=0, \hspace{1cm} t>x_{i+m+1} \hspace{1cm}{\mathrm or} \hspace{1cm} t<x_{i}\\
B^{m}_{i}(t)>0, \hspace{1cm} x_{i}<t<x_{i+m+1}.
\eeq
For a given knot sequence, the set of $m+k+1$ B-splines $\{B^{m}_{-m},\ldots ,B^{m}_{k}\}$ forms a basis (they are 
linearly independent) on [a,b].

The coefficients $a_{j}$ in Eq.~(\ref{bdefn}) can be found by solving the linear system of equations given by:
\beq
\sum_{j=i}^{i+m+1}a_{j}x_{j}^{r}=0, \hspace{1cm} r=0,\ldots, m\\
\sum_{j=i}^{i+m+1}a_{j}x_{j}^{m+1}=(-1)^{m+1}(m+1),
\eeq
which is derived in \cite{splineA}.  

\subsection{Normalized B-Splines and the Recurrence Relation}
\label{norm-splines-recurr}
The normalized B-splines are defined such that the sum of all B-splines at a given point is 1,
\beq
\sum_{i=-m}^{k}N^{m}_{i}(t)=1.
\eeq
The normalized and unnormalized B-splines are related by:
\beq
N^{m}_{i}(t)=\frac{1}{m+1}(x_{i+m+1}-x_{i})B^{m}_{i}(t).
\eeq
The recurrence relation for the normalized B-splines is:
\beq
\label{recurs}
N^{m}_{i}(t)=\frac{t-x_{i}}{x_{i+m}-x_{i}}N^{m-1}_{i}(t)+\frac{x_{i+m+1}-t}{x_{i+m+1}-x_{i+1}}N^{m-1}_{i+1}(t).
\eeq
Finally, the $n^{th}$ derivative (designated by the ``$(n)$'' superscript) of a normalized B-spline is given by the recurrence relation:
\beq
\left(N^{m}_{i}\right)^{(n)}(t)=\frac{m}{x_{i+m}-x_{i}}\left(N_{i}^{m-1}\right) ^{(n-1)}(t)
-\frac{m}{x_{i+m+1}-x_{i+1}}\left(N^{m-1}_{i+1}\right) ^{(n-1)}(t).
\eeq
\subsection{Degenerate Knots and the Recurrence Relation}
\label{degen-knots}
It can be useful to place multiple knots at the same point (degenerate knots).  Although there can be
degenerate knots in the region $[a,b]$ we choose to only use degenerate knots at $a$ and $b$.  This is because degenerate knots create 
discontinuous derivatives at the degenerate knot.  If these discontinuities occur at the boundaries, then one can 
still safely take derivatives in the region of interest, and take the derivative at $a$ or $b$ to be the limit as 
the point is approached from the right or left, respectively.  The recursion relation given in 
Eq.~(\ref{recurs}) must be carefully applied with degenerate knots because the denominators 
can be zero.  However, each term on the right-hand side of Eq.~(\ref{recurs}) is 
finite because in the limit a denominator becomes zero, the product of the denominator with the normalized 
B-spline is finite.

It should be noted that using splines with degenerate knots can lead to a non-Hermitian 
Hamiltonian.  Consider the nonrelativistic kinetic energy term in position space:
\beq
H_{ij}=\int_{a}^{b} dx B_{i}(x) \frac{d^{2}}{dx^{2}} B_{j}(x).
\eeq
If we integrate by parts, this becomes
\beq
\label{herm-check}
H_{ij}=B_{i}(x)\frac{d B_{j}(x)}{dx}\vert_{x=b}-B_{j}(x)\frac{d B_{i}(x)}{dx}\vert_{x=a}-
\int_{a}^{b}\frac{d B_{i}(x)}{dx}\frac{d B_{j}(x)}{dx}.
\eeq
It is obvious the last term is unchanged if we let $i \leftrightarrow j$.

The first ($i=-m$) and last ($i=k$) B-splines do not go to zero at $a$ and $b$, respectively,
\beq
B_{-m}(a) \neq 0, \hspace{1.0 in} B_{k}(b) \neq 0 .
\eeq
This fact prevents $H$ from being hermitian.  Consider the first (left-most) spline ($i=-m$):
\beq
\label{non-herm-1}
H_{-mj}-H_{j-m}=\left[ B_{-m}(x)\frac{d B_{j}(x)}{dx}-B_{j}(x)\frac{d B_{-m}(x)}{dx}\right|_{x=a}.
\eeq
Since $j \neq -m$ we can rewrite Eq. (\ref{non-herm-1}) as:
\beq
H_{-mj}-H_{j-m}= B_{-m}(a)\frac{d B_{j}(a)}{dx},
\eeq
where the expression is evaluated in the limit $x\rightarrow a$.  The only $j$ for which the Hamiltonian 
is not symmetric under interchange of indices is $j=-m+1$.  For all others the derivative at the boundary is zero.  So 
we find for both the first (left-most) and last (right-most) splines:
\beq
\label{extreme-cond}
H_{-m,-m+1}-H_{-m+1,-m}\neq 0 \hspace{.5 in} H_{k,k-1}-H_{k-1,k}\neq 0.
\eeq
Thus, the only way we can maintain Hermiticity in position space is to discard the two splines $B_{-m}$ and $B_{k}$.
The problem can also be avoided by working in momentum space which avoids second derivatives, the source of the 
problem.

\subsection{B-spline Polynomial Generators}

Using the recurrence relations in Section \ref{norm-splines-recurr} is straight-forward analytically and 
numerically.  We can also evaluate the B-splines using the polynomial generator:  
\beq
\label{poly-bspline}
p_{j}(t)=\sum_{r=0}^{m} \frac{1}{r!}s^{(r)}(x_{j})(t-x_{j})^{r}, \hspace{.5in}  t \in [x_{j},x_{j+1}];
\eeq
where,
\beq
s^{(r)}(t)=\sum_{i=-m+r}^{k}a_{i}^{(r)}N_{i}^{m-r}(t),
\eeq
and
\beq 
a_{i}^{(r)}=\left\{ \barr{cc} a_{i} & {\mathrm if} \hspace{.2cm}r=0\\ 
(m+1-r)\frac{a_{i}^{(r-1)}-a_{i-1}^{(r-1)}}{x_{i+m+1-r}-x_{i}} & {\mathrm if} \hspace{.2cm} r>0\earr\right. .
\eeq
Although it is easier to understand the recurrence relations, if we want to repeatedly evaluate a B-spline 
it is faster to use the polynomial generator in Eq.~(\ref{poly-bspline}).
To speed up the calculation further, all the $a_{i}^{(r)}$'s can be calculated in advance.  However, these 
equations, which were taken from \cite{splineA}, are used to represent a spline which is a linear combination 
of B-splines:
\beq
s(t)=\sum_{i=-m}^{k}a_{i}N_{i}^{m}(t).
\eeq
But we are interested in representing only an individual B-spline.  Thus if we want to represent the $I^{th}$ 
B-spline, then we just set all of the $a_{i}$ for $i\ne I$ to $0$ and $a_{I}=1$.  So we can write the 
$I^{th}$ B-spline on the interval $[x_{j},x_{j+1}]$:
\beq
_{j}N_{I}(t)=\sum_{r=0}^{m}\frac{1}{r!}
\left\{ \sum_{i=-m+r}^{k}  a_{i}^{(r)}N_{i}^{m-r}(x_{j})\right\} (t-x_{j})^{r}, \hspace{.5in}  t \in [x_{j},x_{j+1}].
\eeq
\section{Simple B-Spline Applications}
\subsection{Function Approximation}
\begin{figure}
\centerline{\epsfig{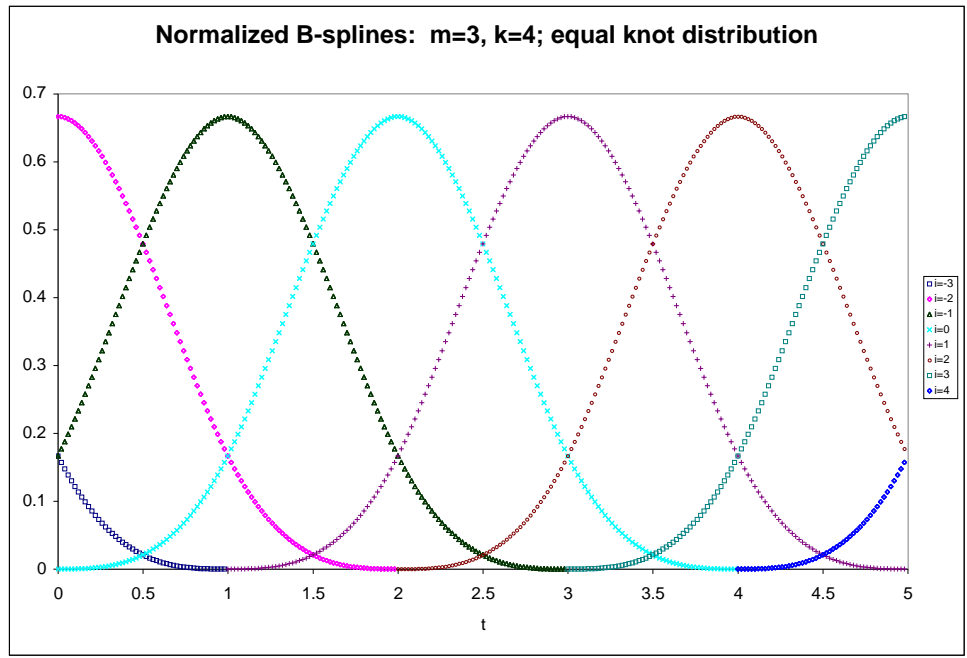}}
\caption[B-spline functions.]{The B-splines $B^{m}_{i}(t)$ for $i$ between $-m$ and $k$, with $m=3$, and $k=4$.  The knots are equally 
spaced with separation $\frac{b-a}{k+1}$ where $a=0$ and $b=5$.}
\label{fig:bspl-plot}
\end{figure}

\begin{figure}
\centerline{\epsfig{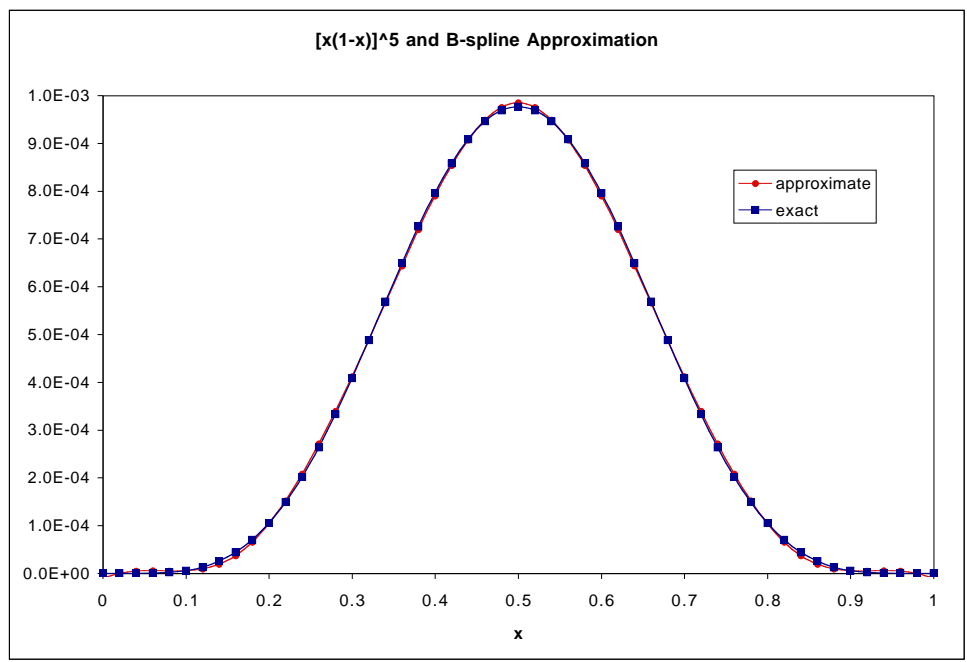}}
\caption[Exact function and B-spline approximation.]{The function $(x(1-x))^{5}$ and the approximation using 
B-splines.  The B-splines are order $m=3$ with $k=5$.}
\label{fig:bsp-err}
\end{figure}

\begin{figure}
\centerline{\epsfig{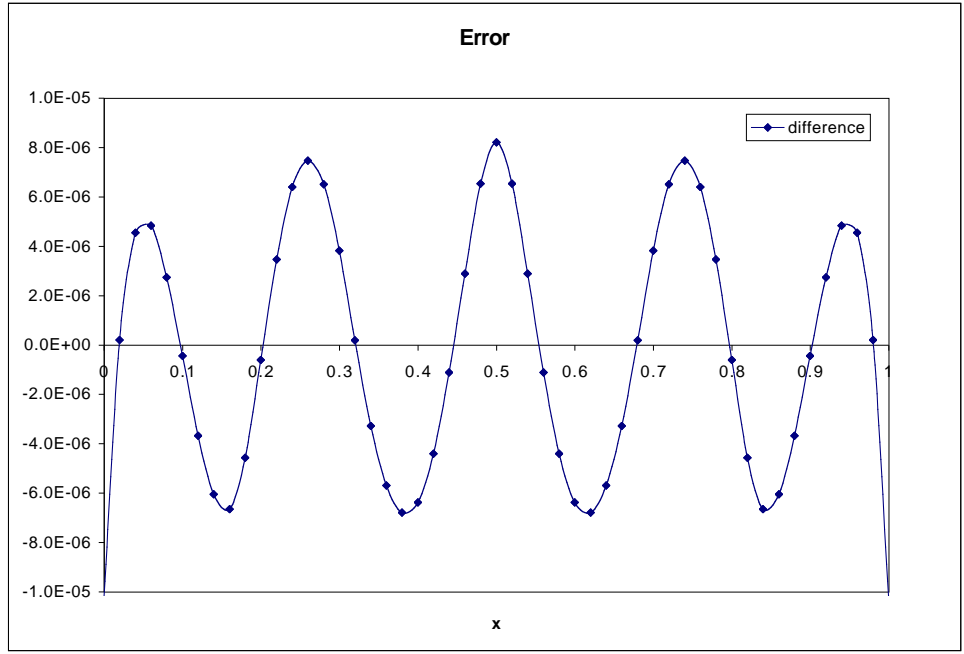}}
\caption[Error in B-spline approximation.]{The difference between the function $(x(1-x))^{3}$ and the 
approximation using B-splines.  
The splines are order $m=3$ with $k=5$.}
\label{fig:bsp-diff}
\end{figure}

A simple problem that illustrates the use of B-splines is function approximation.  We can approximate a 
function $f(x)$ as:
\beq
f(x) \approx \sum_{i} a_{i} B_{i}(x).
\eeq
To solve for the $a_{i}$'s multiply both sides by $B_{j}(x)$ and integrate over $x$:
\beq
\int_{a}^{b} dx B_{j}(x) f(x)=\sum_{i}a_{i}\int_{a}^{b}dx B_{j}(x) B_{i}(x).
\eeq
Note the range of integration is $[a,b]$ since that is the range over which we are going to approximate the 
function and the range over which the B-splines are defined.  This is just a matrix problem of the form:
\beq
a_{i}={\cal O}^{-1}_{ji}f_{j} ,
\eeq
where
\beq
{\cal O}_{ji}=\int_{a}^{b}dx B_{j}(x) B_{i}(x),
\eeq
and
\beq
f_{j}=\int_{a}^{b} dx B_{j}(x) f(x).
\eeq
Figure \ref{fig:bspl-plot} shows the eight normalized B-splines of order $m=3$ with $k=4$ over the range $[a,b]$ 
with $a=0$ and $b=5$.  
At any value of $t$ there are only $m+1=4$ non-zero splines.
We want to know how well the B-splines are going to approximate functions that will come up in QCD.  In the 
longitudinal direction, functions of the form $(x(1-x))^{d}$ where $d$ is positive, are common in light-front QCD.  
Figure \ref{fig:bsp-err} 
shows the function $(x(1-x))^{5}$ and the approximation built from B-splines with $m=3$ and $k=5$.  The knot sequence is 
also uniform with separation $\frac{b-a}{k+1}$ where $a=0$ and $b=1$.  Finally, figure \ref{fig:bsp-diff} shows the 
difference between the exact function and the approximation using B-splines.

\subsection{The One-Dimensional Harmonic Oscillator}
\label{one-d-harmosc}
\begin{figure}
\centerline{\epsfig{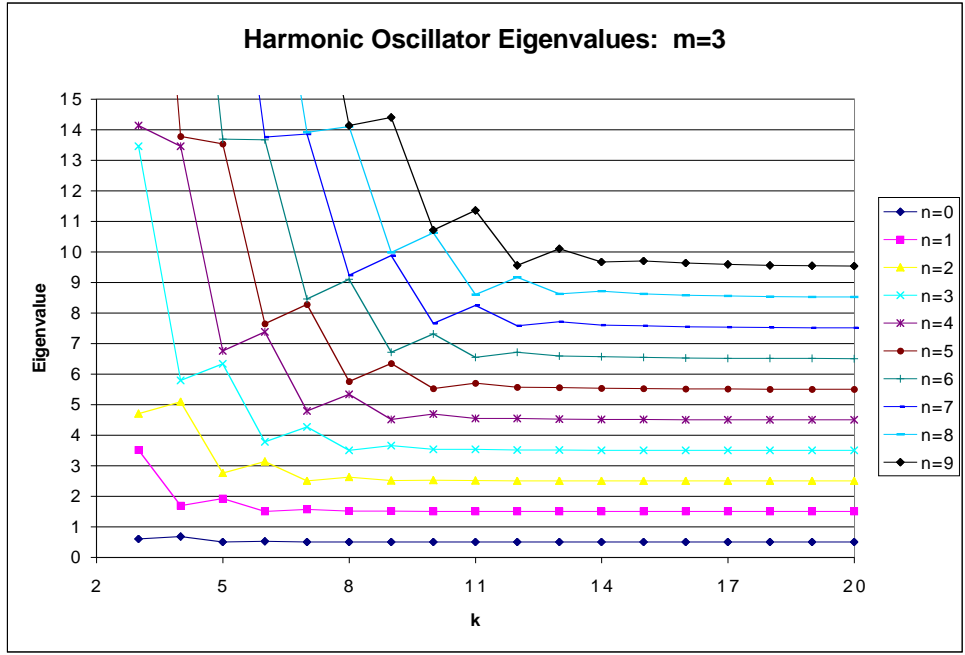}}
\caption[One-dimensional harmonic oscillator eigenvalues.]{Energy eigenvalues of the harmonic oscillator in units of $\hbar\omega$.  The total number of states is $m+k+1$.}
\label{harmeval}
\end{figure}

To test the B-splines as a useful set of basis functions, we use them to solve the one-dimensional harmonic 
oscillator problem.  The harmonic oscillator Hamiltonian is:
\beq
H=-\frac{\hbar^{2}}{2m}\frac{d^{2}}{dx^{2}}+\half m \omega^{2}x^{2}.
\eeq
If we write out the eigenvalue equation for a state $\phi$ that we approximate with B-splines we get:
\beq
H\vert\phi\rangle =E\vert\phi\rangle\rightarrow\sum_{i}a_{i}H\vert B_{i}\rangle=E\sum_{i}a_{i}\vert B_{i}\rangle.
\eeq
Looking at a particular matrix element and dropping the explicit sum, we get:
\beq
\langle B_{j}\vert H\vert B_{i}\rangle a_{i}=\langle B_{j}\vert B_{i}\rangle E a_{i},
\eeq
where $\langle B_{j}\vert H\vert B_{i}\rangle$ is the Hamiltonian matrix ($H$) and $\langle B_{j}\vert B_{i}\rangle$ is 
the overlap matrix ($O$).  This problem can be solved in its current 
form, as a generalized eigensystem problem
\beq
H {\bf x}=E O {\bf x} ,
\eeq
or we can rewrite it as a simple eigensystem problem
\beq
O^{-1} H {\bf x}=E {\bf x} .
\eeq
Since a B-spline of order $m$ has only $m-1$ continuous derivatives, it is necessary to use at least third-order 
B-splines.  Although higher order B-splines may speed the convergence, they also add to the total number of 
states and matrix elements.  For simplicity we use $m=3$ and only adjust $k$.\footnote{Using higher order 
B-splines does not produce convergent results in noticeably less time.}  Figure \ref{harmeval} shows the eigenvalues for the 
ten lowest states in units of $\hbar\omega$.  Note the correct value of the energy is $E_{n}=(n+\half )\hbar 
\omega$, so the eigenvalues plotted approach the correct values.  The total number of splines used is $m+k+1$, or 
for $m=3$, $k+4$.  The knots are equally spaced over the range $[a,b]$ with $a=-5$ and $b=5$.  $a$ 
and $b$ are determined by finding the smallest values that, when increased, do not change the eigenvalues. 

\begin{figure}
\centerline{\epsfig{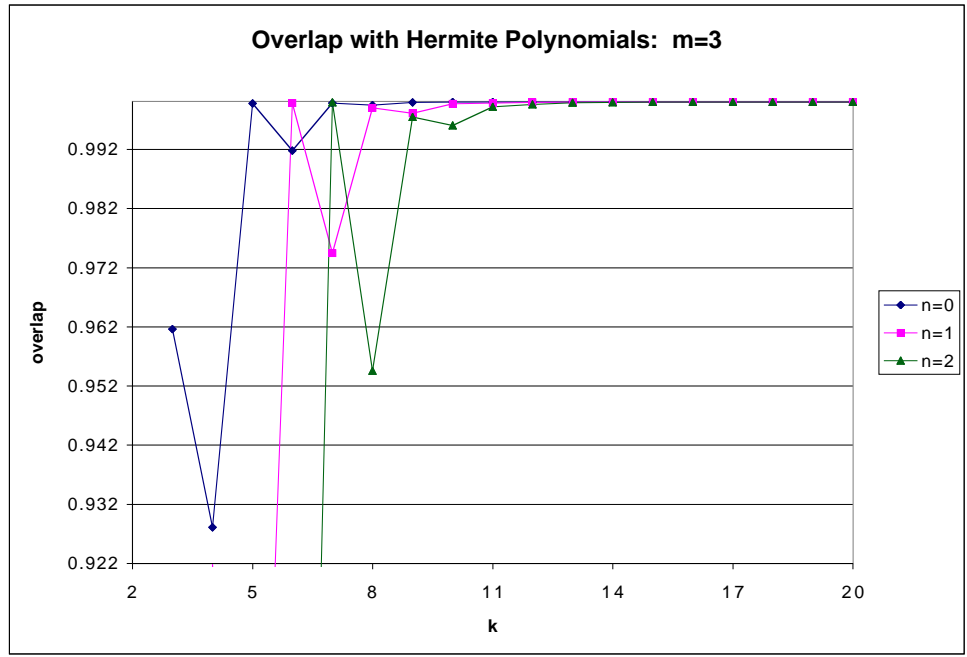}}
\caption{Overlap of the harmonic oscillator eigenstates with the Hermite polynomials.}
\label{harmovr}
\end{figure}

The harmonic oscillator can be solved analytically.  The eigenfunctions are Hermite 
polynomials.  We can see how well the B-splines approximate the solution by considering the overlap between the 
approximate and exact solutions.  The overlap is defined as:
\beq
\frac{\int_{a}^{b} dx \phi(x) H(x)}{\int_{a}^{b} dx H^{2}(x)},
\eeq
where $\phi(x)$ is the approximate solution built from B-splines and $H(x)$ is a Hermite polynomial.  The overlap 
approaches $1$ as a better approximation is made.
Figure \ref{harmovr} shows the overlap of the first three approximate eigenstates determined by B-spline approximation with 
the known solution to the harmonic oscillator problem, the Hermite polynomials.
\subsection{The Two-Dimensional Harmonic Oscillator}
\label{two-d-harmosc}
The two-dimensional harmonic oscillator is a non-trivial extension of the one-dimensional harmonic oscillator.  
The B-spline basis used for the two-dimensional problem is a product of the one-dimensional B-spline basis 
$B_{xy}=B_{x}\otimes B_{y}$ where $B_{x}$ and $B_{y}$ are one-dimensional B-spline basis.
The energy for the two-dimensional case is simply the sum from each individual direction:
\beq
E_{(n_{x},n_{y})}=\half (\omega_{x}+\omega_{y})+(n_{x}\omega_{x}+n_{y}\omega_{y}),
\eeq
where $\hbar =1$.  

\begin{figure}
\centerline{\epsfig{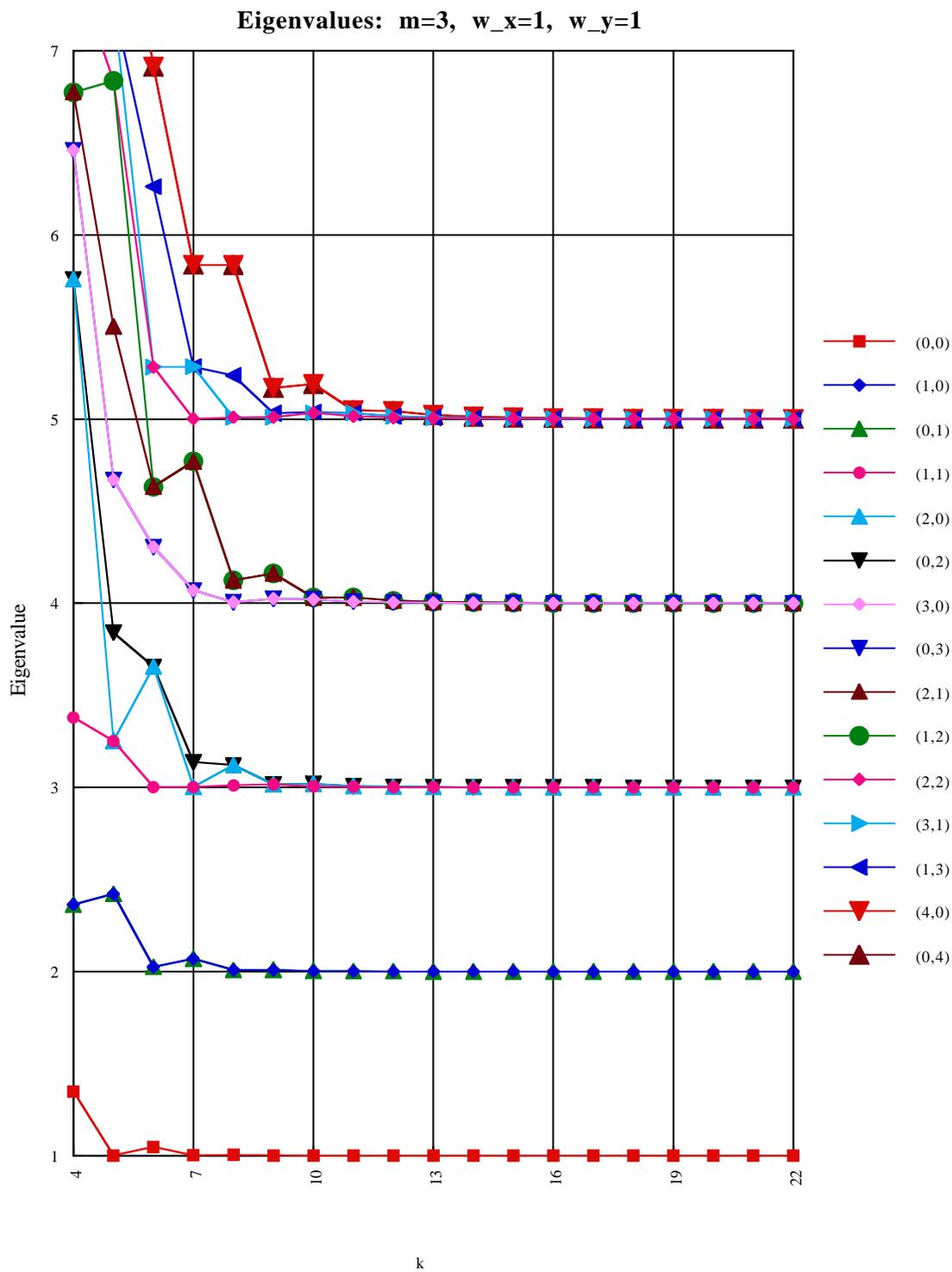}}
\caption[Two-dimensional isotropic harmonic oscillator eigenvalues.]{Eigenvalues for the two-dimensional harmonic oscillator, with $\omega_{x}=\omega_{y}=1$, and
states labeled by $(n_{x},n_{y})$. There are k+m+1 states in each direction.}
\label{2d-eq-evals}
\end{figure}

Figure \ref{2d-eq-evals} shows the eigenvalues for the two-dimensional harmonic oscillator using 
third order B-splines in both the $x$ and $y$ directions.  For convenience we use the same, equally spaced, knot distribution is used in 
both directions.  The frequency in the $y$ direction is the same as in the $x$ direction which gives rise to the 
particular degeneracies shown.

\begin{figure}
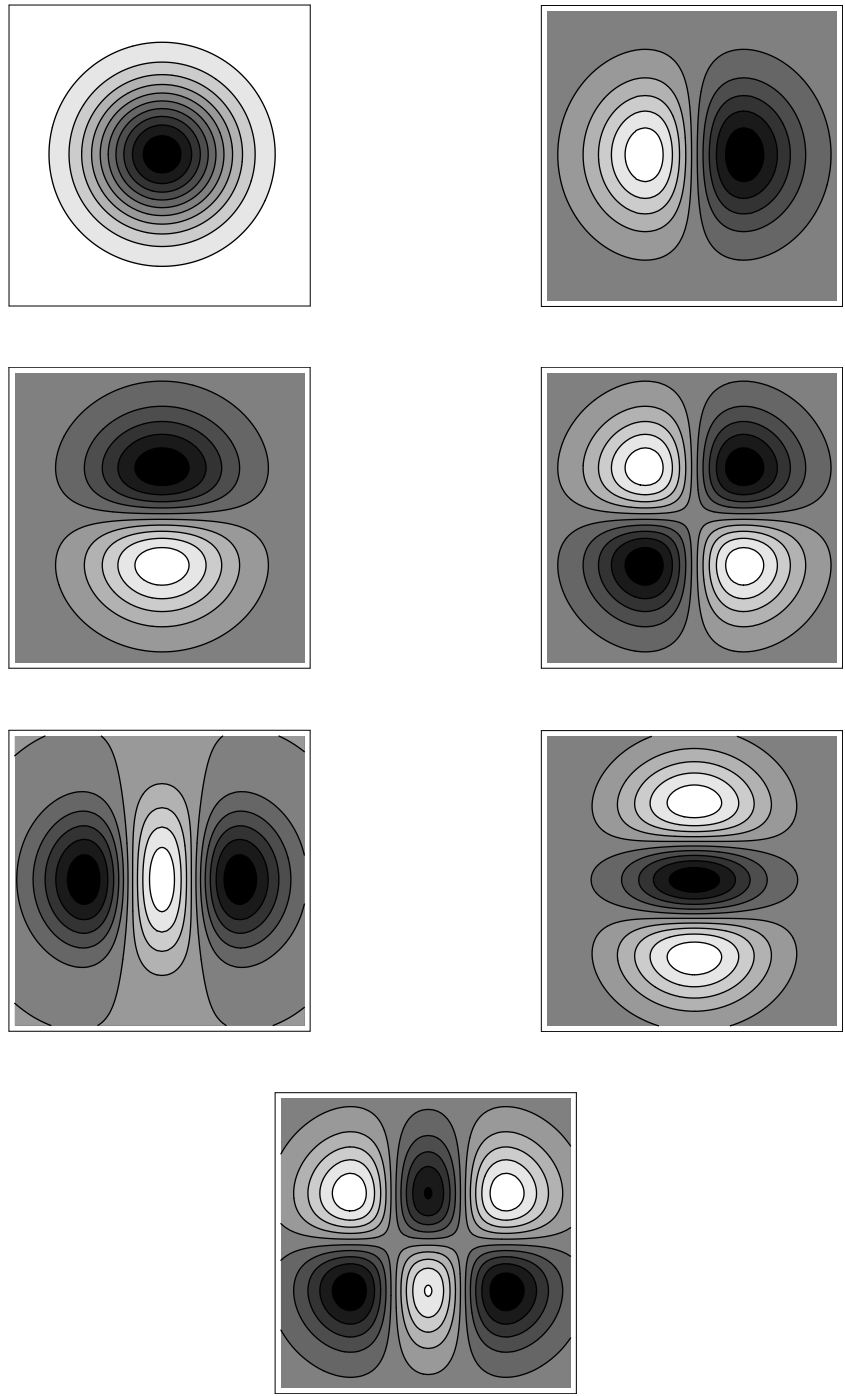

\centerline{\epsfig{file=\bsplpath equal-one.epsf,width=2in}\hspace{2cm}\epsfig{file=\bsplpath equal-two.epsf,width=2in}}
\centerline{\epsfig{file=\bsplpath equal-three.epsf,width=2in}\hspace{2cm}\epsfig{file=\bsplpath equal-four.epsf,width=2in}}
\centerline{\epsfig{file=\bsplpath equal-five.epsf,width=2in}\hspace{2cm}\epsfig{file=\bsplpath equal-six.epsf,width=2in}}
\centerline{\epsfig{file=\bsplpath equal-seven.epsf,width=2in}}
\caption[Two-dimensional isotropic harmonic oscillator wavefunctions.]{The approximation of the seven of the lowest energy wave functions for the two-dimensional harmonic oscillator with 
$\omega_{y}=\omega_{x}=1$.  The labels are : $(0,0)$, $(1,0)$, $(0,1)$, $(1,1)$, $(2,0)$, $(0,2)$ and $(2,1)$.
 There are k+m+1 states in each direction.}
\label{2d-eq-funcs}
\end{figure}

\begin{figure}
\centerline{\epsfig{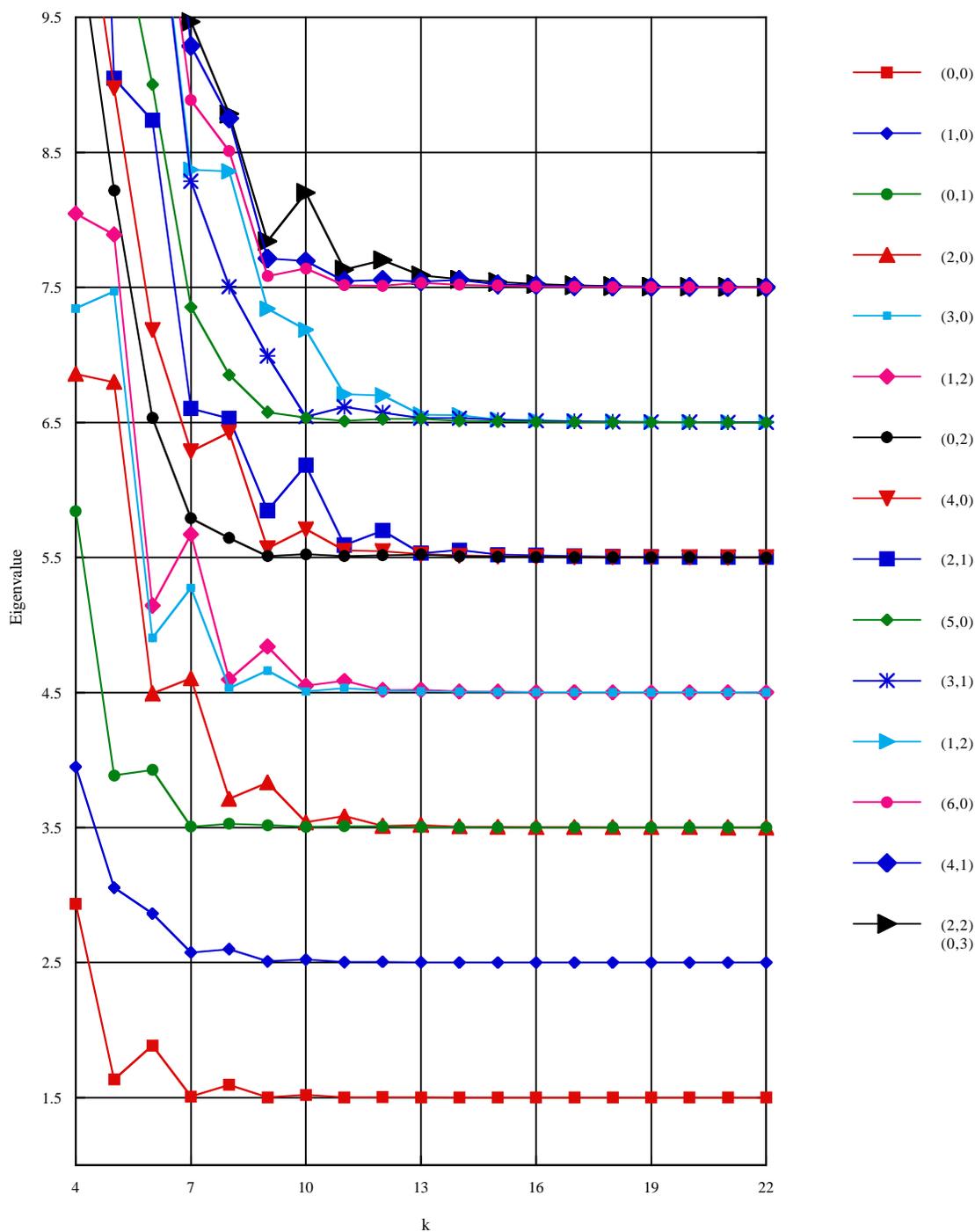}}
\caption[Two-dimensional anisotropic harmonic oscillator eigenvalues.]{Eigenvalues for the two-dimensional 
anisotropic harmonic oscillator, with $\omega_{y}=2\omega_{x}=2$, and
states labeled by $(n_{x},n_{y})$.  There are k+m+1 states in each direction.}
\label{2d-ueq-evals}
\end{figure}

\begin{figure}
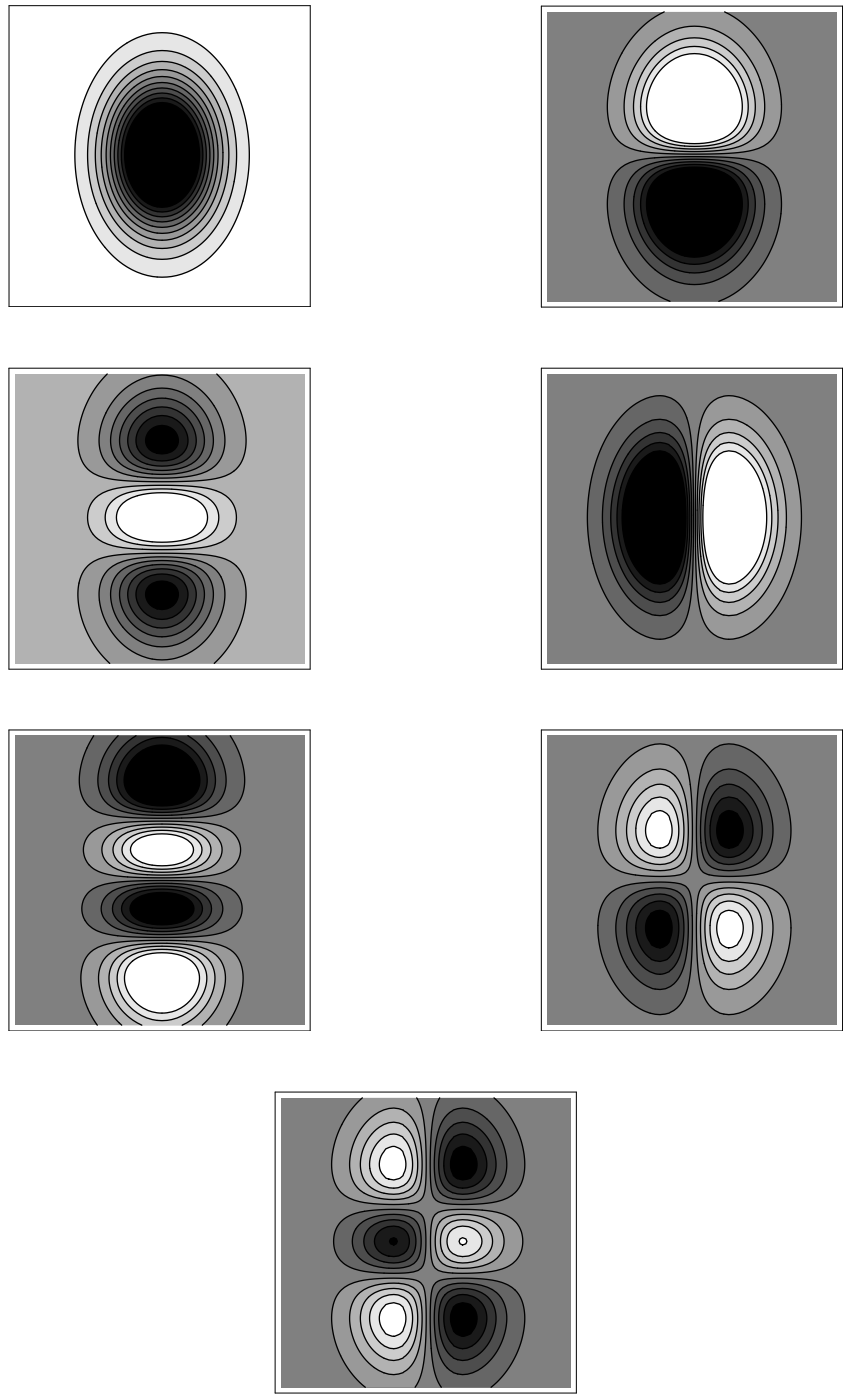

\centerline{\epsfig{file=\bsplpath unequal-one.epsf,width=2in}\hspace{2cm}\epsfig{file=\bsplpath unequal-two.epsf,width=2in}}
\centerline{\epsfig{file=\bsplpath unequal-three.epsf,width=2in}\hspace{2cm}\epsfig{file=\bsplpath unequal-four.epsf,width=2in}}
\centerline{\epsfig{file=\bsplpath unequal-five.epsf,width=2in}\hspace{2cm}\epsfig{file=\bsplpath unequal-six.epsf,width=2in}}
\centerline{\epsfig{file=\bsplpath unequal-seven.epsf,width=2in}}
\caption[Two-dimensional anisotropic harmonic oscillator wavefunctions.]{The approximation of the seven of the lowest energy wave functions for the two-dimensional harmonic oscillator with 
$\omega_{y}=2\omega_{x}=2$.  The labels are : $(0,0)$, $(1,0)$, $(2,0)$, $(0,1)$, $(3,0)$, $(1,1)$ and $(2,1)$. 
 There are k+m+1 states in each direction.}
\label{2d-ueq-funcs}
\end{figure}

The approximation of the lowest seven states are plotted in figure \ref{2d-eq-funcs}.  The ranges shown are limited 
to $-3< x,y < 3$ because the harmonic oscillator states fall off exponentially.  Note also the 
phase of the wave function is arbitrary (note the overall negative sign in the lowest state).
Figure \ref{2d-ueq-evals} shows the eigenvalues for a two-dimensional harmonic oscillator without rotational 
symmetry using 
third order B-splines in both the $x$ and $y$ directions.  For simplicity we use the same, equally spaced, knot distribution is used in 
both directions.  The frequency in the $y$ direction is twice that in the $x$ direction which gives rise to the 
particular degeneracies shown.  The approximation of the lowest seven states are plotted in figure \ref{2d-ueq-funcs}.  The ranges shown are limited 
to $-3< x,y < 3$ because the harmonic oscillator states fall off exponentially.  Note also the 
phase of the wave function is arbitrary (note the overall negative sign in the lowest state).

\chapter{Matrix Elements of the Light-Front QCD Hamiltonian}
\label{chapt:ham-me}

In this chapter we quantize the Hamiltonian.  
We begin with the canonical light-front Hamiltonian \cite{zhang2} as a starting point.  Splitting the 
Hamiltonian density into a free and interacting part, we have:
\beq
\cH&=&\cH_{\mathrm{free}}+\cH_{\mathrm{int}} \\
\cH_{\mathrm{int}}&=&\cH_{qqg}+\cH_{ggg}+\cH_{qqgg}+\cH_{qqqq}+\cH_{gggg},
\eeq
where
\beq
\label{H-qqg}
\cH_{qqg}&=&
-g\left\{-2\psi^{\dagger}_{+}\left[ \inpar\left( \partial^{\bot}A^{\bot}\right) \right] \psi_{+}
+\psi^{\dagger}_{+}\left( \sigma\cdot A^{\bot}\right) 
\right. \n && \hspace{2cm} \left.
\left[ \inpar
\left( \sigma\cdot\partial^{\bot}+m\right) \psi_{+}\right] 
\right. \n &&\hspace{3cm}\left. 
+\psi^{\dagger}_{+}\left( \sigma\cdot\partial^{\bot}-m\right) 
\left[ \inpar\left( \sigma\cdot A_{\bot}\right) \psi_{+}\right] \right\}, \\
\cH_{ggg}&=&-gf^{abc}\left\{ A_{a}^{i}A_{b}^{j}\left( \partial^{i}A_{c}^{j}\right)
+\left( \partial^{i}A_{a}^{i}\right) \inpar
\left[ A_{b}^{j}\left( \partial^{+}A_{c}^{j}\right) \right] \right\}, \\
\cH_{qqgg}&=&g^{2}\left\{\psi^{\dagger}_{+}\left( \sigma\cdot A^{\bot}\right) \left[ \left( \frac{1}{i\partial^{+}}\right) 
\sigma\cdot A^{\bot}\psi_{+}\right] 
\right. \n &&\hspace{2cm} \left.
+2\inpar\left[ f^{abc}A_{b}^{i}\left( \partial^{+}A_{c}^{i}\right) \right] 
\inpar\left( \psi^{\dagger}_{+}\bT^{a}\psi_{+}\right) \right\},\\
\label{H-qqqq}
\cH_{qqqq}&=&2g^{2}\left\{\inpar\left( \psi^{\dagger}_{+}\bT_{a}\psi_{+}\right) 
\inpar\left( \psi^{\dagger}_{+}\bT_{a}\psi_{+}\right) \right\},\\
\cH_{gggg}&=&\frac{g^{2}}{4}f^{abc}f^{ade}\left\{A_{b}^{i}A_{c}^{j}A_{d}^{i}A_{e}^{j}
\right. \n && \hspace{2cm} \left.
+2\inpar\left[ A_{b}^{i}\left( \partial^{+}A_{c}^{i}\right) \right] 
\inpar\left[ A_{d}^{j}\left( \partial^{+}A_{e}^{j}\right) \right] \right\},
\eeq
where the gluon and quark fields are defined in Appendix \ref{lf-formalism}.

\section{Conventions for the States}
We order quark creation and annihilation operators, $b$ and $b^{\dagger}$ before antiquark creation and 
annihilation operators,  $d$ and $d^{\dagger}$.  If there is an initial state that contains a quark and an anti-quark, 
then the ordering will be $\bd \dd \vert 0 \rangle$.  However, since the final state will be the complex conjugate of an 
initial state, the ordering in a final state will be $\langle 0 \vert d b$.  Similarly, if a state contains two 
quarks, then we have:  $\vert q_{1} q_{2} \rangle = \bd_{1} \bd_{2} \vert 0 \rangle$ and $\vert q_{1} 
q_{2} \rangle ^{\dagger} = \langle q_{1} q_{2} \vert = \langle 0 \vert b_{2} b_{1}$.

The results for the matrix elements of the Hamiltonian presented in Sections \ref{qqbg-me} 
and \ref{qqbqqb-me} are analogous to \cite{brentC}.

\section{Quantization of the Hamiltonian}

The full Hamiltonian can be written:
\beq
H=\int dx^{-} d^{2}x_{\bot}\left( \cH_{\mathrm{free}}+\cH_{\mathrm{int}}\right) .
\eeq
We find:
\beq
H_{\mathrm{free}}=\int D_{1}\left\{\delta_{i,j}\frac{\left( p_{1}^{i}\right) ^{2}}{p_{1}^{+}}a^{\dagger}_{1}a_{1}
+\left( \frac{p^{2}_{\bot}+m^{2}}{p^{+}}\right) \left[ b_{1}^{\dagger}b_{1}+d_{1}^{\dagger}d_{1}\right] \right\},
\eeq
where $D_{1}$ is defined in Eq. (\ref{bigD-defn})

The only parts of the interacting Hamiltonian we need to calculate for an ${\cal O}(g^{2})$ calculation are $\cH_{qqg}$ 
and $\cH_{qqqq}$.  We label the transverse momentum of a particle $k_{a}^{(i)}$, where $a$ is the particle label and $(i)$ is the 
direction of the transverse momentum component.  We define:
\beq
\label{k-tilde-defn}
\tilde{k}_{3}^{s_{3}}=(s_{3})k_{3}^{(1)}+ik_{3}^{(2)},
\eeq
and substitute the expressions for the gluon, quark and antiquark fields from 
\ref{gluon-field-expr}, \ref{quark-field-expr}, and \ref{antiquark-field-expr} into Eqs.~(\ref{H-qqg}) and 
(\ref{H-qqqq}) giving:
\beq
\label{qqbarg-vertex}
H_{q\bar{q}g}&=&32\pi^{3}g\int D_{1} D_{3} D_{2}
\sqrt{\frac{k_{1}^{+}\hspace{.1cm}k_{2}^{+}}{2}}\langle c_{1}\vert{\bf T}^{c_{3}}\vert c_{2}\rangle\times
\n &&\hspace{-1cm}\left\{
\delta^{3}(k_{1}+k_{2}-k_{3}) \times
\right. \n && \hspace{.5cm}\left. 
\left[\bd_{1}\dd_{2}a_{3}\left(
\delta_{s_{1},\bar{s}_{2}} \left\{ -\frac{\tilde{k}_{3}^{s_{3}}}{k_{3}^{+}}
+\delta_{\bar{s}_{2},s_{3}}\frac{\tilde{k}_{2}^{s_{3}}}{k_{2}^{+}}
+\delta_{s_{2},s_{3}}\frac{\tilde{k}_{1}^{s_{3}}}{k_{1}^{+}} \right\}
\right. \right. \right. \n && \hspace{4cm} \left. \left. \left.
-ims_{2}\delta_{s_{1},s_{2}}\delta_{s_{2},s_{3}}\left\{\frac{1}{k_{2}^{+}}+\frac{1}{k_{1}^{+}} \right\} \right)
\right. \right. \n && \hspace{2cm}\left. \left.
-\ad_{3}b_{2}d_{1}\left(
\delta_{s_{1},\bar{s}_{2}} \left\{ -\frac{\tilde{k}_{3}^{\ast s_{3}}}{k_{3}^{+}}
+\delta_{\bar{s}_{2},s_{3}}\frac{\tilde{k}_{2}^{\ast s_{3}}}{k_{2}^{+}}
+\delta_{s_{2},s_{3}}\frac{\tilde{k}_{1}^{\ast s_{3}}}{k_{1}^{+}} \right\}
\right. \right. \right. \n && \hspace{5cm} \left. \left. \left.
+ims_{2}\delta_{s_{1},s_{2}}\delta_{s_{2},s_{3}}\left\{\frac{1}{k_{2}^{+}}+\frac{1}{k_{1}^{+}} \right\}
\right) \right]
\right. \n &&\hspace{-1cm}\left. +
\delta^{3}(k_{1}-k_{2}-k_{3})\times 
\right. \n && \hspace{.5cm}\left. 
\left[\dd_{2}\ad_{3}d_{1}\left(
\delta_{s_{1},s_{2}} \left\{ \frac{\tilde{k}_{3}^{\ast s_{3}}}{k_{3}^{+}}
-\delta_{s_{2},s_{3}}\frac{\tilde{k}_{2}^{\ast s_{3}}}{k_{2}^{+}}
-\delta_{s_{2},\bar{s}_{3}}\frac{\tilde{k}_{1}^{\ast s_{3}}}{k_{1}^{+}} \right\}
\right. \right. \right. \n && \hspace{4cm} \left. \left. \left.
-ims_{2}\delta_{s_{1},\bar{s}_{2}}\delta_{s_{2},\bar{s}_{3}}\left\{\frac{1}{k_{2}^{+}}-\frac{1}{k_{1}^{+}} \right\} 
\right)
\right. \right. \n && \hspace{2cm}\left. \left.
+\bd_{1}b_{2}a_{3}\left(
\delta_{s_{1},s_{2}} \left\{ -\frac{\tilde{k}_{3}^{s_{3}}}{k_{3}^{+}}
+\delta_{s_{2},s_{3}}\frac{\tilde{k}_{2}^{s_{3}}}{k_{2}^{+}}
+\delta_{s_{2},\bar{s}_{3}}\frac{\tilde{k}_{1}^{s_{3}}}{k_{1}^{+}} \right\}
\right. \right. \right. \n && \hspace{5cm} \left. \left. \left.
-ims_{2}\delta_{s_{1},\bar{s}_{2}}\delta_{s_{2},\bar{s}_{3}}\left\{\frac{1}{k_{2}^{+}}-\frac{1}{k_{1}^{+}} \right\}
\right) \right]
\right. \n &&\hspace{-1cm}\left. +
\delta^{3}(k_{1}-k_{2}+k_{3}) \times
\right. \n && \hspace{.5cm}\left. 
\left[ \bd_{1}\ad_{3}b_{2}\left(
\delta_{s_{1},s_{2}} \left\{ -\frac{\tilde{k}_{3}^{\ast s_{3}}}{k_{3}^{+}}
+\delta_{s_{2},\bar{s}_{3}}\frac{\tilde{k}_{2}^{\ast s_{3}}}{k_{2}^{+}}
+\delta_{s_{2},s_{3}}\frac{\tilde{k}_{1}^{\ast s_{3}}}{k_{1}^{+}} \right\}
\right. \right. \right. \n && \hspace{4cm} \left. \left. \left.
+ims_{2}\delta_{s_{1},\bar{s}_{2}}\delta_{s_{2},s_{3}}\left\{\frac{1}{k_{2}^{+}}-\frac{1}{k_{1}^{+}} \right\} \right)
\right. \right. \n && \hspace{2cm}\left. \left.
+\dd_{2}d_{1}a_{3}\left(
\delta_{s_{1},s_{2}} \left\{ \frac{\tilde{k}_{3}^{s_{3}}}{k_{3}^{+}}
-\delta_{s_{2},\bar{s}_{3}}\frac{\tilde{k}_{2}^{s_{3}}}{k_{2}^{+}}
-\delta_{s_{2},s_{3}}\frac{\tilde{k}_{1}^{s_{3}}}{k_{1}^{+}} \right\}
\right. \right. \right. \n && \hspace{5cm} \left. \left. \left.
+ims_{2}\delta_{s_{1},\bar{s}_{2}}\delta_{s_{2},s_{3}}\left\{\frac{1}{k_{2}^{+}}-\frac{1}{k_{1}^{+}} \right\}
\right) \right]
\right\} 
\eeq
and
\newpage
\beq
\label{hqqqq}
H_{qqqq}&=&-32g^{2}\pi^{3}\int D_{1}D_{2}D_{3}D_{4}
\langle c_{1}\vert{\bf T}_{a}\vert c_{2}\rangle\langle c_{3}\vert{\bf T}_{a}\vert c_{4}\rangle
\sqrt{k_{1}^{+}k_{2}^{+}k_{3}^{+}k_{4}^{+}}
\times\nonumber\\&&\hspace{0cm}
\left\{\delta^{3}(k_{1}-k_{2}+k_{3}-k_{4})
\frac{\delta_{s_{1},s_{2}}\delta_{s_{3},s_{4}}}{(k_{1}^{+}-k_{2}^{+})^{2}}
\left[ \bd_{1}\bd_{3}b_{2}b_{4}+\dd_{2}\dd_{4}d_{1}d_{3}\right] 
\right.\nonumber\\&&\hspace{0cm}\left.
+\delta^{3}(k_{1}+k_{2}+k_{3}-k_{4})
\frac{\delta_{s_{1},-s_{2}}\delta_{s_{3},s_{4}}}{(k_{1}^{+}+k_{2}^{+})^{2}}
\left[ \bd_{1}\bd_{3}\dd_{2}b_{4}-\dd_{4}b_{2}d_{1}d_{3}\right] 
\right.\nonumber\\&&\hspace{0cm}\left.
-\delta^{3}(k_{1}-k_{2}+k_{3}+k_{4})
\frac{\delta_{s_{1},s_{2}}\delta_{s_{3},-s_{4}}}{(k_{1}^{+}-k_{2}^{+})^{2}}
\left[ \bd_{1}\bd_{3}\dd_{4}b_{2}-\dd_{2}b_{4}d_{1}d_{3}\right] 
\right.\nonumber\\&&\hspace{0cm}\left.
+\delta^{3}(k_{1}+k_{2}-k_{3}+k_{4})
\frac{\delta_{s_{1},-s_{2}}\delta_{s_{3},s_{4}}}{(k_{1}^{+}+k_{2}^{+})^{2}}
\left[ \bd_{1}\dd_{2}\dd_{4}d_{3}-\bd_{3}b_{2}b_{4}d_{1}\right] 
\right.\nonumber\\&&\hspace{0cm}\left.
-\delta^{3}(k_{1}-k_{2}-k_{3}+k_{4})
\frac{\delta_{s_{1},s_{2}}\delta_{s_{3},s_{4}}}{(k_{1}^{+}-k_{2}^{+})^{2}}
\left[ \bd_{1}\dd_{4}b_{2}d_{3}+\bd_{3}\dd_{2}b_{4}d_{1}\right] 
\right.\nonumber\\&&\hspace{0cm}\left.
+\delta^{3}(k_{1}+k_{2}-k_{3}-k_{4})
\frac{\delta_{s_{1},-s_{2}}\delta_{s_{3},-s_{4}}}{(k_{1}^{+}+k_{2}^{+})^{2}}
\left[ \bd_{1}\dd_{2}b_{4}d_{3}+\bd_{3}\dd_{4}b_{2}d_{1}\right] 
\right.\nonumber\\&&\hspace{0cm}\left.
+\delta_{3}(k_{1}-k_{2}-k_{3}-k_{4})
\frac{\delta_{s_{1},s_{2}}\delta_{s_{3},-s_{4}}}{(k_{1}^{+}-k_{2}^{+})^{2}}
\left[ \bd_{1}b_{2}b_{4}d_{3}-\bd_{3}\dd_{2}\dd_{4}d_{1}\right] 
\right\}.
\eeq

\section{$q\bar{q}g$ Matrix Elements}
\label{qqbg-me}

We need to find the matrix elements of all $q\bar{q}g$ vertices.
\newpage

\subsection{$\langle q_{a} \bar{q}_{b} \vert H \vert g_{c}\rangle$}

\centerline{\epsfig{file=\diagrampath gtoqqb.epsf}}

This gives us:
\beq
\label{gtoqqb-me}
&&\langle q_{a} \bar{q}_{b} \vert H \vert g_{c}\rangle = 
\n && \hspace{2cm}
16 \pi^{3} g \sqrt{2k_{a}^{+}k_{b}^{+}}
\langle c_{a}\vert{\bf T}^{c_{c}}\vert c_{b}\rangle\delta^{3}(k_{a}+k_{b}-k_{c})  \times
\n &&\hspace{4cm}
\left(
\delta_{s_{a},\bar{s}_{b}} \left\{ -\frac{\tilde{k}_{c}^{s_{c}}}{k_{c}^{+}}
+\delta_{\bar{s}_{b},s_{c}}\frac{\tilde{k}_{b}^{s_{c}}}{k_{b}^{+}}
+\delta_{s_{b},s_{c}}\frac{\tilde{k}_{a}^{s_{c}}}{k_{a}^{+}} \right\}
\right. \n && \hspace{6cm} \left.
-ims_{b}\delta_{s_{a},s_{b}}\delta_{s_{b},s_{c}}\left\{\frac{1}{k_{b}^{+}}+\frac{1}{k_{a}^{+}} \right\} \right) .
\eeq

\subsection{$\langle g_{c} \vert H \vert q_{a}\bar{q}_{b}\rangle$}

\centerline{\epsfig{file=\diagrampath qqbtog.epsf}}

\beq
\label{qqbtog-me}
&&\langle g_{c} \vert H \vert q_{a}\bar{q}_{b}\rangle = 
\n && \hspace{2cm}
16 \pi^{3} g  \sqrt{2k_{a}^{+}k_{b}^{+}}
\langle c_{b}\vert{\bf T}^{c_{c}}\vert c_{a}\rangle\delta^{3}(k_{a}+k_{b}-k_{c}) \times
\n &&\hspace{4cm}
\left(
\delta_{s_{b},\bar{s}_{a}} \left\{ -\frac{\tilde{k}_{c}^{\ast s_{c}}}{k_{c}^{+}}
+\delta_{\bar{s}_{a},s_{c}}\frac{\tilde{k}_{a}^{\ast s_{c}}}{k_{a}^{+}}
+\delta_{s_{a},s_{c}}\frac{\tilde{k}_{b}^{\ast s_{c}}}{k_{b}^{+}} \right\}
\right. \n && \hspace{6cm} \left.
+ims_{a}\delta_{s_{b},s_{a}}\delta_{s_{a},s_{c}}\left\{\frac{1}{k_{a}^{+}}+\frac{1}{k_{b}^{+}} \right\}
\right) . 
\eeq

\subsection{$\langle \bar{q}_{b} g_{c} \vert H \vert \bar{q}_{a} \rangle$}

\centerline{\epsfig{file=\diagrampath qbtoqbg.epsf}}

\beq
\label{qbtoqbg-me}
&&\langle \bar{q}_{b} g_{c} \vert H \vert \bar{q}_{a} \rangle = 
\n && \hspace{2cm}
16 \pi^{3} g \sqrt{2k_{a}^{+}k_{b}^{+}}
\langle c_{a}\vert{\bf T}^{c_{c}}\vert c_{b}\rangle\delta^{3}(k_{a}-k_{b}-k_{c})\times
\n &&\hspace{4cm}
\left(
\delta_{s_{a},s_{b}} \left\{ \frac{\tilde{k}_{c}^{\ast s_{c}}}{k_{c}^{+}}
-\delta_{s_{b},s_{c}}\frac{\tilde{k}_{b}^{\ast s_{c}}}{k_{b}^{+}}
-\delta_{s_{b},\bar{s}_{c}}\frac{\tilde{k}_{a}^{\ast s_{c}}}{k_{a}^{+}} \right\}
\right. \n && \hspace{6cm} \left.
-ims_{b}\delta_{s_{a},\bar{s}_{b}}\delta_{s_{b},\bar{s}_{c}}\left\{\frac{1}{k_{b}^{+}}-\frac{1}{k_{a}^{+}} \right\} 
\right) .
\eeq

\subsection{$\langle q_{a} \vert H \vert q_{b} g_{c}\rangle$}

\centerline{\epsfig{file=\diagrampath qgtoq.epsf}}

\beq
\label{qgtoq-me}
&&\langle q_{a} \vert H \vert q_{b} g_{c}\rangle = 
\n && \hspace{2cm}
16 \pi^{3} g \sqrt{2k_{a}^{+}k_{b}^{+}}
\langle c_{a}\vert{\bf T}^{c_{c}}\vert c_{b}\rangle\delta^{3}(k_{a}-k_{b}-k_{c})\times
\n &&\hspace{4cm}
\left(
\delta_{s_{a},s_{b}} \left\{ -\frac{\tilde{k}_{c}^{s_{c}}}{k_{c}^{+}}
+\delta_{s_{b},s_{c}}\frac{\tilde{k}_{b}^{s_{c}}}{k_{b}^{+}}
+\delta_{s_{b},\bar{s}_{c}}\frac{\tilde{k}_{a}^{s_{c}}}{k_{a}^{+}} \right\}
\right. \n && \hspace{6cm} \left.
-ims_{b}\delta_{s_{a},\bar{s}_{b}}\delta_{s_{b},\bar{s}_{c}}\left\{\frac{1}{k_{b}^{+}}-\frac{1}{k_{a}^{+}} \right\}
\right) .
\eeq

\subsection{$\langle q_{a} g_{c} \vert H \vert q_{b} \rangle$}

\centerline{\epsfig{file=\diagrampath qtoqg.epsf}}

\beq
\label{qtoqg-me}
&&\langle q_{a} g_{c} \vert H \vert q_{b} \rangle = 
\n && \hspace{2cm}
16 \pi^{3} g \sqrt{2k_{a}^{+}k_{b}^{+}}
\langle c_{a}\vert{\bf T}^{c_{c}}\vert c_{b}\rangle\delta^{3}(k_{a}-k_{b}+k_{c}) \times
\n &&\hspace{4cm}
\left(
\delta_{s_{a},s_{b}} \left\{ -\frac{\tilde{k}_{c}^{\ast s_{c}}}{k_{c}^{+}}
+\delta_{s_{b},\bar{s}_{c}}\frac{\tilde{k}_{b}^{\ast s_{c}}}{k_{b}^{+}}
+\delta_{s_{b},s_{c}}\frac{\tilde{k}_{a}^{\ast s_{c}}}{k_{a}^{+}} \right\}
\right. \n && \hspace{6cm} \left.
+ims_{b}\delta_{s_{a},\bar{s}_{b}}\delta_{s_{b},s_{c}}\left\{\frac{1}{k_{b}^{+}}-\frac{1}{k_{a}^{+}} \right\} 
\right) .
\eeq

\subsection{$\langle \bar{q}_{a} \vert H \vert \bar{q}_{b} g_{c} \rangle$}

\centerline{\epsfig{file=\diagrampath qbgtoqb.epsf}}

\beq
\label{qbgtoqb-me}
&&\langle \bar{q}_{a} \vert H \vert \bar{q}_{b} g_{c} \rangle = 
\n && \hspace{2cm}
16 \pi^{3} g \sqrt{2k_{a}^{+}k_{b}^{+}}
\langle c_{b}\vert{\bf T}^{c_{c}}\vert c_{a}\rangle\delta^{3}(k_{b}-k_{a}+k_{c}) \times
\n &&\hspace{4cm}
\left(
\delta_{s_{b},s_{a}} \left\{ \frac{\tilde{k}_{c}^{s_{c}}}{k_{c}^{+}}
-\delta_{s_{a},\bar{s}_{c}}\frac{\tilde{k}_{a}^{s_{c}}}{k_{a}^{+}}
-\delta_{s_{a},s_{c}}\frac{\tilde{k}_{b}^{s_{c}}}{k_{b}^{+}} \right\}
\right. \n && \hspace{6cm} \left.
+ims_{a}\delta_{s_{b},\bar{s}_{a}}\delta_{s_{a},s_{c}}\left\{\frac{1}{k_{a}^{+}}-\frac{1}{k_{b}^{+}} \right\}
\right) .
\eeq

\section{$q\bar{q}q\bar{q}$ Matrix Elements}
\label{qqbqqb-me}

To calculate a $q\bar{q}$ state to second order, it is only necessary to calculate the instantaneous exchange
matrix element.

\centerline{\epsfig{file=\diagrampath qqbtoqqb-exch.epsf}}

There are two different terms for the instantaneous exchange which give us:
\newpage
\beq
\langle q_{a}\bar{q}_{b} \vert H \vert q_{c}\bar{q}_{d} \rangle _{\mathrm{IN}}&=&
-32g^{2}\pi^{3}\int D_{1}D_{2}D_{3}D_{4}
\langle c_{1}\vert{\bf T}_{a}\vert c_{2}\rangle\langle c_{3}\vert{\bf T}_{a}\vert c_{4}\rangle
\sqrt{k_{1}^{+}k_{2}^{+}k_{3}^{+}k_{4}^{+}}
\times\nonumber\\&&\hspace{-2cm}
\delta^{3}(k_{1}-k_{2}-k_{3}+k_{4})
\frac{\delta_{s_{1},s_{2}}\delta_{s_{3},s_{4}}}{(k_{1}^{+}-k_{2}^{+})^{2}}
\left[ \delta_{a,1}\delta_{b,4}\delta_{c,2}\delta_{d,3}+\delta_{a,3}\delta_{b,2}\delta_{c,4}\delta_{d,1}\right] .
\eeq
The two different sets of delta functions give identical contributions when we use the momentum conserving delta 
function.  Thus, we get:
\beq
\label{inst-exch-me}
\langle q_{a}\bar{q}_{b} \vert H \vert q_{c}\bar{q}_{d} \rangle  &=&
-64g^{2}\pi^{3}
\langle c_{d}\vert{\bf T}_{f}\vert c_{b}\rangle\langle c_{a}\vert{\bf T}_{f}\vert c_{c}\rangle
\sqrt{k_{a}^{+}k_{b}^{+}k_{c}^{+}k_{d}^{+}} \times
\n && \hspace{2cm}
\delta^{3}(k_{a}-k_{c}-k_{d}+k_{b})
\frac{\delta_{s_{a},s_{c}}\delta_{s_{b},s_{d}}}{(k_{a}^{+}-k_{c}^{+})^{2}} .
\eeq

\chapter{Matrix Elements of the Renormalized Hamiltonian}
\label{chapt:me-renorm-ham}

We now determine the matrix elements of our renormalized Hamiltonian in a plane-wave basis.  
Section \ref{mass-o2-begin} shows which matrix elements are needed to calculate the mass-squared operator to second order.
These matrix elements are calculated in Sections \ref{v2nc} through \ref{v2nc-tmatrix}.

We find matrix elements of the IMO which is very closely related to the Hamiltonian matrix elements we have 
already calculated.  Specifically, in the center-of-momentum frame\footnote{Light-front 
formalism is discussed in Section \ref{lf-overview} and Appendix \ref{lf-formalism}.}
\beq
{\cal M}^{2}={\cal P}^{+} {\cal H},
\eeq
where ${\cal P}^{+}$ is the total light-front longitudinal momentum.
\section{Cutoff Dependent, Non-Canonical Contributions}
\label{v2nc}
The non-canonical contributions to the mass matrix are contained in \newline
$\Vcdme{q_{3}\bar{q}_{4}}{q_{1}\bar{q}_{2}}{2}$, or more explicitly in $\dVome{2}$ [Eq. (\ref{deltav})].  For a 
second order calculation, we can write: 
\beq
\dVome{2}=\half\sum_{K}\Vme{q_{3}\bar{q}_{4}}{K}\Vme{K}{q_{1}\bar{q}_{2}}T_{2}^{(\Lambda,\Lambda')}(F,K,I), 
\eeq
where the sum is over a complete set of (intermediate) states and $V^{(1)}$ is the first-order canonical 
Hamiltonian.  The only term in $V^{(1)}$ we need contains a $q \bar{q} g$ vertex so 
$\vert K\rangle$ can only be a  $q\bar{q}g$ state, and the 
interaction is either a gluon exchange or a quark self-energy.

In this section, we make the following momentum definitions for various momenta in the self-energy and exchange diagrams:
\beq
k_{1}&=&(x \p^{+},x \vec{\p}+\vec{q_{\bot}}),\hspace{.5cm}k_{2}=([1-x]\p^{+},[1-x]\vec{\p}-\vec{q_{\bot}}),\n
k_{3}&=&(y \p^{+},y \vec{\p}+\vec{p_{\bot}}),\hspace{.5cm}k_{4}=([1-y]\p^{+},[1-y]\vec{\p}-\vec{p_{\bot}}),\n
k_{5}&=&(z k_{1}^{+},z 
\vec{k}_{1}+\vec{r_{\bot}}),\hspace{.5cm}k_{7}=([1-z]k_{1}^{+},[1-z]\vec{k}_{1}-\vec{r_{\bot}}),
\eeq
where $k_{1}$ and $k_{3}$ are the momentum of the incoming and outgoing quark and anti-quark, respectively.  $k_{5}$ is the 
momentum of the internal quark.  Using these definitions, we find for the free mass squared of the states:
\beq
M_{f}^{2}\vert F \rangle &=&M_{f}^{2}\vert q_{3}\qb_{4}\rangle=\frac{\vec{p}_{\perp}^{\hspace{.05cm}2}+m^{2}}{y(1-y)},\\
M_{f}^{2}\vert I \rangle &=&M_{f}^{2}\vert 
q_{1}\qb_{2}\rangle=\frac{\vec{q}_{\perp}^{\hspace{.05cm}2}+m^{2}}{x(1-x)},\\
M_{f}^{2}\vert K \rangle &=&M_{f}^{2}\vert q_{5}\qb_{6}g_{7}\rangle \n
&=&\frac{m^{2}(1-z)(1-x+xz)+z(1-z) \vec{q}_{\perp}^{\hspace{.05cm}2}+(1-x)\vec{r}_{\perp}^{\hspace{.05cm}2}}{x(1-x)z(1-z)}.
\eeq

\subsection{Self-Energy}
%
The quark self-energy contribution to $\dVome{2}$ is:
\beq
\half \int D_{5}D_{6}D_{7}\langle q_{3}\bar{q}_{4} \vert V^{(1)}_{qg \rightarrow q} \vert q_{5}\bar{q}_{6}g_{7} \rangle
\langle q_{5}\bar{q}_{6}g_{7} \vert V^{(1)}_{q \rightarrow qg} \vert q_{1}\bar{q}_{2} \rangle T_{2}(F,K_{\mathrm{SE}},I), 
\eeq
where the subscript on $V^{(1)}$ indicates which of the vertices calculated in Section \ref{qqbg-me} are used.
For the quark self-energy we will need the matrix elements given in Eqs.~(\ref{qgtoq-me}) and (\ref{qtoqg-me}).
However, these are not the complete matrix elements, since there is a spectator quark.  We use:
\beq
&&\langle q_{5} \bar{q}_{6} g_{7} \vert V^{(1)} \vert \bar{q}_{2} q_{1} \rangle = 
\n && \hspace{1cm}
16 \pi^{3} \p^{+}
 \int D_{1}D_{2}D_{7}D_{5}D_{6} \sqrt{2k_{5}^{+}k_{1}^{+}}
\T{7}{5,1} \delta^{3}(k_{5}-k_{1}+k_{7}) \delta_{2,6}\times
\n &&\hspace{3cm}
\left(
\delta_{s_{5},s_{1}} \left\{ -\frac{\tilde{k}_{7}^{\ast s_{7}}}{k_{7}^{+}}
+\delta_{s_{1},\bar{s}_{7}}\frac{\tilde{k}_{1}^{\ast s_{7}}}{k_{1}^{+}}
+\delta_{s_{1},s_{7}}\frac{\tilde{k}_{5}^{\ast s_{7}}}{k_{5}^{+}} \right\}
\right. \n && \hspace{6.5cm} \left.
+ims_{1}\delta_{s_{5},\bar{s}_{1}}\delta_{s_{1},s_{7}}\left\{\frac{1}{k_{1}^{+}}-\frac{1}{k_{5}^{+}} \right\} 
\right) ,
\eeq
and
\beq
&&\langle \bar{q}_{4}q_{3} \vert V^{(1)} \vert q_{5} \bar{q}_{6} g_{7}\rangle =
\n && \hspace{1cm}
16 \pi^{3} \p^{+}
 \int D_{3}D_{4}D_{7}D_{5}D_{6} \sqrt{2k_{3}^{+}k_{5}^{+}}
\T{7}{3,5} \delta^{3}(k_{3}-k_{5}-k_{7}) \delta_{4,6}\times
\n &&\hspace{3cm}
\left(
\delta_{s_{3},s_{5}} \left\{ -\frac{\tilde{k}_{7}^{s_{7}}}{k_{7}^{+}}
+\delta_{s_{5},s_{7}}\frac{\tilde{k}_{5}^{s_{7}}}{k_{5}^{+}}
+\delta_{s_{5},\bar{s}_{7}}\frac{\tilde{k}_{3}^{s_{7}}}{k_{3}^{+}} \right\}
\right. \n && \hspace{6.5cm} \left.
-ims_{5}\delta_{s_{3},\bar{s}_{5}}\delta_{s_{5},\bar{s}_{7}}\left\{\frac{1}{k_{5}^{+}}-\frac{1}{k_{3}^{+}} \right\}
\right) .
\eeq
Thus the contribution from the quark self-energy diagram can be written:
\beq
&&
(16\pi^{3} {\cal P}^{+})^{2}\delta_{2,4}\int D_{5}D_{7}  k_{5}^{+}\sqrt{k_{1}^{+}k_{3}^{+}} \T{7}{3,5} \T{7}{5,1}
\delta^{3}(k_{3}-k_{5}-k_{7}) \delta^{3}(k_{5}-k_{1}+k_{7}) \n
&&\hspace{2cm} \times 
\left(
\delta_{s_{5},s_{1}} \left\{ -\frac{\tilde{k}_{7}^{\ast s_{7}}}{k_{7}^{+}}
+\delta_{s_{1},\bar{s}_{7}}\frac{\tilde{k}_{1}^{\ast s_{7}}}{k_{1}^{+}}
+\delta_{s_{1},s_{7}}\frac{\tilde{k}_{5}^{\ast s_{7}}}{k_{5}^{+}} \right\}
\right. \n && \hspace{6cm} \left.
+ims_{1}\delta_{s_{5},\bar{s}_{1}}\delta_{s_{1},s_{7}}\left\{\frac{1}{k_{1}^{+}}-\frac{1}{k_{5}^{+}} \right\} 
\right) \n
&&\hspace{3cm} \times
\left(
\delta_{s_{3},s_{5}} \left\{ -\frac{\tilde{k}_{7}^{s_{7}}}{k_{7}^{+}}
+\delta_{s_{3},s_{7}}\frac{\tilde{k}_{5}^{s_{7}}}{k_{5}^{+}}
+\delta_{s_{3},\bar{s}_{7}}\frac{\tilde{k}_{3}^{s_{7}}}{k_{3}^{+}} \right\}
\right. \n && \hspace{7cm} \left.
+ims_{3}\delta_{s_{3},\bar{s}_{5}}\delta_{s_{5},\bar{s}_{7}}\left\{\frac{1}{k_{5}^{+}}-\frac{1}{k_{3}^{+}} \right\}
\right) \n
&& \hspace{4cm} \times T_{2}^{(\Lambda,\Lambda')}(F,K,I).
\eeq
After integrating over everything but the momentum of the internal quark, and completing all sums, this becomes:
\beq
\frac{\delta_{1,3}\delta_{2,4}}{12 \pi^{3}}\int \frac{dz d^{2}\vec{r}_{\bot}}{(1-z)^{3}z^{2}}
\frac{m^{2}(z-1)^{4}+\vec{r}_{\bot}^{\hspace{.1cm}2}(1+z^{2})}{x^{2}}\hspace{.05cm}T_{2}^{(\Lambda,\Lambda')}(F,K,I).
\eeq
Recall that $z=k^{+}_{5}/k^{+}_{1}$.  For the self-energy, the final and initial states have the same energy giving:
\beq
\Delta_{FK}=-\Delta_{KI}=-\frac{\vec{r}^{\hspace{.05cm}2}+m^{2}(1-z)^{2}}{xz(1-z)},
\eeq
and
\beq
T_{2}^{(\la,\la')}(F,K,I)=-\frac{2}{\Delta_{KI}}\left(e^{-2\la'^{-4}\Delta_{KI}^{2}}-e^{-2\la^{-4}\Delta_{KI}^{2}}\right).
\eeq
Using the definitions:
\beq
u&=&\frac{m^{2}(1-z)^{2}+r^{2}}{xz(1-z)},\hspace{1cm}\beta=\frac{m^{2}(1-z)}{xz}, \n
\gamma(a,b)&=&\sqrt{2}\frac{m^{2}}{\Lambda^{2}}\frac{(1-b)}{ab}, \hspace{1cm}
\gamma'(a,b)=\sqrt{2}\frac{m^{2}}{\Lambda'^{2}}\frac{(1-b)}{ab}, \nonumber
\eeq
and changing variables from $r$ to $u$ and integrating over $u$ gives:
\beq
&&-\la'^{2}\frac{\delta_{1,3}\delta_{2,4}}{24\pi^{2}} \left[ \int \frac{z\hspace{.1cm}dz}{1-z}
\left\{ \frac{1+z^{2}}{z}\sqrt{2\pi}\left[1-\mathrm{erf}\left(\gamma'(x,z)\right)\right] \right. \right.\n
&&\hspace{4cm}\left. \left.
-2\sqrt{2}\gamma'(x,z) \mathrm{Ei}\left(1,\gamma'^{2}(x,z)\right)\right\}\right]-\left[ \la' \rightarrow \la\right]
\eeq
for the contribution to $\dVome{2}$ from the quark self-energy, where `erf' is the error function and `Ei' is the 
exponential integral.
%
The anti-quark self-energy contribution to $\dVome{2}$ is:
\beq
\half \int D_{5}D_{6}D_{7}\langle q_{3}\bar{q}_{4} \vert V^{(1)}_{\bar{q}g \rightarrow \bar{q}} \vert q_{5}\bar{q}_{6}g_{7} \rangle
\langle q_{5}\bar{q}_{6}g_{7} \vert V^{(1)}_{\bar{q} \rightarrow \bar{q}g} \vert q_{1}\bar{q}_{2} \rangle T_{2}(F,K_{\mathrm{SE}},I), 
\eeq
where the subscript on $V^{(1)}$ indicates which of the vertices calculated in Section \ref{qqbg-me} is used.

The contribution from the anti-quark self-energy is the same as that from the quark self-energy with $x\rightarrow 
1-x$.  So we find the total contribution to $\delta V$ from the self-energy diagrams is:
\beq
\label{self-energy}
&&\frac{\delta_{1,3}\delta_{2,4}}{24\pi^{2}}\la'^{2}\left[ \int \frac{z\hspace{.1cm}dz}{1-z}
\left\{ \frac{1+z^{2}}{z}\sqrt{2\pi}\left[\mathrm{erf}\left(\gamma'(x,z)\right)
+\mathrm{erf}\left(\gamma'(1-x,z)\right)-2\right]
\right.\right.\n &&\hspace{3cm}\left.\left.
+\sqrt{8} \left[ \gamma'(x,z) \mathrm{Ei}\left(1,\gamma'^{2}(x,z)\right)
\right. \right. \right. \n && \hspace{5cm} \left. \left. \left.
+\gamma'(1-x,z) \mathrm{Ei}\left(1,\gamma'^{2}(1-x,z)\right) \right ] \right\}\right]
\n && \hspace{2cm}
-\left[ \la' \rightarrow \la \right] .
\eeq
To find $\langle q_{3}\bar{q}_{4} \vert V^{(2)}_{\mathrm{SE}}(\Lambda)\vert q_{1}\bar{q}_{2}\rangle$ 
we have to keep only the $\Lambda$ terms and remove the cutoff independent part.  
We remove the cutoff independent part by expanding our answer in a generalized series of 
powers and logs of $\frac{m}{\Lambda}$ and remove the part proportional to $m^{2}$.

Keeping only the terms that contain $\Lambda$ gives us:
\beq
\label{cor-ci}
&&-\frac{\delta_{1,3}\delta_{2,4}}{24\pi^{2}}\la^{2}\left[ \int \frac{z\hspace{.1cm}dz}{1-z}
\left\{ \frac{1+z^{2}}{z}\sqrt{2\pi}\left[\mathrm{erf}\left(\gamma(x,z)\right)
+\mathrm{erf}\left(\gamma(1-x,z)\right)-2\right]\right.\right.\n
&&\hspace{1cm}\left.\left.
+\sqrt{8} \left[ \gamma(x,z) \mathrm{Ei}\left(1,\gamma^{2}(x,z)\right)
+\gamma(1-x,z) \mathrm{Ei}\left(1,\gamma^{2}(1-x,z)\right) \right] \right\}\right].
\eeq
Finally, we have to remove the terms that are independent of the cutoff, since they can not be determined by 
calculating $\delta V$.  The only part of this expression that contains terms proportional to $m^{2}$ are the 
error functions.  We find the cutoff-independent part of this expression to be:
\beq
\label{se-ci}
-4\frac{\delta_{1,3}\delta_{2,4}}{24\pi^{2}}m^{2}\int dz\frac{1+z^{2}}{zx(1-x)}.
\eeq
This term is divergent, however the entire term will be cancelled when calculating the cutoff independent part of 
the reduced interaction (Sec. \ref{v2nc-tmatrix}).  Thus the complete expression for 
$\langle q_{3}\bar{q}_{4} \vert V^{(2)}_{\mathrm{SE}}(\Lambda)\vert q_{1}\bar{q}_{2}\rangle$ is:
\beq
&&\langle q_{3}\bar{q}_{4} \vert V^{(2)}_{\mathrm{SE}}(\Lambda)\vert q_{1}\bar{q}_{2}\rangle
=-\la^{2}\frac{\delta_{1,3}\delta_{2,4}}{24\pi^{2}}\times\n
&&\hspace{1cm}
\left[ \int \frac{z\hspace{.1cm}dz}{1-z}
\left\{ \frac{1+z^{2}}{z}\sqrt{2\pi}\left[\mathrm{erf}\left(\gamma(x,z)\right)
+\mathrm{erf}\left(\gamma(1-x,z)\right)-2\right]\right.\right.\n
&&\hspace{3cm}\left.\left.
+\sqrt{8} \left( \gamma(x,z) \mathrm{Ei}\left[ 1,\gamma^{2}(x,z)\right]
\right. \right. \right. \n && \hspace{6cm} \left. \left. \left.
+\gamma(1-x,z) \mathrm{Ei}\left[ 1,\gamma^{2}(1-x,z)\right] \right) \right\}\right.\n
&&\hspace{5cm}\left.
-4m^{2}\int dz \frac{1+z^{2}}{zx(1-x)}\right].
\eeq

\subsection{Exchange}
The first exchange diagram contributes to $\dVome{2}$:
\beq
\half \int D_{5}D_{6}D_{7}\langle q_{3}\bar{q}_{4} \vert V^{(1)}_{\bar{q}g \rightarrow \bar{q}} \vert q_{5}\bar{q}_{6}g_{7} \rangle
\langle q_{5}\bar{q}_{6}g_{7} \vert V^{(1)}_{q \rightarrow qg} \vert 
q_{1}\bar{q}_{2} \rangle T_{2}(F,K_{\mathrm{EX}},I), 
\eeq
where the subscript on $V^{(1)}$ indicates which of the vertices calculated in Section \ref{qqbg-me} is used.
For this exchange diagram, we need to use the matrix elements derived from Eqs.~(\ref{qtoqg-me}) and (\ref{qbgtoqb-me}):
\beq
\langle q_{5} \bar{q}_{6} g_{7} \vert V^{(1)} \vert q_{1} \bar{q}_{2} \rangle
&=& 16 \pi^{3} \p^{+} \sqrt{2k_{5}^{+}k_{1}^{+}}
\langle c_{5}\vert{\bf T}^{c_{7}}\vert c_{1}\rangle \delta_{2,6} \delta^{3}(k_{5}-k_{1}+k_{7}) \times
\n &&\hspace{-1cm} \left(
\delta_{s_{5},s_{1}} \left\{ -\frac{\tilde{k}_{7}^{\ast s_{7}}}{k_{7}^{+}}
+\delta_{s_{1},\bar{s}_{7}}\frac{\tilde{k}_{1}^{\ast s_{7}}}{k_{1}^{+}}
+\delta_{s_{1},s_{7}}\frac{\tilde{k}_{5}^{\ast s_{7}}}{k_{5}^{+}} \right\}
\right. \n && \hspace{3cm} \left.
+ims_{1}\delta_{s_{5},\bar{s}_{1}}\delta_{s_{1},s_{7}}\left\{\frac{1}{k_{1}^{+}}-\frac{1}{k_{5}^{+}} \right\} 
\right) ,
\eeq
\beq
\langle q_{3}\bar{q}_{4} \vert V^{(1)} \vert q_{1}\bar{q}_{2} g_{7} \rangle
&=& 16 \pi^{3} \p^{+} \sqrt{2k_{4}^{+}k_{6}^{+}}
\langle c_{6}\vert{\bf T}^{c_{7}}\vert c_{4}\rangle \delta_{3,5}\delta^{3}(k_{6}-k_{4}+k_{7})\times
\n &&\hspace{-1cm} \left(
\delta_{s_{6},s_{4}} \left\{ \frac{\tilde{k}_{7}^{s_{7}}}{k_{7}^{+}}
-\delta_{s_{4},\bar{s}_{7}}\frac{\tilde{k}_{4}^{s_{7}}}{k_{4}^{+}}
-\delta_{s_{4},s_{7}}\frac{\tilde{k}_{6}^{s_{7}}}{k_{6}^{+}} \right\}
\right. \n && \hspace{3cm} \left.
+ims_{4}\delta_{s_{6},\bar{s}_{4}}\delta_{s_{4},s_{7}}\left\{\frac{1}{k_{4}^{+}}-\frac{1}{k_{6}^{+}} \right\}
\right) .
\eeq
The total contribution from this exchange diagram is:
\newpage
\beq
&-&(16 \pi^{3} \p^{+})^{2}\int D_{7}\sqrt{k_{1}^{+}k_{2}^{+}k_{3}^{+}k_{4}^{+}}
\T{7}{5,1} \T{7}{6,4}\delta^{3}(k_{2}-k_{4}+k_{7}) \delta^{3}(k_{3}-k_{1}+k_{7}) 
\n &&\hspace{2cm}
\times \left(\delta_{s_{3},s_{1}} \left\{ -\frac{\tilde{k}_{7}^{\ast s_{7}}}{k_{7}^{+}}
+\delta_{s_{1},\bar{s}_{7}}\frac{\tilde{k}_{1}^{\ast s_{7}}}{k_{1}^{+}}
+\delta_{s_{1},s_{7}}\frac{\tilde{k}_{3}^{\ast s_{7}}}{k_{3}^{+}} \right\}
\right. \n && \hspace{5cm} \left.
+ims_{1}\delta_{s_{3},\bar{s}_{1}}\delta_{s_{1},s_{7}}\left\{\frac{1}{k_{1}^{+}}-\frac{1}{k_{3}^{+}} \right\} 
\right) 
\n &&\hspace{3cm}
\times \left(\delta_{s_{2},s_{4}} \left\{ -\frac{\tilde{k}_{7}^{s_{7}}}{k_{7}^{+}}
+\delta_{s_{4},\bar{s}_{7}}\frac{\tilde{k}_{4}^{s_{7}}}{k_{4}^{+}}
+\delta_{s_{4},s_{7}}\frac{\tilde{k}_{2}^{s_{7}}}{k_{2}^{+}} \right\}
\right. \n && \hspace{6cm} \left.
-ims_{4}\delta_{s_{2},\bar{s}_{4}}\delta_{s_{4},s_{7}}\left\{\frac{1}{k_{4}^{+}}-\frac{1}{k_{2}^{+}} \right\}
\right) 
\n && \hspace{4cm}
\times T_{2}^{(\Lambda,\Lambda')}(F,K,I).
\eeq
For the first exchange diagram we find:
\beq
\label{ex1-deltas}
\Delta_{KI}^{(1)}&=&\frac{m^{2}(x-y)^{2}+(x\vec{p}-y\vec{q})^{2}}{xy(x-y)} ,\\
\Delta_{FK}^{(1)}&=&-\frac{m^{2}(x-y)^{2}+\left[(1-x)\vec{p}-(1-y)\vec{q}\hspace{.1cm}\right]^{2}}{(1-x)(1-y)(x-y)}.
\eeq
Finally we use
\beq
\sum_{c_{7}} \T{7}{2,4} 
\T{7}{3,1}=-\frac{1}{6}\delta_{c_{2},c_{4}}\delta_{c_{1},c_{3}}+\frac{1}{2}\delta_{c_{1},c_{2}}\delta_{c_{3},c_{4}},
\eeq 
and find for this exchange graph:
\beq
&&-8 \pi^{3} {\cal P}^{+}\frac{\sqrt{x(1-x)y(1-y)}}{x-y} \delta^{3}({k_{1}+k_{2}-k_{3}-k_{4}})
\n && \hspace{1cm} \times
\left( \delta_{c_{1},c_{2}}\delta_{c_{3},c_{4}}-\frac{1}{3}\delta_{c_{2},c_{4}} \delta_{c_{1},c_{3}} \right)
T_{2}^{(\Lambda,\Lambda')(1)}(F,K,I)
\n &&\hspace{2cm}
\times \left[ \delta_{s_{1},s_{3}}\delta_{s_{2},s_{4}}B_{++}
 +m^{2}\delta_{s_{1},\bar{s}_{3}}\delta_{s_{2},\bar{s}_{4}}\delta_{s_{1},s_{4}}B_{--}
 \right. \n && \hspace{4cm} \left.
+im \left(s_{2}\delta_{s_{1},s_{3}}\delta_{s_{2},\bar{s}_{4}}B_{+-} 
               +s_{1}\delta_{s_{1},\bar{s}_{3}}\delta_{s_{2},s_{4}}B_{-+} \right) \right] ,
\eeq
where the extra parameter on $T_{2}$ simply specifies it is for the first exchange diagram and is defined in 
terms of the $\Delta$'s in Eq. \ref{ex1-deltas}.  The $B$'s are given by:
\newpage
\beq
B_{++}&=&2\frac{(\vec{q}-\vec{p})^{2}}{(x-y)^{2}} 
-\delta_{s_{1},s_{2}}\vec{p}\cdot\vec{q}(s_{1})\frac{x+y-2xy}{x(1-x)y(1-y)}
 \n && \hspace{1cm} 
-\delta_{s_{1},\bar{s}_{2}}\left[ \frac{q^{2}}{x(1-x)}+\frac{p^{2}}{y(1-y)}\right]
\n && \hspace{3cm}
-\frac{1}{x-y}\left[ \frac{2y-1}{y(1-y)}p^{2}-\frac{2x-1}{x(1-x)}q^{2} \right]
\n && \hspace{5cm}
-\frac{\vec{p}\cdot\vec{q}(s_{1})}{xy} - \frac{\vec{p}\cdot\vec{q}(s_{2})}{(1-x)(1-y)}, \\
B_{+-}&=&\frac{x-y}{(1-x)(1-y)}
\left( \frac{\tilde{p}^{s_{2}}-\tilde{q}^{s_{2}}}{x-y}+\delta_{s_{1},s_{2}}\frac{\tilde{q}^{s_{2}}}{x}
+\delta_{s_{1},\bar{s}_{2}}\frac{\tilde{p}^{s_{2}}}{y}\right), \\
B_{-+}&=&\frac{x-y}{xy}
\left( \frac{\tilde{q}^{s_{1}}-\tilde{p}^{s_{1}}}{x-y}+\delta_{s_{1}s,_{2}}\frac{\tilde{q}^{s_{1}}}{1-x}
+\delta_{\bar{s}_{1},s_{2}}\frac{\tilde{p}^{s_{1}}}{1-y} \right), \\
B_{--}&=&\frac{(x-y)^{2}}{x(1-x)y(1-y)} .
\eeq
For the second exchange diagram, we can use the result of the first diagram and switch particles $1$ and $2$, 
as well $3$ and $4$ (See figure \ref{exch-diagrams}).  This means we will have:
\beq
&&x\rightarrow 1-x, \hspace{1cm} \vec{p}\rightarrow -\vec{p}, \hspace{1cm} s_{1}\leftrightarrow s_{2},
\n
&&y\rightarrow 1-y, \hspace{1cm} \vec{q}\rightarrow -\vec{q}, \hspace{1cm} s_{3}\leftrightarrow s_{4},
\eeq
but the color indices do not change.  So for the second exchange diagram, we get:
\beq
\label{ex2-deltas}
\Delta_{FK}^{(2)}&=&\frac{m^{2}(x-y)^{2}+(x\vec{p}-y\vec{q})^{2}}{xy(x-y)}=\Delta_{KI}^{(1)},\\
\Delta_{KI}^{(2)}&=&-\frac{m^{2}(x-y)^{2}+\left[(1-x)\vec{p}-(1-y)\vec{q}\hspace{.1cm}\right]^{2}}{(1-x)(1-y)(x-y)}
=\Delta_{FK}^{(1)},
\eeq
making the contribution from the second diagram:
\beq
&&8 \pi^{3} {\cal P}^{+}\frac{\sqrt{x(1-x)y(1-y)}}{x-y} \delta^{3}({k_{1}+k_{2}-k_{3}-k_{4}})
\n && \hspace{1cm} \times
\left( \delta_{c_{1},c_{2}}\delta_{c_{3},c_{4}}-\frac{1}{3}\delta_{c_{2},c_{4}} \delta_{c_{1},c_{3}} \right)
T_{2}^{(\Lambda,\Lambda')(2)}(F,K,I)
\n &&\hspace{2cm}
\times \left[ \delta_{s_{1},s_{3}}\delta_{s_{2},s_{4}}B_{++}
 +m^{2}\delta_{s_{1},\bar{s}_{3}}\delta_{s_{2},\bar{s}_{4}}\delta_{s_{1},s_{4}}B_{--}
 \right. \n && \hspace{4cm} \left.
+im \left(s_{2}\delta_{s_{1},s_{3}}\delta_{s_{2},\bar{s}_{4}}B_{+-} 
               +s_{1}\delta_{s_{1},\bar{s}_{3}}\delta_{s_{2},s_{4}}B_{-+} \right) \right] .
\eeq
The extra parameter on $T_{2}$ simply specifies it is for the second exchange diagram and is defined in 
terms of the $\Delta$'s in Eq.~(\ref{ex2-deltas}).   $B_{++}$, $B_{+-}$, $B_{-+}$, and $B_{--}$ are 
the same ones defined for the first exchange diagram.  
Eq.~(\ref{ex2-deltas}) shows that 
\beq
T_{2}^{(\Lambda,\Lambda')(1)}(F,K,I)=-T_{2}^{(\Lambda,\Lambda')(2)}(F,K,I)\equiv T_{2}^{(\Lambda,\Lambda')}(F,K,I).
\eeq
There is an implicit cutoff on the exchanged gluon momenta, which we have omitted for clarity, that prevents an infrared 
divergence.  Explicitly, 
in the contribution from the first diagram, the cutoff is on $\vert x-y \vert$, and in the second 
contribution, the cutoff is on $\vert y-x \vert$.

The form of the exchange contribution to the approximate states will look like:
\beq
\int d^{2}\vec{q} d^{2}\vec{p}\int dx dy \theta (\vert x-y \vert -\epsilon ) E(x,\vec{q},y,\vec{p})
\chi_{s_{1}s_{2}}(x,\vec{q}) \chi_{s_{3}s_{4}}(y,\vec{p}) ,
\eeq
where $ E(x,\vec{q},y,\vec{p})$ is a function only of momentum.
If we split this into two terms, one where $x>y$, the other where $x<y$, and note that the exchange 
function is symmetric under particle exchange, we can rewrite the exchange integration variables in the term 
where $x<y$ to get:
\beq
&&\int d^{2}\vec{q} d^{2}\vec{p}\int dx dy \theta (x-y) \theta ( x-y -\epsilon ) E(x,\vec{q},y,\vec{p}) \times \n
&&\hspace{2cm} \left[ \chi_{s_{1}s_{2}}(x,\vec{q}) \chi_{s_{3}s_{4}}(y,\vec{p})
+ \chi_{s_{2}s_{1}}(1-x,-\vec{q}) \chi_{s_{4}s_{3}}(1-y,-\vec{p}) \right] .
\eeq
If both of the spin-momentum wave functions are either symmetric or anti-symmetric under exchange of particles, 
then the two terms are the same.  If one is symmetric and the other anti-symmetric, then the integral is zero.  Thus we 
get:
\beq
2 \int d^{2}\vec{q} d^{2}\vec{p}\int dx dy \theta (x-y) \theta ( x-y -\epsilon ) E(x,\vec{q},y,\vec{p})
\chi_{s_{1}s_{2}}(x,\vec{q}) \chi_{s_{3}s_{4}}(y,\vec{p}) .
\eeq
Therefore the complete contribution is:
\newpage
\beq
&&-16\pi^{3} {\cal P}^{+}\theta ( x-y )\theta ( x-y -\epsilon)\delta^{3}(k_{1}+k_{2}-k_{3}-k_{4})
\n && \hspace{1cm} \times
\left( -\frac{1}{3}\delta_{c_{2},c_{4}}\delta_{c_{1},c_{3}}+\delta_{c_{1},c_{2}}\delta_{c_{3},c_{4}}\right)
\frac{\sqrt{x(1-x)y(1-y)}}{x-y}
\n &&\hspace{2cm}
\times\left[\delta_{s_{1},s_{3}}\delta_{s_{2},s_{4}}B_{++}
+m^{2}\delta_{s_{1},\bar{s}_{3}}\delta_{s_{2},\bar{s}_{4}}B_{--}
\right. \n && \hspace{4cm} \left.
+im\left( s_{2}\delta_{s_{1},s_{3}}\delta_{s_{2},\bar{s}_{4}}B_{+-}
+s_{1}\delta_{s_{1},\bar{s}_{3}}\delta_{s_{2},s_{4}}B_{-+}\right) 
\right]
\n &&\hspace{3cm}
\times \left(\frac{1}{\Delta_{FK}^{(1)}}-\frac{1}{\Delta_{KI}^{(1)}}\right)
\left[e^{2\Lambda'^{-4}\Delta_{FK}^{(1)}\Delta_{KI}^{(1)}}-e^{2\Lambda^{-4}\Delta_{FK}^{(1)}\Delta_{KI}^{(1)}}\right] .
\eeq
To get the ``$\Lambda$ terms'' we can let $\Lambda'\rightarrow\infty$.  This gives us:
\beq
\label{ex-defn}
&&\langle q_{3}\bar{q}_{4} \vert V^{(2)}_{\mathrm{EX}}(\Lambda)\vert q_{1}\bar{q}_{2}\rangle=\n
&&\hspace{1cm}-16\pi^{3} {\cal P}^{+}\theta ( x-y )\theta ( x-y -\epsilon)\delta^{3}(k_{1}+k_{2}-k_{3}-k_{4})
\n && \hspace{2cm}
\times\left( -\frac{1}{3}\delta_{c_{2},c_{4}}\delta_{c_{1},c_{3}}+\delta_{c_{1},c_{2}}\delta_{c_{3},c_{4}}\right)
\sqrt{x(1-x)y(1-y)}
\n && \hspace{3cm}
\times\left[\delta_{s_{1},s_{3}}\delta_{s_{2},s_{4}}\bar{B}_{++}
+m^{2}\delta_{s_{1},\bar{s}_{3}}\delta_{s_{2},\bar{s}_{4}}\bar{B}_{--}
\right. \n && \hspace{5cm} \left.
+im\left( s_{2}\delta_{s_{1},s_{3}}\delta_{s_{2},\bar{s}_{4}}\bar{B}_{+-}
+s_{1}\delta_{s_{1},\bar{s}_{3}}\delta_{s_{2},s_{4}}\bar{B}_{-+}\right) 
\right]
\n &&\hspace{4cm}
\times\left(\frac{1}{\Delta_{FK}^{(1)}}-\frac{1}{\Delta_{KI}^{(1)}}\right)
\left[1-e^{2\Lambda^{-4}\Delta_{FK}^{(1)}\Delta_{KI}^{(1)}}\right],
\eeq
where $B_{\pm \pm}=(x-y)\bar{B}_{\pm \pm}$.

Since there are four interacting particles in this diagram, a cutoff-independent contribution must be independent of 
both $\frac{m}{\Lambda}$ and $\frac{p_{\bot}}{\Lambda}$.  It is also obvious that the only cutoff dependence in the 
expression is from the exponential.  The $1$ subtracted from the exponential removes all the cutoff-independent 
terms, so this is the complete expression for 
$\langle q_{3}\bar{q}_{4} \vert V^{(2)}_{\mathrm{NC_{ex}}}(\Lambda)\vert q_{1}\bar{q}_{2}\rangle$.

\section{$V^{(2)}_{\mathrm{C}}$:  Instantaneous Diagram}

This is a matrix element of the canonical Hamiltonian, and can easily be derived from Eq.~(\ref{inst-exch-me}):
\beq
\langle q_{3}\bar{q}_{4} \vert V_{\mathrm{C}}^{(2)} \vert q_{1}\bar{q}_{2} \rangle
&=&-64\pi^{3} \p^{+}\T{a}{2,4}\T{a}{3,1}\delta^{3}(k_{3}-k_{1}-k_{2}+k_{4})
\n && \hspace{3cm}
\times \sqrt{k_{3}^{+}k_{4}^{+}k_{1}^{+}k_{2}^{+}}
\frac{\delta_{s_{3},s_{1}}\delta_{s_{4},s_{2}}}{(k_{3}^{+}-k_{1}^{+})^{2}} \n
&=&-64\pi^{3} \p^{+}\T{a}{2,4}\T{a}{3,1}\delta^{3}(k_{3}-k_{1}-k_{2}+k_{4})
\n && \hspace{3cm}
\times \sqrt{x(1-x)y(1-y)} \frac{\delta_{s_{3},s_{1}}\delta_{s_{4},s_{2}}}{(x-y)^{2}}.
\eeq
So the complete contribution from the instantaneous diagrams is:
\beq
&&\langle q_{3} \bar{q}_{4}\vert V^{(2)}_{\mathrm{IN}}(\Lambda) \vert q_{1} \bar{q}_{2}\rangle=
\langle q_{3}\bar{q}_{4} \vert V_{\mathrm{C}}^{(2)} \vert q_{1}\bar{q}_{2} \rangle
\n && \hspace{2cm}
-64\pi^{3} {\cal P}^{+}\delta^{3}(k_{1}+k_{2}-k_{3}-k_{4})\sqrt{x(1-x)y(1-y)}\n
&&\hspace{4cm}\times\theta ( \vert x-y \vert -\epsilon)
\left[ \frac{\T{a}{3,1}\T{a}{2,4}\delta_{s_{1},s_{3}}\delta_{s_{4},s_{2}}}{(x-y)^{2}}\right] .
\eeq
We can use the symmetry of the external states to rewrite this as (like in the exchange diagram):
\beq
\label{in-defn}
&&\langle q_{3} \bar{q}_{4}\vert V^{(2)}_{\mathrm{IN}}(\Lambda) \vert q_{1} \bar{q}_{2}\rangle=
\n && \hspace{2cm}
-128\pi^{3} {\cal P}^{+}\delta^{3}(k_{1}+k_{2}-k_{3}-k_{4})\sqrt{x(1-x)y(1-y)}\n
&&\hspace{3cm}\times\theta ( x-y )\theta ( x-y -\epsilon)
\left[ \frac{\T{a}{3,1}\T{a}{2,4}\delta_{s_{1},s_{3}}\delta_{s_{4},s_{2}}}{(x-y)^{2}}\right] .
\eeq

\section{$V^{(2)}_{\mathrm{NC}}$:  The $T$-matrix method}
\label{v2nc-tmatrix}
In this section we derive the second-order, non-canonical, cutoff-independent part of the reduced interaction.  We 
show that $V^{(2)}_{\mathrm{NC}}$ is the same as the cutoff-independent part of the self-energy 
[Eq.~(\ref{se-ci})] with the opposite sign.  This means the sum of the self-energy contribution and the 
cutoff-independent, non-canonical contributions yield the self-energy contribution before the cutoff-independent 
piece is removed [Eq.~(\ref{cor-ci})].

Two particles that do not interact give a zero forward $T$-matrix element.  
We will set the ``forward part'' of the $T$-matrix to zero to fix the second-order non-canonical matrix elements.  
This means we want a particle of mass $m$ in a state to propagate as a particle of mass $m$.  
In a 
light-front theory, we can write the $T$-matrix to second order as:
\beq
T\left( p^{-}_{1}+p^{-}_{2}\right) =H_{\mathrm{int}}(\Lambda)+H_{\mathrm{int}}(\Lambda)
\frac{1}{p^{-}_{1}+p^{-}_{2}-h}H_{\mathrm{int}}(\Lambda),
\eeq
where:
\beq
H_{\mathrm{int}}(\Lambda)=\frac{{\cal M}^{2}_{\mathrm{int}}(\Lambda)}{\p^{+}},
\eeq
and $h$ is the total free energy of the intermediate state.  So explicitly, we want:
\beq
0=\langle q_{3}\bar{q}_{4}\vert V^{(2)}(\la)\vert q_{1}\bar{q}_{2}\rangle
+\langle q_{3}\bar{q}_{4}\vert V^{(1)}\frac{1}{\p^{+}\left (p_{1}^{-}+p_{2}^{-}-h\right) }
V^{(1)}\vert q_{1}\bar{q}_{2}\rangle.
\eeq
If we write:
\beq
V^{(2)}(\la)=V^{(2)}_{\mathrm{NC}}(\la)+V^{(2)}_{\mathrm{C}}=V^{(2)}_{\mathrm{NC}}(\la)+V^{(2)}_{\mathrm{NC}}
+V^{(2)}_{\mathrm{C}},
\eeq
we see that the first term on the RHS is what we calculated in Sec.~\ref{v2nc}, whereas the second term is what we are 
trying to fix, and the last term contains the instantaneous diagrams.  However, since we are fixing the forward 
$T$-matrix, we only want to include diagrams that have the same delta-function structure as a self-energy.  So more 
explicitly, what we want is:
\beq
0&=&\langle q_{3}\bar{q}_{4}\vert V^{(2)}_{\mathrm{SE}}(\la)\vert q_{1}\bar{q}_{2}\rangle
+\langle q_{3}\bar{q}_{4}\vert V^{(2)}_{\mathrm{NC}}\vert q_{1}\bar{q}_{2}\rangle
\n && \hspace{3cm}
+\langle q_{3}\bar{q}_{4}\vert V^{(1)}\frac{1}{\p^{+}\left( p_{1}^{-}+p_{2}^{-}-h\right)}V^{(1)}\vert q_{1}\bar{q}_{2}\rangle
|_{\mathrm{SE}},
\eeq
where $\langle q_{3}\bar{q}_{4}\vert V^{(2)}_{\mathrm{NC}}\vert q_{1}\bar{q}_{2}\rangle$ is the second-order, 
non-canonical, cutoff-independent reduced interaction.
The third term is similar to the first (before the cutoff-independent part is removed), 
except that relating the first to the third will show that:
\beq
&&\half T_{2}\rightarrow  \frac{e^{-\frac{\Delta_{FK}^{2}}{\la^{4}}}e^{-\frac{\Delta_{KI}^{2}}{\la^{4}}}}
{\p^{+}\left( p_{1}^{-}+p_{2}^{-}-h\right)}=
\frac{e^{-\frac{\Delta_{FK}^{2}}{\la^{4}}}e^{-\frac{\Delta_{KI}^{2}}{\la^{4}}}}
{\p^{+}\left( p_{1}^{-}+p_{2}^{-}-p_{5}^{-}-p^{-}_{6}-p^{-}_{7}\right)}\n
&&\hspace{3cm}
=-\frac{e^{-\frac{\Delta_{FK}^{2}}{\la^{4}}}e^{-\frac{\Delta_{KI}^{2}}{\la^{4}}}}{\Delta_{KI}}
=-\frac{e^{-2\frac{\Delta_{KI}^{2}}{\la^{4}}}}{\Delta_{KI}}=-\half T_{2}|_{\Lambda \hspace{.1cm}\mathrm{terms}}
\eeq
This means the third term will cancel the first term, except for the cutoff independent part in 
Eq. (\ref{se-ci}).  This leaves us with:
\beq
0=\langle q_{3}\bar{q}_{4}\vert V^{(2)}_{\mathrm{NC}}\vert q_{1}\bar{q}_{2}\rangle
+4\frac{\delta_{1,3}\delta_{2,4}}{24\pi^{2}}m^{2}\int dz\frac{1+z^{2}}{zx(1-x)},
\eeq
giving:
\beq
\label{vncse}
&&\langle q_{3}\bar{q}_{4}\vert V^{(2)}_{\mathrm{SE}}(\la)\vert q_{3}\bar{q}_{4}\rangle=
\langle q_{3}\bar{q}_{4}\vert V^{(2)}_{\mathrm{NC_{se}}}(\la)+V^{(2)}_{\mathrm{NC}}\vert q_{3}\bar{q}_{4}\rangle=\n
&&-\frac{\delta_{1,3}\delta_{2,4}}{24\pi^{2}}\la^{2}\left[ \int \frac{z\hspace{.1cm}dz}{1-z}
\left\{ \frac{1+z^{2}}{z}\sqrt{2\pi}\left[\mathrm{erf}\left(\gamma(x,z)\right)
+\mathrm{erf}\left(\gamma(1-x,z)\right)-2\right]\right.\right.\n
&&\hspace{3cm}\left.\left.
+\sqrt{8} \left[ \gamma(x,z) \mathrm{Ei}\left(1,\gamma^{2}(x,z)\right)
\right. \right. \right. \n && \hspace{6cm} \left. \left. \left.
+\gamma(1-x,z) \mathrm{Ei}\left(1,\gamma^{2}(1-x,z)\right) \right] \right\}\right].
\eeq

\chapter{The Basis for the Expansion of Real Meson States}
\label{physical-states}
In this chapter we define the basis we use to approximate the real meson states in Chapter \ref{chapt:me-physical}.  
Before we use these basis states to determine the meson spectrum we 
test them by calculating the glueball spectrum in Section \ref{chapt:glue-calc}.  The basis we use to approximate 
the glueball states is 
so closely related to the basis for the meson states that we do not discuss it in detail.

We start by assuming the meson will contain one quark and one anti-quark.  In this case, an 
expansion of the meson state in the free basis looks like:
\beq
\label{wave-exp}
\left| \Psi^{jn}(P)\right> ={\bf 1}\left| \Psi^{jn}(P)\right> 
\simeq \int D_{1}D_{2}\langle q_{1}\bar{q}_{2}\vert \Psi^{jn}(P)\rangle\vert q_{1}\bar{q}_{2}\rangle,
\eeq
$P$ is the three-momentum of the state, $j$ is the projection of the total spin along the 3-axis, $D_{i}$ is 
defined in Eq.~(\ref{bigD-defn}),  
and $n$ labels the mass eigenvalue of the state.  The derivation of the approximate state culminates with the final 
expression in Eq.~(\ref{phys-state-final}).

Since the color part of the wave function can be separated from 
the spin and momentum parts, we write:
\beq
\left| \Psi^{jn}(P)\right> =\left| \Upsilon_{n}\right> \otimes \left| \Gamma^{jn}(P)\right>,
\eeq
where $\left| \Upsilon_{n}\right> $ is the color part of the state and $\left| \Gamma^{jn}(P)\right> $ is the 
momentum/spin part.  Now we can write:
\beq
\langle q_{1}\bar{q}_{2}\vert \Psi^{jn}(P)\rangle = \langle c_{1},c_{2}\vert \Upsilon_{n}\rangle
\langle p_{1},s_{1},p_{2},s_{2}\vert \Gamma^{jn}(P)\rangle.
\eeq
\section{The Color Basis}
We define the color part of the scalar product as:
\beq
M^{n}_{c_{1},c_{2}}=\langle c_{1},c_{2}\vert \Upsilon_{n}\rangle .
\eeq
Using this, the most general color state can be written:
\beq
\left| \Upsilon_{n}\right> =\sum_{c_{1},c_{2}}M^{n}_{c_{1},c_{2}}\vert q_{c_{1}}\bar{q}_{c_{2}}\rangle.
\eeq
If we apply a color rotation to a color singlet state (which will not effect the scalar part), we see:
\beq
\label{color-rot}
\left| \Upsilon'_{n}\right> &=&\sum_{c_{1},c_{2}}M^{n}_{c_{1},c_{2}}\vert q'_{c_{1}}\bar{q}\hspace{.05cm}'_{c_{2}}\rangle\n
&=&\left| \Upsilon_{n}\right>.
\eeq
From \cite{greiner} a color-rotated quark looks like:
\beq
\vert q'_{c_{1}}\rangle =U\vert q_{c_{1}}\rangle
=\sum_{c_{3}}\vert q_{c_{3}}\rangle\langle q_{c_{3}}\vert U\vert q_{c_{1}}\rangle
=\sum_{c_{3}}U_{c_{3}c_{1}}\vert q_{c_{3}}\rangle.
\eeq
$U=\mathrm{exp}[-i\theta_{b}F^{b}]$ is the quark color rotation operator where $F^{b}=\lambda^{b}/2=T^{b}$ are 
the SU$(3)$ generators for the quark representation, and the $\theta_{b}$ are the color rotation parameters.  The 
rotation operator for the anti-quark representation is given by $U^{\ast}$.  So now we can rewrite Eq. 
(\ref{color-rot}) as:
\beq
\left| \Upsilon'_{n}\right> =\sum_{c_{1},c_{2},c_{3},c_{4}}M^{n}_{c_{1},c_{2}}U_{c_{3}c_{1}}U^{\ast}_{c_{4}c_{2}}
\vert q_{c_{3}}\bar{q}_{c_{4}}\rangle .
\eeq
Now, $U^{\ast}=U^{\dagger^{T}}$, so:
\beq
\left| \Upsilon'_{n}\right> =\sum_{c_{1},c_{2},c_{3},c_{4}}M^{n}_{c_{1},c_{2}}U_{c_{3}c_{1}}U^{\dagger}_{c_{2}c_{4}}
\vert q_{c_{3}}\bar{q}_{c_{4}}\rangle .
\eeq
But $U^{\dagger}=U^{-1}$, so if we use $M^{n}_{c_{1},c_{2}}=\delta_{c_{1},c_{2}}$ then:
\beq
\left| \Upsilon'_{n}\right> &=&\sum_{c_{1},c_{3},c_{4}}U_{c_{3}c_{1}}U^{\dagger}_{c_{1}c_{4}}
\vert q_{3}\bar{q}_{4}\rangle\n
&=&\sum_{c_{4}}\vert q_{c_{4}}\bar{q}_{c_{4}}\rangle.
\eeq
Now we normalize the color wavefunction:
\beq
\left< \Upsilon'_{n} \vert \Upsilon_{n} \right>=N^{2}\sum_{c_{4},c'_{4}}
\left< q_{c'_{4}}\bar{q}_{c'_{4}}\vert q_{c_{4}}\bar{q}_{c_{4}}\right>
=N^{2}\sum_{c_{4},c'_{4}}\delta_{c_{4},c'_{4}}\delta_{c_{4},c'_{4}}=N^{2}\sum_{c_{4}}\delta_{c_{4},c_{4}}=
N_{c}N^{2}.
\eeq
Therefore, the normalized color wave function is:
\beq
\left| \Upsilon_{n} \right> =\frac{1}{\sqrt{N_{c}}}\sum_{c_{1},c_{2}}\delta_{c_{1},c_{2}}\vert q_{c_{1}}\bar{q}_{c_{2}}\rangle.
\eeq

\section{Momentum Conservation and Plane Wave Normalization}

Since the $\Psi$'s are mass eigenstates, and since the mass operator conserves momentum, $\langle q_{1}\bar{q}_{2}\vert 
\Psi^{jn}(P)\rangle$ must be proportional to a momentum-conserving delta function.  We write:
\beq
\langle q_{1}\bar{q}_{2}\vert \Psi^{jn}(P)\rangle = &&
\n && \hspace{-2.5cm}
16\pi^{3}\delta^{(3)}(P-p_{1}-p_{2})\frac{1}{\sqrt{16\pi^{3}}}
\frac{1}{\sqrt{N_{c}}}\theta_{\epsilon}\delta_{c_{1},c_{2}}16\pi^{3}\sqrt{p_{1}^{+}p_{2}^{+}}
\Phi^{jn}_{s_{1},s_{2}}(p_{1},p_{2}),
\eeq
where $\Phi^{jn}_{s_{1},s_{2}}(p_{1},p_{2})$ is the spin/momentum wavefunction and
\beq
\theta_{\epsilon}=\theta(x-\epsilon) \theta(1-x-\epsilon),
\eeq
which is used to cut off infrared divergences.
Now we can rewrite Eq. (\ref{wave-exp}) as:
\beq
\frac{1}{\sqrt{16\pi^{3}}}\frac{1}{\sqrt{N_{c}}}\sum_{s_{1},s_{2},c_{1},c_{2}}
\int \frac{d^{2}\vec{q}\hspace{.1cm}dx}{\sqrt{x(1-x)}}\theta_{\epsilon}\delta_{c_{1},c_{2}}
\Phi^{jn}_{s_{1},s_{2}}(P,x,\vec{q}\hspace{.1cm})\vert q_{1}\bar{q}_{2}\rangle,
\eeq
where we have used the Jacobi variables:
\beq
p_{1}=\left( xP^{+},xP_{\bot}+\vec{q}\hspace{.1cm}\right), \hspace{1cm}
p_{2}=\left( [1-x]P^{+},[1-x]P_{\bot}+\vec{q}\hspace{.1cm}\right),
\eeq
and integrated over $p_{2}$.  $\theta_{\epsilon}$ is used to regulate the light-front infrared divergences.  
These divergences will be explicitly cancelled allowing us to let $\epsilon \rightarrow 0$ before the Hamiltonian 
matrix is diagonalized.

Next, we want the state to have a boost-invariant plane-wave normalization:
\beq
\label{plane-wave-norm}
\left< \Psi ^{j' n'}(P')\vert \Psi^{jn}(P)\right>=16\pi^{3}P^{+}\delta^{3}(P-P')\delta_{j,j'}\delta_{n,n'} .
\eeq
We can explicitly write out the left-hand side as:
\beq
&&-\frac{1}{16\pi^{3}}\frac{1}{N_{c}}
\sum_{s_{1},s_{2},c_{1},c_{2}s'_{1},s'_{2},c'_{1},c'_{2}}
\int\frac{d^{2}\vec{q}\hspace{.1cm}dx\theta_{\epsilon}}{\sqrt{x(1-x)}}
\frac{d^{2}\vec{q}\hspace{.05cm}'\hspace{.1cm}dx'\theta'_{\epsilon}}{\sqrt{x'(1-x')}}
\delta_{c_{1},c_{2}}\delta_{c'_{1},c'_{2}}\n
&&\hspace{3cm} \times
\Phi^{\ast j'n'}_{s'_{1},s'_{2}}(P',x',\vec{q'}\hspace{.1cm})\Phi^{jn}_{s_{1},s_{2}}(P,x,\vec{q}\hspace{.1cm})
\langle q'_{1}\bar{q}'_{2}\vert q_{1}\bar{q}_{2}\rangle,
\eeq
where we use:
\beq
\langle \bar{q}'_{2} q'_{1}\vert q_{1}\bar{q}_{2}\rangle=-\langle q'_{1}\bar{q}'_{2}\vert q_{1}\bar{q}_{2}\rangle .
\eeq
We can write:
\beq
-\langle q'_{1}\bar{q}\hspace{.05cm}'_{2}\vert q_{1}\bar{q}_{2}\rangle&=&\left[ 16\pi^{3}\right] ^{2} p_{1}^{+}
\left( P^{+}-p_{1}^{+}\right) \delta^{3}\left( P-P'\right) \n
&&\hspace{3cm} \times
\delta^{3}\left( p_{1}-p'_{1}\right)
\delta_{s_{1},s'_{1}}\delta_{s_{2},s'_{2}}\delta_{c_{1},c'_{1}}\delta_{c_{2},c'_{2}}\n
&=&\left[ 16\pi^{3}\right] ^{2} P^{+} x(1-x) \delta^{3}\left( P-P'\right)\delta\left( x-x'\right)
\delta^{2}\left( \vec{q}\hspace{.05cm}'-\vec{q}\right) \n
&&\hspace{5cm} \times
\delta_{s_{1},s'_{1}}\delta_{s_{2},s'_{2}}\delta_{c_{1},c'_{1}}\delta_{c_{2},c'_{2}}.
\eeq
So now if we integrate over $\vec{q}\hspace{.05cm}'$ and $x$, and complete all the sums except $s_{1}$ and $s_{2}$, we get:
\beq
16\pi^{3}P^{+}\delta^{3}(P-P')\sum_{s_{1},s_{2}}\int d^{2}\vec{q}\hone dx \hone\theta_{\epsilon}
\Phi^{\ast j'n'}_{s_{1},s_{2}}(P,x,\vec{q}\hspace{.1cm})\Phi^{jn}_{s_{1},s_{2}}(P,x,\vec{q}\hspace{.1cm}).
\eeq
This implies:
\beq
\delta_{j,j'}\delta_{n,n'}=\sum_{s_{1},s_{2}}\int d^{2}\vec{q}\hone dx \hone\theta_{\epsilon}
\Phi^{\ast j'n'}_{s_{1},s_{2}}(P,x,\vec{q}\hspace{.1cm})\Phi^{jn}_{s_{1},s_{2}}(P,x,\vec{q}\hspace{.1cm}).
\eeq
Since $\left| \Psi^{jn}(P)\right>$ is uniquely determined by the equation:
\beq
\p^{-}(\la)\left| \Psi^{jn}(P)\right>=\frac{\vec{P}_{\bot}^{\hone 2}+M^{2}_{n}}{P^{+}}\left| \Psi^{jn}(P)\right> ,
\eeq
which can be written:
\beq
{\cal M}^{2}(\la)\left| \Psi^{jn}(P)\right>=M^{2}_{n}\left| \Psi^{jn}(P)\right>,
\eeq
$\Phi^{jn}_{s_{1},s_{2}}(P,x,\vec{q}\hone)$ is uniquely determined by ${\cal M}^{2}(\la)$.  Since the 
free-state matrix elements of ${\cal M}^{2}(\la)$ are independent of the total momentum, so is 
$\Phi^{jn}_{s_{1},s_{2}}(P,x,\vec{q}\hone)$.  Then the normalization condition is:
\beq
\label{norm-condition}
\delta_{j,j'}\delta_{n,n'}=\sum_{s_{1},s_{2}}\int d^{2}\vec{q}\hone dx \hone\theta_{\epsilon}
\Phi^{\ast j'n'}_{s_{1},s_{2}}(x,\vec{q}\hspace{.1cm})\Phi^{jn}_{s_{1},s_{2}}(x,\vec{q}\hspace{.1cm}),
\eeq
and the state is given by:
\beq
\left| \Psi^{jn}(P)\right>=\frac{1}{\sqrt{16\pi^{3}}}\frac{1}{\sqrt{N_{c}}}\sum_{s_{1},s_{2},c_{1},c_{2}}
\int \frac{d^{2}\vec{q}\hspace{.1cm}dx}{\sqrt{x(1-x)}}\theta_{\epsilon}\delta_{c_{1},c_{2}}
\Phi^{jn}_{s_{1},s_{2}}(x,\vec{q}\hspace{.1cm})\vert q_{1}\bar{q}_{2}\rangle.
\eeq

\section{Fermions, Charge Conjugation, and Exchange Symmetry}
\label{meson-particle-exchange}

Charge conjugation is a good quantum number because it is a symmetry of the strong interaction.  We show \cite{weinberg}
that applying the charge conjugation operator, $C$, gives the eigenvalue $\pm 1$, depending on whether or not the spin/momentum 
wavefunction is symmetric or anti-symmetric under particle exchange.

Using some shorthand, we can write the meson state as:
\beq
\vert \Psi \rangle = \delta_{f_{1},f_{2}}\delta_{c_{1},c_{2}}
\sum_{s_{1} s_{2}}\int d\vec{k} d\vec{k'}\hspace{.1cm}\Phi_{s_{1} s_{2}}(\vec{k},\vec{k'})
\hspace{.1cm} \bd_{1} \dd_{2} \vert 0 \rangle ,
\eeq
where $\Phi_{s_{1} s_{2}}(\vec{k},\vec{k'})$ is the spin/momentum part of the wavefunction.  $f_{i}$ and $c_{i}$ 
are the flavor and color indices, respectively.  Next, we act on the state 
with $C$, giving:
\beq
C \vert \Psi \rangle &=& \delta_{f_{1},f_{2}}\delta_{c_{1},c_{2}}\sum_{s_{1} s_{2}}\int d\vec{k} d\vec{k'}
\hspace{.1cm}\Phi_{s_{1} s_{2}}(\vec{k},\vec{k'}) \hspace{.1cm} \dd_{1} \bd_{2} \vert 0 \rangle \n
&=& -\delta_{f_{1},f_{2}}\delta_{c_{1},c_{2}}\sum_{s_{1} s_{2}}\int d\vec{k} d\vec{k'}
\hspace{.1cm}\Phi_{s_{1} s_{2}}(\vec{k},\vec{k'}) \hspace{.1cm} \bd_{2} \dd_{1} \vert 0 \rangle \n
&=& -\delta_{f_{1},f_{2}}\delta_{c_{1},c_{2}}\sum_{s_{1} s_{2}}\int d\vec{k'} d\vec{k}
\hspace{.1cm}\Phi_{s_{2} s_{1}}(\vec{k'},\vec{k}) \hspace{.1cm} \bd_{1} \dd_{2} \vert 0 \rangle ,
\eeq
where the particle indices are switched in the last step.  This means that if $\Phi$ is symmetric under 
particle exchange ($s_{1}\leftrightarrow s_{2}$, $\vec{k}\leftrightarrow\vec{k'}$) then the state is odd under charge conjugation, and if $\Phi$ is antisymmetric under particle 
exchange, then the state is even under charge conjugation.

Thus, if we consider symmetric charge conjugation states, then the spin-momentum wavefunction must be 
anti-symmetric under exchange, and an anti-symmetric charge conjugation state must have a symmetric spin-momentum 
wavefunction under exchange.

\section{Momentum and Spin Wavefunction Bases}
The Spin-Momentum Wavefunction is given by:
\beq
\Phi^{jn}_{s_{1}s_{2}}(x,\vec{k}_{\perp})=\sum_{q=1}^{4} \chi^{s_{1}s_{2}}_{q} \Omega^{jn}_{q}(x,\vec{k}_{\bot}).
\eeq
The $\chi$'s are given by:
\beq
\chi^{s_{1}s_{2}}_{1}&=&\delta_{s_{1},\half}\delta_{s_{2},\half}\n
\chi^{s_{1}s_{2}}_{2}&=&\delta_{\bar{s}_{1},\half}\delta_{\bar{s}_{2},\half}\n
\chi^{s_{1}s_{2}}_{3}&=&\frac{1}{\sqrt{2}}[\delta_{s_{1},\half}\delta_{\bar{s}_{2},\half}+\delta_{\bar{s}_{1},\half}\delta_{s_{2},\half}]\n
\chi^{s_{1}s_{2}}_{4}&=&\frac{1}{\sqrt{2}}[\delta_{s_{1},\half}\delta_{\bar{s}_{2},\half}-\delta_{\bar{s}_{1},\half}\delta_{s_{2},\half}],
\eeq
where $\bar{s}=-s$, and they obey the relation
\beq
\sum_{s_{1}s_{2}}\chi^{s_{1}s_{2}}_{q}\chi^{s_{1}s_{2}}_{q'}=\delta_{q,q'}.
\eeq
The momentum wave function is expanded in a complete basis:
\beq
\Omega^{jn}_{q}(x,\vec{k}_{\perp})=\sum_{l=-m_{l}+1}^{k_{l}-1}\sum_{t=-m_{t}+3}^{k_{t}}\sum_{a=-\infty}^{\infty}
\bar{R}^{jn}_{qlta} B_{l}(x) \tilde{B}_{t}(k) A_{a}(\phi),
\eeq
where $m_{t}$, $k_{t}$, $m_{l}$, and $k_{l}$ are the normal B-spline parameters for the transverse and 
longitudinal basis functions.\footnote{The normal range of the B-spline index is from $-m$ to $k$.  However, two 
longitudinal B-splines ($l=-m_{l},k_{l}$) and three transverse B-splines ($t=-m_{t},-m_{t}+1,-m_{t}+2$) are discarded because they produce 
divergent kinetic energies.}  The $A_{a}(\phi)$ are the basis functions for the transverse-angular degree of 
freedom, the $B_{l}(x)$ are the basis functions for the longitudinal-momentum degree of freedom 
(Sec.~\ref{longitudinal-functions}), and the $\tilde{B}_{t}(k)$ are the basis functions for transverse-momentum 
degree of freedom (Sec.~\ref{transverse-functions}).

The transverse-angular basis functions are given by:
\beq
A_{a}(\phi)=\frac{1}{\sqrt{2\pi}}e^{ia\phi}
\eeq
and have the normalization:
\beq
\int_{0}^{2\pi}d\phi A^{*}_{a'}(\phi) A_{a}(\phi)=\delta_{a,a'}.
\eeq

\subsection{Longitudinal Basis Functions}
\label{longitudinal-functions}
The longitudinal states are functions of $x$, the longitudinal momentum fraction carried by one particle.  The symmetry 
of the problem allows us to choose functions that are symmetric or anti-symmetric under particle exchange $(x 
\leftrightarrow 1-x )$.
  B-splines are not symmetric 
functions about $x=\frac{1}{2}$ a priori.  The longitudinal functions must be symmetric and anti-symmetric combinations of the 
B-splines.  We choose knots that are symmetric about $x=\frac{1}{2}$, so that pairs of B-splines are also symmetric.  For example,
\beq
B^{(m,k)}_{-m}(x)&=&B^{(m,k)}_{k}(1-x) , \n
B^{(m,k)}_{-m+1}(x)&=&B^{(m,k)}_{k-1}(1-x).
\eeq
If there are an even number of B-splines they can all be paired in this manner.  If there are an odd number, the 
``middle''spline with index $l=\bar{l} \equiv \frac{m+k+1}{2}-1$ is, by itself, symmetric about $x=\frac{1}{2}$.  We 
choose to always use an even number of splines so there are an equal number of symmetric and anti-symmetric functions.  
Otherwise, we would expect the basis to better approximate symmetric states than anti-symmetric ones.  We will write 
the symmetric B-splines as:
\beq
\bar{B}_{l\hspace{.1cm}\mathrm{symm}}^{(m,k)}(x)=B_{l}^{(m,k)}(x)+B_{k-m-l}^{(m,k)}(1-x) ; \hspace{1cm} l \le \bar{l},
\eeq
and the anti-symmetric ones as:
\beq
\bar{B}_{l\hspace{.1cm}\mathrm{asymm}}^{(m,k)}(x)=B_{l}^{(m,k)}(x)-B_{k-m-l}^{(m,k)}(1-x) ; \hspace{1cm} l > \bar{l}.
\eeq
We drop the `(a)symm' notation, and all 
B-splines are implicitly (anti)symmeterized as determined by their index.  m and k are the usual indices 
associated with the splines used in Chapter \ref{section:bsplines} and $l$ is used in place of $i$ to signify it is the 
index for a longitudinal function.  We use degenerate knots 
such that $t_{-m},\ldots ,t_{0}=a$ and $t_{k+1},\ldots ,t_{k+m+1}=b$, and we will not use the splines $B_{-m}$ 
and $B_{k}$ because they do not have the correct behavior\footnote{The basis functions must vanish at $x=0,1$ to 
ensure finite kinetic energy.} as $x\rightarrow 0,1$.

Finally the longitudinal states obey the normalization:
\beq
\int_{0}^{1}dx \vert B_{l}(x) \vert ^{2}=1.
\eeq

\subsection{Transverse Basis Functions}
\label{transverse-functions}
We use $t$ as the index for the transverse basis functions.  However, 
the knots that determine each B-spline are designated by $t_{i}$.  Therefore, if a $t$ has a subscript, it refers to a 
knot, otherwise it is a transverse function index.

The magnitude of the relative transverse momentum in a state can be in the range $[0,\infty )$.  
However, to compute integrals numerically we need to restrict this to a finite range.  We change variables:
\beq
\label{ch-of-var}
x=\frac{2}{z+1}-1.
\eeq
The range of z is restricted to $[-1,1]$.  This finite range should lend itself well to using B-splines, as should 
the fact that the integrand under this change of variables tends to be reasonably smooth.  We want a set of 
basis states that after this change of variables produces B-splines.  So let us start at the end, and work 
backwards to develop this set of basis states.

We use degenerate knots such that 
\beq
t_{-m},\ldots ,t_{0}=a
\eeq
and 
\beq t_{k+1},\ldots ,t_{k+m+1}=b.
\eeq
We are interested in the range $z \in [-1,1]$.  Choose the knots $t_{-m}, \ldots ,t_{0}=-1, t_{1},\ldots 
,t_{k+1}=1, \ldots ,t_{k+m+1}$, with the corresponding set of B-splines $\{ B_{-m},\ldots ,B_{k}\}$.  Now spread 
these states out over the range $[0, \infty )$ using the change of variables in Eq. (\ref{ch-of-var}).  The 
knot distribution is changed so that
\beq
t_{i}^{\prime}=\frac{2}{1+t_{i}}-1.
\eeq
This smeared set of splines will be 
labeled $\tilde{B}_{t}$, $t \in [-m,k]$.  The kinetic energy part of the 
Hamiltonian after the change of variables is:
\beq
2 \delta_{q,q^{\prime}} \Lambda^{2} \int_{0}^{1} dx \frac{B_{l}^{m,k}(x)B_{l'}^{m,k}(x)}{x(1-x)}
\int_{-1}^{1}\frac{dy}{(y+1)^{2}}\left( \frac{2}{y+1}-1\right) ^{3} 
B_{t}^{m,k}(y) B_{t'}^{m,k}(y),
\eeq
showing that as $\lim_{y\rightarrow -1}$, $B_{t}^{m,k}(y) B_{t'}^{m,k}(y)$ must die faster than 
$(1+y)^{4}$.  This requires the use of splines of order $m \ge 3$.  However we must also 
consider the power series representation of the B-spline, $B_{t}^{m}$ (which is non-zero between $t_{0}$ and $t_{1}$).
We write it as:
\beq
B_{t}^{m}(y)=\sum_{p=m+t}^{m} a_{p}(1+y)^{p},
\eeq
where $t$ can be negative, and this form can best be understood by considering Eq. (\ref{poly-bspline}).
This means that to ensure there is no divergence, the lowest B-spline we can keep is $B_{t \ge 3-m}^{m}(y)$.
The transverse basis functions obey the normalization:
\beq
\int_{0}^{\infty} dk k \vert \tilde{B}_{t}(k) \vert ^{2}=1.
\eeq

We write the approximate meson state, built from our basis functions as:
\beq
\label{phys-state-final}
\left\vert \Psi^{jn}(P)\right \rangle=\frac{1}{N} \sum_{qlt}R^{jn}_{qlt}\vert q,l,t,j \rangle,
\eeq
where
\beq
\label{qltj-def}
\vert q,l,t,j \rangle&=&\frac{1}{\sqrt{16 \pi^{3}}}\frac{1}{\sqrt{N_{c}}}
\sum_{s_{1},s_{2},c_{1},c_{2}}\int \frac{d^{2}\vec{k} \hspace{.1cm} dx}{\sqrt{x(1-x)}}\theta_{\epsilon}\delta_{c_{1},c_{2}}
\chi_{q}^{s_{1}s_{2}}B_{l}(x)\tilde{B}_{t}(k)\n
&&\times A_{j-s_{1}-s_{2}}(\phi) \vert q_{1}\bar{q}_{2}\rangle.
\eeq
Finally, $N$ is used to ensure plane wave normalization (Eq.~\ref{plane-wave-norm}).  

\section{Restricting the Spin-Momentum Wavefunction}
\label{meson-wavefunction-restrictions}

If we consider states with different charge conjugation separately, we can 
restrict the spin-momentum wavefunction to 
be either symmetric or anti-symmetric under exchange.  For a negative charge conjugation state,
it is necessary for the spin/momentum wave function to be symmetric under particle exchange.                                                      
As described in section \ref{longitudinal-functions}, the longitudinal states, $B_{l}(x)$, are symmetric under $x\rightarrow 1-x$                                                      
if $l \le \bar{l}$ and anti-symmetric if $l > \bar{l}$ (recall $\bar{l}$ refers to the ``middle'' spline).  If $a$ is 
even, then the angular state is symmetric under exchange, 
and odd if $a$ is odd.\footnote{By requiring the approximate meson wavefunction to be an eigenstate of rotations about the 3-axis, 
Allen \cite{brentD} proved that $a=j-s_{1}-s_{2}$, where $j$ is the projection of the total spin along the 3-axis.}  Finally, if $q=1,2,3$ in the spin state it is symmetric 
under exchange, and it is anti-symmetric                                                    
if $q=4$.  Thus, for the entire spin/momentum wavefunction to be symmetric we must have one of the following                                                          
conditions: 

For j even:                                                                                                                                                           
\beq                                                                                                                                                                  
q&=&1,2,4 \rightarrow l > \bar{l}, \n                                                                                                                                  
q&=&3 \rightarrow l \le \bar{l} .                                                                                                                                     
\eeq                                                                                                                                                                  
                                                                                                                                                                      
For j odd:                                                                                                                                                            
\beq                                                                                                                                                                  
q&=&1,2,4 \rightarrow l \le \bar{l}, \n                                                                                                                                
q&=&3 \rightarrow l > \bar{l} .                                                                                                                                       
\eeq 
                                                                                                                                                                      
Next for positive charge conjugation states, we want the total spin-momentum wavefunction to be odd under exchange, 
giving:

For j even:                                                                                                                                                           
\beq                                                                                                                                                                  
q&=&1,2,4 \rightarrow l \le \bar{l}, \n                                                                                                                                
q&=&3 \rightarrow l > \bar{l} .                                                                                                                                       
\eeq                                                                                                                                                                  
                                                                                                                                                                      
For j odd:                                                                                                                                                            
\beq                                                                                                                                                                  
q&=&1,2,4 \rightarrow l > \bar{l}, \n                                                                                                                                  
q&=&3 \rightarrow l \le \bar{l} .                                                                                                                                     
\eeq 

\section{Meson Overlap Matrix}

Since the B-splines are non-orthogonal, we need to calculate the overlap matrix.  We use this section to 
carefully derive the form of the overlap matrix and find relations that will be useful in the two-dimensional 
Hamiltonian integrals.

A free state with spin-state index $q$, and spin $j$ will be written:
\beq
\vert q,l,t,j\rangle &=& \frac{1}{\sqrt{16 \pi^{3}}}\frac{1}{\sqrt{N_{c}}}\sum_{s_{1},s_{2},c_{1},c_{2}}\int d^{2}k_{\perp} dx
\theta_{\epsilon} \delta_{c_{1},c_{2}}\chi_{q}^{s_{1}s_{2}}
\n && \hspace{2cm} \times
\frac{B_{l}(x)}{\sqrt{x(1-x)}}\tilde{B}_{t}(k) A_{j-s_{1}-s_{2}}(\phi)
\vert q_{1} \bar{q}_{2}\rangle.
\eeq
Then:
\beq
\langle q',l',t',j' \vert q,l,t,j \rangle &=& \frac{1}{16 \pi^{3}}\frac{1}{N_{c}} 
\sum_{s_{1},s_{2},s_{1'},s_{2'},c_{1},c_{2},c_{1'},c_{2'}} \int d^{2}k_{\perp} d^{2} k'_{\perp} dx dx' 
\theta_{\epsilon} \theta_{\epsilon'}\n
&&\times \delta_{c_{1},c_{2}}\delta_{c_{1'},c_{2'}} \chi_{q}^{s_{1}s_{2}} \chi_{q'}^{s_{1'}s_{2'}}
A^{*}_{j-s_{1}-s_{2}}(\phi) A_{j'-s_{1'}-s_{2'}}(\phi')\n
&& \hspace{0cm}\times \frac{B_{l}(x) B_{l'}(x')}{\sqrt{x(1-x)x'(1-x')}} \tilde{B}_{t}(k) \tilde{B}_{t'}(k') 
\langle q_{1'} \bar{q}_{2'}\vert q_{1} \bar{q}_{2}\rangle
\eeq
where
\beq
\langle q_{1'} \bar{q}_{2'}\vert q_{1} \bar{q}_{2}\rangle =\delta_{1,1'}\delta_{2,2'}.
\eeq
So this gives us:
\beq
&&\frac{1}{16 \pi^{3}}\frac{1}{N_{c}} 
\sum_{s_{1},s_{2},s_{1'},s_{2'},c_{1},c_{2},c_{1'},c_{2'}} \int d^{2}k_{\perp} d^{2} k'_{\perp} dx dx' 
\theta_{\epsilon} \theta_{\epsilon'}\n
&&\hspace{1in}\times \delta_{c_{1},c_{2}}\delta_{c_{1'},c_{2'}} \chi_{q}^{s_{1}s_{2}} \chi_{q'}^{s_{1'}s_{2'}}
A^{*}_{j-s_{1}-s_{2}}(\phi) A_{j'-s_{1'}-s_{2'}}(\phi')\n
&& \hspace{1.5in}\times \frac{B_{l}(x) B_{l'}(x')}{\sqrt{x(1-x)x'(1-x')}} \tilde{B}_{t}(k) \tilde{B}_{t'}(k') \delta_{1,1'}\delta_{2,2'},
\eeq
where:
\beq
\delta_{1,1'}=16 \pi^{3} p_{1}^{+}\delta^{(3)}(p_{1}-p_{1'})\delta_{c_{1},c_{1'}}\delta_{s_{1},s_{1'}}.
\eeq
Our choice of basis functions guarantees finite kinetic energy, so we can drop the infrared regulator, 
$\theta_{\epsilon}$.
Next, we can use:
\beq
\sum_{c_{1},c_{2},c_{1'},c_{2'}} \delta_{c_{1},c_{2}}\delta_{c_{1'},c_{2'}}\delta_{c_{1},c_{1'}}\delta_{c_{2},c_{2'}}
=\sum_{c_{1},c_{1'}}\delta_{c_{1},c_{1'}}\delta_{c_{2},c_{2'}}
=\sum_{c_{1}}\delta_{c_{1},c_{1}}
=N_{c},
\eeq
and
\beq
&&\sum_{s_{1},s_{2},s_{1'},s_{2'}}\chi_{q}^{s_{1}s_{2}}\chi_{q'}^{s_{1'}s_{2'}}\delta_{s_{1},s_{1'}}\delta_{s_{2},s_{2'}}
A^{*}_{j-s_{1}-s_{2}}(\phi) A_{j'-s_{1'}-s_{2'}}(\phi')\n
&&\hspace{1in}=\sum_{s_{1},s_{2}}\chi_{q}^{s_{1}s_{2}}\chi_{q'}^{s_{1}s_{2}}A^{*}_{j-s_{1}-s_{2}}(\phi) A_{j'-s_{1}-s_{2}}(\phi')\n
&&\hspace{1in}=\delta_{q,q'}A_{j-s_{1}-s_{2}}(\phi) A^{*}_{j'-s_{1}-s_{2}}(\phi').
\eeq
Finally, we will also use:
\beq
&&p_{1}^{+}p_{2}^{+}\delta^{(3)}(p_{1}-p_{1'})\delta^{(3)}(p_{2}-p_{2'})=p_{1}^{+}p_{2}^{+}\delta(p_{1}^{+}-p_{1'}^{+})
\delta^{(2)}(p_{1\perp}-p_{1'\perp})\delta^{(3)}(p_{2}-p_{2'})\n
&&={\cal P}^{+}x(1-x)\delta(x-x')\delta^{(2)}([x-x']{\cal P}_{\perp}+[k_{\perp}-k'_{\perp}])
\delta^{(3)}([p_{1}+p_{2}]-[p_{1'}+p_{2'}])\n
&&={\cal P}^{+}x(1-x)\delta(x-x')\delta^{(2)}(k_{\perp}-k'_{\perp})\delta^{(3)}({\cal P}-{\cal P'}).
\eeq
This leaves us with:
\beq
\label{over-bef-change}
&&16 \pi^{3} {\cal P}^{+}\delta_{q,q'}\delta^{(3)}({\cal P}-{\cal P'})
\int d^{2}k_{\perp}d^{2}k'_{\perp}dx dx' \delta(x-x')\delta^{(2)}(k_{\perp}-k'_{\perp})\n
&&\hspace{1in}\times B_{l}(x) B_{l'}(x') \tilde{B}_{t}(k) 
\tilde{B}_{t'}(k') A^{*}_{j-s_{1}-s_{2}}(\phi) A_{j'-s_{1}-s_{2}}(\phi')\n
&&=16 \pi^{3} {\cal P}^{+}\delta_{q,q'}\delta^{(3)}({\cal P}-{\cal P'})\int d^{2}k_{\perp} dx \n
&&\hspace{1in}\times B_{l}(x) B_{l'}(x) \tilde{B}_{t}(k) \tilde{B}_{t'}(k)
A^{*}_{j-s_{1}-s_{2}}(\phi) A_{j'-s_{1}-s_{2}}(\phi)\n
&&=16 \pi^{3} {\cal P}^{+}\delta_{q,q'}\delta_{j,j'}\delta^{(3)}({\cal P}-{\cal P'})
\int dk k dx B_{l}(x) B_{l'}(x) \tilde{B}_{t}(k) \tilde{B}_{t'}(k).
\eeq
There are three parameters with dimensions of mass in the meson problem, the particle mass, $m$, the relative 
transverse momentum, $k$, and the cutoff $\Lambda$.  We want to write the transverse momentum in units of the 
cutoff so the transverse B-splines will have a dimensionless argument.  For the overlap in 
Eq.~(\ref{over-bef-change}) to have the correct dimensions, the B-splines of the dimensionful parameter, $k$, must 
have dimensions $\left[ \frac{1}{\Lambda} \right]$.  If we introduce the dimensionless B-spline $\hat{B}_{t}(\hat{k})$ 
with dimensionless argument $\hat{k}$ so that:
\beq
\frac{1}{\Lambda}\hat{B}_{t}(\hat{k})=B_{t}(k),
\eeq
we can let
\beq
dk k \tilde{B}_{t}(k)\tilde{B}_{t'}(k) \rightarrow d\hat{k} \hat{k} \hat{B}_{t}(\hat{k})\hat{B}_{t'}(\hat{k}).
\eeq
This change adds extra notation that complicates expressions, although the expression has not changed.  We drop the 
`hat' from $\tilde{B}_{t}$ and $k$ and all B-splines are dimensionless and all transverse momenta are in units of the cutoff.
We now write the overlap:
\beq
\langle q',l',t',j' \vert q,l,t,j \rangle &=& \n
&&\hspace{-1in}16 \pi^{3} {\cal P}^{+}\delta_{q,q'}\delta_{j,j'}\delta^{(3)}({\cal P}-{\cal P'})
\int dk k dx B_{l}(x) B_{l'}(x) \tilde{B}_{t}(k) \tilde{B}_{t'}(k) .
\eeq
This also shows that the $\tilde{B}(k)$ have 
dimensions of inverse transverse momentum.  When we make the change of variables:
\beq
k=\frac{2}{y+1}-1,
\eeq
we find:
\beq
\langle q',l',t',j' \vert q,l,t,j \rangle &=& \n
&&\hspace{-1in}16 \pi^{3} {\cal P}^{+}\delta_{q,q'}\delta_{j,j'}\delta^{(3)}({\cal P}-{\cal P'})
(2)\int_{-1}^{1} \frac{dy}{(1+y)^{2}} \left(\frac{2}{1+y}-1\right) \n
&&\hspace{1in}\times B_{t}(y) B_{t'}(y) \int_{0}^{1} dx B_{l}(x) B_{l'}(x). 
\eeq
Finally, it is conventional to drop the factor $16 \pi^{3} {\cal P}^{+}\delta^{(3)}({\cal P}-{\cal P'})$ as it 
appears in all matrix elements.  So this gives us the final form:
\beq
\label{meson-overlap}
\langle q',l',t',j' \vert q,l,t,j \rangle &=& \n
&&\hspace{-1.2in}2\delta_{q,q'}\delta_{j,j'}
\int_{-1}^{1} \frac{dy}{(1+y)^{2}} \left(\frac{2}{1+y}-1\right) B_{t}(y) B_{t'}(y) 
\int_{0}^{1} dx B_{l}(x) B_{l'}(x). 
\eeq
\section{$q\bar{q}$ Eigenvector Normalization}
\label{qqb-evect-norm}
For our state to have the normalization in Eq.~(\ref{plane-wave-norm}), 
it is necessary for the momentum part of the wave function to obey the 
relation:
\beq
\label{meson-evect-norm}
\int d^{2}k_{\perp}dx \theta(x) \theta(1-x) \sum_{s_{1}s_{2}}\left\vert 
\Phi^{jn}_{s_{1}s_{2}}(x,\vec{k}_{\perp})\right\vert^{2}=1,
\eeq
where
\beq
\left\vert \Phi^{jn}_{s_{1}s_{2}}(x,\vec{k}_{\perp})\right\vert ^{2}&=&
\frac{1}{N^{2}} \sum_{q,q'=1}^{4}\hspace{.1cm}\sum_{l,l'=-m_{l}+1}^{k_{l}-1}\hspace{.1cm}
\sum_{t,t'=-m_{t}+3}^{k_{t}}\hspace{.1cm}\sum_{a,a'=0}^{\infty} \chi^{s_{1}s_{2}}_{q}\chi^{s_{1}s_{2}}_{q'}\n
&&\hspace{.5cm}\times \bar{R}^{jn}_{qlta}\bar{R}^{jn\hspace{.1cm}\ast}_{q'l't'a'}B_{l'}(x)B_{l}(x)\tilde{B}_{t}(k)\tilde{B}_{t'}(k)
A^{*}_{a'}(\phi)A_{a}(\phi),
\eeq
and
\beq
\bar{R}^{jn}_{qlta}=R^{jn}_{qlt}\left[\delta_{q,1}\delta_{a,j-1}+\delta_{q,2}\delta_{a,j+1}
+\delta_{q,3}\delta_{a,j}+\delta_{q,4}\delta_{a,j}\right],
\eeq
which is derived in \cite{brentD} based on symmetry about the 3-axis (there is a subtle 
difference for quarks, since the quarks have spin $\pm \frac{1}{2}$ instead of $\pm 1$).  After some algebra, we 
find:  
\beq
N=\left\{\sum_{q=1}^{4}\sum_{l,t,l',t'} R^{jn}_{qlt} R^{jn\hspace{.1cm}\ast}_{ql't'} {\cal O}^{tt'}_{ll'}\right\}^{-\frac{1}{2}},
\eeq
where
\beq
{\cal O}^{tt'}_{ll'}=\int_{0}^{1} dx B_{l}(x) B_{l'}(x)\int_{0}^{\infty} dk k \tilde{B}_{t}(k)\tilde{B}_{t'}(k),
\eeq
and the $R^{jn}_{qlt}$ are the eigenvector elements determined by diagonalizing the Hamiltonian matrix.  The 
calculation of $N$ also leads to the expression for the spin-averaged dimensionless probability density:
\beq
\label{meson-prob-density}
\Pi(x,k/\Lambda)&=&2\pi \Lambda k \sum_{s_{1}s_{2}}\left \vert \Phi^{jn}_{s_{1}s_{2}}(x,\vec{k}_{\perp})\right\vert ^{2}\n
&=&\frac{\Lambda k}{N^{2}} \sum_{q=1}^{4}\sum_{l,t}\sum_{l't'}R^{jn}_{qlt}R^{jn\hspace{.1cm}\ast}_{ql't'}B_{l}(x)B_{l'}(x)\tilde{B}_{t}(k)\tilde{B}_{t'}(k)\n
&=&\frac{\Lambda k}{N^{2}} \sum_{q=1}^{4}\left \vert \sum_{l,t} R^{jn}_{qlt} B_{l}(x) \tilde{B}_{t}(k)\right \vert ^{2}.
\eeq

\section{Testing the Basis:  Glueball Masses}
\label{chapt:glue-calc}

The calculation of the glueball spectrum in \cite{brentB,brentD} uses oscillating, orthogonal 
basis functions that may be a source of round-off error in the numerical calculation of matrix elements.  In addition 
it is likely their oscillations slow the convergence of the Monte-Carlo integration.  Our approximate meson states 
use B-spline basis functions (Chapter \ref{section:bsplines}) for the 
longitudinal- and transverse-momentum degrees of freedom because we believe they will avoid round-off error and 
their simple structure may speed the convergence of the Monte-Carlo integration.  Before we commit to deriving the 
necessary meson matrix 
elements in our basis, we check to see if there is any improvement in the calculation of the 
glueball spectrum using our longitudinal- and transverse-momentum basis functions defined in Sections 
\ref{longitudinal-functions} and \ref{transverse-functions}, respectively. 

We briefly describe the calculation of the glueball mass spectrum in pure-glue QCD.  
Then in Section \ref{glue-results}, we write the orthogonal functions used 
in the previous glueball calculation, followed by a comparison of our results with those in \cite{brentB,brentD}. 

The second-order pure-glue QCD Hamiltonian matrix with $gg$ external states is similar to the second-order full 
QCD Hamiltonian matrix with $q\bar{q}$ external states.  They are so similar that the same diagrams and 
interactions shown in Section \ref{mass-o2-begin}, with quarks and antiquarks replaced with gluons, 
describe the glueball system.

The coupling is the only fundamental parameter in pure-glue QCD.  Thus we only need to determine the coupling and 
cutoff that most accurately reproduce the glueball spectrum.\footnote{There is no real experimental data for 
glueballs.  Lattice gauge calculations are the most trusted source of data for the glueball system, so we compare 
our results with Ref. \cite{lattice-glue}.}  
We make several calculations with different values of the coupling.  Each calculation produces eigenvalues 
of the renormalized invariant-mass operator, ${\cal 
M}^{2}(\Lambda)$.  These eigenvalues are used to determine the coupling and cutoff. Our 
procedure should produce approximate cutoff-independent results.  The optimal value of the coupling is determined 
by first finding the range of couplings where the results are the most cutoff-independent and then within this range, 
finding the value of the coupling that reproduces the expected degeneracies for different values of $j$, the 
projection of the total spin along the 3-axis.

Once the coupling is determined, we find the cutoff by fixing the first excited glueball state mass using the relation:  
\beq
\Lambda^{2}=\frac{m^{2}_{\mathrm{measured}}}{\langle i \vert {\cal M}^{2} (\Lambda) \vert i \rangle},
\eeq
where $m^{2}_{\mathrm{measured}}$ is the measured mass-squared for the state.

\section{Glueball Results}
\label{glue-results}
In this section we show that our results for the glueball mass spectrum is consistent with \cite{brentB,brentD}.  
We begin by simply stating the basis functions used in the previous calculation (Section \ref{orthog-funcs}) 
followed by a presentation of the results.   We use $k1$ to 
designate the B-spline parameter $k$ for the longitudinal B-splines and $k2$ for the transverse B-splines.
Section \ref{convergence-testing} shows that all desired results have converged when the basis 
parameters are $k1=k2=6$.  This means there are a total of 7 transverse states, 8 longitudinal states (4 symmetric 
and 4 antisymmetric) and 4 spin states, for a 
total of 112 basis states.  All plots except for the ones showing this convergence use these 112 basis states.   Some 
matrix elements are zero by definition due to spin-flip restrictions.\footnote{For example, there are no up-up to 
down-down interactions.}  The four spin states split the Hamiltonian into sixteen blocks for each combination of 
incoming and outgoing spin.  If $j$, the total spin along the $z$ axis is not zero, there are only two blocks that 
are zero.  However, if $j$ is zero, then there are eight more blocks that are either zero or identical to another 
block.  Thus, only six of the sixteen blocks may need to be determined.\footnote{The Hamiltonian is also Hermitian 
which further decreases the number of unique elements.}

We want to make clear that although we do not include as many results as are given in \cite{brentB,brentD}, we are 
able to reproduce all of the previous results. 

\subsection{Orthogonal Basis Functions}
\label{orthog-funcs}

The difference between the basis functions we use and those found in \cite{brentB,brentD} are the functions used 
to represent the longitudinal- and transverse-momentum degrees of freedom.  We use functions based on B-splines 
that are defined in  Sections \ref{longitudinal-functions} and \ref{transverse-functions}.  
The orthogonal longitudinal functions used by Allen are given by:
\beq
L^{e}_{l}(x)=[x(1-x)]^{e}\sum_{m=0}^{l}\lambda_{l,m}^{(e)}x^{m},
\eeq
where the $\lambda_{l,m}^{(e)}$ are determined by Gramm-Schmidt orthogonalization.  The orthogonal transverse 
functions 
are functions of the relative transverse momentum $k$ with the range $[0,\infty )$.  They also 
depend on the parameter $d$ with dimension $\left[ \frac{1}{k} \right]$ which is fixed to help minimize the ground 
state energy.  They are given by:
\beq
T_{t}^{(d)}(k)&=&d\hspace{.1cm} \mathrm{exp}(-k^{2}d^{2}) \sum_{s=0}^{t} \sigma_{t,s} k^{s}d^{s}\\
\bar{T}_{t}(kd)&=&\frac{1}{d}T_{t}^{(d)}(k).
\eeq

\subsection{Convergence Testing}
\label{convergence-testing}

Section \ref{variational-method} introduces the ideas of convergence used in this section.  The first important point is 
that the real ground state energy is always lower than the lowest eigenvalue of the approximate Hamiltonian.  
Conversely, the lowest eigenvalue provides an upper limit to the real ground state energy.  The second important 
point is that for a ``standard'' set of basis functions, as more functions are added, a better approximation to the 
real 
state is achieved, lowering the eigenvalue at each step.  However, each time a new B-spline is added, the other splines 
in the set are rearranged.  This makes it possible for eigenvalues to increase.  This is typically a small effect, and has 
not increased the eigenvalue for more than one function addition.

\begin{figure}
\centerline{\epsfig{file=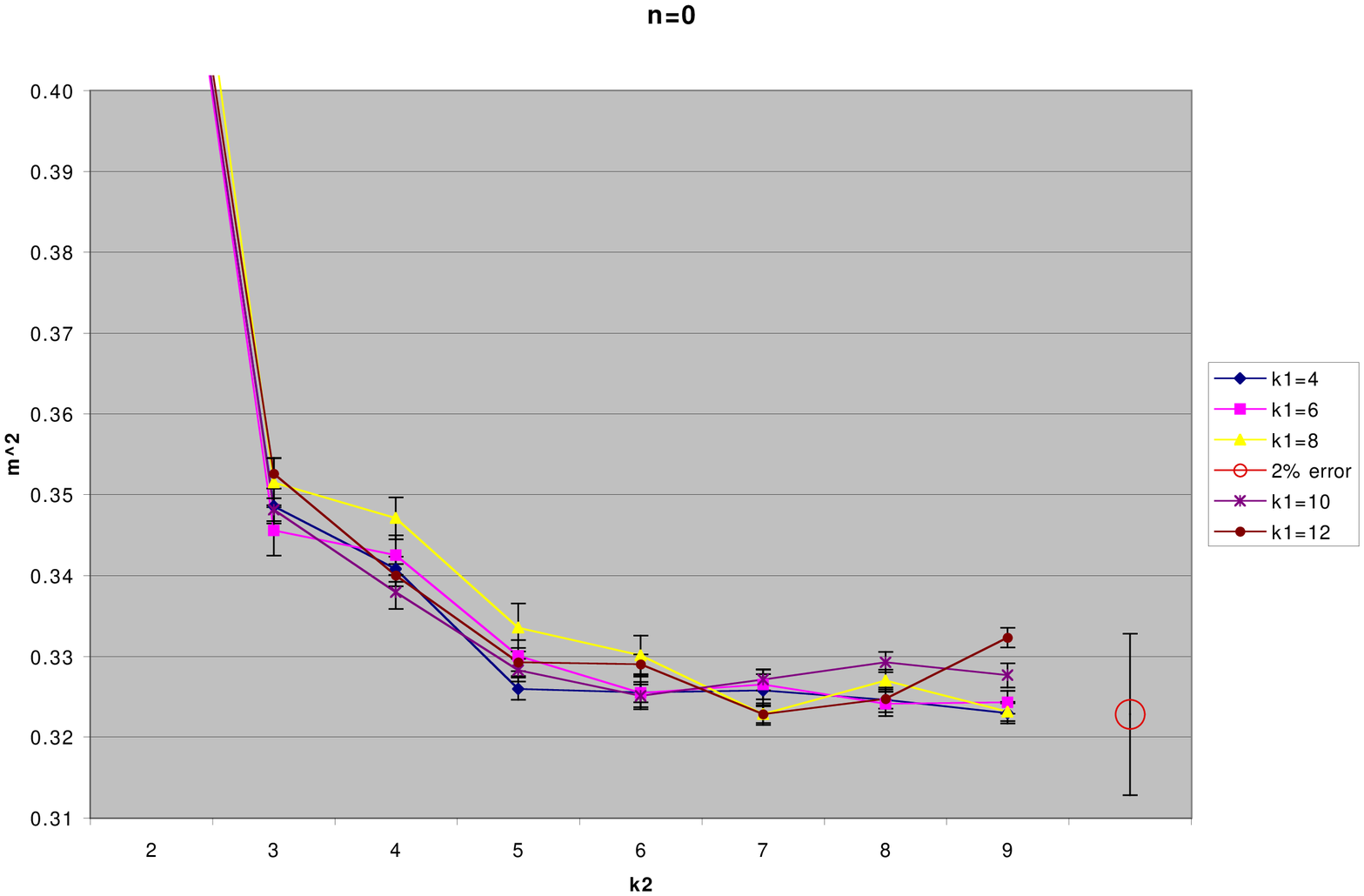,height=3in}}
\caption[Ground state glueball state convergence.]{Ground state eigenvalues of the IMO for different numbers of longitudinal and transverse basis states 
with zero spin and the coupling equal to one-half.  $k1$ and $k2$ are the B-spline parameter $k$ for the 
longitudinal and transverse B-splines, respectively.}
\label{n0-convergence}
\end{figure}

\begin{figure}
\label{n1-convergence}
\centerline{\epsfig{file=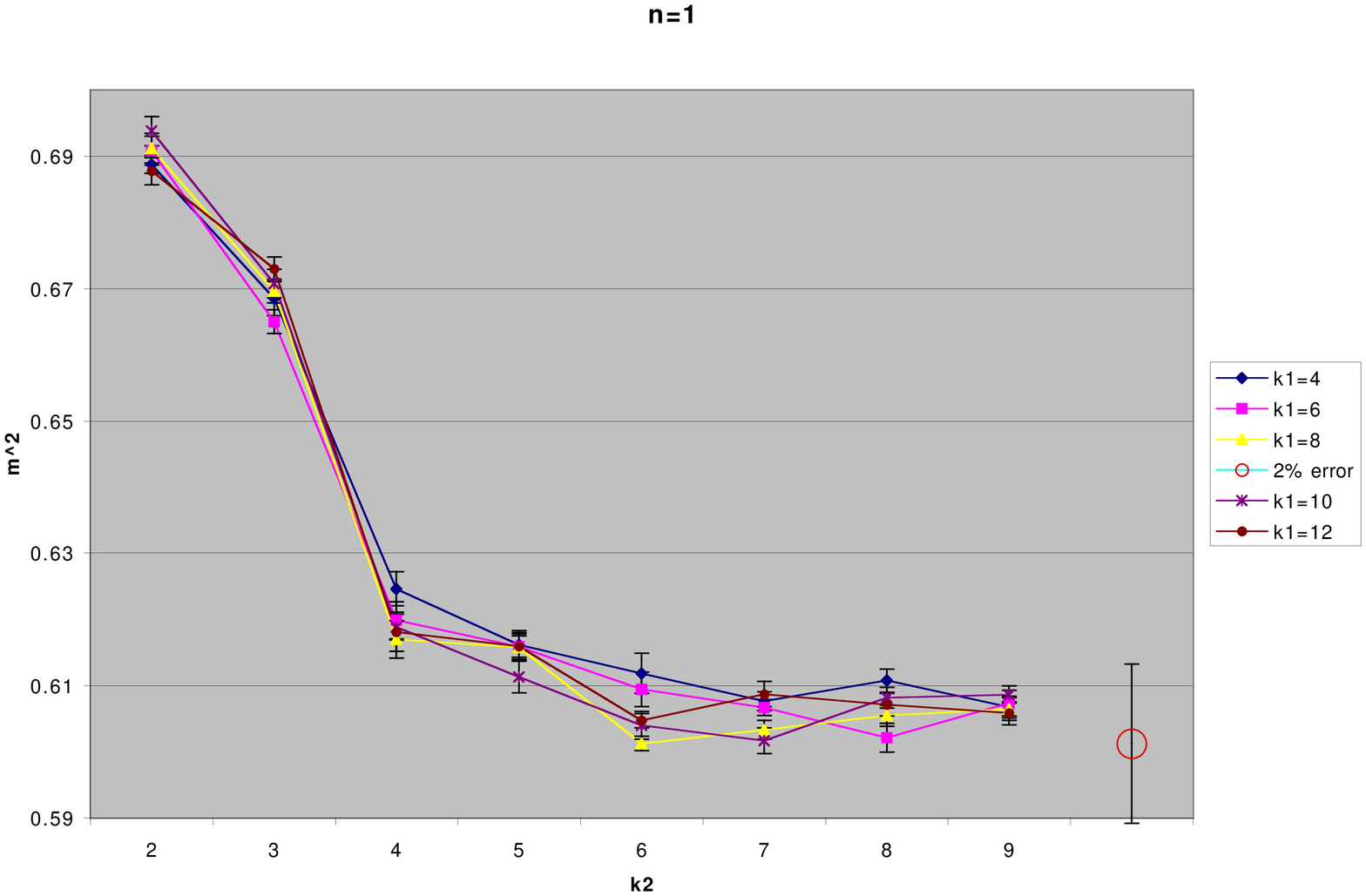,height=3in}}
\caption[First excited glueball state convergence.]{First excited state eigenvalues of the IMO for 
different numbers of longitudinal and transverse basis states with zero spin and the coupling equal to one-half.  $k1$ and $k2$ are the B-spline parameter $k$ for the 
longitudinal and transverse B-splines, respectively.}
\end{figure}

\begin{figure}
\centerline{\epsfig{file=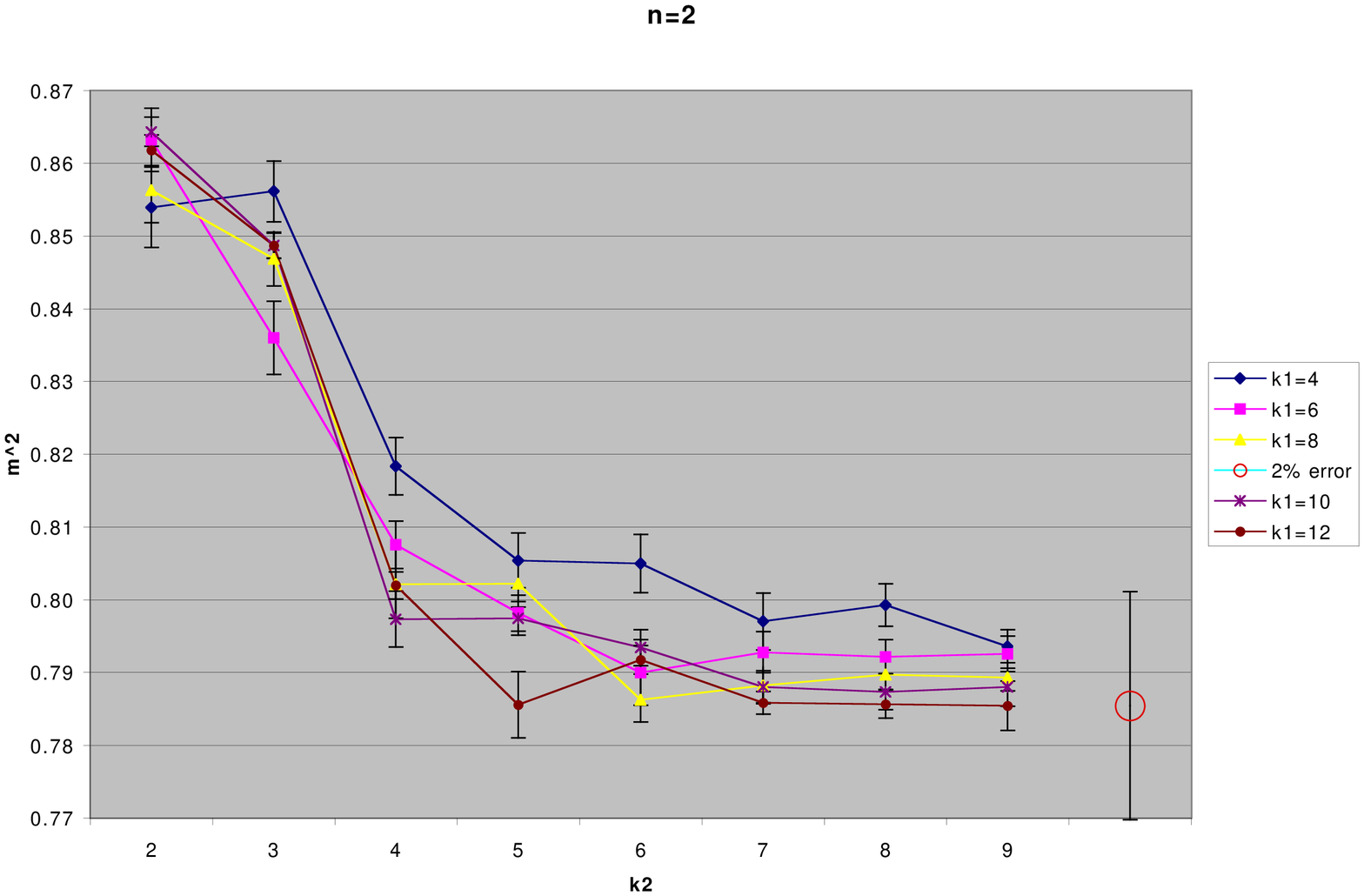,height=3in}}
\caption[Second excited glueball state convergence.]{Second excited state eigenvalues of the 
IMO for different numbers of longitudinal and transverse basis states with zero spin and the coupling equal to one-half.  
$k1$ and $k2$ are the B-spline parameter $k$ for the longitudinal and transverse B-splines, respectively.}
\label{n2-convergence}
\end{figure}

This section shows how the eigenvalues of the IMO converge as we increase the number of transverse and 
longitudinal basis states.  In the plots, the number of transverse basis states is $k2+m+1-3$ where 
three are removed to ensure finite kinetic energy (see Section \ref{transverse-functions}).  The total number of longitudinal 
basis (symmetric plus antisymmetric) states is 
$m+k1+1-2$ where two that do not have the proper $x\rightarrow 0,1$ behavior, are removed.  Increasing the number of longitudinal 
states has very little effect on the eigenvalues, and increasing the number of transverse states leads to rapid 
convergence for low-lying states.  This 
suggests the B-splines in the longitudinal direction are very efficient basis functions for that direction.  Figures 
\ref{n0-convergence} through \ref{n2-convergence} show how the eigenvalues of the mass-squared operator converge as $k1$ and $k2$ 
increase when $\alpha = \frac{1}{2}$ for the ground state through the second excited state, respectively.  These 
figures use $m=3$, $\alpha=\frac{1}{2}$ and $j=0$.  This choice of parameters is arbitrary because we just want to see 
how many B-splines are needed in the transverse and longitudinal directions to give converged results.  The ``2$\%$ error'' point plots the lowest eigenvalue and has 
2$\%$ error bars.  
Thus any lines within these error bars are consistent with the lowest eigenvalue.\footnote{The eigenvalues are 
calculated to only 2\% accuracy.  Any combination of $k1$ and $k2$ that 
yield results within 2\% of the lowest eigenvalue produce converged results.}

\subsection{Determining the Cutoff}
\label{determine-cutoff}

\begin{figure}
\centerline{\epsfig{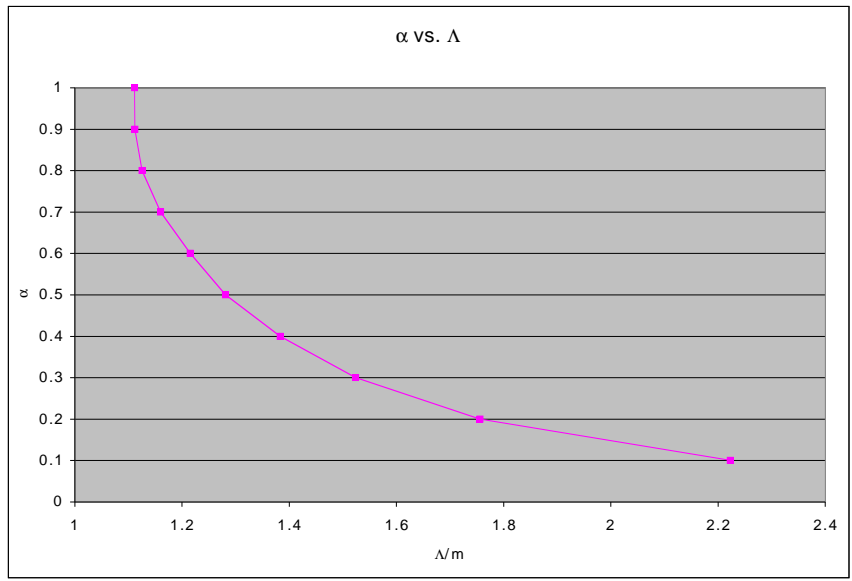}}
\caption[Coupling versus cutoff.]{The coupling $\alpha$ as a function of the cutoff $\Lambda$, in units of the first excited state mass when the first excited state is 
fixed.}
\label{cutoff-vs-coupling}
\end{figure}

In this theory, the only dimensionful parameter is the cutoff, $\Lambda$.  The cutoff can be determined by fitting 
the mass of one state.  Explicitly, if we fix the cutoff using the $i^{th}$ state we find:
\beq
\Lambda^{2}=\frac{m^{2}_{\mathrm{measured}}}{\langle i \vert {\cal M}^{2} (\Lambda) \vert i \rangle},
\eeq
where $m^{2}_{\mathrm{measured}}$ is the measured mass-squared for the state.  There is no real experimental data for 
glueballs, so we use lattice results \cite{lattice-glue} for the ``measured'' 
values.  Figure \ref{cutoff-vs-coupling} shows how the coupling and cutoff are related when we fix the first excited 
state.

\begin{figure}
\centerline{\epsfig{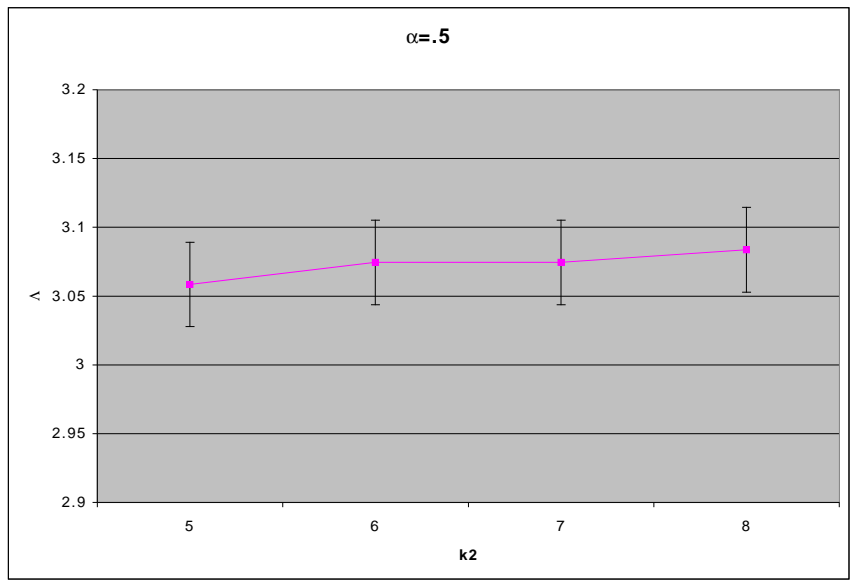}}
\caption[Cutoff convergence.]{The cutoff for a few values of $k2$ when $\alpha=.5$.  This shows the result is converged.}
\label{cutoff-alph.5}
\end{figure}

Figure \ref{cutoff-alph.5} shows that when we fix the first excited 
state, for $\alpha=.5$, $k1=6$ and $k2=6$ the value of the cutoff has converged.

\subsection{Masses versus Coupling}

In this section we show how the masses depend on the coupling, $\alpha$, when the first excited state is fixed.  However in these plots rather than showing the mass's 
dependence on the coupling, we show its dependence on the cutoff.  These plots help determine for what range of cutoffs the masses have a small cutoff dependence to 
determine where the theory is least cutoff dependent.  In agreement with Allen's results \cite{brentB,brentD}, we 
find the slowest cutoff dependence for $.5 \le \alpha \le .7$.

\begin{figure}
\centerline{\epsfig{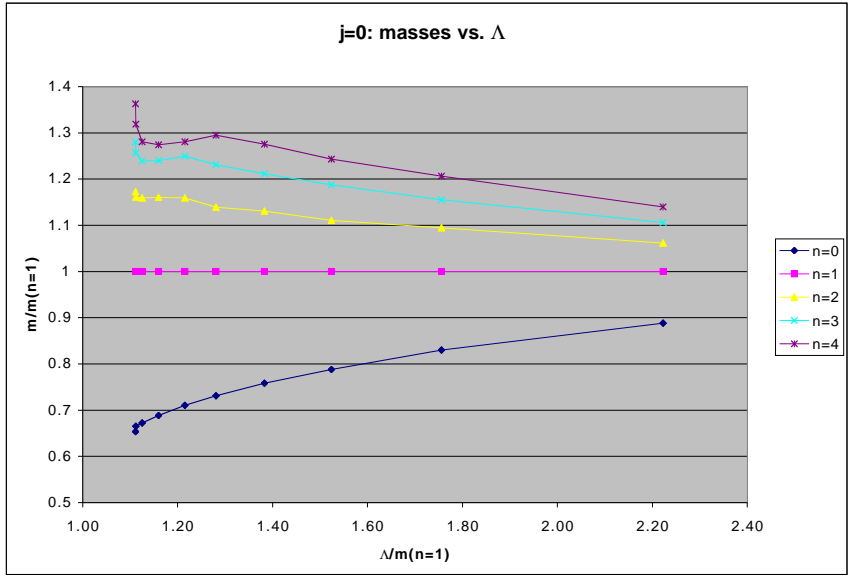}}
\caption{Masses for j=0 states as a function of the cutoff when the first excited state is fixed.}
\label{j0-mass-vs-cut-n1}
\end{figure}

\begin{figure}
\centerline{\epsfig{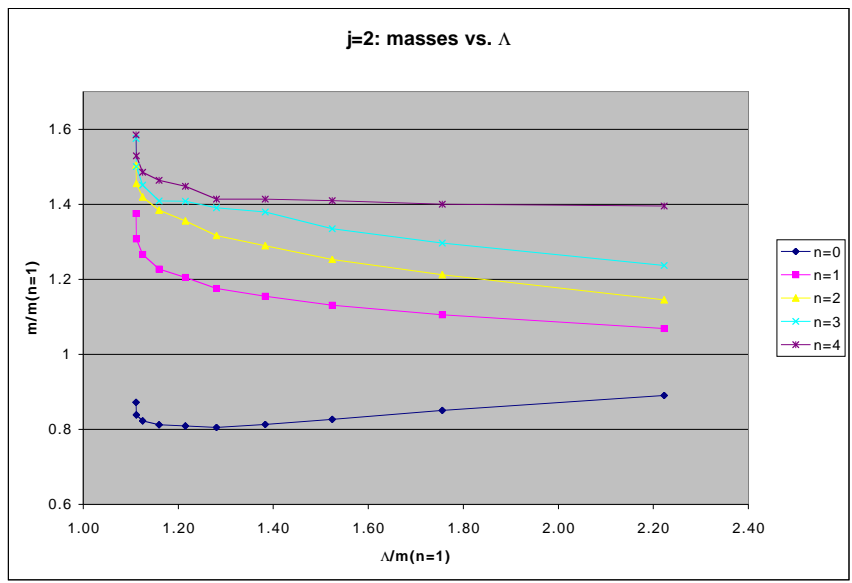}}
\caption{Masses for j=2 states as a function of the cutoff when the first excited state is fixed.}
\label{j2-mass-vs-cut-n1}
\end{figure}

In figures \ref{j0-mass-vs-cut-n1} through \ref{j2-mass-vs-cut-n1} the masses are plotted as a function of the cutoff 
when the first excited state is fixed.  This is equivalent to plotting the masses as a function of the coupling.

\begin{figure}
\centerline{\epsfig{file=\gluepath j1-mass-vs-cutoff-fix-n1.epsf,height=3in}}
\caption{Masses for j=1 states as a function of the cutoff when the first excited state is fixed.}
\label{j1-mass-vs-cut-n1}
\end{figure}

\subsection{Glueball Spectrum}

For completeness we present the glueball spectrum from \cite{brentD} in this section.  Our results are identical to 
within statistical errors to these previous results.  

States are identified by $J^{PC}_{j}$ where $J$ is the total spin of the state, $j$ is the spin projection along 
the 3-axis, $P$ is the parity of the state, and 
$C$ is the state's charge conjugation eigenvalue.   An asterisk in the state notation next to the value of 
$C$ denotes an excited state with the given quantum numbers.  Finally, we need to
distinguish states with identical $J$'s and $P$'s and different $j$'s because we do not have manifest rotational symmetry.  If $J=0$, we
omit the subscript $j$ in the state notation. 

Table \ref{glueball-mass-table} lists the glueball 
masses for $\alpha=.5$, in units of the mass of the ground state (the $0^{++}$ state).  It also gives the average of
lattice results from a number of different calculations for the sake of comparison \cite{lattice-glue}.  The 
uncertainties listed are only statistical errors due to the Monte-Carlo integration.  Finally, the three masses 
listed for the $J=2$ states correspond to $j=0,1,2$, where $j$ is the spin projection along the 3-axis.

\begin{table}
\centerline{\begin{tabular}{|c|c|c|} \hline
State & $M/M_{0^{++}}$ & Lattice \cite{lattice-glue}\\
\hline\hline
$0^{-+}$ & $1.38 \pm 0.02$ & $1.34 \pm 0.18$\\ \hline
& $1.58 \pm 0.01$ & \\
$2^{++}$ & $1.58 \pm 0.02$ & $1.42 \pm 0.06$ \\
& $1.11 \pm 0.01$ & \\
\hline
& $1.70 \pm 0.01$ & \\
$2^{++*}$ & $1.68 \pm 0.02$ & $1.85 \pm 0.20$ \\
& $1.62 \pm 0.02$ & \\ \hline
$0^{++*}$ & $1.77 \pm 0.02$ & $1.78 \pm 0.12$ \\ \hline
\end{tabular}}
\caption[Glueball masses.]{\label{glueball-mass-table}The glueball masses from \cite{brentD}, which are identical 
to our masses within statistical errors, compared to an average of
lattice results from a number of different calculations \cite{lattice-glue}.  
Masses are in units of the mass of the $0^{++}$ state.  The 
uncertainties are only the statistical uncertainties
associated with the Monte Carlo evaluation of the matrix elements of $\M$.  The three values
of the masses for the $2^{++}$ and $2^{++*}$ states for our calculation correspond to $j=0,1,2$.  We use $\alpha=0.5$, with 
8 longitudinal (4 symmetric and 4 antisymmetric) basis functions, 7 transverse-magnitude basis functions, and 4 spin
basis functions, for a total of 112 basis functions.}
\end{table}


\subsection{Wavefunction Plots}

In this section we plot the spin-averaged probability density for the glueballs which is defined in Eq.~(\ref{meson-prob-density}) 
for states some of the low-lying states with $j=0$ and $j=1$.  We only show a few functions since they are similar 
to those found in \cite{brentB,brentD}.  The variable ``$x\_plt$'' is the longitudinal momentum fraction for one of the 
gluons, and the variable ``$k\_plt$'' is the relative transverse momentum of the state in units of the cutoff.

\begin{figure}
\centerline{\epsfig{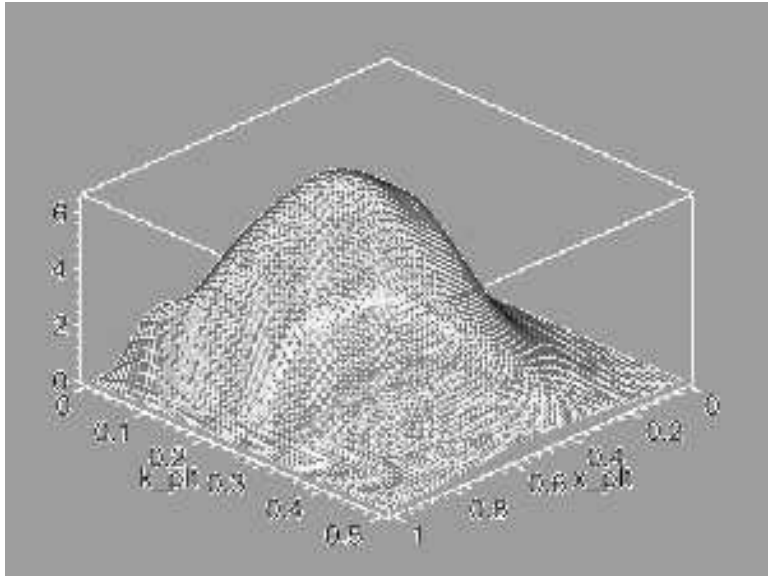}}
\caption{Wavefunction for the $0^{++}$ glueball with $\alpha =\frac{1}{2}$.}
\label{j0-wave-n0}
\end{figure}

\begin{figure}
\centerline{\epsfig{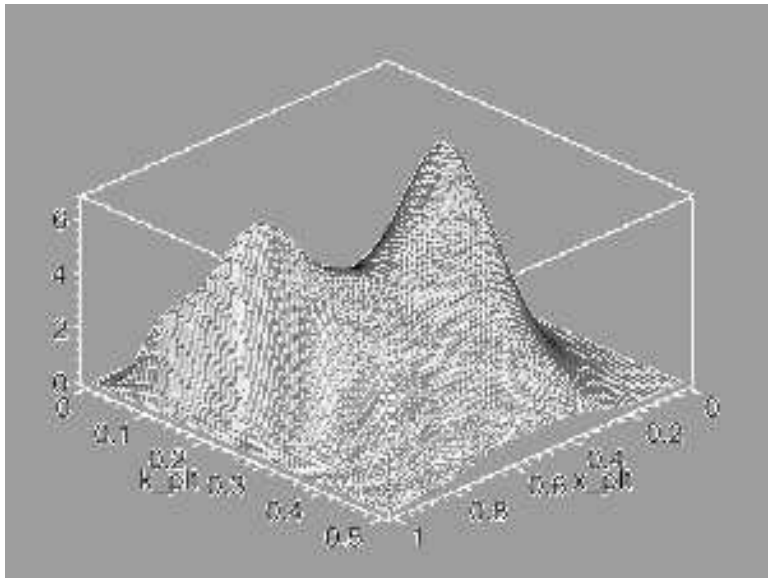}}
\caption{Wavefunction for the $2^{++}_{0}$ glueball with $\alpha =\frac{1}{2}$.}
\label{j1-wave-n0}
\end{figure}

Figures \ref{j0-wave-n0} through \ref{j1-wave-n0} show the probability density for four of the the low-lying 
glueball states.  Each wavefunction is built from 8 longitudinal (4 symmetric and 4 antisymmetric) basis functions 
and 7 transverse functions.  Including the four spin states, there are 112 basis states.

\begin{figure}
\centerline{\epsfig{file=\gluepath j0n1-wave.epsf,height=3in}}
\caption{Wavefunction for the $0^{-+}$ glueball with $\alpha =\frac{1}{2}$.}
\label{j0-wave-n1}
\end{figure}
\begin{figure}
\centerline{\epsfig{file=\gluepath j0n2-wave.epsf,height=3in}}
\caption{Wavefunction for the $2^{++}_{0}$ glueball with  $\alpha =\frac{1}{2}$.}
\label{j0-wave-n2}
\end{figure}
%
%
%

\chapter{Matrix Elements of the Approximate Meson States}
\label{chapt:me-physical}

We calculate all of the matrix elements of the approximate IMO (introduced in Section~\ref{mass-o2-begin}) necessary 
for a second-order calculation.  The matrix elements are divided into four types of contributions; 
kinetic energy (KE), self-energy (SE), exchange (EX) and instantaneous exchange (IN).  The four contributions are 
combined to explicitly cancel divergences (discussed in Section~\ref{comb-inter}).  With this division the full 
expression can be written:
\beq
\langle q',l',t',j' \vert {\cal M}^{2}(\Lambda)\vert q,l,t,j \rangle &=&
\langle q',l',t',j' \vert {\cal M}^{2}(\Lambda)\vert q,l,t,j \rangle_{\mathrm{KE}} \n &+&
\langle q',l',t',j' \vert {\cal M}^{2}(\Lambda)\vert q,l,t,j \rangle_{\mathrm{SE}}^{\mathrm{F}} \n &+&
\langle q',l',t',j' \vert {\cal M}^{2}(\Lambda)\vert q,l,t,j \rangle_{\mathrm{EX}}^{\mathrm{F}}\n &+&
\langle q',l',t',j' \vert {\cal M}^{2}(\Lambda)\vert q,l,t,j \rangle_{\mathrm{IN+EX}} \n &+&
\langle q',l',t',j' \vert {\cal M}^{2}(\Lambda)\vert q,l,t,j \rangle_{\mathrm{IN}}^{\mathrm{B,F}},
\eeq
where the free-state functions $\vert q,l,t,j \rangle$ are defined in Eq.~(\ref{qltj-def}).
In the remainder of this Chapter we determine the matrix elements that need to be calculated by numerical 
integration.  The integrals are divided into two groups:  two-dimensional and five-dimensional.

\section{Two-Dimensional Integrals}
\label{physical-state-me-2d}
\subsection{The Kinetic Energy}

The two-dimensional integral for the kinetic energy is very similar to the overlap integral.  The only difference 
in the derivation is:
\beq
\langle q'_{1} \bar{q}'_{2} \vert q_{1}\bar{q}_{2} \rangle \rightarrow \frac{\vec{k}^{2}+m^{2}}{x(1-x)} 
\langle q'_{1} \bar{q}'_{2} \vert q_{1}\bar{q}_{2} \rangle .
\eeq
Thus, we have:
\beq
&&\langle q',l',t',j' \vert {\cal M}^{2}(\Lambda) \vert q,l,t,j \rangle_{\mathrm{KE}}=
2 \Lambda^{2}\delta_{q,q'}\delta_{j,j'}\int_{-1}^{1}\frac{dy_{k}}{(1+y_{k})^{2}}
\left( \frac{2}{1+y_{k}}-1\right)\n
&&\hspace{2cm}\times \left[ \left( \frac{2}{1+y_{k}}-1\right)^{2}+m^{2}\right] 
B_{t}(y_{k}) B_{t'}(y_{k}) \int_{0}^{1} dx \frac{B_{l}(x) B_{l'}(x)}{x(1-x)}.
\eeq
\subsection{The Self Energy}

We need to separate Eq. (\ref{vncse}) into divergent and finite pieces:
\beq
&&\langle q_{3}\bar{q}_{4}\vert V^{(2)}_{\mathrm{SE}}(\la)\vert q_{1}\bar{q}_{2}\rangle=\n
&&\hspace{1cm}-\frac{\delta_{1,3}\delta_{2,4}}{24\pi^{2}}\la^{2}\left[ \int \frac{z\hspace{.1cm}dz}{1-z}
\left\{ \frac{1+z^{2}}{z}\sqrt{2\pi}\left[\mathrm{erf}\left(\gamma(x,z)\right)
+\mathrm{erf}\left(\gamma(1-x,z)\right)-2\right]\right.\right.\n
&&\hspace{1.5cm}\left.\left.
+\sqrt{8} \left[\gamma(x,z) \mathrm{Ei}\left(1,\gamma^{2}(x,z)\right)
+\gamma(1-x,z) \mathrm{Ei}\left(1,\gamma^{2}(1-x,z)\right) \right] \right\}\right].
\eeq
The divergent part of the self-energy is contained in the term:
\beq
-\frac{\delta_{1,3}\delta_{2,4}}{24\pi^{2}}\la^{2}\int dz \frac{1+z^{2}}{1-z}\left(-2 \sqrt{2\pi}\right)=
-2\sqrt{2 \pi}\frac{\delta_{1,3}\delta_{2,4}}{24\pi^{2}}\Lambda^{2} \left[ 2 \log(\epsilon) +\frac{3}{2}\right].
\eeq
So now we can split the self-energy contribution into a finite and divergent piece:
\newpage
\beq
&&\langle q_{3}\bar{q}_{4}\vert V^{(2)}_{\mathrm{SE}}(\la)\vert q_{1}\bar{q}_{2}\rangle ^{\mathrm{F}}
=-\frac{\delta_{1,3}\delta_{2,4}}{24\pi^{2}}\la^{2}\n
&&\hspace{1cm}\left[ \int \frac{z\hspace{.1cm}dz}{1-z}
\left\{ \frac{1+z^{2}}{z}\sqrt{2\pi}\left[\mathrm{erf}\left(\gamma(x,z)\right)
+\mathrm{erf}\left(\gamma(1-x,z)\right)\right]\right.\right.\n
&&\hspace{3cm}\left.\left.
+\sqrt{8} \left[\gamma(x,z) \mathrm{Ei}\left(1,\gamma^{2}(x,z)\right)
\right. \right. \right. \n && \hspace{4cm} \left. \left. \left.
+\gamma(1-x,z) \mathrm{Ei}\left(1,\gamma^{2}(1-x,z)\right) \right] \right\}+3\sqrt{2\pi}\right],\\
&&\langle q_{3}\bar{q}_{4}\vert V^{(2)}_{\mathrm{SE}}(\la)\vert q_{1}\bar{q}_{2}\rangle ^{\mathrm{D}}=
-\frac{\delta_{1,3}\delta_{2,4}}{24\pi^{2}}\la^{2}4\sqrt{2 \pi}\log(\epsilon),
\eeq
with
\beq
\langle q',l',t',j' \vert {\cal M}^{2}(\Lambda) \vert q,l,t,j \rangle_{\mathrm{SE}}^{\mathrm{F}}
=g_{\Lambda}^{2}e^{-\Lambda^{-4}\Delta^{2}_{FI}}
\langle q',l',t',j' \vert V^{(2)}_{\mathrm{SE}}(\la) \vert q,l,t,j \rangle^{\mathrm{F}},\\
\langle q',l',t',j' \vert {\cal M}^{2}(\Lambda) \vert q,l,t,j \rangle_{\mathrm{SE}}^{\mathrm{D}}
=g_{\Lambda}^{2}e^{-\Lambda^{-4}\Delta^{2}_{FI}}
\langle q',l',t',j' \vert V^{(2)}_{\mathrm{SE}}(\la) \vert q,l,t,j \rangle^{\mathrm{D}}.
\eeq
We find for the finite part (see Eq.~\ref{qltj-def}):
\beq
\label{self-energy-before-cov}
&&\langle q',l',t',j {\cal M}^{2}(\la) \vert q,l,t,j \rangle^{\mathrm{F}}_{\mathrm{SE}}=
16 \pi^{3} g_{\Lambda}^{2}{\cal P}^{+}\delta^{(3)}({\cal P}-{\cal P}') \delta_{q,q'}\delta_{j,j'}\Lambda^{2}\n
&&\hspace{2cm} 
\times \left( -\frac{1}{24\pi^{2}}\right) \int_{0}^{1}dx B_{l}(x)B_{l'}(x) {\mathrm I } (x)
\int_{0}^{\infty}dq \hspace{.1cm} q \tilde{B}_{t}(q)\tilde{B}_{t'}(q),
\eeq
where
\beq
&&\mathrm{I}(x)=3\sqrt{2\pi}+\int \frac{z\hspace{.1cm}dz}{1-z}
\left\{ \frac{1+z^{2}}{z}\sqrt{2\pi}\left[\mathrm{erf}\left(\gamma(x,z)\right)
+\mathrm{erf}\left(\gamma(1-x,z)\right)\right]\right.\n
&&\hspace{4cm}\left.
+\sqrt{8} \left[ \gamma(x,z) \mathrm{Ei}\left(1,\gamma^{2}(x,z)\right)
\right. \right. \n && \hspace{6cm} \left. \left.
+\gamma(1-x,z) \mathrm{Ei}\left(1,\gamma^{2}(1-x,z)\right) \right] \right\}.
\eeq
The divergent part of the self-energy is identical, except:
\beq
\mathrm{I}(x) \rightarrow 4\sqrt{2 \pi} \log(\epsilon),
\eeq
giving
\beq
&&\langle q',l',t',j \vert {\cal M}^{2}(\la) \vert q,l,t,j \rangle^{\mathrm{D}}_{\mathrm{SE}}=
-\frac{4}{3} (2\pi)^{\frac{3}{2}}
\Lambda^{2} \log(\epsilon)\hspace{.1cm} g_{\Lambda}^{2}{\cal P}^{+}\delta^{(3)}({\cal P}-{\cal P}') \n
&&\hspace{3cm} \times
\delta_{q,q'}\delta_{j,j'}\int_{0}^{1}dx B_{l}(x)B_{l'}(x)\int_{0}^{\infty}dq \hspace{.1cm} q \tilde{B}_{t}(q)\tilde{B}_{t'}(q).
\eeq

\section{Five-Dimensional Integrals}
\label{physical-state-me-5d}
\subsection{The Instantaneous Interaction}
\label{inst-inter}
The matrix element of the instantaneous interaction in the approximate basis is given by:
\beq
&&\langle q',l',t',j' \vert {\cal M}^{2}(\la) \vert q,l,t,j \rangle_{\mathrm{IN}}=
g_{\Lambda}^{2}e^{-\Lambda^{-4}\Delta_{FI}^{2}} \times\n
&&\hspace{1cm}
\frac{1}{16\pi^{3}}\frac{1}{N_{c}} \sum_{s_{1\rightarrow 4},c_{1\rightarrow 4}}
\int \frac{d^{2}\vec{q} dx}{\sqrt{x(1-x)}}  \int \frac{d^{2}\vec{p} dy}{\sqrt{y(1-y)}}
\theta_{\epsilon}\theta_{\epsilon'} \delta_{c_{1},c_{2}}\delta_{c_{3},c_{4}}
\chi_{q}^{s_{1}s_{2}}\chi_{q'}^{s_{3}s_{4}}\times\n
&&\hspace{1cm} 
B_{l'}(y)B_{l}(x)\tilde{B}_{t}(q)\tilde{B}_{t'}(p) A^{\ast}_{j'-s_{3}-s_{4}}(\phi') A_{j-s_{1}-s_{2}}(\phi)
\langle q_{3}\bar{q}_{4}\vert V_{\mathrm{IN}}^{(2)}(\Lambda) \vert q_{1}\bar{q}_{2}\rangle ,
\eeq
where the matrix element of the reduced interaction is defined in Eq. (\ref{in-defn}).  This can be reduced to:
\beq
&&\langle q',l',t',j' \vert {\cal M}^{2}(\la) \vert q,l,t,j \rangle_{\mathrm{IN}}=
\n && \hspace{1cm}
-\frac{32}{N_{c}} g_{\Lambda}^{2} {\cal P}^{+}\delta({\cal P}-{\cal P}')\delta_{j,j'}\int d\gamma_{\phi} dq qdx  \int dp p dy
\theta_{\epsilon}\theta_{\epsilon'} \n
&&\hspace{2cm} \times \theta (x-y-\epsilon) B_{l'}(y)B_{l}(x)\tilde{B}_{t}(q)\tilde{B}_{t'}(p)
\frac{e^{-\Lambda^{-4}\Delta_{FI}^{2}}}{(x-y)^{2}}
\n && \hspace{3cm} \times
\left[
\delta_{q,1}\delta_{q',1}\cos ( [j-1]\gamma_{\phi})
+\delta_{q,2}\delta_{q',2}\cos ( [j+1]\gamma_{\phi})
\right.\n &&\hspace{6cm}\left.
+\left(\delta_{q,3}\delta_{q',3}+\delta_{q,4}\delta_{q',4}\right) \cos ( j\gamma_{\phi})
\right] .
\eeq
Finally, as discussed in Section \ref{comb-inter}, we split the instantaneous interaction into two parts, a part above the cutoff that vanishes in 
the limit $\Lambda \rightarrow \infty$, and a part below the cutoff that remains in this limit:
\beq
\label{inst-above}
\langle q',l',t',j' \vert {\cal M}^{2}(\la) \vert q,l,t,j \rangle^{\mathrm{A}}_{\mathrm{IN}}
&=&\left[ 1-e^{2 \Lambda^{-4}\Delta_{FK}^{(1)}\Delta_{KI}^{(1)}}\right]
\n && \hspace{1cm} \times
\langle q',l',t',j' \vert {\cal M}^{2}(\la) \vert q,l,t,j \rangle _{\mathrm{IN}},\\
\langle q',l',t',j' \vert {\cal M}^{2}(\la) \vert q,l,t,j \rangle^{\mathrm{B}}_{\mathrm{IN}}
&=&e^{2 \Lambda^{-4}\Delta_{FK}^{(1)}\Delta_{KI}^{(1)}}
\n && \hspace{1cm} \times
\langle q',l',t',j' \vert {\cal M}^{2}(\la) \vert q,l,t,j \rangle_{\mathrm{IN}}.
\eeq

\subsection{The Finite Part of the Exchange Interaction}
The matrix element of the finite part of the exchange interaction is:
\beq
\label{finite-exchange-start}
&&\langle q',l',t',j' \vert {\cal M}^{2}(\la) \vert q,l,t,j \rangle_{\mathrm{EX}}=
g_{\Lambda}^{2}e^{-\Lambda^{-4}\Delta_{FI}^{2}} \times\n
&&\hspace{1cm}
\frac{1}{16\pi^{3}}\frac{1}{N_{c}} \sum_{s_{1\rightarrow 4},c_{1\rightarrow 4}}
\int \frac{d^{2}\vec{q} dx}{\sqrt{x(1-x)}}  \int \frac{d^{2}\vec{p} dy}{\sqrt{y(1-y)}}
\theta_{\epsilon}\theta_{\epsilon'} \delta_{c_{1},c_{2}}\delta_{c_{3},c_{4}}
\chi_{q}^{s_{1}s_{2}}\chi_{q'}^{s_{3}s_{4}}\times\n
&&\hspace{1cm} 
B_{l'}(y)B_{l}(x)\tilde{B}_{t}(q)\tilde{B}_{t'}(p) A^{\ast}_{j'-s_{3}-s_{4}}(\phi') A_{j-s_{1}-s_{2}}(\phi)
\langle q_{3}\bar{q}_{4}\vert V_{\mathrm{EX}}^{(2)}(\Lambda) \vert q_{1}\bar{q}_{2}\rangle ,
\eeq
where the matrix element is given in Eq. (\ref{ex-defn}).  After many pages of algebra the contribution from the
finite part of the exchange interaction is given by:
\beq
\label{finite-exchange-before-change}
&&\langle q',l',t',j' \vert {\cal M}^{2}(\la) \vert q,l,t,j \rangle^{\mathrm{F}}_{\mathrm{EX}}=
- {\cal P}^{+}\delta^{3}({\cal P}-{\cal P}')\left( N_{c}-\frac{1}{3}\right) \theta (x-y-\epsilon) \times\n
&&\hspace{.5in}g_{\Lambda}^{2}e^{-\Lambda^{-4}\Delta_{FI}^{2}} \int d\gamma_{\phi} dq q dx \int dp p dy 
\frac{1}{x-y}\frac{B_{l'}(y)B_{l}(x)\tilde{B}_{t}(q)\tilde{B}_{t'}(p)}{x(1-x)y(1-y)} \n
&&\hspace{4cm}\times\left(\frac{1}{\Delta_{FK}^{(1)}}-\frac{1}{\Delta_{KI}^{(1)}}\right)
\left[ 1-e^{2\Lambda^{-4}\Delta_{FK}^{(1)}\Delta_{KI}^{(1)}}\right] \delta_{j,j'}S_{q,q'},
\eeq
where the $S_{q,q'}$ are given by:
\beq
S_{1,1}&=&-\frac{1}{x-y} \left[ q^{2}y(1-y)(1-2x) -p^{2}x(1-x)(1-2y)\right] \cos ([j-1]\gamma_{\phi})\n
&&\hspace{3cm}-pq\left[ y^{2}(1-2x)-x^{2}(1-2y)-2y(1-x)\right] \cos(j \gamma_{\phi}), \n
S_{2,2}&=&-\frac{1}{x-y} \left[ q^{2}y(1-y)(1-2x) -p^{2}x(1-x)(1-2y)\right] \cos ([j+1]\gamma_{\phi})\n
&&\hspace{3cm}-pq\left[ y^{2}(1-2x)-x^{2}(1-2y)-2y(1-x)\right] \cos(j \gamma_{\phi}), \n
S_{3,3}&=&-\cos (j\gamma_{\phi})\left[ pq \cos \gamma_{\phi}(1-x-y+2xy) 
\right. \n &&\hspace{2cm} \left.
-\frac{1-x-y}{x-y}\left( y(1-y)q^{2}-x(1-x)p^{2}\right)-m^{2} (x-y)^{2} \right],\n
S_{4,4}&=&-\cos (j\gamma_{\phi})\left[ pq \cos \gamma_{\phi}(1-x-y+2xy)
\right. \n && \hspace{2cm} \left.
-\frac{1-x-y}{x-y}\left( y(1-y)q^{2}-x(1-x)p^{2}\right) +m^{2} (x-y)^{2} \right],\n
S_{1,2}&=&0,\n
S_{2,1}&=&0,\n
S_{1,3}&=& \frac{im}{\sqrt{2}} \left[(1-2y)q\cos( j \gamma_{\phi})-(1-x-y)p\cos( [j-1]\gamma_{\phi})\right],\n
S_{3,1}&=& \frac{im}{\sqrt{2}} \left[(1-x-y)q\cos([j-1]\gamma_{\phi})-(1-2x)p\cos(j\gamma_{\phi})\right],\n
S_{1,4}&=& \frac{im}{\sqrt{2}} \left[-(1-2y+2y^{2})q\cos( j\gamma_{\phi})+(1-y-x+2xy)p\cos( [j-1]\gamma_{\phi})\right],\n
S_{4,1}&=& \frac{im}{\sqrt{2}} \left[ (1-2x+2x^{2})p\cos( j\gamma_{\phi})-(1-x-y+2xy)q\cos( [j-1]\gamma_{\phi})\right],\n
S_{2,3}&=& \frac{im}{\sqrt{2}} \left[(1-2y)q\cos(j\gamma_{\phi})-(1-x-y)p\cos([j+1]\gamma_{\phi})\right],\n
S_{3,2}&=& \frac{im}{\sqrt{2}} \left[(1-x-y)q\cos([j+1]\gamma_{\phi})-(1-2x)p\cos(j\gamma_{\phi})\right], \n
S_{2,4}&=& \frac{im}{\sqrt{2}} \left[(1-2y+2y^{2})q\cos(j\gamma_{\phi})-(1-y-x+2xy)p\cos([j+1]\gamma_{\phi})\right],\n
S_{4,2}&=& \frac{im}{\sqrt{2}} 
\left[(1-x-y+2xy)q\cos([j+1]\gamma_{\phi})-(1-2x+2x^{2})p\cos(j\gamma_{\phi})\right],\n
S_{3,4}&=& \sqrt{2}pq\sin \gamma_{\phi} \sin (j\gamma_{\phi})(1-x-y), \n
S_{4,3}&=& \sqrt{2}pq\sin \gamma_{\phi} \sin (j\gamma_{\phi})(1-x-y) .\nonumber
\eeq

\subsection{The Divergent Part of the Exchange Interaction}

The complete contribution from the divergent part of the 
exchange diagrams is:
\beq
\label{exch-div}
&&\langle q',l',t',j' \vert {\cal M}^{2}(\la) \vert q,l,t,j \rangle^{\mathrm{D}}_{\mathrm{EX}}=
\n && \hspace{1cm}
-2g_{\Lambda}^{2}e^{-\Lambda^{-4}\Delta_{FI}^{2}} {\cal P}^{+} 
\delta^{3}({\cal P}-{\cal P}')\delta_{j,j'}\left( N_{c}-\frac{1}{3}\right) 
\n &&\hspace{1.5cm}\times
\int d\gamma_{\phi} dq q dx \int dp p dy 
\frac{\theta( x-y -\epsilon )}{x-y}B_{l'}(y)B_{l}(x)\tilde{B}_{t}(q)\tilde{B}_{t'}(p) \n
&&\hspace{2cm}\times 
\left(\frac{1}{\Delta_{FK}^{(1)}}-\frac{1}{\Delta_{KI}^{(1)}}\right)
\left[ 1-e^{2\Lambda^{-4}\Delta_{FK}^{(1)}\Delta_{KI}^{(1)}}\right] 
\left\{\frac{p^{2}+q^{2}-2pq \cos (\gamma_{\phi})}{(x-y)^{2}}\right\}
\n &&\hspace{2cm}\times
\left[
\delta_{q,1}\delta_{q',1}\cos ([j-1]\gamma_{\phi})+\delta_{q,2}\delta_{q',2}\cos ([j+1]\gamma_{\phi})
\right. \n && \hspace{5cm} \left.
+\delta_{q,3}\delta_{q',3}\cos (j\gamma_{\phi})+\delta_{q,4}\delta_{q',4}\cos (j\gamma_{\phi})\right] .
\eeq
\subsection{Combining the Divergent Part of the Exchange Interaction and the Instantaneous Interaction Above the Cutoff}

If we consider the contribution from the instantaneous interaction above the cutoff (Eq. \ref{inst-above}) and the 
divergent part of the exchange interaction (Eq. \ref{exch-div}), we see that in the limit $x=y$ the two 
divergent contributions cancel.  Thus, we can combine the two interactions and write them as:
\beq
\label{in-plus-exch-before-change}
&&\langle q',l',t',j' \vert {\cal M}^{2}(\la) \vert q,l,t,j \rangle_{\mathrm{IN+EX}}=
\n && \hspace{1cm}
-\frac{16}{3}g_{\Lambda}^{2}{\cal P}^{+}\delta({\cal P}-{\cal P}')\delta_{j,j'}
\n && \hspace{2cm} 
\int d\gamma_{\phi} dq qdx  \int dp p dy
\theta_{\epsilon}\theta_{\epsilon'} \theta (x-y-\epsilon) e^{-\Lambda^{-4}\Delta_{FI}^{2}}
\n && \hspace{1.5cm} \times
B_{l'}(y)B_{l}(x)\tilde{B}_{t}(q)\tilde{B}_{t'}(p)
\left[ 1-e^{2\Lambda^{-4}\Delta_{FK}^{(1)}\Delta_{KI}^{(1)}}\right] \n
&&\hspace{1.5cm} \times
\left\{
\frac{W_{q,q'}}{(x-y)^{2}} 
\left[ \left(\frac{1}{\Delta_{FK}^{(1)}}-\frac{1}{\Delta_{KI}^{(1)}}\right)
\left( \frac{p^{2}+q^{2}-2\vec{p}\cdot\vec{q}}{x-y}\right)+2 \right]\right\} ,
\eeq
where
\beq
W_{q,q'}&=&\delta_{q,1}\delta_{q',1}\cos ([j-1]\gamma_{\phi})+\delta_{q,2}\delta_{q',2}\cos ([j+1]\gamma_{\phi})
\n && \hspace{5cm}
+\left(\delta_{q,3}\delta_{q',3}+\delta_{q,4}\delta_{q',4}\right) \cos (j\gamma_{\phi}).
\eeq
In the limit $\vert x-y \vert \rightarrow 0$ the term in square brackets (inside the braces) vanishes, leaving the entire term finite 
in this limit so we no longer need the infrared cutoff for this term.

\subsection{Combining the Divergent Part of the Self-Energy and the Instantaneous Interaction Below the Cutoff}

The combination of the instantaneous interaction below the cutoff and the divergent part of the self-energy is 
algebraically complex.  The details of the calculation can be found in Appendix \ref{appx:se-inst}.  We find:
\beq
\label{before-subtraction}
&&\hspace{-1cm}\langle q',l',t',j' \vert {\cal M}^{2}(\la) \vert q,l,t,j \rangle_{\mathrm{IN}}^{\mathrm{B,F}} =
-g_{\Lambda}^{2}\frac{32}{3} \delta_{j,j'} {\cal P}^{+}\delta({\cal P-P'})\theta(x-y) \theta(x-y-\epsilon) \n
&&\hspace{0cm}
 \int_{2\epsilon}^{1-\epsilon} dx
\int_{\epsilon}^{x-\epsilon} dy \log(x-y) \int d^{2}\vec{q} d^{2}\vec{p}
B_{l}(x)\tilde{B}_{t}(q) \tilde{B}_{t'}(p)
W_{q,q'} e^{-\Lambda^{-4}(\Delta^{2}_{FK}+\Delta^{2}_{IK})}
\n && \hspace{2cm} \times
\left[ \frac{B_{l'}(y)}{(x-y)^{2}}+\frac{B'_{l'}(y)}{x-y}
-2\frac{B_{l'}(y)}{x-y}\left(\Delta_{FK}\Delta_{FK}'+\Delta_{KI}\Delta_{KI}' \right) \right] ,
\eeq
where,
\beq
\Delta_{KI}&=&\frac{m^{2}(x-y)^{2}+(x\vec{p}-y\vec{q})^{2}}{xy(x-y)} \n
&=&\frac{m^{2}(x-y)-yq^{2}+xp^{2}}{xy}+\frac{(\vec{p}-\vec{q})^{2}}{x-y},\n
\Delta_{FK}&=&-\frac{m^{2}(x-y)^{2}+\left[(1-x)\vec{p}-(1-y)\vec{q}\hspace{.1cm}\right]^{2}}{(1-x)(1-y)(x-y)} \n
&=&-\frac{m^{2}(x-y)+(1-y)q^{2}-(1-x)p^{2}}{(1-x)(1-y)}-\frac{(\vec{p}-\vec{q})^{2}}{x-y},\n
\Delta_{KI}'&=&\frac{(\vec{p}-\vec{q})^{2}}{(x-y)^{2}}-\frac{m^{2}+p^{2}}{y^{2}}, \n
\Delta_{FK}'&=&\frac{m^{2}+p^{2}}{(1-y)^{2}}-\frac{(\vec{p}-\vec{q})^{2}}{(x-y)^{2}} . \nonumber
\eeq
However, the integrand has an apparent divergence as $x \rightarrow y$.  The integrals over transverse momenta, however, 
integrate to zero in this case, killing the divergence.  We can subtract a term that integrates to zero and that 
explicitly cancels the false divergence from the above contribution.  We make the following change of variables to 
aid in the calculation of the term we want to subtract:

\beq
\label{first-change-begin}
\vec{r}=\half (\vec{p}+\vec{q}), \hspace{1cm} \vec{w}=\half \frac{\vec{q}-\vec{p}}{\sqrt{\eta}},
\eeq
where
\beq
\vec{r}=r\left[ \cos \alpha \hat{x} +\sin \alpha \hat{y} \right], \hspace{1cm}
\vec{w}=w\left[ \cos \delta \hat{x} +\sin \delta \hat{y} \right]
\eeq
and
\beq
\beta=\alpha-\delta, \hspace{1cm} \vec{r}\cdot\vec{w}=rw\cos \beta .
\eeq
Then
\beq
\vec{q}=\vec{r}+\eta \vec{w}, \hspace{1cm} \vec{p}=\vec{r}-\eta \vec{w},
\eeq
and
\beq
q=\sqrt{r^{2}+\eta w^{2}+2rw\sqrt{\eta} \cos \beta }, \hspace{1cm}
p=\sqrt{r^{2}+\eta w^{2}-2rw\sqrt{\eta} \cos \beta } .
\eeq
For convenience, we define:
\beq
r_{\pm}\equiv \sqrt{r^{2}+\eta w^{2}\pm 2rw \sqrt{\eta} \cos \beta} .
\eeq
Finally, 
\beq
\label{first-change-end}
\vec{q}\cdot\vec{p}=r^{2}-\eta w^{2}, \hspace{1cm}  \cos \gamma_{\phi}=\frac{r^{2}-\eta w^{2}}{r_{+}r_{-}}, 
\hspace{1cm} \sin \gamma_{\phi}=\frac{-2rw\sqrt{\eta}}{r_{+}r_{-}}\sin \beta,
\eeq 
are useful relations.
The details of this subtraction are in Appendix \ref{appx:se-inst}.  After making the subtraction, we have:
\beq
&&\langle q',l',t',j' \vert {\cal M}^{2}(\la) \vert q,l,t,j \rangle_{\mathrm{IN}}^{\mathrm{B,F}}=
\n && \hspace{1cm}
-g_{\Lambda}^{2}\frac{128}{3} \delta_{j,j'} {\cal P}^{+}\delta({\cal P-P'})
 \int_{0}^{1} dx B_{l}(x) \int_{0}^{x} dy \hspace{.1cm} \eta\log\eta
\int d^{2}\vec{r} d^{2}\vec{w}
\n && \hspace{2cm} \times
\left\{ e^{-(\Delta^{2}_{FK}+\Delta^{2}_{IK})} 
 W_{q,q'}\tilde{B}_{t}(r_{+}) \tilde{B}_{t'}(r_{-})\left[ \frac{B_{l'}(y)}{\eta^{2}}+\frac{B'_{l'}(y)}{\eta}
 \right. \right. \n &&\hspace{5cm} \left. \left.
-2\frac{B_{l'}(y)}{\eta}\left(\Delta_{FK}\Delta_{FK}'+\Delta_{KI}\Delta_{KI}' \right) \right] 
\right. \n &&\hspace{4cm} \left.
-\tilde{B}_{t}(r) \tilde{B}_{t'}(r)  B_{l'}(y) \frac{\delta_{q,q'}}{\eta^{2}}e^{-32w^{4}}\left[ 1-64 w^{4} \right] \right\}.
\eeq
The integral is explicitly finite, but still sharply peaked around $x \approx y$.  We can speed convergence of the 
numerical integration by smoothing the integral using the change of variables:
\beq
\label{second-change-begin}
\eta=xe^{-p},\hspace{1cm} dy=\eta dp,
\eeq
and
\beq
\label{second-change-end}
y_{w}=\frac{2}{1+w}-1, \hspace{1cm} y_{r}=\frac{2}{1+r}-1, \hspace{1cm} y_{p}=\frac{2}{1+p}-1 .
\eeq
Finally, we get:
\beq
\label{instant-after-change}
&&\langle q',l',t',j' \vert {\cal M}^{2}(\la) \vert q,l,t,j \rangle_{\mathrm{IN}}^{\mathrm{B,F}}=
\n && \hspace{1cm}
-8g_{\Lambda}^{2}\frac{128}{3} \delta_{j,j'} {\cal P}^{+}\delta({\cal P-P'})
\n &&\hspace{2cm} \times
 \int_{0}^{1} dx \int_{0}^{2\pi} 
d\gamma_{\phi}\int_{-1}^{1} \frac{dy_{p}}{(1+y_{p})^{2}} \int_{-1}^{1} \frac{dy_{r}}{(1+y_{r})^{2}} \int_{-1}^{1} 
\frac{dy_{w}}{(1+y_{w})^{2}}
\n && \hspace{2cm} \times
\eta\log\eta \hspace{.1cm} r \hspace{.1cm} w B_{l}(x)
\n && \hspace{2cm} \times 
\left\{ e^{-(\Delta^{2}_{FK}+\Delta^{2}_{IK})} W_{q,q'}\tilde{B}_{t}(r_{+}) \tilde{B}_{t'}(r_{-})
\right. \n && \hspace{3cm} \left. \times
\left[ \frac{B_{l'}(y)}{\eta}+B'_{l'}(y)
-2B_{l'}(y)\left(\Delta_{FK}\Delta_{FK}'+\Delta_{KI}\Delta_{KI}' \right) \right] 
\right. \n &&\hspace{4.5cm} \left.
-\tilde{B}_{t}(r) \tilde{B}_{t'}(r) B_{l'}(y) \frac{\delta_{q,q'}}{\eta}e^{-32w^{4}}\left[ 1-64 w^{4} \right] \right\},
\eeq
where,
\beq
\Delta_{KI}&=&\frac{m^{2}\eta-yr_{+}^{2}+xr_{-}^{2}}{xy}+4w^{2},\n
\Delta_{FK}&=&-\frac{m^{2}\eta+(1-y)r_{+}^{2}-(1-x)r_{-}^{2}}{(1-x)(1-y)}-4w^{2},\n
\Delta_{KI}'&=&\frac{4w^{2}}{\eta}-\frac{m^{2}+r_{-}^{2}}{y^{2}}, \n
\Delta_{FK}'&=&\frac{m^{2}+r_{-}^{2}}{(1-y)^{2}}-\frac{4w^{2}}{\eta}. \nonumber
\eeq

\chapter{Numerical Issues and Parallelization}
\label{chapt:numerical-parallel}

With a problem of this nature the analytic calculation is only part of the solution.  To get results that can be 
used to check the theory or make predictions, it is necessary to further refine the calculation for efficient 
numerical calculations.  Furthermore we can obtain results more rapidly from a well-designed algorithm that makes 
the best use of computer resources.

In this chapter we discuss some of the refinements that help make the Monte-Carlo integration more efficient.  We 
also describe the algorithm and how it is parallelized.

\section{Numerical Calculation of Meson Mass Spectra}
In order to aid the numerical calculations, we will use the change of variables given in 
Eq.~(\ref{first-change-begin}) through Eq.~(\ref{first-change-end}) and Eq.~(\ref{second-change-begin}) through 
Eq.~(\ref{second-change-end}).  The complete change of variables gives:
\beq
\int_{0}^{x} dy \int_{0}^{\infty} dq \int_{0}^{\infty}dp
\rightarrow 32 \eta^{2} \int_{-1}^{1} \frac{dy_{s}}{(1+y_{s})^{2}} \int_{-1}^{1} \frac{dy_{w}}{(1+y_{w})^{2}} \int_{-1}^{1} 
\frac{dy_{r}}{(1+y_{r})^{2}}.
\eeq

We divide out the common factor $16\pi^{3}{\cal P}^{+}\delta^{3}({\cal P}-{\cal P}')$ from all the matrix 
elements and use the following definitions for all of the integrals:
\newpage
\beq
&&W_{q,q'}=\delta_{q,1}\delta_{q',1}\cos ([j-1]\gamma_{\phi})
+\delta_{q,2}\delta_{q',2}\cos ([j+1]\gamma_{\phi})
\n && \hspace{6cm}
+\left(\delta_{q,3}\delta_{q',3}+\delta_{q,4}\delta_{q',4}\right) \cos (j\gamma_{\phi}) ,
\eeq
\beq
\Delta_{FI}&=&\frac{m^{2}+r_{-}^{2}}{y(1-y)}-\frac{m^{2}+r_{+}^{2}}{x(1-x)} \n
&=&\frac{1}{x(1-x)y(1-y)} \left \{ m^{2}\left[ x(1-x)-y(1-y)\right]
\right. \n &&\hspace{1cm} \left.
 +\eta (1-x-y) \left( r^{2}+\eta w^{2}\right) 
\right. \n && \hspace{2cm} \left.
-2rw\sqrt{\eta} \cos\beta \left[ x(1-x)+y(1-y) \right] \right\}, \n
\Delta_{KI}&=&\frac{m^{2}\eta-yr_{+}^{2}+xr_{-}^{2}}{xy}+4w^{2}\n
&=&\frac{\eta m^{2}+\eta \left(r^{2}+\eta w^{2}\right) -2rw\sqrt{\eta}\cos \beta (x+y)}{xy}+4w^{2},\n
\Delta_{FK}&=&-\frac{m^{2}\eta+(1-y)r_{+}^{2}-(1-x)r_{-}^{2}}{(1-x)(1-y)}-4w^{2}\n
&=&-\frac{\eta m^{2}+\eta \left(r^{2}+\eta w^{2}\right) +2rw\sqrt{\eta}\cos\beta \left( 1-x+1-y\right) 
}{(1-x)(1-y)}-4w^{2},\n
\bar{\Delta}_{KI}'&=&4w^{2}-\eta\frac{m^{2}+r_{-}^{2}}{y^{2}}, \n
\bar{\Delta}_{FK}'&=&\eta\frac{m^{2}+r_{-}^{2}}{(1-y)^{2}}-4w^{2} ,
\eeq
where we have rewritten some of the definitions to reduce round-off error.
When calculating the cosine and sine functions of $\gamma_{\phi}$, we will need to use recursion 
relations since we can really only calculate $\sin\gamma_{\phi}$ and $ \cos\gamma_{\phi}$ [see 
Eq.~(\ref{first-change-end})].
\subsection{The Finite Part of the Exchange Interaction}
From Eq. (\ref{finite-exchange-before-change}) we get:
\newpage
\beq
&&\langle q',l',t',j' \vert {\cal M}^{2}(\la) \vert q,l,t,j \rangle^{\mathrm{F}}_{\mathrm{EX}}=
\n && \hspace{1cm}
-\frac{ 1}{6\pi^{3}} g_{\Lambda}^{2} \int d\gamma_{\phi} dq q dx \int dp p dy 
\frac{e^{-\Delta_{FI}^{2}}}{x-y}  \delta_{j,j'}S_{q,q'} 
\n &&\hspace{2cm}\times\frac{B_{l'}(y)B_{l}(x)\tilde{B}_{t}(q)\tilde{B}_{t'}(p)}{x(1-x)y(1-y)}
\left(\frac{1}{\Delta_{FK}}-\frac{1}{\Delta_{KI}}\right)
\left[ 1-e^{2\Delta_{FK}\Delta_{KI}}\right]\n \n \n
&&\hspace{0cm}
=-\frac{16}{3\pi^{3}} g_{\Lambda}^{2} \delta_{j,j'} \int_{0}^{1} dx \int_{0}^{2\pi} d\gamma_{\phi} \int_{-1}^{1} \frac{dy_{s}}{(1+y_{s})^{2}} 
\int_{-1}^{1} \frac{dy_{w}}{(1+y_{w})^{2}} \int_{-1}^{1} \frac{dy_{r}}{(1+y_{r})^{2}} \n
&&\hspace{2cm}\times\frac{B_{l'}(y)B_{l}(x)B_{t}(y_{r_{+}})B_{t'}(y_{r_{-}})}{x(1-x)y(1-y)}
r w e^{-\Delta_{FI}^{2}} \left(\frac{1}{\Delta_{FK}}-\frac{1}{\Delta_{KI}}\right)
\n && \hspace{4cm} \times
\left[ 1-e^{2\Delta_{FK}\Delta_{KI}}\right] \bar{S}_{q,q'} ,
\eeq
where
\beq
y_{r_{\pm}}=\frac{2}{1+r_{\pm}}-1 ,
\eeq
and:
\beq
\bar{S}_{1,1}&=&- \left[ r_{+}^{2}y(1-y)(1-2x) -r_{-}^{2}x(1-x)(1-2y)\right] \cos ([j-1]\gamma_{\phi})\n
&&\hspace{1cm}-\eta \hspace{.1cm} r_{-}r_{+}\left[ y^{2}(1-2x)-x^{2}(1-2y)-2y(1-x)\right] \cos(j \gamma_{\phi}), \n
\bar{S}_{2,2}&=&- \left[ r_{+}^{2}y(1-y)(1-2x) -r_{-}^{2}x(1-x)(1-2y)\right] \cos ([j+1]\gamma_{\phi})\n
&&\hspace{1cm}-\eta \hspace{.1cm} r_{-}r_{+}\left[ y^{2}(1-2x)-x^{2}(1-2y)-2y(1-x)\right] \cos(j \gamma_{\phi}), \n
\bar{S}_{3,3}&=&-\cos (j\gamma_{\phi})\left[ \eta r_{-}r_{+} \cos \gamma_{\phi}(1-x-y+2xy)
\right. \n && \hspace{2cm} \left.
-(1-x-y)\left( y(1-y)r_{+}^{2}-x(1-x)r_{-}^{2}\right) -m^{2}\eta^{3}\right],\n
\bar{S}_{4,4}&=&-\cos (j\gamma_{\phi})\left[ \eta r_{-}r_{+} \cos \gamma_{\phi}(1-x-y+2xy)
\right. \n && \hspace{2cm} \left.
-(1-x-y)\left( y(1-y)r_{+}^{2}-x(1-x)r_{-}^{2}\right) +m^{2}\eta^{3}\right],\n
\bar{S}_{1,2}&=&0,\n
\bar{S}_{2,1}&=&0,\n
\bar{S}_{1,3}&=& \frac{im\eta}{\sqrt{2}} \left[(1-2y)r_{+}\cos(j \gamma_{\phi})-(1-x-y)r_{-}\cos([j-1]\gamma_{\phi})\right],\n
\bar{S}_{3,1}&=& \frac{im\eta}{\sqrt{2}} \left[(1-x-y)r_{+}\cos([j-1]\gamma_{\phi})-(1-2x)r_{-}\cos(j\gamma_{\phi})\right],\n
\bar{S}_{1,4}&=& \frac{im\eta}{\sqrt{2}} \left[-(1-2y+2y^{2})r_{+}\cos(j\gamma_{\phi})
\right.\n && \hspace{3cm} \left.
+(1-y-x+2xy)r_{-}\cos([j-1]\gamma_{\phi})\right],\n
\bar{S}_{4,1}&=& \frac{im\eta}{\sqrt{2}} \left[ (1-2x+2x^{2})r_{-}\cos(j\gamma_{\phi})
\right. \n && \hspace{3cm} \left.
-(1-x-y+2xy)r_{+}\cos([j-1]\gamma_{\phi})\right],\n
\bar{S}_{2,3}&=& \frac{im\eta}{\sqrt{2}} \left[(1-2y)r_{+}\cos(j\gamma_{\phi})-(1-x-y)r_{-}\cos([j+1]\gamma_{\phi})\right],\n
\bar{S}_{3,2}&=& \frac{im\eta}{\sqrt{2}} \left[(1-x-y)r_{+}\cos([j+1]\gamma_{\phi})-(1-2x)r_{-}\cos(j\gamma_{\phi})\right], \n
\bar{S}_{2,4}&=& \frac{im\eta}{\sqrt{2}} \left[(1-2y+2y^{2})r_{+}\cos(j\gamma_{\phi})
\right. \n && \hspace{3cm} \left.
-(1-y-x+2xy)r_{-}\cos([j+1]\gamma_{\phi})\right],\n
\bar{S}_{4,2}&=& \frac{im\eta}{\sqrt{2}} \left[(1-x-y+2xy)r_{+}\cos([j+1]\gamma_{\phi})
\right. \n && \hspace{3cm} \left.
-(1-2x+2x^{2})r_{-}\cos(j\gamma_{\phi})\right],\n
\bar{S}_{3,4}&=&\eta\sqrt{2}r_{-}r_{+}\sin \gamma_{\phi} \sin (j\gamma_{\phi})(1-x-y), \n
\bar{S}_{4,3}&=&\eta\sqrt{2}r_{-}r_{+}\sin \gamma_{\phi} \sin (j\gamma_{\phi})(1-x-y) . \nonumber
\eeq
\subsection{The Instantaneous and Exchange Interaction}
Eq. (\ref{in-plus-exch-before-change}) gives:
\newpage
\beq
&&\langle q',l',t',j' \vert {\cal M}^{2}(\la) \vert q,l,t,j \rangle_{\mathrm{IN+EX}}= {\cal P}^{+}
g_{\Lambda}^{2}e^{-\Lambda^{-4}\Delta_{FI}^{2}} \times\n
&&\hspace{.5cm}
-\frac{16}{3}\delta({\cal P}-{\cal P}')\delta_{j,j'}\int d\gamma_{\phi} dq qdx  \int dp p dy
\theta_{\epsilon}\theta_{\epsilon'}B_{l'}(y)B_{l}(x)\tilde{B}_{t}(q)\tilde{B}_{t'}(p)
\n && \hspace{2cm} \times 
\left[ 1-e^{2\Lambda^{-4}\Delta_{FK}^{(1)}\Delta_{KI}^{(1)}}\right] \n
&&\hspace{3cm} \times
\frac{W_{q,q'}}{(x-y)^{2}} 
\left[ \left(\frac{1}{\Delta_{FK}^{(1)}}-\frac{1}{\Delta_{KI}^{(1)}}\right)
\left( \frac{p^{2}+q^{2}-2\vec{p}\cdot\vec{q}}{x-y}\right)+2 \right] \n \n \n
&&\hspace{1cm}
=-\frac{32}{3\pi^{3}}g_{\Lambda}^{2}\delta_{j,j'}\int_{0}^{1} dx \int_{0}^{2\pi} d\gamma_{\phi} 
\int_{-1}^{1} \frac{dy_{s}}{(1+y_{s})^{2}} 
\int_{-1}^{1} \frac{dy_{w}}{(1+y_{w})^{2}} \int_{-1}^{1} \frac{dy_{r}}{(1+y_{r})^{2}} r w  \n
&&\hspace{2cm}\times
e^{-\Delta_{FI}^{2}} B_{l'}(y)B_{l}(x)B_{t}(y_{r_{+}})B_{t'}(y_{r_{-}})\left[ 1-e^{2\Delta_{FK}\Delta_{KI}}\right] \n
&&\hspace{5cm} \times
W_{q,q'}
\left[ 4w^{2}\left(\frac{1}{\Delta_{FK}}-\frac{1}{\Delta_{KI}}\right)+2 \right] ,
\eeq
\subsection{The Instantaneous Interaction}
The final five-dimensional contribution comes from Eq. (\ref{before-subtraction}):
\beq
&&\hspace{-1cm}\langle q',l',t',j' \vert {\cal M}^{2}(\la) \vert q,l,t,j \rangle_{\mathrm{IN}}^{\mathrm{B,F}} 
=-8g_{\Lambda}^{2}\frac{128}{3} \delta_{j,j'} {\cal P}^{+}\delta({\cal P-P'})\n
&&\hspace{0cm}
 \int_{0}^{1} dx \int_{0}^{2\pi} 
d\gamma_{\phi}\int_{-1}^{1} \frac{dy_{p}}{(1+y_{p})^{2}} \int_{-1}^{1} \frac{dy_{r}}{(1+y_{r})^{2}} \int_{-1}^{1} 
\frac{dy_{w}}{(1+y_{w})^{2}} \hspace{.1cm} \eta\log\eta \hspace{.1cm} r \hspace{.1cm} w
B_{l}(x) \times \n && \hspace{0cm} \left\{
e^{-(\Delta^{2}_{FK}+\Delta^{2}_{IK})} W_{q,q'}\tilde{B}_{t}(r_{+}) \tilde{B}_{t'}(r_{-})
\right. \n && \hspace{2cm} \left. \times
\left[ \frac{B_{l'}(y)}{\eta}+B'_{l'}(y)
-2B_{l'}(y)\left(\Delta_{FK}\Delta_{FK}'+\Delta_{KI}\Delta_{KI}' \right) \right] 
\right. \n &&\hspace{2cm} \left.
-\tilde{B}_{t}(r) \tilde{B}_{t'}(r) B_{l'}(y) \frac{\delta_{q,q'}}{\eta}e^{-32w^{4}}\left[ 1-64 w^{4} \right] 
\right\} \nonumber 
\eeq
\newpage
\beq
&&\hspace{0cm}=
-\frac{64}{3\pi^{3}}g_{\Lambda}^{2} \delta_{j,j'} \int_{0}^{1} dx \int_{0}^{2\pi} 
d\gamma_{\phi}\int_{-1}^{1} \frac{dy_{p}}{(1+y_{p})^{2}} \int_{-1}^{1} \frac{dy_{r}}{(1+y_{r})^{2}} \int_{-1}^{1} 
\frac{dy_{w}}{(1+y_{w})^{2}} 
\n && \hspace{1cm} \times
\log\eta \hspace{.1cm} r \hspace{.1cm} w
B_{l}(x)  
\left\{
e^{-(\Delta^{2}_{FK}+\Delta^{2}_{IK})} W_{q,q'}\tilde{B}_{t}(r_{+}) \tilde{B}_{t'}(r_{-})
\right. \n && \hspace{2cm} \left. \times
\left[ B_{l'}(y)+\eta B'_{l'}(y)
-2B_{l'}(y)\left(\Delta_{FK}\bar{\Delta}_{FK}'+\Delta_{KI}\bar{\Delta}_{KI}' \right) \right] 
\right. \n &&\hspace{2cm} \left.
-\tilde{B}_{t}(r) \tilde{B}_{t'}(r) B_{l'}(y) \delta_{q,q'}e^{-32w^{4}}\left[ 1-64 w^{4} \right] 
\right\} .
\eeq

\subsection{The Self-Energy}
The self-energy is given from Eq. (\ref{self-energy-before-cov}):
\beq
&&\langle q',l',t',j' {\cal M}^{2}(\la) \vert q,l,t,j \rangle^{\mathrm{F}}_{\mathrm{SE}}=
\n && \hspace{1cm}
16 \pi^{3} g_{\Lambda}^{2}{\cal P}^{+}\delta^{(3)}({\cal P}-{\cal P}') \delta_{q,q'}\delta_{j,j'}\Lambda^{2}\n
&&\hspace{3cm} \times
\frac{-1}{24\pi^{2}}\int_{0}^{1}dx B_{l}(x)B_{l'}(x)\mathrm{I}(x)
\int_{0}^{\infty}dq \hspace{.1cm} q \tilde{B}_{t}(q)\tilde{B}_{t'}(q) \n \n \n
&&\hspace{0cm}=
-\frac{1}{24\pi^{2}}g_{\Lambda}^{2}\delta_{q,q'}\delta_{j,j'}\Lambda^{2}
\int_{0}^{1}dx B_{l}(x)B_{l'}(x)\mathrm{I}(x)
\int_{0}^{\infty}dq \hspace{.1cm} q \tilde{B}_{t}(q)\tilde{B}_{t'}(q) \n \n
&&\hspace{0cm}=
-\frac{1}{12\pi^{2}}g_{\Lambda}^{2}\delta_{q,q'}\delta_{j,j'}\Lambda^{2}
\int_{0}^{1}dx B_{l}(x)B_{l'}(x)\mathrm{I}(x)
\n && \hspace{5cm} \times
\int_{-1}^{1} \frac{dy_{q}}{(1+y_{q})^{2}} \hspace{.1cm} q B_{t}(y_{q})B_{t'}(y_{q}) .
\eeq

\subsection{The Complex Hamiltonian}

One substantial difference between the meson and glueball Hamiltonians is that the meson Hamiltonian is complex.  
However, each matrix element is either real or pure imaginary, so there is still only one 
integral per matrix element.  This means we can treat the problem the same way we treated the glueball numerically. 
Specifically, if $q=q'$ or ($q=3$ and $q'=4$) or ($q=4$ and $q'=3$) the matrix element is real, otherwise it is 
imaginary.

\subsection{Reducing the Number of Matrix Elements to Calculate}

We can take advantage of certain features of the Hamiltonian to reduce the number of matrix elements we actually 
need to calculate.  First consider the exchange interaction since it contains the only contributions when $q \ne q'$.  

The exchange interaction contains non-zero contributions only when $q\ne q'$.  The ``main'' part of the exchange diagram 
(everything except the $\bar{S}$'s) is obviously symmetric when we switch final and initial states.  This switch is accomplished 
simply by letting $q \leftrightarrow q'$, $l \leftrightarrow l'$, $t \leftrightarrow t'$, $x \leftrightarrow y$, 
$q\hspace{.1cm}(r_{+}) \leftrightarrow p\hspace{.1cm}(r_{-})$, and $\gamma_{\phi} \rightarrow -\gamma_{\phi}$.  

We see that the real $\bar{S}$'s are also symmetric under interchange of final and initial states, and 
because they are real this part of the Hamiltonian is already Hermitian.  The imaginary $\bar{S}$'s 
are also antisymmetric under interchange of final and initial states.  This guarantees the entire Hamiltonian is 
Hermitian and cuts almost in half the number of matrix elements that must be calculated.  Of course Hermiticity is 
necessary to obtain real eigenvalues and it provides a check on the analytic work up to this point.

More redundancies in the Hamiltonian can be found if we consider the case $j=0$.  First, consider the $\bar{S}$'s in this 
case:
\beq
\bar{S}_{1,1}&=&- \left[ r_{+}^{2}y(1-y)(1-2x) -r_{-}^{2}x(1-x)(1-2y)\right] \cos (\gamma_{\phi})\n
&&\hspace{2cm}-\eta \hspace{.1cm} r_{-}r_{+}\left[ y^{2}(1-2x)-x^{2}(1-2y)-2y(1-x)\right], \n
\bar{S}_{2,2}&=&- \left[ r_{+}^{2}y(1-y)(1-2x) -r_{-}^{2}x(1-x)(1-2y)\right] \cos (\gamma_{\phi})\n
&&\hspace{2cm}-\eta \hspace{.1cm} r_{-}r_{+}\left[ y^{2}(1-2x)-x^{2}(1-2y)-2y(1-x)\right],  \n
\n
\bar{S}_{3,3}&=&-\left[ \eta r_{-}r_{+} \cos \gamma_{\phi}(1-x-y+2xy)
\right. \n && \hspace{2cm} \left.
-(1-x-y)\left( y(1-y)r_{+}^{2}-x(1-x)r_{-}^{2}\right) -m^{2}\eta^{3} \right],\n
\bar{S}_{4,4}&=&-\left[ \eta r_{-}r_{+} \cos \gamma_{\phi}(1-x-y+2xy)
\right. \n && \hspace{2cm} \left.
-(1-x-y)\left( y(1-y)r_{+}^{2}-x(1-x)r_{-}^{2}\right) +m^{2}\eta^{3} \right],\n
\bar{S}_{1,2}&=&0,\n
\bar{S}_{2,1}&=&0,\n
\bar{S}_{1,3}&=& \frac{im\eta}{\sqrt{2}} \left[(1-2y)r_{+}-(1-x-y)r_{-}\cos(\gamma_{\phi})\right],\n
\bar{S}_{3,1}&=& \frac{im\eta}{\sqrt{2}} \left[(1-x-y)r_{+}\cos(\gamma_{\phi})-(1-2x)r_{-}\right],\n
\bar{S}_{1,4}&=& \frac{im\eta}{\sqrt{2}} \left[-(1-2y+2y^{2})r_{+}+(1-y-x+2xy)r_{-}\cos(\gamma_{\phi})\right],\n
\bar{S}_{4,1}&=& \frac{im\eta}{\sqrt{2}} \left[ (1-2x+2x^{2})r_{-}-(1-x-y+2xy)r_{+}\cos(\gamma_{\phi})\right],\n
\bar{S}_{2,3}&=& \frac{im\eta}{\sqrt{2}} \left[(1-2y)r_{+}-(1-x-y)r_{-}\cos(\gamma_{\phi})\right],\n
\bar{S}_{3,2}&=& \frac{im\eta}{\sqrt{2}} \left[(1-x-y)r_{+}\cos(\gamma_{\phi})-(1-2x)r_{-}\right], \n
\bar{S}_{2,4}&=& \frac{im\eta}{\sqrt{2}} \left[(1-2y+2y^{2})r_{+}-(1-y-x+2xy)r_{-}\cos(\gamma_{\phi})\right],\n
\bar{S}_{4,2}&=& \frac{im\eta}{\sqrt{2}} \left[(1-x-y+2xy)r_{+}\cos(\gamma_{\phi})-(1-2x+2x^{2})r_{-}\right],\n
\bar{S}_{3,4}&=& 0, \n
\bar{S}_{4,3}&=& 0 . \nonumber
\eeq
From this we see that $\bar{S}_{1,3}=\bar{S}_{2,3}$, $\bar{S}_{1,4}=-\bar{S}_{2,4}$, and $\bar{S}_{3,4}=0$.

The only redundancy when $q=q'$ is when $q=1,q'=1$ and $q=2,q'=2$.  So we have the following symmetries:
\beq
\langle 2,l',t',0 \vert {\cal M}^{2} \vert 2,l,t,0 \rangle &=&\langle 1,l',t',0 \vert {\cal M}^{2} \vert 1,l,t,0 
\rangle , \n
\langle 1,l',t',0 \vert {\cal M}^{2} \vert 3,l,t,0 \rangle &=&\langle 2,l',t',0 \vert {\cal M}^{2} \vert 3,l,t,0 
\rangle , \n
\langle 1,l',t',0 \vert {\cal M}^{2} \vert 4,l,t,0 \rangle &=&-\langle 2,l',t',0 \vert {\cal M}^{2} \vert 4,l,t,0 
\rangle , \n
\langle 3,l',t',0 \vert {\cal M}^{2} \vert 4,l,t,0 \rangle &=& 0 .
\eeq

\subsection{Rotational Symmetry: $j\rightarrow -j$}

Reference \cite{brentD} shows that the basis $\vert q,l,t,-j\rangle$ is related to the basis $\vert 
q,l,t,j\rangle$ by swapping the states $\vert 1,l,t,j\rangle$ and $\vert 2,l,t,j\rangle$, and changing the sign of 
$\vert 4,l,t,j\rangle$.  Renaming the basis states and changing their phases has no effect on the eigenvalues of 
the matrix, so the eigenvalues of states with the same absolute value of $j$ are the same.

\section{Hamiltonian Matrix with Parallel Processing}

Fundamental physics research is dominated by analytical calculations.  However, as theories become 
more complex, comparing theory to experiment becomes more numerically intensive.  Fortunately the computational performance 
to price ratio has been increasing at such a rate that many numerical problems previously considered impractical 
can be done on readily available desktop computers.

Although computers have and will continue to get faster, algorithms written to run on parallel processors almost 
always perform better than their single processor counterparts.  The amount of work that is required to write a 
program using such standards as  the Message Passing Interface (MPI) may seem inhibiting, but with certain algorithms, like the one used to numerically 
determine Hamiltonian matrix elements, there can be a tremendous benefit.

\subsection{Thread Independence}

Algorithms used in fields such as Lattice Gauge Theory use program threads that directly depend on each other.  
For instance, the lattice may be equally divided among the threads, and then boundary information is passed between 
them.  The efficiency of this type of algorithm, whose threads communicate frequently, can be largely influenced by 
the network environment.  However, when calculating matrix elements, each matrix element can be calculated almost 
independently of the others.  Thus the communication between threads is very infrequent and does not greatly 
decrease the performance of the program.  This fact allows our algorithm to run efficiently on many different 
machines over a large, inefficient network.

The majority of our algorithm used to calculate the Hamiltonian matrix is embarrasingly parallel, each 
worker process only being told which matrix element to calculate.  However, there are parts of the program 
(Sec.~\ref{simple-algorithm}) that use a parallelized version of the numerical integrator (Sec.~\ref{parallel-vegas}) to 
limit idle worker processes.

\subsection{Determining Accuracy of Matrix Elements}
\label{me-accuracy}

The longer a program runs, the more important it is to make 
sure the source code is written well to reduce run times.  Similarly, it is important that the program is not 
performing calculations that do not affect the result.

Each matrix element in the Hamiltonian is determined by numerically calculating a five-dimensional integral using 
VEGAS, an adaptive Monte-Carlo integrator.  Monte-Carlo integrators are much more efficient then 
nested single dimensional integrators for a large enough number of dimensions.\footnote{For three or 
more dimensions Monte-Carlo is generally faster.}  However, the results converge 
slowly\footnote{Monte-Carlo results converge like $\frac{1}{\sqrt{N}}$ for N function calls.}.  
Thus, each matrix element should only be calculated as accurately as needed to 
produce results within the desired error.  

If all matrix elements are calculated to the same precision, the program will run for many times longer than 
necessary.  Although it is dependent on the value of input parameters, to achieve two-percent accuracy in the 
eigenvalues only about ten percent of the matrix elements 
are important enough to require an error less than fifty percent.  Unfortunately which matrix elements are 
important can 
not be known a priori.  The method used to calculate each matrix element to within the required precision is 
discussed in Section \ref{simple-algorithm}.

\subsection{The Simplified Algorithm}
\label{simple-algorithm}

The purpose of the algorithm is to calculate all Hamiltonian matrix elements to sufficient accuracy to give 
eigenvalues within a pre-determined error.  Two questions must be answered before we can determine how accurately to 
calculate each matrix element.   How should the error in the eigenvalues be determined from a matrix in which every matrix 
element has a known uncertainty?  How does the uncertainty in a matrix element translate into errors in the 
eigenvalues?  

Before answering these questions we give a little bit of detail about how the matrix element is actually 
determined.  Each matrix element has a contribution from a two-dimensional and a five-dimensional integral.  The 
two-dimensional contributions are determined by nesting two one-dimensional integrations.  This first contribution 
can be calculated easily and quickly to a much higher accuracy than the remaining contribution from the 
five-dimensional integral.  VEGAS saves statistical information about the integral, so if the 
desired accuracy is not reached, the routine can be re-entered, and the calculation will proceed from where it left 
off.  If the routine did not have the ability to be restarted with previously obtained information, the desired 
precision would need to be known a priori.  Since matrix diagonalization is a highly non-linear problem, 
understanding how each matrix element affects the answer is nearly impossible. 

With no reasonable analytic way of understanding how the uncertainties in the matrix elements translate to errors 
in the eigenvalues, we calculate the error statistically.  Each matrix element is independently and randomly 
varied about its average value using a gaussian distribution determined by the standard deviation,
then a set of eigenvalues is determined.  This is repeated to generate a large 
number of eigenvalue sets.  Since we only expect the lowest several eigenvalues to be correctly determined by our 
renormalized Hamiltonian, we calculate the average and standard deviation of the lowest ten eigenvalues.  The largest 
uncertainty in any of these lowest eigenvalues is defined as the error in the eigenvalues.  Although this is a good 
statistical way to determine the error, it can be very cpu-intensive for large matrices, and thus is not an 
efficient way to determine the deviation produced in the eigenvalues for each matrix element individually.

To approximate the deviation in the eigenvalues caused by a single matrix element we use a much less thorough method that 
only requires calculating the eigenvalues three times (versus a hundred or more).  The eigenvalues are determined 
when the element under question is at its average value, and then when it is its average value plus and minus 
its standard deviation.  The deviation is defined as the largest change in an eigenvalue when the element is 
shifted up or down by its standard deviation.

\begin{figure}
    \centerline{\bf Simplified Algorithm for the 5D Contribution to a Matrix Element}
    \vspace{1cm}
    \centerline{\epsfig{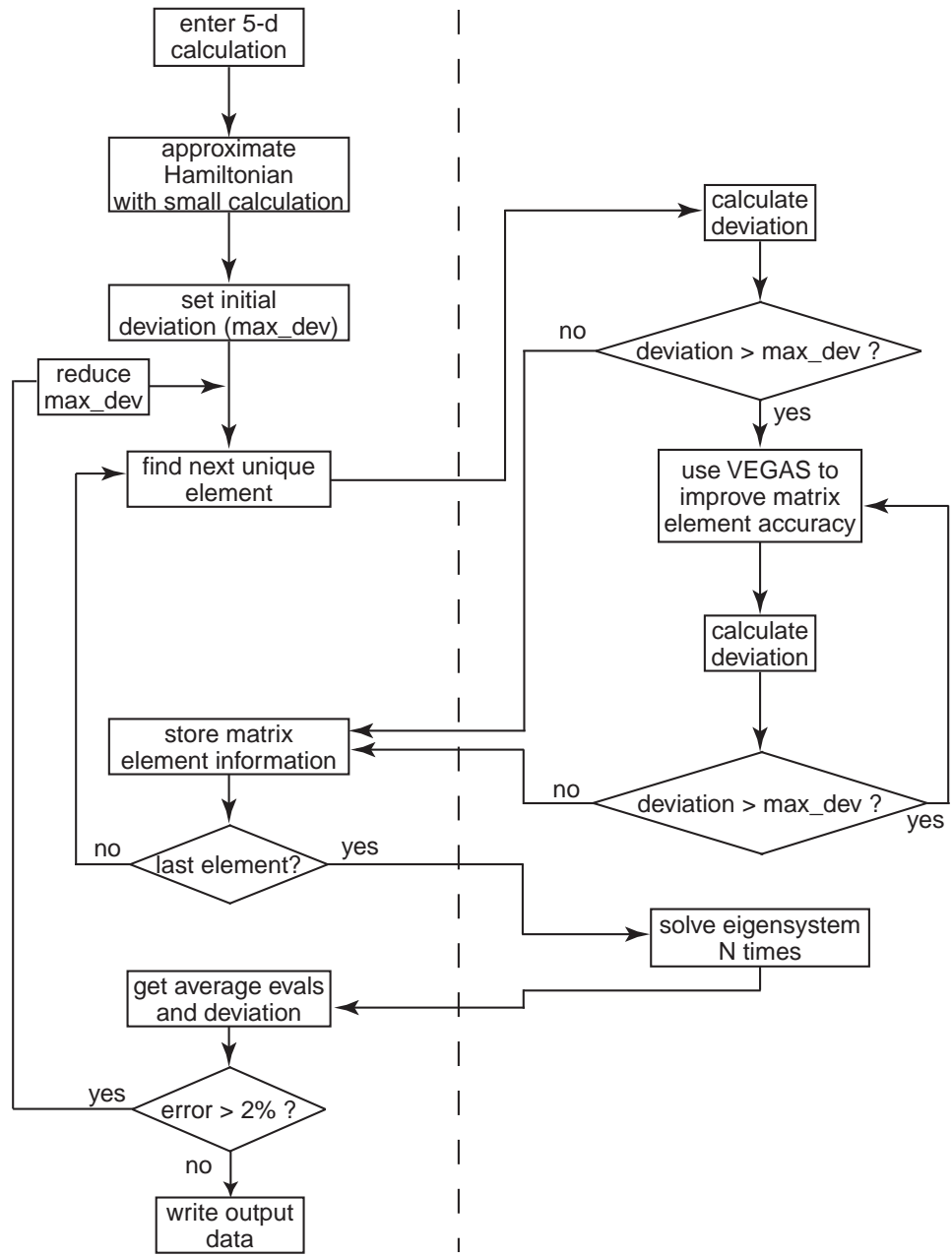}}
    \caption[Simple Algorithm Flowchart]{Flowchart for the simple algorithm.  The dashed line can be ignored to understand the basic flow, 
    however in the parallelized version (Sec.~\ref{parallel-algorithm}) it separates the dispatcher's flow (left) 
    from the worker's flow (right).}
    \label{fig:simp-algo}
\end{figure}

With this understanding of the deviation produced by a single matrix element, and the error in the 
eigenvalues from the entire matrix, the basic algorithm is straightforward.  After the two-dimensional integrals 
have been calculated, the algorithm cycles through each matrix element reducing the deviation produced by each element 
to a predetermined value.  At the end of an iteration, the error in the eigenvalues is determined.  If the error is 
less than the desired error, the program exits, otherwise, it cycles through the matrix elements again, this 
time requiring the deviation to be smaller than it was for the previous iteration.  The program iterates through 
the matrix in this manner until the desired accuracy is reached.  Figure \ref{fig:simp-algo} shows the flow chart for 
the basic algorithm.

\subsection{Saving Data}
\label{saving-data}

Saving basic data like the Hamiltonian and its error can be very useful because 
output can be rapidly regenerated or generated in a different format if needed.  However there is much more important data that 
can be saved.  The most cpu-expensive data generated while running this program is the statistical information generated 
by VEGAS for each integral.  

In its basic form VEGAS only saves the information for the one 
integral it is calculating, and discards that information when it moves on to the next integral.  The problem is that 
the
same integral may need to be calculated more accurately in a later iteration, thus the integration 
would have to start from the beginning.  Although saving the VEGAS information for each matrix element uses a lot of 
storage (30 integers, 1649 double precision), it is important to save all of it.  
If there is not enough memory available, the algorithm uses a scratch file to store the VEGAS information for the 
entire matrix.  During most of a new calculation ($\approx 70-90\%$)\footnote{A calculation that is restarted or one 
that uses saved data may go through several iterations reducing the deviation until there are integrals that 
need more precision.  Until this point is reached, the VEGAS information is accessed frequently.  If the information is 
stored on disk instead of memory, there is a significant 
but temporary slowdown in the calculation.}, there is little slowdown from disk access.

Despite the performance gain in modern computers, they are still subject to crashes.  Also, if the code is running 
in parallel on 
many different computers, the chance of the code stalling due to a single computer failure increases.  This problem 
can be remedied in a straightforward manner by occasionally saving the VEGAS information from memory to a file which 
can later be read in for code restart.

One final benefit of saving data, in particular the data for the five-dimensional integrals, is that the coupling is 
an overall factor multiplying all five-dimensional integrals.  If the parameters vary only in the coupling from a previous run, the old 
matrix element estimates can be used as a starting point, which speeds searches through our parameter space.

\subsection{Parallelizing The Algorithm}
\label{parallel-algorithm}

We first define the nomenclature we use when discussing the parallelized algorithm.  The Message Passing 
Interface (MPI) uses one dispatcher process and several worker processes.\footnote{There is nothing in MPI that forces 
the distinction between the dispatcher and the worker.}  In this algorithm the dispatcher is the organizer.  
It sends jobs for the workers to do and receives their results.  In addition the dispatcher process 
does not use a lot of computer cycles and therefore does not require its own processor.  There is an important distinction between processes and 
processors.  The processor(s) is(are) the computer's physical cpu(s), whereas the processes are simply jobs running 
on the machine.  For example, on a particular machine, we typically run four processes on two processors.  One process 
is the dispatcher and the remaining three are all workers.\footnote{The reason we run more than one worker process per 
processor is to artificially slow the workers on faster machines so the cluster is more homogeneous.}

There are many different ways this problem can be parallelized.  For example, the matrix can be divided 
equally between the processors, each responsible for one submatrix.  Another method is to let each process 
always work independently, moving from element to element until the desired error in the eigenvalues is reached.  
These two methods are examples of something that is too simple, and too complex, respectively.  The first method 
assumes that each section of the matrix is equally important.  However, the far off-diagonal elements are very 
small and calculating them to within more than an order of magnitude is often unnecessary.  The second approach 
requires a very complicated method of tracking which process is working on which matrix element, trying to reach a 
particular deviation.  Also it is not clear that this method would be efficient since only a handful of matrix 
elements are very important, and they take a lot of time to calculate to sufficient accuracy.

The philosophy we use in parallelizing the program is one of efficient simplicity.  The method is simple 
because the calculation is broken up into steps and because no process continues to the next step until all other 
processes are ready to proceed.  Efficiency is achieved by having processes that have finished help the processes 
that are still working.

Although the entire code is parallelized, we only discuss two sections in detail.  The first 
is the parallelization of the error calculation and the second is the parallelization of the 5-d integral calculation.  
At the end of each iteration (reducing the deviation in the five-dimensional integrals) the 
error in the eigenvalues needs to be calculated.  However, particularly for large matrices, this calculation can take a 
long time because it requires the complete eigensystem analysis.  If only the dispatcher is used to do the error 
calculation, it may take several minutes, leaving all of the worker processes idle.  The error calculation is parallelized by having each 
worker process generate an equal number of eigenvalue sets, then sending the results to the dispatcher, who then 
determines the averages and standard deviations.

The part of the calculation that benefits the most from parallel processing is the calculation of the five-dimensional 
integral contribution to each matrix element.  Each matrix element can be calculated independently so there is almost 
no overhead due to communication between processes.  A worker is given an assignment from the dispatcher.  It then 
calculates it and returns the result, the process repeats with a new assignment, and an updated Hamiltonian.  Figure 
\ref{fig:simp-algo}, the simple algorithm, also shows which part of the flow is executed by the dispatcher process 
(left of dotted line) and the worker processes (right of dotted line).
The only time this is not true is at the end of each iteration.  
Once all of the matrix elements have been assigned, as individual worker processes finish, the dispatcher directs 
them to help another worker process complete its assignment.  The implementation of this ``helping'' algorithm is the first 
important and difficult addition to the parallelized algorithm.  

\begin{figure}
    \centerline{\bf Helping Algorithm For a Worker Process}
    \vspace{1cm}
    \centerline{\epsfig{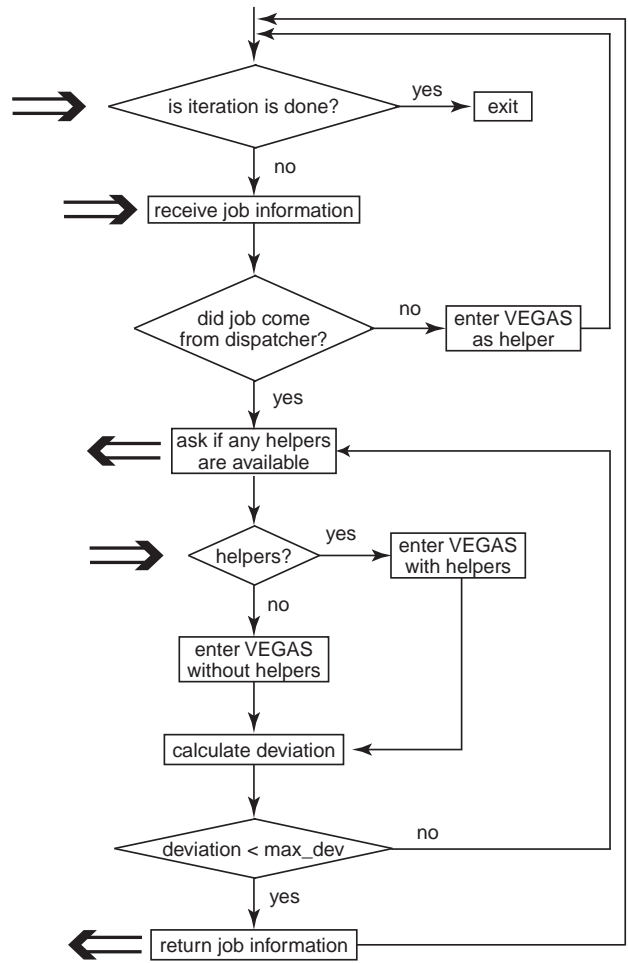}}
    \caption[Helping Algorithm Flowchart.]{Flowchart for a worker process during the five-dimensional calculation, 
    including the steps to help other processes.  The double-lined arrows represent communication to and from 
    the dispatcher.}
    \label{fig:help-algo}
\end{figure}

If there is only one matrix element that requires a long time to calculate, then as the other workers finish, they 
sit idle waiting for the last element to be determined.  This is a major drawback of the 
simple parallelization (independent matrix elements) since, except at the end of each iteration, the program can 
efficiently run using many processes.  Thus, all but one process can be sitting idle while the last one continues to work.  
This can be remedied by having the 
dispatcher keep track of who is working and who is available to help.  This increases the  
communication between the dispatcher and the worker at the time of assignment.  The worker must be 
notified if it will be helping another process, or if it will get help, and who will be helping it.  When the worker is 
finished it must let the dispatcher know if it was helping or getting help, etc.  Figure \ref{fig:help-algo} shows 
the flowchart for a worker process with the option to help other processes.

The last complication that requires additional communication between processes is saving VEGAS information.  
Not only should the VEGAS information be saved for the completed matrix elements, it 
should also be saved for calculations in progress.  Otherwise, information from an integral that has consumed 
hours of cpu time might be lost.  Thus the worker may 
also send updated VEGAS information, without sending a result.  It is non-trivial to work out the communication 
between the worker and dispatcher process to allow for a wide range of information transfer.

\subsection{Parallelized VEGAS}
\label{parallel-vegas}

The parallelization of the numerical integrator is done in an attempt to limit the total number of 
communications between the worker in charge of a particular integral, and the processes helping it.  Also, we do 
not want to get into the details of the adaptive nature of the integrator but do want a simple way of 
speeding the convergence of the integral.  

The worker calls VEGAS with, among other things, the number of points to use and the number of iterations to 
attempt.  After the desired number of iterations is completed, the routine statistically combines each of the 
iterations to get the result and its uncertainty.  Thus, it is very straightforward to include information from 
the helper routines as extra iterations in the main worker.

The information that is lost using this method is that which helps refine the grid to make VEGAS an adaptive 
integrator.  However, according to the documentation \cite{num-recips}, the grid does not get significantly refined after a 
few iterations of a small number of points.  Thus discarding the grid information should be unimportant.

\chapter{Results}
\label{results}

In this Chapter we present our fully relativistic results for the meson mass spectrum.  The results are presented along with 
experimental data \cite{particle-data-table} and predictions from a light-front calculation using a similar 
renormalization method with a nonrelativistic reduction \cite{martinaA,martinaB}.

We first present important information about the experimental results we use.  
Our results begin with a brief description of the steps required to calculate the meson spectrum followed by two 
applications.  
We apply the method to the $b\bar{b}$ system as a check of the theory to verify that it agrees with 
nonrelativistic results and the $c\bar{c}$ system is investigated in detail.
Finally, we show the spin-averaged meson probability densities defined in Eq. (\ref{meson-prob-density}) to show 
what type of wavefunctions are produced.

\section{Experimental Results}

Heavy quarkonia states can be produced by colliding electrons and positrons.\footnote{It is possible that some 
states couple very weakly to this production channel which could lead to missing states in the Particle Data 
Table.  There is important ongoing work seeking new hadrons using new beams and targets.}  As the center-of-mass energy is 
changed, an increase in scattering events signals the existence of a resonance.  The resonance is associated with a 
particle being created and then decaying into other particles.  The mass of the initial particle is determined 
simply by finding the center-of-mass energy at which the resonance occurs.  Daughter particle masses are 
determined by subtracting the energy of emitted photons or pions from the parent particle.  Identifying
the particles generated in a detector is not simple.  One must infer the quantum numbers by knowing the 
quantum numbers of the beam and the daughter particles that are actually detected.  

Although particle accelerators have produced a large amount of data that has been analyzed to give a plethora of 
information on spectroscopy of systems and transitions between particles, not all theoretically predicted particles 
have been seen.  This fact is important when comparing new theoretical results to experimental data.

The charmonium system contains two (1S) states, the $J/\psi$ and the $\eta_{c}$.  The $\eta_{c}$ is not directly 
produced in colliding $e^{+}e^{-}$ experiments but it is observed through transitions from the $J/\psi$ and other 
heavier states.  The bottomonium system is also expected to have two (1S) states but the $\eta_{b}$ has not been 
seen experimentally because the transition from the $\Upsilon$ is greatly suppressed relative to the $J/\psi 
\rightarrow \gamma \eta_{c}$ because of the different quark masses, charges and photon energies.

\section{Procedure}

It should first be noted that there is no simple way to guarantee correct state identification 
when performing this calculation because, for instance, the spin operator is dynamic on the light-front as is 
equal-time parity.  To simplify state identification, the quantum numbers we use for each data set include the 
z-component of the total spin ($j$) and charge conjugation (C).\footnote{We can also (usually) determine 
the intrinsic spin of the system and the state's symmetry in the longitudinal momentum direction}  The drawback is 
that there are fewer low-lying states associated with each set of quantum numbers.  For instance, the lowest five 
$b\bar{b}$ states contain two with negative charge conjugation ($n=0$ and $n=4$), and three with positive charge 
conjugation ($n=1,2,3$).  We must use the $\mathrm{C}=+$ states to try to determine the mass ratios because we need  
at least two ratios to fix $m$ and $\alpha$.\footnote{In addition since the second $\mathrm{C}=-$ state is 
actually $n=4$, we can not trust the accuracy of such an excited state.}

There are two fundamental QCD parameters that need to be fixed, and the scale (cutoff) needs to be determined.  The 
calculation produces eigenvalues of the renormalized invariant-mass operator, ${\cal M}^{2}(m_{_{\Lambda}},\Lambda)$.  
These eigenvalues are used to determine the fundamental QCD parameters and the cutoff.  The steps to 
determine these values are described next.

Eigenvalues are generated for a set of $\alpha$ and $\frac{m}{\Lambda}$.  If the renormalization was entirely 
non-perturbative and the Hamiltonian was calculated to all orders, the spectrum would be independent of the cutoff; 
however, we expect some cutoff dependence because we renormalize perturbatively and keep only two partons in our 
states.  The renormalization is most 
trustworthy where there is little cutoff dependence.  

The first step in finding the correct value of $\alpha$ and \mol is to determine the cutoff by fixing one 
mass\footnote{We fix the lowest mass state for the given quantum numbers.} using the relation:
\beq
\Lambda^{2}=\frac{m^{2}_{measured}}{\langle i \vert {\cal M}^{2}(\Lambda ) \vert i \rangle},
\eeq
where $m^{2}_{measured}$ is taken from \cite{particle-data-table} and $i$ refers to the state used to fix the 
cutoff.  Next the rest of the meson spectrum can be calculated using this cutoff:
\beq
m^{2(j)}_{calculated}=\Lambda^{2}\langle j \vert {\cal M}^{2}(\Lambda ) \vert j \rangle,
\eeq
where $j$ refers to the state being calculated.  

Secondly we analyze the spectrum for a range of $\alpha$ and \mol to find the values that give the correct values 
for the lowest and second lowest mass states above the state we fix.  Finding which values of $\alpha$ and \mol 
correctly determine these two observables gives us our best values of $\alpha$ and \mol.

\section{Convergence Testing}

We must be certain we are using an adequate number of basis functions when taking final data.  We verify our 
results have converged for various values of \mol with $\alpha=\frac{1}{2}.$\footnote{We expect to need more 
longitudinal basis functions as \mol increases because the wavefunction should become more sharply peaked about 
$x=\half$ for heavy quarks.}  We reiterate that it is not necessary for the ground state eigenvalue to decrease as each 
additional function is added since the previous functions are shifted \footnote{See Chapter \ref{section:bsplines} for 
details.}.  

\begin{figure}
    \centerline{\epsfig{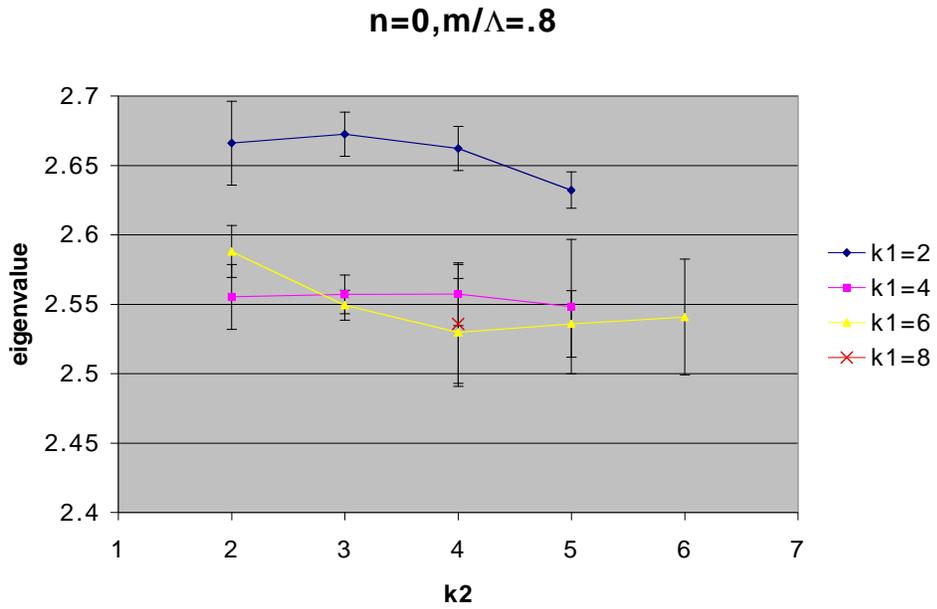}}
    \caption{\label{ccbar-conv-m08-n0}Convergence of the ground state for \mol$=.8$ and $\alpha=.5$.}
\end{figure}

\begin{figure}
    \centerline{\epsfig{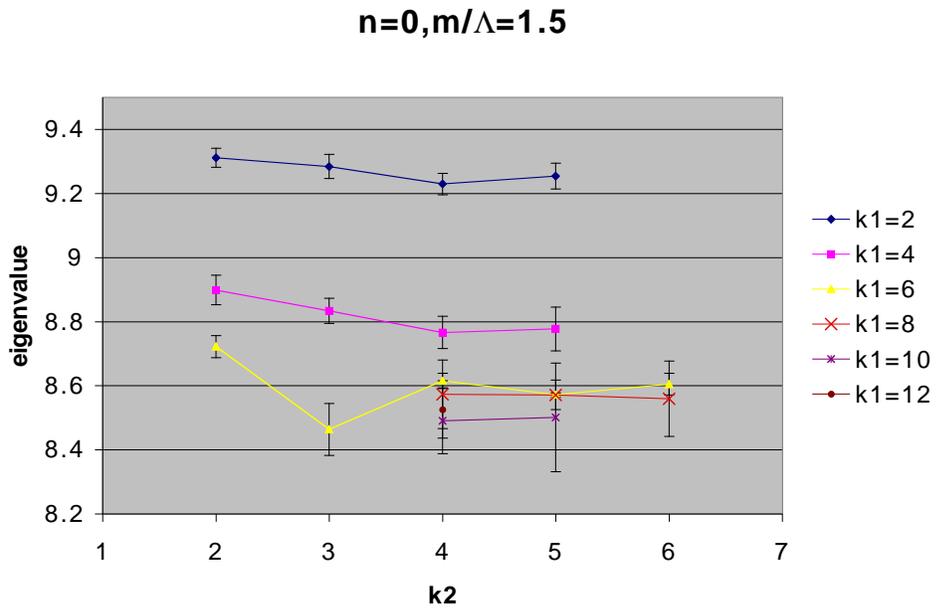}}
    \caption{\label{ccbar-conv-m15-n0}Convergence of the ground state for \mol$=1.5$ and $\alpha=.5$.}
\end{figure}

Increasing the order of the B-splines ($m$ in B-spline notation) beyond third order does not improve convergence, 
but increases the number of nonzero matrix elements.  Therefore all B-splines we use are third order.
Figures \ref{ccbar-conv-m02-n0} through \ref{ccbar-conv-m15-n2} show the convergence of the lowest three states 
for $\frac{m}{\Lambda}=.2$, $.8$, and $1.5$ as we increase the number of longitudinal and transverse states.  
The number of B-splines in the longitudinal direction is:
\beq
k_{1}+m+1-2 = k_{1}+2,
\eeq
where two states are not used because they have the incorrect behavior as $x \rightarrow 0,1$.  We pair 
the splines into symmetric and antisymmetric functions under particle exchange 
(see Sec.~\ref{longitudinal-functions}).  Thus the total number of splines 
in the longitudinal direction is:
\beq
\label{number-long}
\mathrm{N}_{l}=\frac{k_{1}}{2}+1.
\eeq
Similarly the number of B-splines in the transverse direction is:
\beq
\label{number-trans}
\mathrm{N}_{t} = k_{2}+m+1-3 = k_{2}+1,
\eeq
where three states are not used to ensure finite kinetic energy due to small x singularities.

\begin{figure}
    \centerline{\epsfig{file=\mesonpath ccbar-conv-m02-n0.epsf,width=5in}}
    \caption{\label{ccbar-conv-m02-n0}Convergence of the ground state for \mol$=.2$ and $\alpha=.5$.}
\end{figure}
\begin{figure}
    \centerline{\epsfig{file=\mesonpath ccbar-conv-m02-n1.epsf,width=5in}}
    \caption{\label{ccbar-conv-m02-n1}Convergence of the first exited state for \mol$=.2$ and $\alpha=.5$.}
\end{figure}
\begin{figure}
    \centerline{\epsfig{file=\mesonpath ccbar-conv-m02-n2.epsf,width=5in}}
    \caption{\label{ccbar-conv-m02-n2}Convergence of the second excited state for \mol$=.2$ and $\alpha=.5$.}
\end{figure}
\begin{figure}
    \centerline{\epsfig{file=\mesonpath ccbar-conv-m08-n1.epsf,width=5in}}
    \caption{\label{ccbar-conv-m08-n1}Convergence of the first exited state for \mol$=.8$ and $\alpha=.5$.}
\end{figure}
\begin{figure}
    \centerline{\epsfig{file=\mesonpath ccbar-conv-m08-n2.epsf,width=5in}}
    \caption{\label{ccbar-conv-m08-n2}Convergence of the second excited state for \mol$=.8$ and $\alpha=.5$.}
\end{figure}
\begin{figure}
    \centerline{\epsfig{file=\mesonpath ccbar-conv-m15-n1.epsf,width=5in}}
    \caption{\label{ccbar-conv-m15-n1}Convergence of the first exited state for \mol$=1.5$ and $\alpha=.5$.}
\end{figure}
\begin{figure}
    \centerline{\epsfig{file=\mesonpath ccbar-conv-m15-n2.epsf,width=5in}}
    \caption{\label{ccbar-conv-m15-n2}Convergence of the second excited state for \mol$=1.5$ and $\alpha=.5$.}
\end{figure}

As \mol increases a larger number of longitudinal states 
are needed for the eigenvalues to converge.  Although we could have used fewer basis functions for smaller values 
of \mol, all of our calculations use $k_{1}=8$ and $k_{2}=5$.  Using equations (\ref{number-long}),  
(\ref{number-trans}), and noting that we have four spin states, the number of basis states we use is:
\beq
4 \times N_{l} \times N_{t} = 4 \times 5 \times 6 = 120 .
\eeq
Since the Hamiltonian is Hermitian, at most 7260 of the 14400 matrix elements are unique.  Since not all of the 
B-splines spatially overlap all of the other B-splines, there are 1680 elements that are zero.  For our choice 
of $k_{1}$ and $k_{2}$ we have reduced the number of elements that need to be calculated by $23\%$.

\section{Error Estimation}

There are three sources of errors in our meson spectra.  Since we do not use an infinite number of basis functions 
we can not exactly determine the eigenvalues of our Hamiltonian.  Based on our convergence testing, we can choose the 
smallest number of basis functions that produce converged eigenvalues to within two percent.  These eigenvalues are of the 
Hamiltonian (IMO), so a two percent error in the eigenvalues will produce approximately a one 
percent error in the masses.

The second source of errors is statistical.  Each matrix element is calculated by Monte-Carlo integration which 
suffers from statistical uncertainty.  We calculate the error in the eigenvalues by allowing each matrix element to vary 
randomly within its gaussian distribution.  This is done many times to generate a large number of eigenvalue 
sets that 
are used to calculate the average eigenvalue and the errors in the eigenvalues.  The largest error allowed in any 
eigenvalue is two percent, giving a one percent error in the masses.

The final and most significant source of error is in our approximation of the theory:  limiting the approximate states 
to a quark-antiquark pair and truncating the Hamiltonian at second order in the coupling.  We cannot 
reliably estimate the error produced by our approximation to the real state without first solving the meson 
spectrum with $\vert q\bar{q} g \rangle$ states.  We can estimate the error from 
approximating the Hamiltonian.  

Our Hamiltonian contains a contribution from the kinetic energy and second-order interactions, so we should expect 
errors of order $g^{4}$ ($\alpha^{2}$); however, for the heaviest mesons the errors in the masses should be very 
small because they are dominated by the kinetic energy.  We can write
\beq
{\cal H}(1+\delta) = \mathrm{KE} + V (1+\epsilon)
\eeq
where $\delta$ is the fractional error in the Hamiltonian and $\epsilon$ is the fractional error in the 
interactions ($\frac{g^{4}}{g^{2}}=g^{2} \propto \alpha$).  However, what we actually solved is 
\beq
{\cal H} = \mathrm{KE} + V .
\eeq
We can now solve for the fractional error in the binding energy by considering the expectation values of ${\cal H}$ 
and KE:
\beq
\delta = \epsilon \frac{\langle{\cal H}\rangle - \langle\mathrm{KE}\rangle}{\langle{\cal H}\rangle} 
\propto \alpha \frac{\langle{\cal H}\rangle - \langle\mathrm{KE}\rangle}{\langle{\cal H}\rangle},
\eeq
where we used the fact that the corrections to the interactions are ${\cal O}(\alpha)$.  This means the error in the 
eigenvalues is roughly $\alpha$ times the fraction of the eigenvalue that comes from the interaction.

\section{$b\bar{b}$ Spectrum}

We calculate the $b\bar{b}$ spectrum to verify that our method gives similar results to \cite{martinaA,martinaB}, which 
uses nonrelativistic reduction (NR Reduction).  The 
splittings in the $b\bar{b}$ system are so small that our method can not determine them in a 
reasonable amount of time.  However the comparison is useful to make sure that we get similar results for the quark 
mass, cutoff and meson masses.  With the small splittings, our method also has little predictive power for this system.

We use the values 
$\alpha=.4$ and $\frac{m}{\Lambda}=1.38$ from \cite{martinaA} and determine the cutoff to be $3.6$ GeV by fixing the mass of 
$\chi_{b_{0}}$ (the lowest $\mathrm{C}=+$ state) to data.  Table \ref{bbbar-spectrum} shows our results (Fully 
Relativistic) along with the 
experimental results as well as those calculated using NR Reduction.  $3.5 \%$ of mass of the $\chi_{b_{0}}$ 
is due to interactions.  Thus we believe the errors in our bottomonium spectrum due to our approximate Hamiltonian 
to be about $1.4\%$.  Our spectrum agrees with the other results within errors.  This method passes the simplest test.

\begin{table}
    \begin{center}
	\begin{tabular}{|c||c|c|c|}
	    \hline
	    State & Experimental & NR Reduction & Fully Relativistic \\ [0.5ex]
	    \hline
	    \hline
	    $\Upsilon_{a}$ & $9.460$ & $9.4$ & $9.64 \pm .1 \pm .1 \pm .14$ \\
	    \hline
	    $\chi_{b_{0}}$ & $9.860$ & $9.9$ & N/A \footnotemark \\
	    \hline
	    $\chi_{b_{1}}$ & $9.893$ & $9.9$ & $9.87 \pm .1 \pm .1 \pm .14$ \\
	    \hline
	    $\chi_{b_{2}}$ & $9.913$ & $9.9$ & $9.88 \pm .1 \pm .1 \pm .14$ \\
	    \hline
	    $\Upsilon_{b}$ & $10.023$ & N/A & $9.86 \pm .1 \pm .1 \pm .14$ \\
	    \hline
	\end{tabular}
    \end{center}
    \caption[Bottomonium masses]{\label{bbbar-spectrum}Bottomonium masses in GeV from experimental results, 
    nonrelativistic reduction and our fully relativistic calculation.  The NR Reduction and Fully Relativistic calculations use 
    $\alpha=.4$ and $m=4.9$ GeV.  The errors given for 
    the relativistic calculation are from using a finite number of basis states, 
    statistics and our second-order approximation to the Hamiltonian.}
\end{table}
\footnotetext{This is the particle used to find the cutoff.}

\begin{figure}
	\centerline{\epsfig{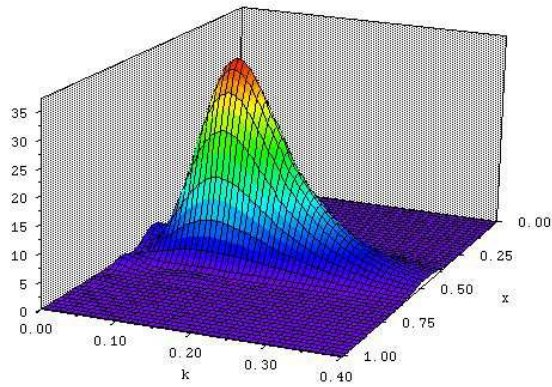}}
    \caption[$\chi_{b_{2}}$ wavefunction.]{\label{bbbar-cf+j0n2-wave}$\chi_{b_{2}}$ wavefunction.  $x$ is the longitudinal momentum fraction carried by one particle and 
    $k$ is the relative transverse momentum in units of the cutoff.}
\end{figure}

    The wavefunctions for the $\Upsilon_{a}$, $\chi_{b_{0}}$, $\chi_{b_{1}}$, and $\chi_{b_{2}}$, with $\alpha=.4$ and $j=0$ are plotted in figures 
\ref{bbbar-cf-j0n0-wave}, \ref{bbbar-cf+j0n0-wave}, \ref{bbbar-cf+j0n1-wave}, and \ref{bbbar-cf+j0n2-wave} respectively.
\begin{figure}
	\centerline{\epsfig{file=\mesonpath bbbar-cf-j0-n0.epsf,height=2.5in}}
    \caption[$\Upsilon_{a}$ wavefunction.]{\label{bbbar-cf-j0n0-wave}$\Upsilon_{a}$ wavefunction.  $x$ is the longitudinal momentum fraction carried by one particle and 
    $k$ is the relative transverse momentum in units of the cutoff.}
\end{figure}
\begin{figure}
	\centerline{\epsfig{file=\mesonpath bbbar-cf-j0-n0.epsf,height=2.5in}}
    \caption[$\chi_{b_{0}}$ wavefunction.]{\label{bbbar-cf+j0n0-wave}$\chi_{b_{0}}$ wavefunction.  $x$ is the longitudinal momentum fraction carried by one particle and 
    $k$ is the relative transverse momentum in units of the cutoff.}
\end{figure}
\begin{figure}
	\centerline{\epsfig{file=\mesonpath bbbar-cf+j0-n1.epsf,height=2.5in}}
    \caption[$\chi_{b_{1}}$ wavefunction.]{\label{bbbar-cf+j0n1-wave}$\chi_{b_{1}}$ wavefunction.  $x$ is the longitudinal momentum fraction carried by one particle and 
    $k$ is the relative transverse momentum in units of the cutoff.}
\end{figure}

\section{$c\bar{c}$ Spectrum}

Our approximations may work best for the $c\bar{c}$ system.  We find a good 
representation of the mass spectrum, but find unusual values for the coupling and cutoff.  If 
we use the values of the coupling and \mol from \cite{martinaB} we do not fit the spectrum within errors.

\begin{figure}
	\centerline{\epsfig{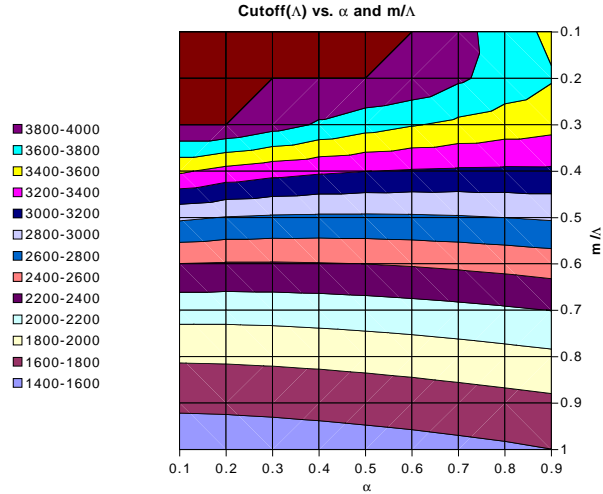}}
    \caption[The cutoff versus the coupling and the charm quark mass.]{\label{ccbar-cut-vs-a-m}The cutoff (MeV) versus the coupling and the charm quark mass divided by the cutoff.}
\end{figure}

We first generate eigenvalues for $.1 \le \alpha \le .9$ and $.1 \le \frac{m}{\Lambda} \le 1$ for states with 
positive charge conjugation since only one of the five lightest states in the $c\bar{c}$ spectrum has negative charge 
conjugation.  $\eta_{c}$ is used to determine the cutoff which is plotted (in MeV) against $\alpha$ and \mol in Figure 
\ref{ccbar-cut-vs-a-m}.

\begin{figure}
    \centerline{\epsfig{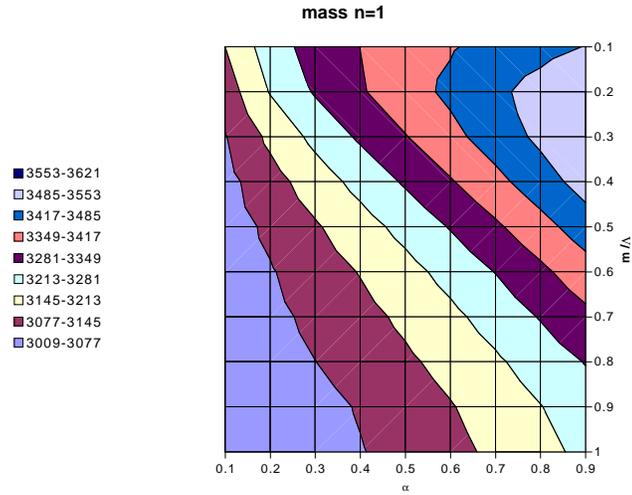}}
    \caption[Mass of $\chi_{c_{0}}$ as a function of $\alpha$ and \mol .]
    {\label{ccbar-m-n1-vs-a-m}Mass of the first excited positive-charge-conjugation state, $\chi_{c_{0}}$ as a function of $\alpha$ 
    and \mol .  The experimental value is 3417 MeV and each color band represents a mass range equal to two 
    percent of the experimental mass.}
\end{figure}

\begin{figure}
    \centerline{\epsfig{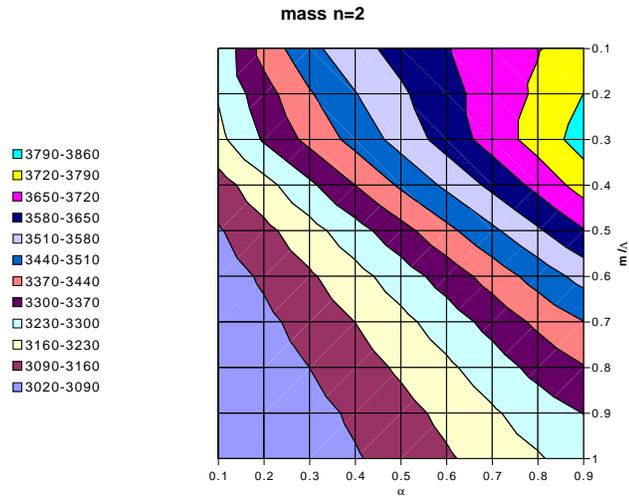}}
    \caption[Mass of $\chi_{c_{1}}$ as a function of $\alpha$ and \mol .]
    { \label{ccbar-m-n2-vs-a-m}Mass of the second excited positive-charge-conjugation state, $\chi_{c_{1}}$ as a function of $\alpha$ 
    and \mol .  The experimental value is 3510 MeV and each color band represents a mass range equal to two 
    percent of the experimental mass.}
\end{figure}

Using the corresponding value of $\Lambda$ for $\alpha$ and \mol, we plot the mass of the first (n=1) and second 
(n=2) excited positive-charge-conjugation states in figures \ref{ccbar-m-n1-vs-a-m} and \ref{ccbar-m-n2-vs-a-m} 
respectively.

\begin{figure}
    \centerline{\epsfig{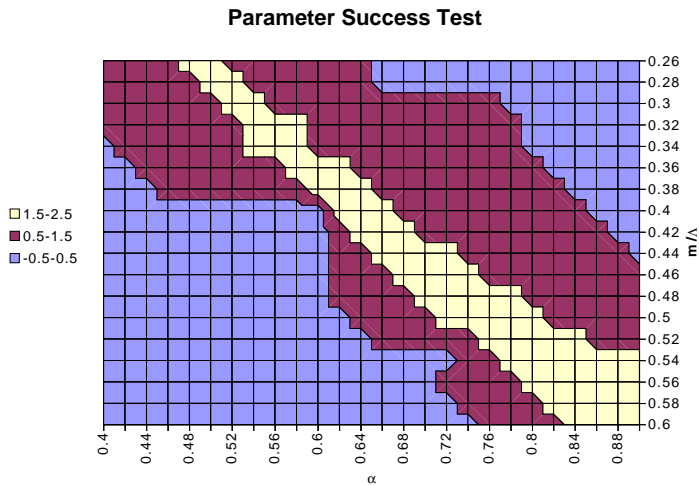}}
    \caption[Parameter Success]{ \label{ccbar-param-success}This figure displays the success for choices of 
    $\alpha$ and \mol at predicting the first and second excited state masses.  We define a correct prediction to 
    be within two percent of the actual value.  The value is zero (purple) if neither mass 
    is correctly determined, one (red) if one mass is correctly predicted and two (yellow) if both are correct.}
\end{figure}

We find the values $\alpha$ and \mol that correctly produce the ratios of the first and second exited states by 
finding the region where the mass of the first and second excited states are predicted to within two percent.  
Figure \ref{ccbar-param-success} shows where both masses are predicted within the error.
Unfortunately fixing these two masses is not sufficient to determine $\alpha$ and \mol because there is a range of 
$\alpha$ and \mol that correctly predict the spectrum.  Also as \mol is lowered the eigenvalues become smaller, 
increasing the cutoff.  All of the combinations of $\alpha$ and \mol that successfully predict the spectrum 
produce a cutoff and/or coupling that is too large  to be consistent with our ${\cal O} (g^{2})$ approximation.  
If we consider $\alpha = \frac{1}{2}$ and \mol$=.28$, we get a cutoff of $3.8$ GeV and the mass of the 
charm quark equal to $1.05$ GeV.  It is clear the errors generated by our approximations are too large to 
successfully predict the charmonium spectrum.

\begin{figure}
	\centerline{\epsfig{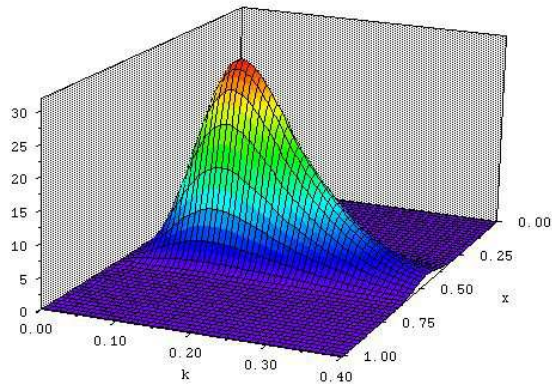}}
    \caption[$\chi_{c_{1}}$ wavefunction.]{\label{ccbar-bris-cf+j0-n2}$\chi_{c_{1}}$ wavefunction.  $x$ is the longitudinal momentum fraction carried by one particle and 
    $k$ is the relative transverse momentum in units of the cutoff.}
\end{figure}

If we use $\alpha=.5$ and \mol$=.88$ from \cite{martinaB} we find the spectrum in Table 
\ref{ccbar-spectrum} and determine the cutoff to be $1.7$ GeV, and the charm quark mass to be $1.5$ GeV.  $5\%$ of 
the $\eta_{c}$ mass is due to interactions, making the error from the approximate Hamiltonian in our spectrum 
about $2.5\%$.  
\begin{table}
    \begin{center}
	\begin{tabular}{|c||c|c|c|}
	    \hline
	    State & Experimental & NR Reduction & Fully Relativistic \\ [0.5ex]
	    \hline
	    \hline
	    $\eta_{c}$ & $2.9798 $ & $3.0$ & N/A \footnotemark  \\
	    \hline
	    $J/\psi$ & $3.097$ & $3.0$ & $ 2.981 \pm 0.03 \pm 0.03 \pm 0.075$ \\
	    \hline
	    $\chi_{c_{0}}$ & $3.415$ & 3.5 & $3.114 \pm 0.03 \pm 0.03 \pm 0.078 $ \\
	    \hline
	    $\chi_{c_{1}}$ & $3.510$ & $3.5$ & $3.142 \pm 0.03 \pm 0.03 \pm 0.079 $ \\
	    \hline
	    $\chi_{c_{2}}$ & $3.556$ & $3.5$ & $3.145 \pm 0.03 \pm 0.03 \pm 0.079 $ \\
	    \hline
	\end{tabular}
    \end{center}
    \caption[Charmonium masses]{\label{ccbar-spectrum}Charmonium masses in GeV from experimental results, nonrelativistic 
    reduction and fully relativistic calculations 
    with $\alpha=.5$ and \mol$=.88$.  The errors given for the Fully Relativistic calculation are from using a finite number of basis states, 
    statistics and our second-order approximation to the Hamiltonian.}
\end{table}
\footnotetext{This was the particle used to find the cutoff.}
The spin-averaged probability densities for the $\eta_{c}$, $J/\psi$, $\chi_{c_{0}}$ and $\chi_{c_{1}}$ with $\alpha=.5$ and 
\mol$=.88$ are shown in figures \ref{ccbar-bris-cf+j0-n0} through \ref{ccbar-bris-cf+j0-n2}, respectively.
\begin{figure}
	\centerline{\epsfig{file=\mesonpath ccbar-bris-cf+j0-n0.epsf,height=2.5in}}
    \caption[$\eta_{c}$ wavefunction.]{\label{ccbar-bris-cf+j0-n0}$\eta_{c}$ wavefunction.  $x$ is the longitudinal momentum fraction carried by one particle and 
    $k$ is the relative transverse momentum in units of the cutoff.}
\end{figure}
\begin{figure}
	\centerline{\epsfig{file=\mesonpath ccbar-bris-cf-j0-n0.epsf,height=2.5in}}
    \caption[$J/\psi$ wavefunction.]{\label{ccbar-bris-cf-j0-n0}$J/\psi$ wavefunction.  $x$ is the longitudinal momentum fraction carried by one particle and 
    $k$ is the relative transverse momentum in units of the cutoff.}
\end{figure}
\begin{figure}
	\centerline{\epsfig{file=\mesonpath ccbar-bris-cf+j0-n1.epsf,height=2.5in}}
    \caption[$\chi_{c_{0}}$ wavefunction.]{\label{ccbar-bris-cf+j0-n1}$\chi_{c_{0}}$ wavefunction.  $x$ is the longitudinal momentum fraction carried by one particle and 
    $k$ is the relative transverse momentum in units of the cutoff.}
\end{figure}

\chapter{Conclusions and Future}
\label{conclusions}

The method we use to regulate and renormalize the light-front QCD Hamiltonian does not remove any physics, 
add extra parameters or fundamentally limit itself to particular problems (like 
heavy quark systems).  The renormalized, second-order, light-front Hamiltonian 
is reasonably successful at determining the glueball spectrum.  This same 
approximation is much less successful at determining the heavy meson spectra.

The success of the glueball calculation shows our method can be successful.  It is clear that to more successfully 
determine (with predictive power) the heavy meson spectra we must either calculate the renormalized Hamiltonian to 
higher order in the strong coupling and/or allow for more complicated external states ($q\bar{q}g$).  In addition 
to these improvements using basis functions that are more sharply peaked should improve the accuracy of the 
heaviest meson system ($b\bar{b}$)  where the difference in the masses are fractions of a percent.

We determined the light-front Hamiltonian to second 
order in the strong coupling while approximating the real states as a quark-antiquark pair.  The results from 
the glueball and meson mass calculations show our method can reproduce experimental results with varying degrees of 
success.  The 
renormalized Hamiltonian we determined is the building block of higher order calculations and will be a necessary 
part of future calculations.

It is unclear if B-splines are an efficient basis to approximate the real states.  They reproduced the glueball spectrum with 
fewer basis functions than an earlier calculation but appear inefficient at representing the sharply peaked heavy 
meson wave functions.  Although the full flexibility of the B-splines was not explored, it is likely more efficient to 
use basis functions better tailored to the problem at hand.
%

The analytic calculation of the renormalized Hamiltonian requires a large time investment and great care.  However the 
analytic result is not very useful without an effective and efficient method to numerically calculate the integrals the 
analytic calculation yields.  We developed an algorithm to calculate these integrals without wasting time calculating any 
matrix element more accurately than needed.  We also generalized the algorithm to run on a single processor or in a 
parallel processor environment, including clusters of desktop machines.  This is a significant development because the 
algorithm is independent of the details of the integrand so higher order analytic calculations can utilize it with very little change.

We successfully added to the systematic determination of the renormalized light-front Hamiltonian by including 
masses. We explored the effectiveness of B-spline basis functions and 
developed an algorithm for calculating the renormalized Hamiltonian matrix.

%
%

\appendix
\chapter{Light-Front Conventions}
\label{lf-formalism}

In this Appendix we list the light-front conventions used in this dissertation.  For a more general introduction see the 
light-front review by A. Harindranath \cite{pedestrians}.

\setlength{\arraycolsep}{.1cm}
\section{Coordinates}
\beq
\setlength{\arraycolsep}{.5cm}
\barr{ll}
x^{\pm}=x^{0}\pm x^{3}&
x\cdot y=\half x^{+}y^{-}+\half x^{-}y^{+}-x^{i}_{\bot}y^{i}_{\bot}\\
\partial^{\pm}=2\frac{\partial}{\partial x^{\mp}}&
\partial\cdot x=\half\partial^{+}x^{-}+\half\partial^{-}x^{+}-\partial^{i}_{\bot}x^{i}_{\bot}\\
\partial^{i}_{\bot}=-\frac{\partial}{\partial x^{i}_{\bot}}\nonumber
\earr
\eeq
\beq
\barr{lll}
x^{\mu}=\left( \barr{c} x^{+}\\ x^{-}\\ x^{1}\\ x^{2}\earr\right) &
\hspace{.5cm}g^{\mu \nu}=\left( \barr{rrrr} 
0&2&0&0\\
2&0&0&0\\
0&0&-1&0\\
0&0&0&-1 
\earr\right) &
\hspace{.5cm} g_{\mu \nu}=\left( \barr{rrrr}
0&\half&0&0\\
\half&0&0&0\\
0&0&-1&0\\
0&0&0&-1
\earr\right) \nonumber
\earr
\eeq
\section{Gamma Matrices}
\beq
\setlength{\arraycolsep}{.1cm}
\barr{lll}
\gp=\left( \barr{cc} 0&0\\ 2i&0\earr\right) &
\hspace{.5cm}\gamma^{-}=\left( \barr{cc} 0&-2i\\ 0&0\earr\right) &
\hspace{.5cm}\gamma^{0}=\left( \barr{cc} 0&-i\\ i&0\earr\right) \\
\gamma^{5}=\left( \barr{cc} \sigma^{3}&0\\ 0&-\sigma^{3}\earr\right) &
\hspace{.5cm}\gamma^{i}=\left( \barr{cc} -i\sigma^{i}&0\\ 0&i\sigma^{i}\earr\right) &
\hspace{.5cm}\alpha^{i}_{\bot}=\left( \barr{cc} 0&\sigma^{i}_{\bot}\\ \sigma^{i}_{\bot}&0 \earr\right) \nonumber
\earr
\eeq
\section{Pauli Matrices}
\beq
\sigma^{1}=\left( \barr{cc}0&1\\ 1&0\earr\right) \hspace{1cm}
\sigma^{2}=\left( \barr{cc}0&-i\\ i&0\earr\right) \hspace{1cm}
\sigma^{3}=\left( \barr{cc}1&0\\ 0&-1\earr\right) \nonumber
\eeq
\section{Projection Operators}
\label{conv-proj}
\beq
\Lpm&=&\half \gz \gpm\nonumber\\
\Lp&=&\half\left( \barr{rr} 0&-i\\ i&0\earr\right) 
\left( \barr{rr} 0&0\\ 2i&0\earr\right) =\left( \barr{rr} 1&0\\ 0&0\earr\right) \nonumber\\
\Lm&=&\half\left( \barr{rr} 0&-i\\ i&0\earr\right) 
\left( \barr{rr} 0&-2i\\ 0&0\earr\right) =\left( \barr{rr} 0&0\\ 0&1\earr\right) \nonumber
\eeq
{\large Note:}
\beq
\setlength{\arraycolsep}{.5cm}
\barr{lll}
\Lp+\Lm=1&\Lp^{2}=\Lp&\Lm^{2}=\Lm\\
\Lpm\Lmp=0&\gp\gp=\gm\gm=0&(\gpm)^{\dagger}=\gmp\\
\Lpm^{\dagger}=\Lpm&\gz\gpm=\gmp\gz&
\gamma^{i}_{\bot}\gpm=-\gpm\gamma^{i}_{\bot}\\
\gmp\gpm=4\Lpm&\alpha^{i}_{\bot}\Lpm=\Lmp\alpha^{i}_{\bot}&
\gz\Lpm=\Lmp\gz\nonumber
\earr
\eeq
\section{Lagrangian}
\beq
{\cal L}_{\mathrm{QCD}}=-\frac{1}{4}F_{\mu\nu}^{c}F^{\mu\nu}_{c}+
\bar{\psi}(iD\hspace{-.25cm}\slash-m_{F})\psi,
\eeq
where
\beq
F^{\mu\nu}_{c}=\partial^{\mu}A_{c}^{\nu}-\partial^{\nu}A_{c}^{\mu}-gA^{\mu}_{c_{1}}A^{\nu}_{c_{2}}f^{c_{1}c_{2}c}
\eeq
and
\beq
{\bf D}_{\mu}\psi=(\partial_{\mu}+igA^{a}_{\mu}\bT_{a})\psi.
\eeq
Greek indices are Lorentz indices, $c$'s are color indices, $\alpha$ is the color index, repeated indices are 
summed over, and the $f$'s are the SU($N_{c}$) structure constants.  The $\bT_{a}$'s are the $SU(3)$ Gell-Mann 
matrices which obey the commutation relation $\left[ \bT_{a},\bT_{b}\right] =if_{abc}\bT_{c}$ and are given by:
\beq
\barr{llll}
\bT_{1}=\half\left( \barr{ccc} 0&1&0\\ 1&0&0\\ 0&0&0\earr\right) \nonumber&
\bT_{2}=\half\left( \barr{ccc} 0&-i&0\\ i&0&0\\ 0&0&0\earr\right) \nonumber&
\bT_{3}=\half\left( \barr{ccc} 1&0&0\\ 0&-1&0\\ 0&0&0\earr\right) \nonumber&
\bT_{4}=\half\left( \barr{ccc} 0&0&1\\ 0&0&0\\ 1&0&0\earr\right) \nonumber\\
\bT_{5}=\half\left( \barr{ccc} 0&0&-i\\ 0&0&0\\ i&0&0\earr\right) \nonumber&
\bT_{6}=\half\left( \barr{ccc} 0&0&0\\ 0&0&1\\ 0&1&0\earr\right) \nonumber&
\bT_{7}=\half\left( \barr{ccc} 0&0&0\\ 0&0&-i\\ 0&i&0\earr\right) \nonumber&
\bT_{8}=\half\left( \barr{ccc} \frac{1}{\sqrt{3}}&0&0\\ 0&\frac{1}{\sqrt{3}}&0\\ 
0&0&\frac{-2}{\sqrt{3}}\earr\right). \nonumber
\earr
\eeq
\section{The Gluon Field}
\label{gluon-field}
The expansion of the transverse gluon field takes the form:
\beq
\label{gluon-field-expr}
\vec{A}_{c\bot}(x^{-},\vec{x_{\bot}})=\int D_{i}\delta_{c,c_{i}}\left[ a_{i}\vec{\epsilon}_{\bot s_{i}}e^{-ip_{i}\cdot x}
+a^{\dagger}_{i}\vec{\epsilon}_{\bot s_{i}}^{\hspace{.1cm}\ast}e^{ip_{i}\cdot x}\right] |_{x^{+}=0},
\eeq
where
\beq
\label{bigD-defn}
D_{i}=\sum_{c_{i}=1}^{N_{c}}\sum_{s_{i}=-1,1}\frac{d^{2}p_{i\bot}dp_{i}^{+}}{16\pi^{3}p_{i}^{+}}
\theta (p_{i}^{+}-\epsilon{\cal P}^{+}),
\eeq
${\cal P}$ is the four-momentum operator, $\epsilon$ is a positive infinitesimal, and
\beq
\vec{\epsilon}_{\bot s}=\frac{-1}{\sqrt 2}(s,i).
\eeq
The creation and annihilation operators follow the convention
\beq
a_{i}=a(p_{i},s_{i},c_{i}),
\eeq
and have the commutation relations
\beq
[a_{i},a_{j}^{\dagger}]=16\pi^{3}p_{i}^{+}\delta^{(3)}(p_{i}-p_{j})\delta_{s_{i},s_{j}}\delta_{c_{i},c_{j}}
\eeq
and
\beq
[a_{i},a_{j}]=[a_{i}^{\dagger},a_{j}^{\dagger}]=0,
\eeq
where
\beq
\delta^{(3)}(p_{i}-p_{j})=\delta(p_{i}^{+}-p_{j}^{+})\delta^{(2)}(\vec{p}_{i\bot}-\vec{p}_{j\bot}).
\eeq
The gluon field commutation relation is:
\beq
\left[ A^{ic_{1}}_{\bot}(x),\partial^{+}A^{jc_{2}}_{\bot}(y)\right] _{x^{+}=y^{+}=0}
=i\delta^{ij}\delta_{c_{1}c_{2}}\delta^{3}(x-y).
\eeq
\section{Gluon Polarization Vector}
The gluon polarization vector is
\beq
\epsilon^{+}=0,\hspace{.5cm}\epsilon^{-}=\frac{2{\bf q}_{\bot}\cdot\epsilon_{\bot}}{q^{+}},
\eeq
and
\beq
\sum_{\lambda}\epsilon^{\mu}_{\bot}(\lambda)\epsilon^{\ast\hspace{.05cm}\nu}_{\bot}(\lambda)=-g^{\mu \nu}_{\bot},
\eeq
so that
\beq
\sum_{\lambda}\epsilon^{\mu}(\lambda)\epsilon^{\ast\hspace{.05cm}\nu}(\lambda)
=g^{\mu\nu}_{\bot}+\frac{1}{q^{+}}\left( \eta^{\mu}q^{\nu}_{\bot}+\eta^{\nu}q^{\mu}_{\bot}\right) 
+\frac{{\bf q}_{\bot}^{2}}{(q^{+})^{2}}\eta^{\mu}\eta^{\nu},
\eeq
where $\eta^{+}=\eta^{1}=\eta^{2}=0$ and $\eta^{-}=2$.  Also
\beq
\vec{\epsilon}_{\bot\lambda}\cdot\vec{\epsilon}^{\hspace{.1cm}\ast}_{\bot\lambda^{\prime}}=\delta_{\lambda\lambda^{\prime}}
\hspace{1cm}
\vec{\epsilon}_{\bot\lambda}\cdot\vec{\epsilon}_{\bot\lambda^{\prime}}=-\delta_{\lambda\lambda^{\prime}}
.
\eeq
We can write them out:
\beq
\epsilon^{\mu}_{\bot 1}=\left( \barr{c}0\\ 0\\ -\frac{1}{\sqrt{2}}\\ -\frac{i}{\sqrt{2}}\earr\right) \hspace{2cm}
\epsilon^{\mu}_{\bot -1}=\left( \barr{c}0\\ 0\\ \frac{1}{\sqrt{2}}\\ -\frac{i}{\sqrt{2}}\earr\right) .
\eeq
We also find:
\beq
\sigma_{\bot}\cdot\epsilon^{\bot}_{-1}=-\sigma_{\bot}\cdot\epsilon^{\bot\ast}_{1}
=\sqrt{2}\left( \barr{cc}0&0\\ 1&0\earr\right) \hspace{.5cm}
\sigma_{\bot}\cdot\epsilon^{\bot\ast}_{-1}=-\sigma_{\bot}\cdot\epsilon_{1}^{\bot}
=\sqrt{2}\left( \barr{cc}0&1\\ 0&0\earr\right) 
\eeq
\section{The Fermion Field}
The dynamical fermion degree of freedom is $\psi_{+}$ and can be expanded in terms of plane wave creation and 
annihilation operators at $x^{+}=0$,
\beq
\label{quark-field-expr}
\psi_{+}^{c}\left( x^{-},\vec{x}_{\bot}\right) =\int D_{1}\delta_{c,c1}
\left[ b_{1}u_{+}(k,\sigma)e^{-ik\cdot x}+d^{\dagger}_{1}v_{+}(k,\sigma)e^{ik\cdot x}\right] ,
\eeq
\beq
\label{antiquark-field-expr}
\psi_{+}^{c\dagger}\left( x^{-},\vec{x}_{\bot}\right) =\int D_{1}\delta_{c,c1}
\left[ b^{\dagger}_{1}u_{+}^{\dagger}(k,\sigma)e^{ik\cdot x}+d_{1}v_{+}^{\dagger}(k,\sigma)e^{-ik\cdot x}\right] ,
\eeq
where the creation and annihilation operators are labeled the same as in Section \ref{gluon-field}.  The field 
operators satisfy
\beq
\left\{\psi^{c_{1}}_{+}(x),\psi^{c_{2}\hspace{.05cm}\dagger}_{+}(y)\right\}_{x^{+}=y^{+}=0}
=\Lambda_{+}\delta_{c_{1},c_{2}}\delta^{3}(x-y),
\eeq
and the creation and annihilation operators satisfy:
\beq
\left\{b_{i},b^{\dagger}_{j}\right\}=\left\{d_{i},d^{\dagger}_{j}\right\}
=16\pi^{3}p^{+}\delta_{c_{i},c_{j}}\delta_{s_{i},s_{j}}\delta^{3}(p_{i}-p_{j}).
\eeq
\section{Dirac Spinors}
The dirac spinors $u(p,\sigma)$ and $v(p,\sigma)$ satisfy
\beq
\left( \psl-m\right) u(p,\sigma)=0,\hspace{.5cm}
\left( \psl+m\right) v(p,\sigma)=0,
\eeq
and
\beq
\bar{u}(p,\sigma)u(p,\sigma^{\prime})&=&-\bar{v}(p,\sigma)v(p,\sigma^{\prime})=2m\delta_{\sigma \sigma^{\prime}},\\
\bar{u}(p,\sigma)\gamma^{\mu}u(p,\sigma^{\prime})&=&\bar{v}(p,\sigma)\gamma^{\mu}v(p,\sigma^{\prime})
=2p^{\mu}\delta_{\sigma \sigma^{\prime}},
\eeq
\beq
\sum_{\sigma =\pm\half} u(p,\sigma)\bar{u}(p,\sigma)=\psl +m,\hspace{.5cm}
\sum_{\sigma =\pm\half} v(p,\sigma)\bar{v}(p,\sigma)=\psl -m.
\eeq
We use the normalizations:
\beq
u^{\dagger}_{+k\lambda^{\prime}}\hspace{.1cm}u_{+k\lambda}=v^{\dagger}_{+k\lambda^{\prime}}\hspace{.1cm}v_{+k\lambda}
=k^{+}\delta_{\lambda,\lambda^{\prime}}.
\eeq
We can define:
\beq
u_{+k\lambda}\equiv\sqrt{k^{+}}\hspace{.1cm}\chp_{\lambda}\hspace{1cm}
v_{+k\lambda}\equiv\sqrt{k^{+}}\hspace{.1cm}\chp_{\bar{\lambda}}.
\eeq
Use the eigenvalue equation:
\beq
\Lambda_{+}\chp_{\lambda}=\chp_{\lambda}
\eeq
to find the eigenvector with eigenvalue $1$.  This leads to two solutions which we write:
\beq
\chp_{\half}=\left( \barr{c}1\\ 0\\ 0\\ 0\earr\right) \hspace{2cm}
\chp_{-\half}=\left( \barr{c}0\\ 1\\ 0\\ 0\earr\right) \hspace{2cm}.
\eeq
We define $\chi_{\lambda}$ as the upper $2$ components of $\chp_{\lambda}$. Finally, we discover that:
\beq
\label{fermion-spinor}
u^{\dagger}_{s_{1}}(p_{1})\hspace{.1cm}v_{s_{2}}(p_{2})=v^{\dagger}_{s_{1}}(p_{1})\hspace{.1cm}u_{s_{2}}(p_{2})
&=&\sqrt{p_{1}^{+}\hspace{.1cm}p_{2}^{+}}\hspace{.2cm}\delta_{s_{1},-s_{2}}\n
u^{\dagger}_{s_{1}}(p_{1})\hspace{.1cm}u_{s_{2}}(p_{2})=v^{\dagger}_{s_{1}}(p_{1})\hspace{.1cm}v_{s_{2}}(p_{2})
&=&\sqrt{p_{1}^{+}\hspace{.1cm}p_{2}^{+}}\hspace{.2cm}\delta_{s_{1},s_{2}}.
\eeq
\chapter{Details of the Combination of the Divergent Part of the Self-Energy and the Instantaneous Interaction Below the Cutoff}
\label{appx:se-inst}

The non-vanishing part of the instantaneous interaction below the cutoff is 
given by:\footnote{This is an intermediate step in the calculation of the instantaneous interaction in 
Section~\ref{inst-inter} that is not explicitly given.}
\beq
&&-g_{\Lambda}^{2}\frac{32}{3} {\cal P}^{+}\delta({\cal P-P'})\sum_{s_{1},s_{2}}\int d^{2}\vec{q} dx \int d^{2}\vec{p} dy 
\theta_{\epsilon} \theta_{\epsilon'}\chi_{q}^{s_{1}s_{2}} \theta (x-y-\epsilon) 
\n && \hspace{1cm} \times
\chi_{q'}^{s_{1}s_{2}} B_{l}(x)B_{l'}(y)\tilde{B}_{t}(q) \tilde{B}_{t'}(p)
A^{\ast}_{j'-s_{1}-s_{2}}(\phi')A_{j-s_{1}-s_{2}}(\phi) 
\n && \hspace{2cm} \times 
\frac{e^{-\Lambda^{-4}\Delta^{2}_{FI}}e^{-2\Lambda^{-4}\Delta_{FK}\Delta_{IK}}}{(x-y)^{2}}.
\eeq
If we make the following definitions:
\beq
X_{1}&=&-g_{\Lambda}^{2}\frac{32}{3}\sum_{s_{1},s_{2}}\chi_{q}^{s_{1}s_{2}}\chi_{q'}^{s_{1}s_{2}}
 {\cal P}^{+}\delta({\cal P-P'}) \n
F(x,y,\vec{q},\vec{p})&=&B_{l}(x)B_{l'}(y)\tilde{B}_{t}(q) \tilde{B}_{t'}(p)
 A^{\ast}_{j'-s_{1}-s_{2}}(\phi')A_{j-s_{1}-s_{2}}(\phi) 
 \n && \hspace{4cm} \times
e^{-\Lambda^{-4}\Delta^{2}_{FI}}e^{-2\Lambda^{-4}\Delta_{FK}\Delta_{IK}} , \nonumber
\eeq
we get:
\beq
X_{1}\int d^{2}\vec{q} d^{2}\vec{p} \int_{2\epsilon}^{1-\epsilon} dx \int_{\epsilon}^{x-\epsilon} dy
\frac{1}{(x-y)^{2}} F(x,y,\vec{q},\vec{p}) .
\eeq
Now make the change of variables:
\beq
\vec{R}=\vec{q}-\vec{p} &\hspace{1cm}& \vec{Q}=\vec{q}+\vec{p} \n
\vec{q}=\frac{\vec{Q}+\vec{R}}{2} &\hspace{1cm}& \vec{p}=\frac{\vec{Q}-\vec{R}}{2},
\eeq
and find the Jacobian:
\beq
d^{2}p d^{2}q = \frac{1}{4} d^{2}Q d^{2}R .
\eeq
This gives us:
\beq
\frac{1}{4}X_{1}\int_{2\epsilon}^{1-\epsilon} dx \int_{\epsilon}^{x-\epsilon} dy \int d^{2}\vec{Q} d^{2}\vec{R} 
\frac{1}{(x-y)^{2}} F(x,y,\frac{\vec{Q}+\vec{R}}{2},\frac{\vec{Q}-\vec{R}}{2}) .
\eeq
Defining:
\beq
\vec{N}&=&\frac{1}{\sqrt{x-y}}\vec{R}, \n
X_{2}&=&\frac{1}{4}X_{1} \int d^{2}\vec{Q} d^{2}\vec{N}\int_{2\epsilon}^{1-\epsilon} dx, \n
X_{3}&=& F(x,y,\frac{\vec{Q}+\sqrt{x-y}\vec{N}}{2},\frac{\vec{Q}-\sqrt{x-y}\vec{N}}{2}), \nonumber
\eeq
we get:
\beq
X_{2}\int_{\epsilon}^{x-\epsilon} dy \frac{1}{(x-y)} X_{3} .
\eeq
Integrating by parts,
\beq
\int u dv = uv - \int v du
\eeq
with
\beq
dv=\frac{dy}{x-y} \hspace{1cm} v=-\log(x-y) \hspace{1cm} u=X_{3}
\eeq
gives:
\beq
X_{2}\left[ X_{3}\log(x-y) \vert^{\epsilon}_{x-\epsilon} 
+\int_{\epsilon}^{x-\epsilon} dy \log(x-y) \frac{dX_{3}}{dy} \right] .
\eeq
Consider the first term:
\beq
&&X_{2}X_{3}\log(x-y)\vert_{y=\epsilon} \n 
&& \hspace{1cm}=\frac{1}{4}X_{1} \int_{2\epsilon}^{1-\epsilon} dx \int d^{2}\vec{Q} d^{2}\vec{N}
\n && \hspace{2cm} \times
F(x,y,\frac{\vec{Q}+\sqrt{x-y}\vec{N}}{2},\frac{\vec{Q}-\sqrt{x-y}\vec{N}}{2}) \log(x-y)\vert_{y=\epsilon} 
\n && \hspace{1cm}
=\frac{1}{4}X_{1} \int_{2\epsilon}^{1-\epsilon} dx \int d^{2}\vec{Q} d^{2}\vec{R} \frac{1}{x-y}
\n && \hspace{2cm} \times
F(x,y,\frac{\vec{Q}+\vec{R}}{2},\frac{\vec{Q}-\vec{R}}{2}) \log(x-y)\vert_{y=\epsilon} 
\n && \hspace{1cm}
=X_{1} \int_{2\epsilon}^{1-\epsilon} dx \int d^{2}\vec{q} d^{2}\vec{p}\frac{1}{x-y}
\n && \hspace{2cm} \times
F(x,y,\vec{q},\vec{p}) \log(x-y)\vert_{y=\epsilon} 
\n && \hspace{1cm}
=X_{1} \int_{2\epsilon}^{1-\epsilon} dx \int dq q dp p d\phi d\phi' \frac{1}{x-y}
 \n && \hspace{2cm} \times
 A^{\ast}_{j'-s_{1}-s_{2}}(\phi')A_{j-s_{1}-s_{2}}(\phi) 
B_{l}(x)B_{l'}(y)\tilde{B}_{t}(q) \tilde{B}_{t'}(p) 
\n && \hspace{3cm} \times
e^{-\Lambda^{-4}\Delta^{2}_{FI}}e^{-2\Lambda^{-4}\Delta_{FK}\Delta_{IK}} \log(x-y)\vert_{y=\epsilon}.
\eeq
If we let
\beq
X_{4}&=&X_{1} \int d\phi d\phi' A^{\ast}_{j'-s_{1}-s_{2}}(\phi')A_{j-s_{1}-s_{2}}(\phi) \n
s&=&\frac{p}{\sqrt{\epsilon}} , \nonumber
\eeq
we get:
\beq
&& \epsilon X_{4} \int_{2\epsilon}^{1-\epsilon} dx \int dq q ds s 
\frac{1}{x-y}B_{l}(x)B_{l'}(y)\tilde{B}_{t}(q) \tilde{B}_{t'}(s\sqrt{\epsilon}) 
\n &&\hspace{5cm} \times
e^{-\Lambda^{-4}\Delta^{2}_{FI}}e^{-2\Lambda^{-4}\Delta_{FK}\Delta_{IK}} \log(x-y)\vert_{y=\epsilon}.
\eeq
So now we need to consider this in the $\epsilon \rightarrow 0$ limit.
\beq
\Delta_{FK}&=&-\frac{m^{2}(x-y)^{2}+\epsilon s^{2}(1-x)^{2}+q^{2}-2(1-x)(1-y)\sqrt{\epsilon}sq \cos(\gamma)}
{(1-x)(1-y)(x-y)} \n
&\rightarrow & -\frac{m^{2}x^{2}+q^{2}}{x(1-x)} \n
\Delta_{KI}&=&\frac{m^{2}(x-y)^{2}+\epsilon s^{2}x^{2}+\epsilon^{2}q^{2} -2x\epsilon^{\frac{3}{2}}sq\cos(\gamma)}
{xy(x-y)} \n
&\rightarrow & \frac{m^{2}}{\epsilon}+s^{2} \n
\Delta_{FI}&=&\frac{m^{2}+\epsilon s^{2}}{y(1-y)}-\frac{m^{2}+q^{2}}{x(1-x)} \n
&\rightarrow & \frac{m^{2}}{\epsilon}+s^{2}-\frac{m^{2}+q^{2}}{x(1-x)} .
\eeq
The B-splines also go to zero as $\epsilon \rightarrow 0$.  However, it is sufficient to note the exponential 
terms are the dominant factor in this limit, thus this term is zero.

The second term is:
\beq
&&X_{2}X_{3}\log(x-y)\vert_{y=x-\epsilon} \n 
&& \hspace{1cm}=-X_{1} \int_{2\epsilon}^{1-\epsilon} dx \int dq q dp p d\phi d\phi'
\frac{1}{\epsilon}\log(\epsilon) B_{l}(x)B_{l'}(y)\tilde{B}_{t}(q) \tilde{B}_{t'}(p)
 \n && \hspace{2cm} \times
 A^{\ast}_{j'-s_{1}-s_{2}}(\phi')A_{j-s_{1}-s_{2}}(\phi) 
e^{-\Lambda^{-4}\Delta^{2}_{FI}}e^{-2\Lambda^{-4}\Delta_{FK}\Delta_{IK}} \vert_{y=x-\epsilon}.
\eeq
We can use the fact that when the angular integrals are done, we get $\delta_{j,j'}$
and the change of variables; $\gamma_{\phi}=\phi-\phi'$, to get:
\beq
&&X_{2}X_{3}\log(x-y)\vert_{y=x-\epsilon} \n 
&& \hspace{1cm}=-\frac{\delta_{j,j'}}{\sqrt{2\pi}}X_{1} \int_{2\epsilon}^{1-\epsilon} dx \int dq q dp p d\phi d\phi'
B_{l}(x)B_{l'}(y)\tilde{B}_{t}(q) \tilde{B}_{t'}(p)
 \n && \hspace{2cm} \times
\frac{\log(\epsilon)}{\epsilon} A_{j-s_{1}-s_{2}}(\gamma_{\phi}) 
e^{-\Lambda^{-4}\Delta^{2}_{FI}}e^{-2\Lambda^{-4}\Delta_{FK}\Delta_{IK}} \vert_{y=x-\epsilon}.
\eeq
Next, when we integrate over $\gamma$ we get terms coming from the cosine in the angular basis 
function.  We can also use the change of variables we used for the first term to get:
\beq
&&X_{2}X_{3}\log(x-y)\vert_{y=x-\epsilon} \n 
&& \hspace{1cm}=-\frac{\delta_{j,j'}}{8\pi}X_{1} \int_{2\epsilon}^{1-\epsilon} dx \int d^{2} Q d^{2} N
\log(\epsilon)B_{l}(x)B_{l'}(y)\tilde{B}_{t}(Q/2) \tilde{B}_{t'}(Q/2)
 \n && \hspace{2cm} \times
\cos([j-s_{1}-s_{2}]\gamma_{\phi}) 
e^{-\Lambda^{-4}\Delta^{2}_{FI}}e^{-2\Lambda^{-4}\Delta_{FK}\Delta_{IK}} \vert_{y=x-\epsilon}.
\eeq
Use the following definitions:
\beq
\vec{Q}=Q\cos \alpha \hat{x} + Q \sin \alpha \hat{y}, \hspace{1cm}
\vec{N}=N\cos \eta \hat{x} + N \sin \eta \hat{y},
\eeq
\beq
q&=&\half \sqrt{Q^{2}+\epsilon N^{2}+2\sqrt{\epsilon}QN \cos(\alpha-\eta)}, \n
p&=&\half \sqrt{Q^{2}+\epsilon N^{2}-2\sqrt{\epsilon}QN \cos(\alpha-\eta)}, \n
\vec{q}\cdot \vec{p}&=&\frac{1}{4} (Q^{2}-\epsilon N^{2}) .
\eeq
With these, we know that:
\beq
(\vec{p}-\vec{q})^{2}=\epsilon N^{2}, \hspace{1cm} p^{2}-q^{2}={\cal O}(\sqrt{\epsilon}) ,
\eeq
\beq
\Delta_{FK} =-\Delta_{KI} \rightarrow N^{2}, \hspace{1cm} \Delta_{FI}\rightarrow 0 .
\eeq
Also, if we use:
\beq
\vec{p}\cdot \vec{q} = pq \cos \gamma ,
\eeq
we find that $\cos \gamma = 1$.  This means that $\gamma = 0$, so that $\cos([j-s_{1}-s_{2}]\gamma)=1$.  This 
leaves us with:
\beq
&&X_{2}X_{3}\log(x-y)\vert_{y=x-\epsilon} \n 
&& \hspace{-.5cm}=-\frac{\delta_{j,j'}}{8\pi}X_{1} \int_{2\epsilon}^{1-\epsilon} dx \int d^{2} Q d^{2} N
\log(\epsilon)B_{l}(x)B_{l'}(x)\tilde{B}_{t}(Q/2) \tilde{B}_{t'}(Q/2)
e^{-2\Lambda^{-4}N^{4}} \n
&& \hspace{-.5cm}=-\frac{\delta_{j,j'}}{8\pi}X_{1} \int_{2\epsilon}^{1-\epsilon} dx \int d^{2} Q 
\log(\epsilon)B_{l}(x)B_{l'}(x)\tilde{B}_{t}(Q/2) \tilde{B}_{t'}(Q/2) \Lambda^{2}\frac{2\pi}{16}\sqrt{8\pi}.
\eeq
Next, if we do the angular integral in $Q$ and let $Q\rightarrow 2q$:
\beq
\int d^{2}\vec{Q} f(Q/2)=2\pi \int dQ Q f(Q/2) = 8\pi \int dq q f(q),
\eeq
we get:
\beq
&&X_{2}X_{3}\log(x-y)\vert_{y=x-\epsilon} \n 
&& \hspace{-.5cm}=-\frac{\pi\delta_{j,j'}\Lambda^{2}}{8}\sqrt{8\pi}X_{1} \int_{2\epsilon}^{1-\epsilon} dx \int dq q
\log(\epsilon)B_{l}(x)B_{l'}(x)\tilde{B}_{t}(q) \tilde{B}_{t'}(q) \n
&& \hspace{-.5cm}=\delta_{j,j'}\Lambda^{2}\frac{4}{3} (2 \pi)^{\frac{3}{2}}\int_{2\epsilon}^{1-\epsilon} dx \int dq q
\log(\epsilon)B_{l}(x)B_{l'}(x)\tilde{B}_{t}(q) \tilde{B}_{t'}(q) .
\eeq
So after adding the overall momentum-conserving delta function and the total longitudinal momenta which were 
dropped for convenience, we find:
\beq
X_{2}X_{3}\log(x-y)\vert_{y=x-\epsilon}=
-\langle q',l',t',j' \vert V^{(2)}_{\mathrm{SE}}(\Lambda) \vert q,l,t,j \rangle ^{\mathrm{D}}.
\eeq
Thus, the divergence in the self energy is cancelled.

The final term we need to calculate is:
\beq
&&X_{2} \int_{\epsilon}^{x-\epsilon} dy \log(x-y) \frac{dX_{3}}{dy} \n
&&\hspace{1cm}=
\frac{1}{4}X_{1} \int d^{2}\vec{Q} d^{2}\vec{N}\int_{2\epsilon}^{1-\epsilon} dx
\int_{\epsilon}^{x-\epsilon} dy \log(x-y) \frac{dX_{3}}{dy} \n
&&\hspace{1cm}=
\frac{1}{4}X_{1} \int_{2\epsilon}^{1-\epsilon} dx
\int_{\epsilon}^{x-\epsilon} dy \log(x-y) \frac{d}{dy}\int d^{2}\vec{Q} d^{2}\vec{N}X_{3} \n
&&\hspace{1cm}=
X_{1} \int_{2\epsilon}^{1-\epsilon} dx
\int_{\epsilon}^{x-\epsilon} dy \log(x-y) \frac{d}{dy}\int d^{2}\vec{q} d^{2}\vec{p}\frac{X_{3}}{x-y} \n
&&\hspace{1cm}=
X_{1} \int_{2\epsilon}^{1-\epsilon} dx
\int_{\epsilon}^{x-\epsilon} dy \log(x-y) \int d^{2}\vec{q} d^{2}\vec{p}
\n && \hspace{2cm} \times
B_{l}(x)\tilde{B}_{t}(q) \tilde{B}_{t'}(p)
 A^{\ast}_{j'-s_{1}-s_{2}}(\phi')A_{j-s_{1}-s_{2}}(\phi) 
 \n && \hspace{3cm} \times
\frac{d}{dy}\frac{1}{x-y}B_{l'}(y)
e^{-\Lambda^{-4}\Delta^{2}_{FI}}e^{-2\Lambda^{-4}\Delta_{FK}\Delta_{IK}} .
\eeq
We can rewrite $\Delta_{FI}$ in terms of $\Delta_{FK}$ and $\Delta_{IK}$.  After simplifying, we get:
\newpage
\beq
&&\hspace{0cm}=
X_{1} \int_{2\epsilon}^{1-\epsilon} dx
\int_{\epsilon}^{x-\epsilon} dy \log(x-y) \int d^{2}\vec{q} d^{2}\vec{p}
\n && \hspace{2cm} \times
B_{l}(x)\tilde{B}_{t}(q) \tilde{B}_{t'}(p)
 A^{\ast}_{j'-s_{1}-s_{2}}(\phi')A_{j-s_{1}-s_{2}}(\phi) 
 \n && \hspace{3cm} \times
\frac{d}{dy}\frac{1}{x-y}B_{l'}(y) e^{-\Lambda^{-4}(\Delta^{2}_{FK}+\Delta^{2}_{IK})} ,
\eeq
or
\beq
&&\hspace{0cm}=
X_{1} \int_{2\epsilon}^{1-\epsilon} dx
\int_{\epsilon}^{x-\epsilon} dy \log(x-y) \int d^{2}\vec{q} d^{2}\vec{p}
\n && \hspace{1cm} \times
B_{l}(x)\tilde{B}_{t}(q) \tilde{B}_{t'}(p)
 A^{\ast}_{j'-s_{1}-s_{2}}(\phi')A_{j-s_{1}-s_{2}}(\phi) e^{-\Lambda^{-4}(\Delta^{2}_{FK}+\Delta^{2}_{IK})} 
 \n && \hspace{1.5cm} \times
\left[ \frac{B_{l'}(y)}{(x-y)^{2}}+\frac{B'_{l'}(y)}{x-y}
-2\frac{B_{l'}(y)}{x-y}\left(\Delta_{FK}\Delta_{FK}'+\Delta_{KI}\Delta_{KI}' \right) \right] .
\eeq
As we have seen before, when we do the sum over spins and do the integral over $\phi'$, we get:
\beq
\label{appx-eq:before-subtraction}
&&\hspace{-1cm}\langle q',l',t',j' \vert {\cal M}^{2}(\la) \vert q,l,t,j \rangle_{\mathrm{IN}}^{\mathrm{B,F}} =
-g_{\Lambda}^{2}\frac{32}{3} \delta_{j,j'} {\cal P}^{+}\delta({\cal P-P'})\theta(x-y) \theta(x-y-\epsilon) \n
&&\hspace{0cm}
 \int_{2\epsilon}^{1-\epsilon} dx
\int_{\epsilon}^{x-\epsilon} dy \log(x-y) \int d^{2}\vec{q} d^{2}\vec{p}
B_{l}(x)\tilde{B}_{t}(q) \tilde{B}_{t'}(p)
W_{q,q'} e^{-\Lambda^{-4}(\Delta^{2}_{FK}+\Delta^{2}_{IK})}
\n && \hspace{2cm} \times
\left[ \frac{B_{l'}(y)}{(x-y)^{2}}+\frac{B'_{l'}(y)}{x-y}
-2\frac{B_{l'}(y)}{x-y}\left(\Delta_{FK}\Delta_{FK}'+\Delta_{KI}\Delta_{KI}' \right) \right] ,
\eeq
where,
\beq
W_{q,q'}&=&\delta_{q,1}\delta_{q',1}\cos ([j-1]\gamma_{\phi})
+\delta_{q,2}\delta_{q',2}\cos ([j+1]\gamma_{\phi})
\n && \hspace{5cm}
+\left(\delta_{q,3}\delta_{q',3}+\delta_{q,4}\delta_{q',4}\right) \cos (j\gamma_{\phi}) ,
\eeq
\beq
\Delta_{KI}&=&\frac{m^{2}(x-y)^{2}+(x\vec{p}-y\vec{q})^{2}}{xy(x-y)} \n
&=&\frac{m^{2}(x-y)-yq^{2}+xp^{2}}{xy}+\frac{(\vec{p}-\vec{q})^{2}}{x-y},\n
\Delta_{FK}&=&-\frac{m^{2}(x-y)^{2}+\left[(1-x)\vec{p}-(1-y)\vec{q}\hspace{.1cm}\right]^{2}}{(1-x)(1-y)(x-y)} \n
&=&-\frac{m^{2}(x-y)+(1-y)q^{2}-(1-x)p^{2}}{(1-x)(1-y)}-\frac{(\vec{p}-\vec{q})^{2}}{x-y},\n
\Delta_{KI}'&=&\frac{(\vec{p}-\vec{q})^{2}}{(x-y)^{2}}-\frac{m^{2}+p^{2}}{y^{2}}, \n
\Delta_{FK}'&=&\frac{m^{2}+p^{2}}{(1-y)^{2}}-\frac{(\vec{p}-\vec{q})^{2}}{(x-y)^{2}} , \nonumber
\eeq
and we have left all six integrals intact.  This means when we get to the end and are back in the original 
coordinates, we can just remove the integral over $\phi'$ as it has already been done. 
This integral appears divergent because if $\vec{p}=\vec{q}$, as $x\rightarrow y$ the integrand diverges but 
the argument of the outer exponential is zero.  
The divergent part should cancel when we integrate over the transverse momenta.
In the limit that $x-y=\eta:  \eta \rightarrow 0$, the integrals over transverse momenta will bring down a 
factor of $\eta^{2}$. Thus the only term that may be divergent is the last one proportional to 
$\Delta_{FK}\Delta_{FK}'+\Delta_{KI}\Delta_{KI}'$.  However, we will consider the first and third terms:
\beq
&&\hspace{0cm}
-g_{\Lambda}^{2}\frac{32}{3} \delta_{j,j'} {\cal P}^{+}\delta({\cal P-P'}) \int_{2\epsilon}^{1-\epsilon} dx
\int_{\epsilon}^{x-\epsilon} dy \log\eta \int d^{2}\vec{q} d^{2}\vec{p}
B_{l}(x)\tilde{B}_{t}(q) \tilde{B}_{t'}(p)
W_{q,q'} 
\n && \hspace{1cm} \times
e^{-\Lambda^{-4}(\Delta^{2}_{FK}+\Delta^{2}_{IK})} 
\left[ \frac{B_{l'}(y)}{\eta^{2}}
-2\frac{B_{l'}(y)}{\eta}\left(\Delta_{FK}\Delta_{FK}'+\Delta_{KI}\Delta_{KI}' \right) \right] ,
\eeq
Now consider writing this where $x\approx y$:
\beq
&&\hspace{0cm}
-g_{\Lambda}^{2}\frac{32}{3} \delta_{j,j'} {\cal P}^{+}\delta({\cal P-P'}) \int_{2\epsilon}^{1-\epsilon} dx
\int_{\epsilon}^{x-\epsilon} dy \log\eta \int d^{2}\vec{q} d^{2}\vec{p}
B_{l}(x)B_{l'}(y)\tilde{B}_{t}(q) \tilde{B}_{t'}(p)
\n && \hspace{3cm} \times
\delta_{q,q'} e^{-2\eta^{-2}\Lambda^{-4}(\vec{p}-\vec{q})^{4}}
\left[ \frac{1}{\eta^{2}}-4\frac{1}{\eta^{4}}\left(\vec{p}-\vec{q}\right) ^{4} \right] .
\eeq
Note that in this limit, $\gamma_{\phi}\rightarrow 0$, so that $W_{q,q'}\rightarrow \delta_{q,q'}$.  
Now let:
\beq
\vec{r}=\half (\vec{p}+\vec{q}), \hspace{1cm} \vec{w}=\half \frac{\vec{q}-\vec{p}}{\sqrt{\eta}},
\eeq
where
\beq
\vec{r}=r\left[ \cos \alpha \hat{x} +\sin \alpha \hat{y} \right], \hspace{1cm}
\vec{w}=w\left[ \cos \delta \hat{x} +\sin \delta \hat{y} \right],
\eeq
and
\beq
\beta=\alpha-\delta, \hspace{1cm} \vec{r}\cdot\vec{w}=rw\cos \beta .
\eeq
Then
\beq
\vec{q}=\vec{r}+\eta \vec{w}, \hspace{1cm} \vec{p}=\vec{r}-\eta \vec{w},
\eeq
and
\beq
q=\sqrt{r^{2}+\eta w^{2}+2rw\sqrt{\eta} \cos \beta }, \hspace{1cm}
p=\sqrt{r^{2}+\eta w^{2}-2rw\sqrt{\eta} \cos \beta } .
\eeq
Define:
\beq
r_{\pm}\equiv \sqrt{r^{2}+\eta w^{2}\pm 2rw \sqrt{\eta} \cos \beta} .
\eeq
Then
\beq
\vec{q}\cdot\vec{p}=r^{2}-\eta w^{2}, \hspace{1cm}  \cos \gamma_{\phi}=\frac{r^{2}-\eta w^{2}}{r_{+}r_{-}} ,
\hspace{1cm} \sin \gamma_{\phi}=\frac{-2rw\sqrt{\eta}}{r_{+}r_{-}}\sin \beta.
\eeq 
We have:
\beq
d^{2}\vec{q} d^{2}\vec{p}=4\eta d^{2}\vec{r} d^{2}\vec{w},
\eeq
which gives:
\beq
&&\hspace{0cm}=
-g_{\Lambda}^{2}\frac{128}{3} \delta_{j,j'} {\cal P}^{+}\delta({\cal P-P'}) \int_{2\epsilon}^{1-\epsilon} dx
\int_{\epsilon}^{x-\epsilon} dy \frac{\log\eta}{\eta} \int d^{2}\vec{r} d^{2}\vec{w}
B_{l}(x)B_{l'}(y)\tilde{B}_{t}(r) \tilde{B}_{t'}(r)
\n && \hspace{3cm} \times
\delta_{q,q'} e^{-32w^{4}}\left[ 1-64 w^{4} \right] .
\eeq
However,
\beq
\int dw w \left( 1-64w^{4}\right) e^{-32 w^{4}} =0,
\eeq
so the integral is finite.

Now let us try to rewrite the finite part of the apparent divergent integral in the original variables.
\beq
0&=&
-g_{\Lambda}^{2}\frac{128}{3} \delta_{j,j'} \int_{2\epsilon}^{1-\epsilon} dx
\int_{\epsilon}^{x-\epsilon} dy \eta\log\eta \int d^{2}\vec{r} d^{2}\vec{w}
B_{l}(x)B_{l'}(y)\tilde{B}_{t}(r) \tilde{B}_{t'}(r)
\delta_{q,q'} \n && \hspace{3cm} \times
e^{-32w^{4}}\frac{1}{\eta^{2}}\left[ 1-64 w^{4} \right] \n
&=&
-g_{\Lambda}^{2}\frac{32}{3} \delta_{j,j'} \int_{2\epsilon}^{1-\epsilon} dx
\int_{\epsilon}^{x-\epsilon} dy \log\eta \int d^{2}\vec{q} d^{2}\vec{p}
B_{l}(x)B_{l'}(y)\tilde{B}_{t}(r) \tilde{B}_{t'}(r)
\delta_{q,q'} \n && \hspace{3cm} \times
e^{-2\eta^{-2}(\vec{p}-\vec{q})^{4}}\frac{1}{\eta^{2}}\left[ 1-4 \frac{(\vec{p}-\vec{q})^{4}}{\eta^{2}} \right] \n
&=&
-g_{\Lambda}^{2}\frac{32}{3} \delta_{j,j'} \int_{2\epsilon}^{1-\epsilon} dx
\int_{\epsilon}^{x-\epsilon} dy \log\eta \int d^{2}\vec{q} d^{2}\vec{p}
B_{l}(x)B_{l'}(y)\tilde{B}_{t}(r) \tilde{B}_{t'}(r)
\delta_{q,q'} \n && \hspace{3cm} \times
e^{-(\Delta_{FK}^{2}+\Delta_{KI}^{2})}
\frac{e^{-2\frac{(\vec{p}-\vec{q})^{4}}{\eta^{2}}+\Delta_{FK}^{2}+\Delta_{KI}^{2}}}
{\eta^{2}}\left[ 1-4 \frac{(\vec{p}-\vec{q})^{4}}{\eta^{2}} \right] .
\eeq

Before the subtraction we had (Eq. \ref{appx-eq:before-subtraction}):
\beq
&&\langle q',l',t',j' \vert {\cal M}^{2}(\la) \vert q,l,t,j \rangle_{\mathrm{IN}}^{\mathrm{B,F}} =
\n && \hspace{1cm}
-g_{\Lambda}^{2}\frac{32}{3} \delta_{j,j'} {\cal P}^{+}\delta({\cal P-P'})
 \int_{2\epsilon}^{1-\epsilon} dx
\int_{\epsilon}^{x-\epsilon} dy \log(x-y) \int d^{2}\vec{q} d^{2}\vec{p}
\n &&\hspace{1.5cm} \times
B_{l}(x)\tilde{B}_{t}(q) \tilde{B}_{t'}(p)
W_{q,q'} e^{-\Lambda^{-4}(\Delta^{2}_{FK}+\Delta^{2}_{IK})}
\n && \hspace{1.5cm} \times
\left[ \frac{B_{l'}(y)}{(x-y)^{2}}+\frac{B'_{l'}(y)}{x-y}
-2\frac{B_{l'}(y)}{x-y}\left(\Delta_{FK}\Delta_{FK}'+\Delta_{KI}\Delta_{KI}' \right) \right] ,
\eeq
which when we change to the variables $r$ and $w$ gives:
\beq
&&\langle q',l',t',j' \vert {\cal M}^{2}(\la) \vert q,l,t,j \rangle_{\mathrm{IN}}^{\mathrm{B,F}}=
\n && \hspace{1cm}
-g_{\Lambda}^{2}\frac{128}{3} \delta_{j,j'} {\cal P}^{+}\delta({\cal P-P'})
 \int_{2\epsilon}^{1-\epsilon} dx
\int_{\epsilon}^{x-\epsilon} dy \hspace{.1cm} \eta\log\eta \int d^{2}\vec{r} d^{2}\vec{w}
\n &&\hspace{1.5cm} \times
B_{l}(x)\tilde{B}_{t}(r_{+}) \tilde{B}_{t'}(r_{-})
W_{q,q'} e^{-\Lambda^{-4}(\Delta^{2}_{FK}+\Delta^{2}_{IK})}
\n && \hspace{1.5cm} \times
\left[ \frac{B_{l'}(y)}{\eta^{2}}+\frac{B'_{l'}(y)}{\eta}
-2\frac{B_{l'}(y)}{\eta}\left(\Delta_{FK}\Delta_{FK}'+\Delta_{KI}\Delta_{KI}' \right) \right] .
\eeq
So after making the subtraction, we have:
\newpage
\beq
&&\langle q',l',t',j' \vert {\cal M}^{2}(\la) \vert q,l,t,j \rangle_{\mathrm{IN}}^{\mathrm{B,F}}=
\n && \hspace{1cm}
-g_{\Lambda}^{2}\frac{128}{3} \delta_{j,j'} {\cal P}^{+}\delta({\cal P-P'})
 \int_{0}^{1} dx B_{l}(x) \int_{0}^{x} dy \hspace{.1cm} \eta\log\eta
\int d^{2}\vec{r} d^{2}\vec{w}
\n && \hspace{2cm} \times
\left\{ e^{-(\Delta^{2}_{FK}+\Delta^{2}_{IK})} 
 W_{q,q'}\tilde{B}_{t}(r_{+}) \tilde{B}_{t'}(r_{-})\left[ \frac{B_{l'}(y)}{\eta^{2}}+\frac{B'_{l'}(y)}{\eta}
 \right. \right. \n &&\hspace{5cm} \left. \left.
-2\frac{B_{l'}(y)}{\eta}\left(\Delta_{FK}\Delta_{FK}'+\Delta_{KI}\Delta_{KI}' \right) \right] 
\right. \n &&\hspace{4cm} \left.
-\tilde{B}_{t}(r) \tilde{B}_{t'}(r)  B_{l'}(y) \frac{\delta_{q,q'}}{\eta^{2}}e^{-32w^{4}}\left[ 1-64 w^{4} \right] \right\}.
\eeq
Making the final changes of variables:
\beq
\eta=xe^{-p},\hspace{1cm} dy=\eta dp,
\eeq
and
\beq
y_{w}=\frac{2}{1+w}-1, \hspace{1cm} y_{r}=\frac{2}{1+r}-1, \hspace{1cm} y_{p}=\frac{2}{1+p}-1 ,
\eeq
\beq
\label{apdx:instant-after-change}
&&\langle q',l',t',j' \vert {\cal M}^{2}(\la) \vert q,l,t,j \rangle_{\mathrm{IN}}^{\mathrm{B,F}}=
\n && \hspace{1cm}
-8g_{\Lambda}^{2}\frac{128}{3} \delta_{j,j'} {\cal P}^{+}\delta({\cal P-P'})
\n &&\hspace{2cm} \times
 \int_{0}^{1} dx \int_{0}^{2\pi} 
d\gamma_{\phi}\int_{-1}^{1} \frac{dy_{p}}{(1+y_{p})^{2}} \int_{-1}^{1} \frac{dy_{r}}{(1+y_{r})^{2}} \int_{-1}^{1} 
\frac{dy_{w}}{(1+y_{w})^{2}}
\n && \hspace{2cm} \times
\eta\log\eta \hspace{.1cm} r \hspace{.1cm} w B_{l}(x)
\n && \hspace{2cm} \times 
\left\{ e^{-(\Delta^{2}_{FK}+\Delta^{2}_{IK})} W_{q,q'}\tilde{B}_{t}(r_{+}) \tilde{B}_{t'}(r_{-})
\right. \n && \hspace{3cm} \left. \times
\left[ \frac{B_{l'}(y)}{\eta}+B'_{l'}(y)
-2B_{l'}(y)\left(\Delta_{FK}\Delta_{FK}'+\Delta_{KI}\Delta_{KI}' \right) \right] 
\right. \n &&\hspace{4.5cm} \left.
-\tilde{B}_{t}(r) \tilde{B}_{t'}(r) B_{l'}(y) \frac{\delta_{q,q'}}{\eta}e^{-32w^{4}}\left[ 1-64 w^{4} \right] \right\},
\eeq
where,
\beq
&&W_{q,q'}=\delta_{q,1}\delta_{q',1}\cos ([j-1]\gamma_{\phi})
+\delta_{q,2}\delta_{q',2}\cos ([j+1]\gamma_{\phi})
\n && \hspace{5cm} 
+\left(\delta_{q,3}\delta_{q',3}+\delta_{q,4}\delta_{q',4}\right) \cos (j\gamma_{\phi}) ,
\eeq
and
\beq
\Delta_{KI}&=&\frac{m^{2}\eta-yr_{+}^{2}+xr_{-}^{2}}{xy}+4w^{2},\n
\Delta_{FK}&=&-\frac{m^{2}\eta+(1-y)r_{+}^{2}-(1-x)r_{-}^{2}}{(1-x)(1-y)}-4w^{2},\n
\Delta_{KI}'&=&\frac{4w^{2}}{\eta}-\frac{m^{2}+r_{-}^{2}}{y^{2}}, \n
\Delta_{FK}'&=&\frac{m^{2}+r_{-}^{2}}{(1-y)^{2}}-\frac{4w^{2}}{\eta}. 
\eeq



\end{document}